\documentclass[12pt, a4paper]{report}
\usepackage[T1]{fontenc}
\usepackage[latin1]{inputenc}
\usepackage[english]{babel}
\usepackage{siunitx}
\usepackage{graphicx}
\usepackage{tipa} 
\usepackage{graphics} 
\usepackage{times} 
\usepackage{amsmath}
\usepackage{latexsym}
\usepackage{amscd}
\usepackage{mathrsfs} 
\usepackage{relsize}
\usepackage{delarray}
\usepackage{pstricks}
\usepackage{theorem}
\usepackage{changepage}
\usepackage{euscript}
\usepackage{textcomp}
\usepackage{esvect}
\usepackage{parskip}
\usepackage{placeins}
\usepackage{subfigure}
\usepackage{array}
\usepackage{delarray}
\usepackage{stmaryrd}
\usepackage{fancyhdr}
\usepackage{graphpap}
\usepackage{makeidx}
\usepackage{enumerate}
\usepackage{esint}
\usepackage{datetime}
\usepackage{caption}
\usepackage{smartdiagram}
\usesmartdiagramlibrary{additions}
\usepackage{abstract}
\usepackage{color}
\definecolor{igreen}{rgb}{0.0, 0.56, 0.0}
\usepackage{xcolor, colortbl}
\colorlet{gred}{-red!75!green!65!}
\colorlet{mamber}{-red!75!green!15!blue!50!}
\colorlet{grown}{-red!75!blue!20!green}
\colorlet{bled}{-red!85!blue!40!green!45!}
\colorlet{waters}{cyan!25} 
\colorlet{water}{cyan!25!green!20!} 
\definecolor{grin}{HTML}{00F9DE}
\usepackage{rotating}

\usepackage{arydshln}
\usepackage{booktabs}
\usepackage{graphicx}
\usepackage{tikz}
\usepackage{tikz-3dplot}
\usetikzlibrary{
arrows,
arrows.meta,
automata,
backgrounds,
calc,
decorations,
decorations.pathmorphing,
decorations.pathreplacing,
decorations.fractals,
external,
fit,
matrix,
petri,
positioning,
shadows,
shapes,
shapes.multipart,
topaths,
intersections
}
\usepackage{eso-pic}
\def\ba{\begin{array}}
\def\ea{\end{array}}
\def\beann{\begin{eqnarray*}}
\def\eeann{\end{eqnarray*}}
\def\bea{\begin{eqnarray}}
\def\eea{\end{eqnarray}}

\makeatletter
\newlength\qvec@height
\newlength\qvec@depth
\newlength\qvec@width
\newcommand{\qvec}[2][]{
    \settoheight{\qvec@height}{$#2$}
    \settodepth{\qvec@depth}{$#2$}
    \settowidth{\qvec@width}{$#2$}
  \def\qvec@arg{#1}
  \raisebox{.2ex}{\raisebox{\qvec@height}{\rlap{%
    \kern.05em
    \begin{tikzpicture}[scale=1,shorten >=-3pt,shorten <=-3pt]
    \pgfsetroundcap
    \coordinate (Stx) at (.05em,0) ;
		\coordinate (Arx) at (\qvec@width-.05em,0) ;
    \draw[->](Stx) to[bend left] (Arx);
    \end{tikzpicture}
  }}}
  #2
}
\makeatother
\makeatletter
\newlength\pvec@height
\newlength\pvec@depth
\newlength\pvec@width
\newcommand{\pvec}[2][]{
    \settoheight{\pvec@height}{$#2$}
    \settodepth{\pvec@depth}{$#2$}
    \settowidth{\pvec@width}{$#2$}
  \def\pvec@arg{#1}
  \raisebox{.2ex}{\raisebox{\pvec@height}{\rlap{%
    \kern.05em
    \begin{tikzpicture}[scale=1,shorten >=-3pt,shorten <=-3pt]
    \pgfsetroundcap
    \coordinate (Stx) at (.05em,0) ;
		\coordinate (Arx) at (\pvec@width-.05em,0) ;
    \draw[->](Stx) to[bend right] (Arx);
    \end{tikzpicture}
  }}}
  #2
}
\makeatother
\makeatletter
\newlength\vvec@height%
\newlength\vvec@depth%
\newlength\vvec@width%
\newcommand{\vvec}[2][]{%
  \ifmmode%
    \settoheight{\vvec@height}{$#2$}%
    \settodepth{\vvec@depth}{$#2$}%
    \settowidth{\vvec@width}{$#2$}%
  \else 
    \settoheight{\vvec@height}{#2}%
    \settodepth{\vvec@depth}{#2}%
    \settowidth{\vvec@width}{#2}%
  \fi%
  \def\vvec@arg{#1}%
  \def\vvec@dd{:}%
  \def\vvec@d{.}%
  \raisebox{.2ex}{\raisebox{\vvec@height}{\rlap{%
    \kern.05em%
    \begin{tikzpicture}[scale=1]
    \pgfsetroundcap
    \draw (.05em,0)--(\vvec@width-.05em,0);
    \draw (\vvec@width-.05em,0)--(\vvec@width-.15em, .075em);
    \draw (\vvec@width-.05em,0)--(\vvec@width-.15em,-.075em);
    \ifx\vvec@arg\vvec@d%
      \fill(\vvec@width*.45,.5ex) circle (.5pt);%
    \else\ifx\vvec@arg\vvec@dd%
      \fill(\vvec@width*.30,.5ex) circle (.5pt);%
      \fill(\vvec@width*.65,.5ex) circle (.5pt);%
    \fi\fi%
    \end{tikzpicture}%
  }}}%
  #2%
}
\makeatother
\def\ba{\begin{array}}
\def\ea{\end{array}}
\def\beann{\begin{eqnarray*}}
\def\eeann{\end{eqnarray*}}
\def\bea{\begin{eqnarray}}
\def\eea{\end{eqnarray}}

\usepackage{titlesec}
\usepackage{multirow}
\usepackage{hyperref}
\usepackage{apacite}
\usepackage{lipsum}
\usepackage{tikz-cd}
\usepackage{float}
\usepackage{titling}
\usepackage{epigraph}
\usepackage[title, titletoc]{appendix}
\titleformat{\chapter}{\normalfont\LARGE}{\thechapter\,$\vert$}{20pt}{\LARGE}{\setcounter{chapter}{0}}
\titlespacing*{\chapter}{0pt}{-70pt}{40pt} 
\makeatletter
\newcommand\BackgroundPicturea[3]{
	\setlength{\unitlength}{1pt}
	\put(0,\strip@pt\paperheight){
		\parbox[t]{\paperwidth}{
			\vspace{#2}\hspace{#3}
			\mbox{\includegraphics[scale=0.5]{#1}}
}}}
\newcommand\BackgroundPictureb[3]{
	\setlength{\unitlength}{1pt}
	\put(0,\strip@pt\paperheight){
		\parbox[t]{\paperwidth}{
			\vspace{#2}\hspace{#3}
			\mbox{\includegraphics[scale=0.3]{#1}}
}}}
\makeatother

\usepackage{setspace}
\addto\captionsenglish{
	\renewcommand{\contentsname}%
	{Table of Contents}
	\setcounter{tocdepth}{3}
	\setcounter{secnumdepth}{3}
	}

\hypersetup{pdftitle = Thesis, pdfauthor = {First Last}, pdfstartview=FitH, pdfkeywords = essay, pdfpagemode=FullScreen, colorlinks, anchorcolor = red, citecolor = blue, urlcolor=blue, filecolor=green, linkcolor=red, plainpages=false}
\usepackage{lineno}
\usepackage{upgreek}
\usepackage{float, extarrows, tikz-cd}
\usepackage{graphicx}
\usepackage{slashed}
\usepackage{tabularx,ragged2e}
\usepackage{amssymb}
\usepackage{subcaption}
\usepackage{amsmath,amssymb}
\usepackage{slashed}
\usepackage{caption}
\usepackage{xcolor}
\usepackage{dsfont}
\usepackage{verbatim}
\usepackage{mathtools, xcolor,ytableau, amsfonts,tikz}
\usepackage{graphicx}
\usepackage{physics}
\usepackage{simpler-wick}
\usepackage{microtype}
\usepackage{graphicx}
\usepackage{placeins}
\usepackage{hyperref}
\usepackage{fancyhdr}

\fancyheadoffset{9pt}
\fancyfootoffset{9pt}

\numberwithin{equation}{section}
 
\hypersetup{
    colorlinks=true,
    linkcolor=black,
    citecolor=blue,
    urlcolor=blue
}

\newcommand{\mL}{\mathcal{L}}
\newcommand{\mH}{\mathcal{H}}

\newcommand{\mE}{\mathcal{E}}

\newcommand{\oA}{\overline{A}}

\def\le{\left}
\def\ri{\right}
\newcommand\ov{\over}
\newcommand\p{\ensuremath{\partial}}
\newcommand{\es}[2] {\begin{equation} \label{#1} \begin{split} #2 \end{split} \end{equation}}

\def\AreaLaw{{\rm Area}(\p A)}

\def\<{\langle}
\def\>{\rangle}

\newcommand\lam{\lambda}
\newcommand\Lam{\Lambda}

\newcommand\Om{\Omega}

\newcommand\de{{\ensuremath{{\delta}}}}

\def\mop#1{\mathop{\rm #1}\nolimits}
\def\vol{\mop{vol}}

\date{}

\title{On Entanglement Dynamics in Holographic Theories}
\author{\\ \Large{Hanzhi Jiang}
\\ Lady Margaret Hall
\\
\\
\\
\\ University of Oxford
\\
A thesis presented for the degree of \\ \textit{Doctor of Philosophy}
\\ \\
Michaelmas 2025
}

\begin{document}
\AddToShipoutPicture*{\BackgroundPicturea{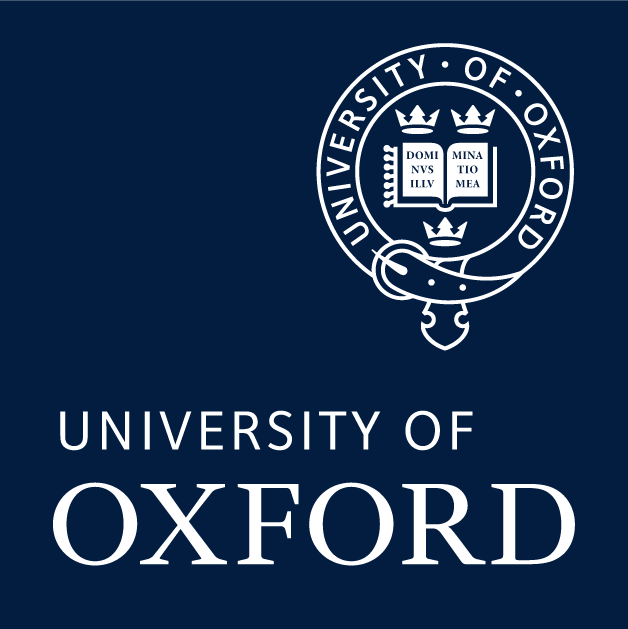}{0.7in}{5.8in}}
\AddToShipoutPicture*{\BackgroundPictureb{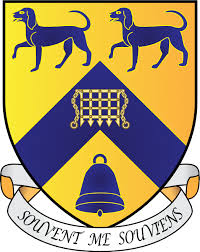}{6.1in}{3.7in}}
\thispagestyle{headings}
	\maketitle
\FloatBarrier
\pagenumbering{roman}



\thispagestyle{empty}
\chapter*{Acknowledgements}
I would like to first thank my advisor, Prof. Ard A. Louis, for accepting me to Oxford, for introducing me to algorithmic information theory, for teaching me to think critically, and for giving me the freedom to explore other fields in the middle of my DPhil. 

I am extremely grateful to Prof. M\'ark Mezei, who kindly agreed to co-supervise me at the beginning of my third year. In my early days, M\'ark found interesting and reasonably doable projects for me to work on, spent hours answering all my elementary questions and guided me through calculations; later on, he encouraged me to be independent, albeit still provided generous help even to projects that he did not actively involved. I cannot imagine what my DPhil would be like without the supervision of M\'ark. 

I am much obliged to Prof. Vijay Balasubramanian, Prof. Mike Blake, Prof. Richard Davison, Prof. Simon F. Ross, Julio Virrueta, Anthony Thompson, Tom Yildirim and Prof. Kamaludin Dingle for fruitful collaborations. I am grateful to Prof. Andrei Starinets and Prof. Benjamin Withers for being my thesis committee and for providing constructive feedback to this thesis. 

I would like to thank all my friends and colleagues at Rudolf Peierls Centre for Theoretical Physics and Mathematical Institute, University of Oxford. I extend my appreciation to Stacey Doyle and Jordan Summers for helping me with all the admin-related issues. I am grateful to my College, Lady Margaret Hall, for financial support. I would also like to acknowledge my college advisor, Prof. Gianluca Gregori, for all the academic advice. 

Last but not least, I would like to express my gratitude to my parents for their love and support. This thesis is dedicated to them.






\thispagestyle{empty}
\begin{abstract}
Entanglement entropy (EE) is a concept of paramount importance in modern theoretical physics. Nevertheless, calculating EE in generic interacting theories with non-trivial time dependence is notoriously difficult. In this thesis, we address the real-time dynamics of EE in certain strongly-coupled quantum field theories (QFT) using the Anti-de Sitter/Conformal Field Theory (AdS/CFT) correspondence. We first study the time evolution of EE in an exactly solvable model of two-dimensional CFT with an inhomogeneous sine-square deformation (SSD). We argue that the non-trivial time evolutions driven by the SSD Hamiltonian can be regarded as changing the foliations of spacetime. Remarkably, this leads to the presence of new horizons in the dual three-dimensional bulk. We then compute the EE in SSD CFT using holography and find agreement with CFT results in all comparable cases. Afterwards, we proceed to entanglement dynamics in generic chaotic quantum many-body systems. In the limit of large subsystems and late times, entanglement dynamics is believed to be described by an entanglement membrane effective theory. For 2d CFTs, a degenerate tension function reproduces several results regarding the dynamics of the entropy; this stands in contrast to higher dimensions where the tension is non-degenerate. We use holography to show that, in order to correctly capture the reflected entropy in 2d CFT, one needs to add an additional degree of freedom to the membrane description. Furthermore, we show that the conventional non-degenerate membrane tension function emerges upon introducing a relevant deformation to the CFT, dual to a planar Banados-Teitelboim-Zanelli (BTZ) black hole with scalar hair and with an interior Kasner universe. Subsequently, we establish a membrane effective theory for the dynamics of reflected entropy in higher dimensions. Lastly, we apply the membrane theories for reflected entropy to study the time evolution of multipartite entanglement signals. We found that scrambling dynamics generate multipartite entanglement with rich structures. 
\vspace{-10mm} 
\end{abstract}
\tableofcontents
\thispagestyle{plain}


\chapter{Introduction}
\pagenumbering{arabic}


A characteristic phenomenon in quantum physics is the presence of entanglement. The entanglement in quantum systems can be quantitatively described by an information-theoretic quantity, the entanglement entropy (EE). EE plays a crucial role in understanding the structures of quantum matters. For example, the ground state of interacting quantum many-body systems can be understood in terms of the entanglement structures of their wavefunctions: EE serves as an order parameter describing the phase structures of these systems~\cite{Kitaev:2005dm,Levin:2006zz}. The study of EE is also of paramount importance in time-dependent cases. Interacting systems thermalise. For isolated systems, while unitary time evolutions map pure states to pure states, the reduced density matrix of a subregion is expected to be very close to a thermal density matrix~\cite{Deutsch:1991msp,Srednicki:1994mfb,Rigol:2007juv}, which has EE proportional to the volume of the subregion. Therefore, the thermalisation of quantum many-body systems are fundamentally characterised by their entanglement dynamics. In addition to theoretical interests, thermalisation as probed by entanglement dynamics can be measured experimentally in cold atom systems under a quench~\cite{Kaufman:2016mif}, i.e. a sudden change of the Hamiltonian. Observations in their setup are consistent with the aforementioned volume law of EE.\footnote{To be more precise, the observable in~\cite{Kaufman:2016mif} is the quantum purity, whose natural logarithm is the 2$^{\rm nd}$ Renyi entropy defined in~\eqref{RenyiDef} that is related to EE.} 

While EE is a profound quantity in modern theoretical physics, the calculation of EE in generic interacting systems is a formidable task, and has been carried out mostly in special systems like two-dimensional conformal field theories (CFTs)~\cite{Holzhey:1994we,Calabrese:2004eu,Calabrese:2005in} and free field theories~\cite{Calabrese:2004eu}. In recent years, nevertheless, important progress has been made thanks to the AdS/CFT correspondence~\cite{Maldacena:1997re,Gubser:1998bc,Witten:1998qj}, the idea that certain quantum matter systems are mathematically equivalent to quantum gravity theories in one dimension higher. The strong-weak nature of the AdS/CFT correspondence allows one to map problems in strongly-interacting quantum systems to weakly coupled gravitational theories that are computationally tractable. In particular, the Ryu-Takayanagi (RT)~\cite{Ryu:2006bv,Ryu:2006ef} and Hubeny-Rangamani-Takayanagi (HRT)~\cite{Hubeny:2007xt} proposals equate EE in field theory to the area of certain co-dimension 2 extremal surfaces (in the former case, minimal surfaces) in the dual gravitational theories. 

The goal of this thesis is to employ the AdS/CFT correspondence to explore the time evolution of entanglement entropy in certain strongly interacting quantum many-body systems with holographic gravity duals. The thesis is organised as follows: 

To proceed, we will first review the essentials of entanglement entropy in section~\ref{EEreview}, and then survey holographic entanglement entropy in the simpler time-symmetric case in section~\ref{HoloRTReview}. We will turn to our main interest, the time evolution of holographic entanglement entropy, in section~\ref{HRTReview}, where we review the basis and point out the open problems that led to the original research in Chapter~\ref{Floquet} and~\ref{Chap3}. Subsequently, in section~\ref{ReflectedEntropyReview} and~\ref{multipartySignalReview}, we will review information-theoretic quantities characterising multipartite correlations that can be regarded as generalisations of the RT/HRT proposals. 

To gain intuitions for the difficult problem of entanglement dynamics, it is often instructive to look into special toy models with nice theoretical properties. In Chapter~\ref{Floquet}, we first consider the time evolution of EE in an exactly solvable setup of a two-dimensional CFT driven by Hamiltonians built out of conformal symmetry generators~\cite{Wen:2018vux}: in this case, the quantum dynamics can be understood using two-dimensional geometry. We investigate how the dynamics is reflected in the holographic dual three-dimensional spacetime and find $\emph{new horizons}$. Our result provides a novel example demonstrating the emergence of horizons when quantum systems are driven by alternative Hamiltonians: a familiar example of this kind is the Rindler horizon in general relativity. We then compute the EE in SSD CFT using holography and found agreement with CFT results in all comparable cases. 

In Chapter~\ref{Chap3}, we move on to exploring entanglement dynamics in generic chaotic many-body systems. We focus on the the limit of late time and large region, i.e. the "hydrodynamic" limit, under which the time evolution of EE has an effective description in terms of a minimal membrane~\cite{Jonay:2018yei,Mezei:2018jco}. For 2d CFTs, a degenerate tension function reproduces several results regarding the dynamics of the EE; this stands in contrast to higher dimensions where the tension is non-degenerate. In Chapter 3, we use holography to propose the $\emph{generalised membrane}$ effective theory for entanglement dynamics in 2d CFT by adding $\emph{an additional degree of freedom}$ to the membrane description. This additional degree of freedom has two important consequences: First, it plays a crucial role in the dynamics of a novel information-theoretic function known as reflected entropy. Second, it is believed to be related to the infinite-dimensional conformal symmetry of 2d CFT, which prevents the emergence of dissipative hydrodynamics. Indeed, as we will also address in Chapter~\ref{Chap3}, when we break this infinite-dimensional conformal symmetry by introducing a relevant deformation, we can recover the ordinary membrane theory with a non-degenerate tension function. We then briefly comment on the consistency of our holographic studies with dual field-theoretic results on universal hydrodynamics in two-dimensional QFTs at high temperature~\cite{Davison:2024msq}.\footnote{In a subsequent work~\cite{Davison:2025xdj}, holography was used to verify the universal hydrodynamics in two-dimensional QFTs, through the phenomenon of "pole skipping"~\cite{Grozdanov:2017ajz,Blake:2017ris,Blake:2018leo}.} Finally, we also found that the dynamics of reflected entropy in higher dimensions are captured by ordinary membrane theory. 

Subsequently, in Chapter~\ref{Chap4}, we apply the (generalised) membrane theories for reflected entropy to study the real-time dynamics of multiparty entanglement signals~\cite{Balasubramanian:2024ysu} in chaotic quantum many-body systems, including holographic gauge theories.  We find that the time evolution profile of multipartite entanglement signals depict rich structures. This indicates that scrambling dynamics often produces non-trivial multipartite entanglement. On the other hand, entanglement dynamics in integrable models are described by the quasi-particle picture~\cite{Calabrese:2005in,Casini:2015zua}. Assuming no initial multipartite entanglement,\footnote{See~\cite{Casini:2015zua} for discussions on the quasiparticle picture with initial multiparty entanglement.} the entanglement dynamics in the quasiparticle picture are bipartite by construction, and involve free-streaming EPR pairs of entangled quasiparticles which can be shared by at most two regions at a time. Compared with the membrane theory result, we can regard generating multiparty entanglement as a common feature of scrambling dynamics. The results in Chapter~\ref{Chap4} then unveil an overlooked aspect of scrambling dynamics with potentially abundant novel properties.  

Technical details of Chapter~\ref{Floquet}-\ref{Chap4} are presented in the Appendices. 

\section{Entanglement entropy}
\label{EEreview}
Suppose we have a quantum system in a pure state $\ket{\psi}$, whose density matrix is $\rho=\ket{\psi}\bra{\psi}$. We divide the system into two subregions $A$ and $\oA$, and denote the artificial boundary $\partial A$ separating them as the entangling surface. The Hilbert space of the system equals the tensor product
\begin{align}
    \mathcal{H}_{A\oA}=\mathcal{H}_A\otimes\mathcal{H}_{\oA}.
\end{align}
The reduced density matrix $\rho_A$ of subregion $A$ is defined by the partial trace
\begin{align}
    \rho_A=\Tr_{\oA}\rho.
\end{align}
Intuitively, $\rho_A$ captures the degrees of freedom in $A$ while assuming complete ignorance of what happens in $\oA$. If $\ket{\psi}$ is factorized, we would have a pure state in $\mathcal{H}_A$; the presence of quantum entanglement, nevertheless, leads to a density matrix, i.e. a list of probabilities for the occurrence of potential states in $\mathcal{H}_A$, characterising the price we pay for our ignorance. 

The entanglement entropy (EE) of the subregion $A$ is given by the von Neumann entropy associated with the reduced density matrix $\rho_A$ of $A$
\begin{align}
\label{EEdef}
    S(A)=-\Tr_A(\rho_A\log\rho_A). 
\end{align}
From this definition, it is evident that $S(A)=S(\oA)$. If $A$ and $\oA$ are unentangled, $S(A)=0$; $S(A)$ reaches its maximum when $A$ and $\oA$ are maximally entangled. 

To illustrate the definition of entanglement entropy, let us consider an elementary example with $A$ and $\oA$ each being a spin-$\frac{1}{2}$ particle. The Hilbert space of the full system is then spanned by $\ket{\uparrow\uparrow}$, $\ket{\uparrow\downarrow}$, $\ket{\downarrow\uparrow}$, $\ket{\downarrow\downarrow}$. Suppose the system is in the following pure state $\ket{\psi}=\cos\theta \ket{\uparrow\downarrow}+\sin\theta \ket{\downarrow\uparrow}$. The reduced density matrix of $A$ is $\rho_A=\cos^2\theta\ket{\uparrow}\bra{\uparrow}+\sin^2\theta\ket{\downarrow}\bra{\downarrow}$, and $S(A)=-2\cos^2\theta\log\cos\theta-2\sin^2\theta\log\sin\theta$. One can easily verify that $S(A)$ takes its maximal value $S(A)=\log 2$ when $\cos^2\theta=\frac{1}{2}$, i.e. when $A$ and $\oA$ are maximally entangled. More generally, if we had $q$ qubits in $A$ and $q$ qubits in $\oA$ that are maximally entangled, $S_A=q\log 2$. Hence, we can intuitively understand $S_A$ as counting the number of bits, and, since $q$ qubits have $2^{q}$ states, $e^{S_A}$ as counting the number of entangled states.

In the above elementary example, $S_A$ is computed by the eigenvalues $\lambda_i$ of $\rho_A$. More generally, in any finite systems, the Schmidt decomposition allows one to diagonalize the operator $\rho_A$ and find its eigenvalues $\lambda_i$. The entanglement entropy is then simply $S_A=-\sum_i \lambda_i\log\lambda_i$.

Instead of finite systems, in physics, we are often interested in entanglement entropy in continuum QFTs. In these cases, we have wavefunctionals $\Psi[\Phi(x)]$ on Cauchy slices, where we define the subregion $A$, its complement $\oA$, as well as the entangling surface $\partial A$.\footnote{Note that for gauge theories, in general $\mathcal{H}_{A\oA}\neq\mathcal{H}_A\otimes\mathcal{H}_{\oA}$. In what follows, we will ignore this subtlety as it won't affect the discussions too badly. See, for example~\cite{Lin:2017uzr,Lin:2018bud} for discussions regarding this issue in holographic theories that this thesis primarily concerns.} Afterwards, one may use~\eqref{EEdef} to calculate $S(A)$: tracing out $\oA$ amounts to integrating out all fields in $x\in\oA$. Once we have the reduced density matrix $\rho_A$, we need to further obtain the operator $\log\rho_A$. This is very subtle, and it is more practical to obtain the entanglement entropy as the appropriate analytic continuation of the $n^{th}$-Renyi entropy $S^{(n)}(A)$
\begin{align}
\label{RenyiDef}
    S^{(n)}(A)=\frac{1}{1-n}\log\Tr_A\rho_A^{n}, && S(A)=\lim_{n\to 1} S^{(n)}(A).
\end{align}
The quantity $\Tr_A\rho_A^{n}$ in the definition of Renyi entropy~\eqref{RenyiDef} admits a replica path integral derivation; in cases with non-trivial time-dependences, the Schwinger-Keldysh formalism is needed, where the forward-backward evolution allows one to cancel out unknown information of the final state in the time evolution. See e.g.~\cite{Rangamani:2016dms} for formalism and~\cite{Calabrese:2004eu} for explicit calculations in two-dimensional conformal field theories. 

\subsection{UV and IR properties}
When working in the continuum, we expect UV singularities. In the UV, any finite energy state is the same as the vacuum state. Therefore, to study the UV divergences, we can restrict to $\rho=\ket{0}\bra{0}$. Physically, the UV divergence arises from the short-distance correlations in the UV. The definition of $\rho_A$ entails bi-partitioning the system across the entangling surface $\partial A$. One then expects the correlations of modes across $\partial A$ to have a leading contribution to EE, and needs to account for these modes within a cutoff $\epsilon$ distance away from $\partial A$. Intuitively, the divergences are proportional to the number of EPR pairs that are shared between the two sides of the entangling surface $\partial A$. Hence, it should scales as the $\emph{area}$ of this entangling surface. Indeed, in $d$-dimensional free field theory, it has been proved that the leading UV divergent of EE obeys the area law~\cite{Bombelli:1986rw,Srednicki:1993im}.
\begin{equation}
\label{areaLaw}
    S_A=\alpha\frac{{\rm Area}(\partial A)}{\epsilon^{d-2}}+...,
\end{equation}
where $d$ denotes the spacetime dimensions, and the power $d-2$ of $\epsilon$ ensures that $S_A$ is dimensionless, $\alpha$ depends on the field theory in question, and the ellipses include less divergent terms. In general, for a $d$-dimensional relativistic QFT, the UV divergence takes the form 
\begin{align}
\label{genericUVdiv}
    S_A=
    \begin{cases}
        a_{d-2}\Big(\frac{L}{\epsilon}\Big)^{d-2}+a_{d-4}\Big(\frac{L}{\epsilon}\Big)^{d-4}+...+a_1\frac{L}{\epsilon}+(-1)^{\frac{d-1}{2}}\mathcal{S}_A+... \ &d\ {\rm odd}, \\ 
        a_{d-2}\Big(\frac{L}{\epsilon}\Big)^{d-2}+a_{d-4}\Big(\frac{L}{\epsilon}\Big)^{d-4}+...+(-1)^{\frac{d-2}{2}}\mathcal{S}_A\log \frac{L}{\epsilon}+... &d\ {\rm even}\,,
    \end{cases}
\end{align}
where $L$ characterises the size of the system. Notice that only even powers appear. This is because the UV divergence is related to the extrinsic curvature $K\sim\nabla n$ of the entangling surface $\partial A$, where $n$ is the unit normal vector. $K$ flips sign if we consider $A$ vs $\oA$. $S_{{\rm div}}(A)=S_{{\rm div}}(\oA)$ then implies that only even powers of $K$ are allowed. The constants $a_n$ are in general theory-dependent. However, $\mathcal{S}_A$ is universal. We note that in $d=2$ there is only a logarithmic divergence. In this case, the entangling surface $\partial A$ are sets of disconnected points. 

The above area law holds generally for QFTs in the ground state. For highly excited or thermal states, however, EE scales differently. To illustrate this behaviour, let us consider a 1-dimensional chain of $N$ spin-$\frac{1}{2}$ particles. We take $A$ to be a spatial region containing $k$ spins ($1\ll k\ll N$), and $\oA$ to be everything else.  The most general state of the system is given by
\begin{align}
\label{volLawEx}
    \ket{\Psi}=\sum_{s_i} c_{s_1...s_N}\ket{s_1}...\ket{s_N}, && s_i\in\{0,1\}.
\end{align}
A highly excited or thermal state is expected to be random, i.e. for such states $c_{s_1...s_N}$ is drawn from a random distribution. Since $k\ll N$, Page's theorem~\cite{Page:1993df} indicates that $\lambda_i~\sim\frac{1}{2^{k}}$. As such, $S(A)\sim-\sum \frac{1}{2^{k}} \log \frac{1}{2^{k}}=k\log 2$, where the sum runs over all $2^k$ eigenvalues. We see that $S(A)$ goes as the number of spins $k$ in $A$. In other words, for this 1-dimensional system, $S(A)\sim \vol(A)$. Intuitively, this volume scaling can be understood as the cumulative effects of long-range correlations between generic points contained in $A$ and those outside. Hence, we expect the volume law  
\begin{align}
\label{volLaw}
    S_{{\rm reg},A}=s_{\rm th}\vol(A)\sim L^{d-1}
\end{align}
to be generic for highly excited or thermal states. The subscript "reg" is due to the following reason: Unlike the above example~\eqref{volLawEx}, which is on a discrete spin chain, for EE in a highly excited or thermal state in a continuous system, there are UV divergent terms~\eqref{genericUVdiv} including the leading area law divergence~\eqref{areaLaw}. To distil the IR behaviour of EE, in~\eqref{volLaw} we regulate $S_A$ by subtracting these UV divergent terms.\footnote{Notice that there is a potential subtlety regarding this regularisation scheme: the UV divergence could be state-dependent~\cite{Marolf:2016dob}.} 

Apart from in specific cases such as two-dimensional conformal field theories~\cite{Calabrese:2004eu,Calabrese:2005in} and free field theory~\cite{Calabrese:2004eu}, entanglement entropy is generally difficult to compute using the replica trick mentioned in the previous subsection. In particular, calculating EE in interacting QFTs is a formidable task. Fortunately, the Ryu-Takayanagi(RT)~\cite{Ryu:2006bv,Ryu:2006ef}/Hubeny-Rangamani-Takayanagi(HRT)~\cite{Hubeny:2007xt} formula allows one to calculate EE in certain strongly-interacting quantum field theories with large degrees of freedom using holography.  

\section{Holographic entanglement entropy}
\label{HoloRTReview}
Intuitively, the AdS/CFT correspondence states that certain conformal field theory in $d$ spacetime dimensions (the "boundary") with a large degree of freedom is equivalent to a quantum gravity theory in asymptotically AdS$_{d+1}\times M$ spacetime (the "bulk"), where the asymptotically AdS$_{d+1}$ metric can be written as
\begin{align}
    ds^2=g_{rr}dr^2+g_{\mu\nu}dx^{\mu}dx^{\nu},
\end{align}
with $r$ being the bulk radial coordinate, and the AdS boundary at $r=\infty$. $M$ is some compact manifold. A more precise statement is as follows. 

A $d$-dimensional conformal field theory has a semiclassical dual if there exists a finite set of CFT primary operators $O_i$, and a bulk effective action $S_{\rm eff}[\phi_i]$, and $\phi_i$ are a finite set of bulk fields such that 
\begin{align}
\label{dictionary}
    \int\mathcal{D}\phi_ie^{i S_{\rm eff}[\phi_i]}O_{i_1}(t_1,\vec{x}_1)...O_{i_n}(t_n,\vec{x}_n)\approx\langle O_{i_1}(t_1,\vec{x}_1)...O_{i_n}(t_n,\vec{x}_n)\rangle_{\rm CFT},
\end{align}
where 
\begin{align}
    \lim_{r\to\infty}r^{\Delta_i}\phi_i(t,r,\vec{x})=O_i(t,\vec{x})
\end{align}
and $\Delta_i$ are the scaling dimensions of $O_i$. Equation~\eqref{dictionary} establishes a quantitative relation between the bulk fields $\phi_i$ and boundary operators $O_i$, and is known as the extrapolate dictionary~\cite{Maldacena:1997re,Gubser:1998bc,Witten:1998qj}. For example, the boundary stress tensor $T_{\mu\nu}$ is dual to bulk metric $g_{\mu\nu}$; the Noether current $J_{\mu}$ of a continuous global symmetry is dual to a gauge field $A_{\mu}$ in the bulk.  

In a conformal field theory with semiclassical gravity dual, the entanglement entropy is holographically computed by the Ryu-Takayanagi(RT)~\cite{Ryu:2006bv,Ryu:2006ef}/Hubeny-Rangamani-Takayanagi(HRT)~\cite{Hubeny:2007xt} formula. The original RT formula concerns only time-independent cases, which is then generalised by HRT to include non-trivial time dependences. The  HRT formula~\cite{Hubeny:2007xt} states that if we consider states in a CFT$_d$ with a holographic dual, the von Neumann entropy of a spatial subregion $A$ lying on some Cauchy slice is computed by the area of a co-dimension 2 $\emph{extremal}$ surface $\mE_A$ in the asymptotically AdS$_{d+1}$ bulk that is $\emph{homologous to}$ $A$, i.e. anchored on $\partial A$ and smoothly retractable to $A$ on the AdS$_{d+1}$ boundary. Among the family of extremal surfaces that satisfy the homological constraints, we should pick the one with the smallest area, 
\begin{equation}
\label{eq:HRT}
 S(A) = \frac{{\rm Area}(\mE_A)}{4G_N}.
\end{equation}
Notice that the condition of the co-dimension 2 surface being extremal is equivalent to the vanishing of the null extrinsic curvatures in the two normal directions to the surface~\cite{Hubeny:2007xt}. 

If the field theory state in question is static, one can uniquely extend the boundary Cauchy slice to the corresponding constant time slice in the asymptotically AdS$_{d+1}$ bulk.\footnote{This is in contrast to the time dependent cases studied above, where Cauchy slices in the bulk are not uniquely specified by Cauchy slices on the boundary.} Consequently, the co-dimension 2 extremal surface $\mE_A$ degenerates to a $\emph{minimal}$ surface that lies on it. This is because we can decompose the null extrinsic curvature into timelike $K^{(t)}$ and spacelike $K^{(s)}$ components. Under the presence of a time reflection symmetry, $K^{(t)}=0$ automatically. Solving the latter $K^{(s)}=0$ is then equivalent to finding minimal surfaces~\cite{Rangamani:2016dms}. \eqref{eq:HRT} in this static special case is the original RT formula~\cite{Ryu:2006bv,Ryu:2006ef}

With the RT/HRT formula, the calculation of $S(A)$ is then $\emph{geometrised}$ to the computation of the area of $\mE_A$. The RT formula has been proved in~\cite{Lewkowycz:2013nqa}. See~\cite{Dong:2016hjy} for progress on deriving the HRT formula. 

Now let us find the area of the HRT surface ${\rm Area}(\mE_A)$ in~\eqref{eq:HRT}. We denote the metric of the asymptotically AdS$_{d+1}$ spacetime as $ds^2=g_{\mu\nu}dx^{\mu}dx^{\nu}$ ($\mu,\nu=1,\dots,d+1$). We parametrise the co-dimension 2 extremal surface $\mE_A$ with coordinate $\vec{\xi}$. The induced metric $\gamma_{ab}$ ($a,b=1,\dots,d-1$) of $\mE_A$ is 
\begin{align}
\label{inducedMetric}
    \gamma_{ab}=g_{\mu\nu}\frac{\partial x^{\mu}}{\partial \xi^a}\frac{\partial x^{\nu}}{\partial \xi^b}.
\end{align}
The area of $\mE_A$ is then simply the determinant of $\gamma_{ab}$
\begin{align}
\label{area}
    S=\frac{1}{4G_N}\int d^{d-1}\xi \sqrt{\det(\gamma_{ab})}.
\end{align}
To find the EE, we $\emph{extremise}$ $S$~\eqref{area}, i.e. find $\mE_A$ such that $\delta S=0$. In the static case, the extremal surfaces reduces to minimal surfaces and the practice $\emph{minimises}$ $S$.

\subsection{CFT$_d$ in the vacuum on $\mathbb{R}^{d-1,1}$}\label{EEVacGenD}
Now, let us consider general $d$-dimensional CFT in the vacuum state on $\mathbb{R}^{d-1,1}$. Since the state is static, we are instructed to use the RT formula and find minimal surfaces in the dual bulk geometry on a constant time slice (which we can, without loss of generality, take to be $t=0$). The dual geometry is Poincare AdS$_{d+1}$, whose metric is 
\begin{align}
\label{pureAdSmetric}
    ds^2=\frac{1}{z^2}(-dt^2+dz^2+d\vec{x}_{d-1}^2),
\end{align}
and the AdS boundary is at $z=0$. In $d>2$, there are different possible shapes for the entangling subregions. Here we pick two most symmetric cases: strip and spherical regions. In the strip case, the RT surfaces have non-trivial shapes only along the $x_1$ direction, and it would be more convenient to adapt Cartesian coordinate for the boundary spatial directions: $d\vec{x}_{d-1}^2=dx_1^2+d\vec{x}_{d-2}^2$. We choose $\xi^{i}=x^{i}$ ($i=1,...,d-1$) to parametrise the co-dimension 2 minimal surface. This surface can then be specified by the function $z(x_1)$. Using~\eqref{inducedMetric}, the induced metric of the minimal surface is 
\begin{align}
    ds^2=\gamma_{ab}d\xi^a d\xi^{b}=\frac{1}{z^2}\Big(1+\Big(\frac{\partial z(x_1)}{\partial x_1}\Big)^2\Big)dx_1^2+\frac{1}{z^2}dx_2^2+...+\frac{1}{z^2}dx_{d-1}^2,
\end{align}
and the EE~\eqref{area} is hence 
\begin{align}
    S_{{\rm strip}}&=\frac{1}{4G_N}L^{d-2}\int dx_1\frac{\sqrt{1+z'(x_1)^2}}{z^{d-1}},\label{EEVacStrip}
\end{align}
where $L$ is the IR regulator of the translationally-invariant directions. To calculate the EE, we need to $\emph{minimise}$ the area functional in~\eqref{EEVacStrip}, i.e. to find $z(x_1)$ such that $\delta S=0$. This is a problem in the calculus of variation like those in classical mechanics. Generally speaking, the EE~\eqref{area} we are minimising can be regarded as an "action", hence $\sqrt{\det(\gamma_{ab})}$ in~\eqref{area} plays the role of the "Lagrangian". In the case of~\eqref{EEVacStrip}, this "Lagrangian" is
\begin{align}
\mathcal{L}\big(z(x_1),z'(x_1),x_1\big)=\frac{\sqrt{1+z'(x_1)^2}}{z^{d-1}}\label{EEVacStripLagrangian},
\end{align}
where the coordinate $x_1$ parametrising the minimal surface serves as "time", and the function $z(x_1)$ describing the minimal surface acts as the "generalised coordinate". In general, one can find $z(x_1)$ from the Euler-Lagrange equation with "Lagrangian"~\eqref{EEVacStripLagrangian}; in practice, however, it is easier to invoke $\emph{conservation laws}$. Since the "Lagrangian"~\eqref{EEVacStripLagrangian} does not contain "time" $x_1$, the "energy" $\mH$ is conserved 
\begin{align}
    \mH=z'(x_1)\frac{\partial \mL}{\partial z'(x_1)}-\mL=-\frac{1}{z^{d-1}\sqrt{1+z'^2}}=-\frac{1}{z_0^{d-1}}\label{EEVacStripH},
\end{align}
where we have evaluate the "energy" at the deepest point $z_0$ the RT surface can reach into the bulk, at which point $z'|_{z_0}=0$. From \eqref{EEVacStripH}, we can solve for $z'(x_1)$
\begin{align}
    z'(x_1)=\frac{\sqrt{z_0^{2(d-1)}-z^{2(d-1)}}}{z^{d-1}}\label{EEVacStripZ}.
\end{align}

To relate $z_0$ with the size $2a$ of the entangling subregion $A$, we can integrate \eqref{EEVacStripZ} to find the $x$ displacement of the RT surface
\begin{align}
    a=\int dx=\int_0^{z_0}\frac{dz}{z'(x)},&&\Rightarrow && z_0=a\frac{\Gamma\big(\frac{1}{2(d-1)}\big)}{\sqrt{\pi}\Gamma\big(\frac{d}{2(d-1)}\big)}.
\end{align}


From the differential equation~\eqref{EEVacStripZ}, we can solve $x_1(z)$ in terms of hypergeometric function
\begin{align}
    x_1(z)=\frac{z^d}{z_0^{d-1}d}{}_2F_{1}\Big(\frac{1}{2},\frac{d}{2(d-1)},1+\frac{d}{2(d-1)},\Big(\frac{z}{z_0}\Big)^{2 (d-1)}\Big)-\frac{\sqrt{\pi}}{d}\frac{\Gamma\big(\frac{3d-2}{2(d-1)}\big)}{\Gamma\big(\frac{2d-1}{2(d-1)}\big)},
\end{align}
which then allows us to compute $S$ in~\eqref{EEVacStrip}
\begin{align}
\label{EEStripResult}
    S=\frac{1}{4G_N}\frac{L^{d-2}}{d-2}\Big(\frac{2}{\epsilon^{d-2}}-\Big(\frac{2}{z_0}\Big)^{d-1}\frac{1}{a^{d-2}}\Big).
\end{align}
In evaluating~\eqref{EEVacStrip}, we have introduced a UV cutoff $z=\epsilon\ll 1$. This is due to the fact that the minimal surfaces extend to the AdS boundary, and have infinite area. Notice that the resulting divergent term scales as $\frac{{\rm Area}(\partial A)}{\epsilon^{d-2}}$, as is expected for EE in the ground state in a QFT~\eqref{areaLaw}.

In the special case of $d=2$, \eqref{EEStripResult} becomes degenerate. Now, let us calculate the EE in this case separately. In $d=2$, the co-dimension 2 minimal surface is a 1-dimensional geodesic. The solution to~\eqref{EEVacStripZ} is a semi-circle $z(x)=\pm \sqrt{a^2-(x+C_1)^2}$. We can adopt an alternative parametrisation $\alpha$, and describe it with two functions
\begin{align}
\label{semiCircle}
    x(\alpha)=a\cos\alpha,& && z(\alpha)=a\sin\alpha.
\end{align}
With this parametrisation, the "action"~\eqref{EEVacStrip} can be written as
\begin{align}
\label{EEVacAdS3}
    S&=\frac{1}{4G_N}\int d\alpha\frac{\sqrt{x'(\alpha)^2+z'(\alpha)}}{z^2}=\frac{c}{6}\cdot 2\int_{\frac{\epsilon}{a}}^{\frac{\pi}{2}}\frac{d\theta}{\sin\theta}=\frac{c}{3}\log\frac{2a}{\epsilon},
\end{align}
where in getting the second equality, we have also used the Brown-Henneaux central charge~\cite{Brown:1986nw} $c=\frac{3\ell_{AdS}}{2G_N}$. The divergence behaves as
$\log\frac{2a}{\epsilon}$, as is expected from~\eqref{genericUVdiv}. The holographic calculation~\eqref{EEVacAdS3} matches with CFT$_2$ results~\cite{Holzhey:1994we,Calabrese:2004eu}.

We can likewise study the case with spherical entangling subregions. For this case, it would be more convenient to adapt spherical coordinates the boundary spatial coordinate: $d\vec{x}_{d-1}^2=dr^2+r^2 d\vec{\theta}_{d-2}^2$. We choose $\xi^i=(r,\theta_{d-2})$ to parametrise the minimal surface, and describe this surface with the function $z(r)$. Using~\eqref{inducedMetric}, we have the induced metric 
\begin{align}
    ds^2=\frac{1}{z^2}\Big(1+\Big(\frac{\partial z(r)}{\partial r}\Big)^2\Big)dr^2+\frac{r^2}{z^2}d\theta_1^2+...+\frac{r^2}{z^2}d\theta_{d-1}^2,
\end{align}
which then leads to the EE
\begin{align}
    S_{{\rm sphere}}&=\frac{1}{4G_N}\omega_{d-2}\int dr\frac{r^{d-2}}{z^{d-1}}\sqrt{1+z'(r)^2}\label{EEVacSphere},
\end{align}
where $\omega_{d-2}=\frac{2 \pi^{\frac{d-1}{2}}}{\Gamma(\frac{d-1}{2})}$ is the area of unit $S^{d-2}$. Now, the "Lagrangian"\\ $\mathcal{L}\big(z(r),z'(r),r\big)=\frac{r^{d-2}}{z^{d-1}}\sqrt{1+z'(r)^2}$ does explicitly contain the "time" $r$, so there's no "energy" conservation any more. One thus needs to solve the Euler-Lagrange equation directly~\cite{Hubeny:2012ry}
\begin{align}
    \frac{d}{d r}\frac{\partial \mathcal{L}\big(z(r),z'(r),r\big)}{\partial z'(r)}-\frac{\partial \mathcal{L}\big(z(r),z'(r),r\big)}{\partial z(r)}&=0,\\
    \Rightarrow z''+(1+z'^2)\Big(\frac{d-2}{r}z'+\frac{d-1}{z}\Big)&=0\label{EEsphereODE}.
\end{align}
One can easily verify that the hemisphere 
\begin{align}
\label{hemisphere}
    z^2+r^2=R^2
\end{align}
obeys the differential equation~\eqref{EEsphereODE}.\footnote{Obviously, \eqref{hemisphere} agrees with~\eqref{semiCircle} when $d=2$. In this case there's no distinction between strip or spherical entangling subregions.}

The EE is more technical to evaluate. It is shown~\cite{Ryu:2006ef} that like in the strip case, the leading contribution is proportional to the area of the boundary of the entangling subregion $A$, $\frac{{\rm Area}(\partial A)}{\epsilon^{d-2}}$. 

\subsection{Holographic UV properties}
We have seen that for both strip and spherical subregions, the EE scales as $\frac{{\rm Area}(\partial A)}{\epsilon^{d-2}}$ ($d>2$). This behaviour is in fact general for RT as well as HRT surfaces. This is because the bulk $z$ coordinate corresponds to the energy scale of the dual field theory~\cite{Susskind:1998dq}, and the UV physics of the field theory is determined by the $z\to 0$ regions of the dual gravitational theories. The $(d+1)$-dimensional bulk metric takes universal form asymptotically. This reflects the fact that in the dual field theories, any finite energy states in the UV are the same as the vacuum. Choosing $z$ as one of the coordinates $\vec{\xi}$ for the RT/HRT surface, the induced metric of the RT/HRT surface at small $z$ can be written as
\begin{align}
    ds^2=\frac{1}{z^2}(dz^2+h_{ab}d\xi^a d\xi^b+...).
\end{align}
The area of the RT/HRT surface is then given by 
\begin{align}
\label{areaLawRT}
    A=\int_{\epsilon}^z\frac{dz'}{z'^{d-1}}\int d^{d-2}\xi \sqrt{h}+...=\frac{{\rm Area}(\partial A)}{\epsilon^{d-2}}+...,
\end{align}
in agreement with field theory considerations~\eqref{areaLaw}.

\subsection{Multiple regions}
\begin{figure}[htbp]
\centering
\includegraphics[width=.6\textwidth]{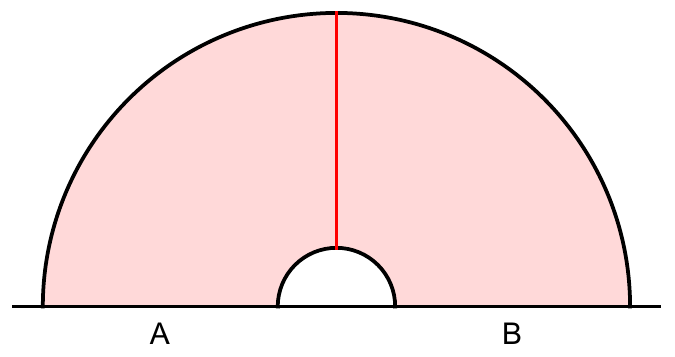}
\includegraphics[width=.6\textwidth]{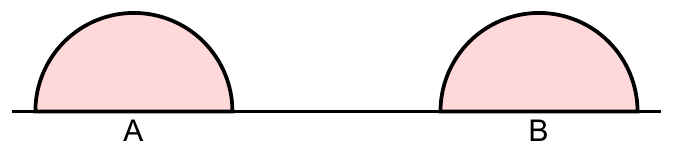}
\qquad
\caption{A demonstration of different RT surface configurations for $AB=A\cup B$. The bulk regions enclosed by the RT surfaces and the boundary, which are depicted as shaded, are the AB entanglement wedges. In the upper figure, the entanglement wedge cross-section computing the reflected entropy is shown in red. \label{fig:EWCSVacConn}} 
\end{figure}

We can use the RT surfaces as fundamental building block to evaluate the EE when $A$ is a disjoint interval $A=\cup_i A_i$. While the calculation in field theory is difficult, the holographic computation turns to be simple: we just minimise all possible configurations of RT surfaces homologous to $A=\cup_i A_i$. Taking vacuum AdS$_3$ as an example, in which case RT surfaces are semi-circle geodesic, we have
\begin{align}
    S_A=\frac{c}{3}\min\Big(\sum_{i,j}\log\frac{|u_i-v_j|}{\epsilon}\Big).
\end{align}
Notice that depending on the subregion sizes, the minimal RT configurations are in general different, see Figure~\eqref{fig:EWCSVacConn}. This strategy of minimizing over all possible RT surface configurations obeying the homology condition keeps to hold in higher dimensions and for thermal states. 

\subsection{Thermal CFT$_d$ on $\mathbb{R}^{d-1,1}$}
Having studied the case of vacuum CFT$_d$ on $\mathbb{R}^{d-1,1}$, we can proceed to studying thermal CFT$_d$ states, which are dual to asymptotically AdS$_{d+1}$ black holes. In the simplest neutral case, the dual geometry is planar AdS$_{d+1}$ black holes, whose metric is given by\footnote{Notice that unlike the vacuum case~\eqref{pureAdSmetric}, the metric~\eqref{AdSSchMetric} is no longer in Fefferman-Graham gauge~\cite{deHaro:2000vlm}, as $g_{zz}=\frac{1}{z^2f(z)}\neq\frac{1}{z^2}$. See~\cite{Hubeny:2012ry} for conversion of~\eqref{AdSSchMetric} to Fefferman-Graham gauge. } 
\begin{align}
\label{AdSSchMetric}
    ds^2=\frac{1}{z^2}\Big(-f(z)dt^2+\frac{dz^2}{f(z)}+d\vec{x}_{d-1}^2\Big), \ \ f(z)=1-z^d,
\end{align}
where $d\geq 2$. Following the same steps in section~\ref{EEVacGenD}, we adopt Cartesian and spherical coordinates for $d\vec{x}_{d-1}^2$, parametrise the minimal surface with $\xi^{i}=x^i$ and $\xi^i=(r,\theta_{d-2})$, and describe them with $z(x_1)$ and $z(r)$, respectively. The EE are then given by
\begin{align}
    S_{{\rm strip}}&=\frac{1}{4G_N}L^{d-2}\int dx_1\frac{1}{z^{d-1}}\sqrt{1+\frac{z'(x_1)^2}{f(z)}}\label{EEThmStrip},\\
    S_{{\rm sphere}}&=\frac{1}{4G_N}\omega_{d-2}\int dr\frac{r^{d-2}}{z^{d-1}}\sqrt{1+\frac{z'(r)^2}{f(z)}}\label{EEThmSphere}.
\end{align}

When the entangling subregion is large, due to the $\frac{1}{z^2}$ factor in \eqref{AdSSchMetric}, the RT surfaces consist of large plateau regions where they skim the horizon. In these plateau regions, $z\approx 1$ and $z'\approx 0$. Plugging these into \eqref{EEThmStrip}, subtracting the universal UV divergent term $\propto\frac{{\rm Area}(\partial A)}{\epsilon^{d-2}}$, we have
\begin{align}
\label{EEThmStrip2}
    S_{{\rm strip}}=\frac{1}{4G_N}L^{d-2}\cdot \Delta x_1=\frac{1}{4G_N}\vol(A),
\end{align}
and into \eqref{EEThmSphere} we have 
\begin{align}
\label{EEThmSphere2}
    S_{{\rm sphere}}=\frac{1}{4G_N}\omega_{d-2}\frac{r^{d-1}}{d-1}\Big|^R_0=\frac{1}{4G_N}\frac{\pi^{\frac{d-1}{2}}}{\Gamma(\frac{d+1}{2})}R^{d-1}=\frac{1}{4G_N}\vol(A).
\end{align}
In both cases, the EE is given by the $\emph{volume}$ of the entangling subregion $A$. This is expected for EE in the thermal state in a QFT~\eqref{volLaw}.

\subsection{Entanglement plateaux}\label{sec:lateauxReview}
\begin{figure}[htbp]
\centering
\includegraphics[width=.35\textwidth]{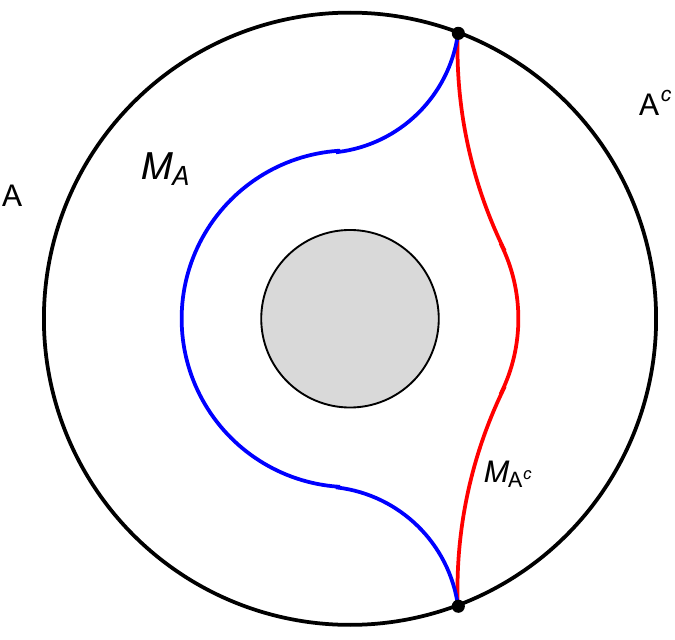}
\qquad
\caption{A demonstration of the minimal surface for a subregion $A$ and its complementary $A^c$. \label{fig:plateaux}}
\end{figure}

So far we have been focusing on planar AdS spacetime. It would also be interesting to study global AdS$_{d+1}$ geometry, dual to CFT$_d$ on $\mathbb{R}\times\mathbb{S}^{d-1}$. When the state of the total system is a thermal density matrix, $S_A$ and $S_{A^c}$ are not equal, as the minimal surfaces $M_A$ and $M_{A^c}$ computing them encircle complementary parts of the black hole horizon \cite{Ryu:2006bv,Ryu:2006ef} (see Figure \ref{fig:plateaux}). Let us study $S_A$ in the $d=2$ case that is relevant to the discussions in Chapter~\ref{Floquet} first. In addition to the connected minimal surface $M_A$, there exists another disconnected minimal surface $M_{A^c}\cup \mathbb{S}_{\rm hor}$, where $\mathbb{S}_{\rm hor}$ is the bifurcation surface of the black hole. See Figure \ref{fig:plateaux}. When $A$ gets large, $M_{A^c}\cup \mathbb{S}_{\rm hor}$ has a smaller area than $M_A$ and hence becomes the correct RT surface. In other words, we have 
\es{2minsurfs}{
    S_A=\frac{1}{4G_N}{\rm min}\{\mathcal{L}(M_A),\mathcal{L}(M_{A^c})+ \mathcal{L}(\mathbb{S}_{\rm hor})\}\,,
}
where $\mathcal{L}(\mathbb{S}_{\rm hor})=2\pi r_+$. This phenomenon is known as $\emph{entanglement plateaux}$~\cite{Hubeny:2013gta}. 

Entanglement plateaux are consistent with the Araki-Lieb inequality~\cite{Araki:1970ba}, which upper bounds the difference between $S_A$ and $S_{A^c}$ in a bipartite system with the entropy of the total density matrix $S_{\rho}$
\begin{align}
    |\Delta S_A|=|S_A-S_{A^c}|\leq S_{A\cup A^c}=S_{\rho}\label{Araki-Lieb}.
\end{align}
If we naively take $M_A$ as the RT surface throughout, for large $A$ we would violate~\eqref{Araki-Lieb}, as $|\Delta S_A|$ diverges in the limit of $A$ becoming the entire system~\cite{Hubeny:2013gta}. Employing~\eqref{2minsurfs} to minimise between $M_A$ and $M_{A^c}\cup \mathbb{S}_{\rm hor}$, on the other hand, manifests~\eqref{Araki-Lieb} for $A$ of all sizes: when $M_{A^c}\cup \mathbb{S}_{\rm hor}$ becomes the RT surface, the inequality~\eqref{Araki-Lieb} simply saturates. 

Entanglement plateaux also exist in $d>2$: in these cases, for sufficiently large $A$ there are no connected minimal surfaces $M_A$ at all~\cite{Hubeny:2013gta}. This leaves no option but to consider the disconnected $M_{A^c}\cup \mathbb{S}_{\rm hor}$ as the RT surface for $S_A$. 



\section{Time evolution of holographic entanglement entropy}
\label{HRTReview}
So far, we have reviewed the basis of entanglement entropy as well as its holographic calculations at fixed time. Now, we are ready to discuss the major question we would like to address in this thesis, the out-of-equilibrium $\emph{dynamics}$ of EE in interacting many-body systems. This is challenging, as there are no conventional small parameters to expand in and numerical simulations are hampered by the exponentially large Hilbert space that the system explores. There are a handful of special setups, where analytical progress can be made: in random, dual unitary, and stabiliser quantum circuits~\cite{Fisher:2022qey}; in models of Sachdev-Ye-Kitaev type~\cite{Rosenhaus:2018dtp,Chowdhury:2021qpy}; in systems with a holographic gravity dual~\cite{Son:2002sd,Balasubramanian:2011ur,Shenker:2013pqa}; and in conformal field theories (CFTs) driven by unconventional Hamiltonians. In this thesis, we will focus on the the latter two cases, both of which describe quantum dynamics in terms of geometry: in terms of gravity in asymptotically AdS spacetimes and in terms of conformal transformations of the complex plane. We will briefly review CFTs driven by unconventional Hamiltonians in section~\ref{FloquetIntro}, and entanglement dynamics in holographic CFTs in section~\ref{HMReview}.

\subsection{CFTs driven by unconventional Hamiltonians}
\label{FloquetIntro}
In this section, we will review 1+1 dimensional CFTs with inhomogeneous deformations. It provides a framework for analytic studies of entanglement dynamics, which we will address in Chapter~\ref{Floquet}. 1+1 dimensional CFTs with inhomogeneous deformations have received considerable attention recently \cite{Wen:2018vux,Wen:2018agb,Fan:2019upv,Wen:2020wee,Han:2020kwp,Fan:2020orx,Wen:2022pyj}, as they provide exactly solvable models that describe quantum critical systems out of equilibrium.  These deformations include the so-called sine square deformation (SSD) \cite{Wen:2018vux}, which is a special case of $SL(2)$ deformations of 2d CFTs. In SSD, one changes the CFT Hamiltonian $H_0=\int_0^l dx\,T_{00}(x)$ on a circle of circumference $l$ to a new Hamiltonian with a sine-square enveloping function
\begin{align}
    H_1=2\int_0^l dx\ \sin ^2\Big(\frac{\pi x}{l}\Big)T_{00}(x).
\end{align}
SSD was first introduced in quantum many-body systems to eliminate boundary effects \cite{Gendiar:2008udd,Hikihara:2011mtb,gendiar2011suppression}. Later on, 2d CFT with SSD Hamiltonian and its generalisation, the Mobius deformed Hamiltonian were studied in \cite{Katsura:2011ss,Ishibashi:2015jba,Ishibashi:2016bey,Okunishi:2016zat,Wen:2016inm,Tamura:2017vbx}, where it was shown \cite{Ishibashi:2015jba,Ishibashi:2016bey} that the spectrum of SSD CFT is continuous. In \cite{Wen:2018vux} 
explicit expressions for entanglement entropy (EE) after a quantum quench were found. It was also pointed out in \cite{Wen:2018vux} that SSD CFT can be interpreted as a CFT on curved spacetime. 

SSD CFT also serves as the building block for Floquet CFT \cite{Wen:2018agb,Fan:2019upv,Wen:2020wee,Han:2020kwp,Fan:2020orx,Wen:2022pyj}. Floquet CFT is constructed by $\emph{periodically}$ driving the CFT with ordinary and SSD Hamiltonians, $H_0$ and $H_1$. The physical motivation is to provide an exactly solvable toy model that captures some of the features of the more realisable periodically driven systems including Floquet topological insulators, Floquet symmetry-protected phases, as well as Floquet time crystals. Floquet CFT describes driven quantum many-body systems at criticality. Floquet CFT was found to have heating and non-heating phases \cite{Wen:2018agb,Fan:2019upv}, corresponding to different types of $SL(2)$ transformations. In the heating phase, the energy is localised in peaks that also dominate entanglement entropy. If initially prepared in a thermal state, the system outside these peaks cools down and behaves like the vacuum state, leading to the name "Floquet's refrigerators" \cite{Wen:2022pyj}. In the non-heating phases, the energy and entanglement show oscillatory behaviour without a fixed period.

In Chapter~\ref{Floquet}, we study entanglement dynamics in the aforementioned SSD CFT from both field-theoretic and holographic points of view.\footnote{The original paper~\cite{Jiang:2024hgt} that Chapter~\ref{Floquet} is based on also contains holographic as well as field-theoretic studies of Floquet CFTs. } There are already holographic studies of SSD and Floquet CFTs in the literature. In \cite{MacCormack:2018rwq}, a $(2+1)$-dimensional bulk metric with SSD boundary is constructed. Holographic dynamics of SSD quench with thermal initial states was investigated in \cite{Goto:2021sqx}, and the time evolution of mutual information under this framework was further explored in \cite{Goto:2023wai}. There are also recent works  \cite{Bernamonti:2024fgx,Kudler-Flam:2023ahk} exploring holographic aspects of SSD or Mobius quench in boundary conformal field theory, whose gravity dual involves end-of-the-world branes. 

\subsection{Time evolution of entanglement entropy from black hole interior}
\label{HMReview}
We would like to understand entanglement dynamics in generic holographic theories, where the time-dependent entanglement entropy is computed by the HRT surfaces that enters the black hole interior. We again consider thermal CFT$_d$ states on $\mathbb{R}^{d-1,1}$, dual to asymptotically AdS$_{d+1}$ black hole. It would be convenient to write the metric as\footnote{In the special case of AdS$_{d+1}$-Schwarzschild~\eqref{AdSSchMetric}, we have~\cite{Hartman:2013qma}
\begin{align*}
    d\rho=\frac{dz}{z\sqrt{1-z^d}}
\end{align*}
}
\begin{align}
    ds^2=-g(\rho)^2dt^2+d\rho^2+h(\rho)^2dx_{d-1}^2,
\end{align}
where there's a horizon at $\rho=0$, near which $g(\rho)=\rho+O(\rho^3)$ and $h(\rho)=C+O(\rho^2)$. In the interior $\rho=i\kappa$, and $t_I=t-i\pi/2$ (the Schwarzschild time $t$ becomes complex in the black hole interior~\cite{Fidkowski:2003nf}).

We consider the entanglement entropy of subregion $A$ consisting of half of the space of each of the two copies of the thermofield double. We evolve time forward on both copies of the field theory. Due to the symmetry of the problem, the HRT surface lies on the $x_1=0$ plane, and extends along the $x_2,...,x_{d-1}$ directions. Noticing that the co-dimension 2 extremal surface no longer lies on a constant time slice, we choose to parametrise it with $\xi^i=(t,\vec{x}_{d-2})$. The HRT surface is then described by the function $\rho(t)$ (or $t(\rho)$). Using~\eqref{inducedMetric} and~\eqref{area}, we find
\begin{align}
\label{HMarea}
    S=\frac{1}{4G_N}L^{d-2}\int dt\ h\big(\rho(t)\big)^{d-2}\sqrt{-g\big(\rho(t)\big)^2+\dot\rho(t)^2}.
\end{align}
To calculate the EE, we need to $\emph{extremise}$ the area functional in~\eqref{HMarea}. Like in the static RT case, it would be easier to apply conservation laws: we notice that the "Lagrangian" $\mathcal{L}\big(\rho(t),\dot\rho(t),t\big)=h\big(\rho(t)\big)^{d-2}\sqrt{-g\big(\rho(t)\big)^2+\dot\rho(t)^2}$ is $t$-independent. Hence, the "energy" is conserved 
\begin{align}
\label{HMconversedE}
    \mathcal{H}=\dot{\rho}(t)\frac{\partial\mathcal{L}}{\partial\dot{\rho}(t)}-\mathcal{L}=\frac{g^2h^{d-2}}{\sqrt{-g^2+\dot{\rho}^2}}=-ig_0h_0^{d-2},
\end{align}
where we again evaluate the "energy" at the deepest point $\rho_0=i\kappa_0$ the HRT surface can reach in the interior, at which point $\dot\rho|_{\rho_0}=0$. From~\eqref{HMconversedE}, we can solve for $\dot{\rho}$. This then allows us to find the integral expression describing the HRT surface\footnote{Notice that the integral has a pole at $\rho=0$, and the integral should be taken along a $\epsilon$-ball around $\rho=0$.} 
\begin{align}
\label{HMtRho}
    t(\rho)=\int \frac{d\rho}{\dot{\rho}}=-i\frac{\pi}{2}-\int_{i \kappa_0}^{\rho}\frac{d\rho'}{\sqrt{1-\frac{g^2h^{2d-4}}{g_0^2h_0^{2d-4}}}}.
\end{align}
Using $\dot{\rho}$, we can also express the area~\eqref{HMarea} integral 
\begin{align}
    S&=L^{d-2}\int\frac{d\rho}{\dot{\rho}}\ h(\rho)^{d-2}\sqrt{-g(\rho)^2+\dot\rho^2},\\
    &=2L^{d-2}\int_{i\kappa_0}^{\infty}d\rho\frac{h^{d-2}}{\sqrt{1-\frac{g_0^2h_0^{2d-4}}{g^2h^{2d-4}}}}.
\end{align}
Now let us study the $t(\rho)$ integral~\eqref{HMtRho}, and denote $a(\rho)=-ig(\rho)h(\rho)^{d-2}$ for convenience. We expand the integrand $\frac{1}{g\sqrt{1-\frac{a^2}{a_0^2}}}$ around $\rho=\rho_0$. If $a$ is not extremized at $\rho_0=i\kappa_0$, i.e. $a'(\rho_0)\neq 0$, we find that the $t(\rho)$ integral~\eqref{HMtRho} goes as $\sqrt{\rho-\rho_0}$ as $\rho\to\rho_0$ and is hence finite; nevertheless, if $a'(\rho_0)= 0$, the integral instead goes as $\log(\rho-\rho_0)$ as $\rho\to\rho_0$ and diverges. We denote this point where $a(\rho)$ reach its extremum (typically a maximum, as can be explicitly checked in AdS$_{d+1}$-Schwarzschild) as $\rho_*$. These mean that the boundary time $t_b=t(\rho=\infty)$ remains finite for $\rho_0<\rho_*$, but becomes large, i.e. $t_b\to\infty$ as $\rho_0\to\rho_*$. In other words, at late time, the HRT surface gets stuck at the constant $\rho$ slice $\rho=\rho_*$ instead of going deeper into the black hole interior. In addition to this portion at $\rho\to\rho_*$, the late-time HRT surface also has a part that exits the horizon and goes to the boundary. For large $t_b$, this part gives a constant contribution of EE (after regularisations). On the other hand, from~\eqref{HMarea}, it is evident that the area of the interior part of the HRT surface at $\rho\to\rho_*$ grows linear in $t_b$: as the HRT surface lies close to $\rho_*$, we have $\dot\rho=0$, and hence 
\begin{align}
\label{HMarea2}
    S=\frac{1}{4G_N}L^{d-2}\int dt\ h(\rho_*)^{d-2}\sqrt{-g(\rho_*)^2}=L^{d-2} h(\rho_*)^{d-2}\sqrt{-g(\rho_*)^2}\cdot t_b.
\end{align}
This means that the entanglement entropy also grows $\emph{linearly}$ with time $t_b$, with the proportional constant $v_E$ specified by the interior geometry of the black hole at $\rho_*=i\kappa_*$.
\begin{align}
    S=\frac{1}{4G_N}L^{d-2}s v_E t_b, && v_E= \frac{\sqrt{-g(\rho_*)^2}h(\rho_*)^{d-2}}{h(0)^{d-1}}.
\end{align}
where $s=\frac{h(0)^{d-1}}{4G_N}$ is the coarse-grained entropy density. The speed $v_E$ at which EE grows linearly is called the $\emph{entanglement velocity}$. For the planar AdS$_{d+1}$-Schwarzschild black hole, we have 
\begin{align}
\label{vE}
    \tan \frac{\kappa_* d}{2}=\sqrt{\frac{d}{d-2}}, && v_E=\frac{\big(\frac{d-2}{d}\big)^{\frac{d-2}{2d}}}{\big(\frac{2(d-1)}{d}\big)^{\frac{d-1}{d}}}.
\end{align}

\subsection{Entanglement membrane from holography}
\label{sec:membraneReview}
\begin{figure}[htbp]
\centering
\includegraphics[width=.5\textwidth]{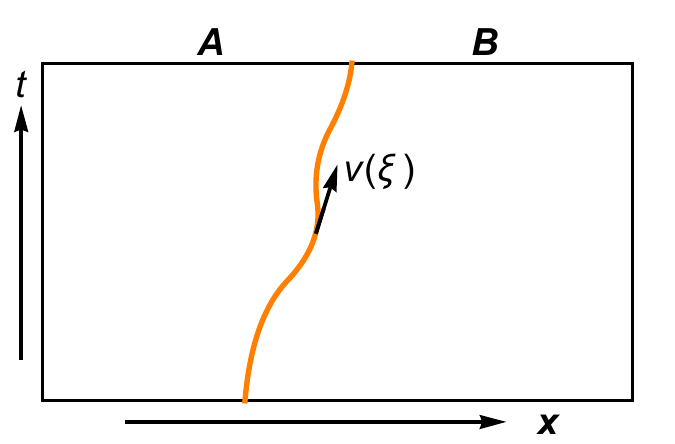}
\qquad
\caption{Cartoon of a minimal membrane (orange) extending in spacetime in $(1+1)$-dimensions. The top and bottom horizontal lines denote the constant time slices at $t$ and $0$, respectively. We are computing the time-dependent EE for a single region $A$ (its complement is denoted as $B$). The minimal membrane is obtained by minimising the action~\eqref{MinMemb} over all possible membranes obeying the boundary condition: that it anchors on $\partial A$ at time $t$. The figure can either represent entropy dynamics in a quench, where $A(t)$ is on the top time slice and the bottom time slice represents the initial state, or the operator entanglement of the time evolution operator $e^{-(\beta/2+it)H}$, where the top region is $A_\text{ket}$ and the bottom one is $A_\text{bra}$. \label{fig:MembraneCartoon}}
\end{figure}

From the derivation of~\eqref{HMarea2},~\eqref{EEThmStrip2} and~\eqref{EEThmSphere2}, it is evident that while the full expressions of EE appear to be somewhat complicated and often have no analytic solutions, major simplifications occur in the limit of late time and large entangling subregions. Under such limits, only certain portions of the full RT/HRT expressions play roles to leading order. This simplification suggests the existence of an $\emph{effective theory}$ that governs the leading dynamical behaviour of EE in the late time and large region limit. Indeed, in chaotic many-body quantum systems, there is mounting evidence that the dynamics of entanglement and R\'enyi entropies in the limit of large region sizes and long times can be captured by an effective description in terms of a minimal membrane~\cite{Nahum:2016muy,Jonay:2018yei,Mezei:2018jco}. This effective description states that the entanglement entropy is computed by the action
\begin{equation}\label{MinMemb}
S=\min_{x(\xi)} \int d^{d-1}\xi\, \sqrt{\abs{\gamma(\xi)}}\, \frac{s_\text{th}\, {\cal E}(v(\xi))}{\sqrt{1-v^2(\xi)}}\,.
\end{equation}
In the above formula, $\xi$ are the membrane worldvolume coordinates,\footnote{In $d=1+1$, the membrane is a curve in spacetime.} $x(\xi)$ is its embedding into spacetime, $\gamma$ is the induced metric on the membrane, $s_\text{th}$ is the coarse grained entropy density, $v(\xi)$ is local transverse velocity of the membrane, and ${\cal E}(v)$ is the membrane tension function that depends the microscopic details of the many-body system. The minimisation is performed over all membrane shapes satisfying boundary conditions determined by the situation of interest. If we wish to study a quench, the membrane connects the time slice of the boundary of the subregion $A(t)$ and the time slice of the initial state; similarly, if we are interested in operator entanglement, the membrane connects the $A_\text{bra}$ and $A_\text{ket}$ subregions of the operator, see Fig.~\ref{fig:MembraneCartoon}.

The membrane effective description can be derived for random quantum circuits~\cite{Nahum:2016muy,Jonay:2018yei,Zhou:2018myl}, Floquet circuits (up to a plausible assumption about the convergence of an RG scheme)~\cite{Zhou:2019pob}, generalised dual-unitary circuits~\cite{Rampp:2023vah}, Brownian Hamiltonians \cite{Vardhan:2024wxb}, holographic CFTs~\cite{Mezei:2018jco,Mezei:2019zyt}, and Brownian SYK chain~\cite{Mezei:2025fcd}. There is also convincing numerical evidence for this description in chaotic spin chains~\cite{Jonay:2018yei}. In this thesis, we will focus on the holographic case. 

\subsubsection{The entanglement membrane for a half space subregion}
The holographic derivation of the entanglement membrane~\cite{Mezei:2018jco} starts with the study of HRT~\cite{Hubeny:2007xt} and RT~\cite{Ryu:2006bv,Ryu:2006ef} surfaces in black brane spacetimes. In the quench setup, at early times we have some non-equilibrium features in the spacetime that get encoded in boundary conditions for the membrane. At later times local thermal equilibrium is reached, which is dual to the black brane. In the operator entanglement setup, we consider an eternal black brane and put the $A_\text{bra}$ and $A_\text{ket}$ subregions on the left and right asymptotic boundaries~\cite{Hartman:2013qma}.

Hence, we consider an 
 asymptotically AdS$_{d+1}$  black brane in infalling coordinates:
\begin{equation}
ds^2 = \frac{1}{z^2}\left(-a(z)du^2 -{2\ov b(z)}dudz+dx^2+ d\vec{y}_{d-2}^2\right)\,,\label{blackBraneInf}
\end{equation}
where the conformal boundary is at $z=0$, and the horizon is at $z=1$; therefore $a(1)=0$. The  AdS$_{d+1}$ asymptotics requires $a(0)=b(0)=1$. For a neutral black brane $a(z)=1-z^d,\, b(z)=1$. 

\begin{figure}[htbp]
\centering
\includegraphics[width=.365\textwidth]{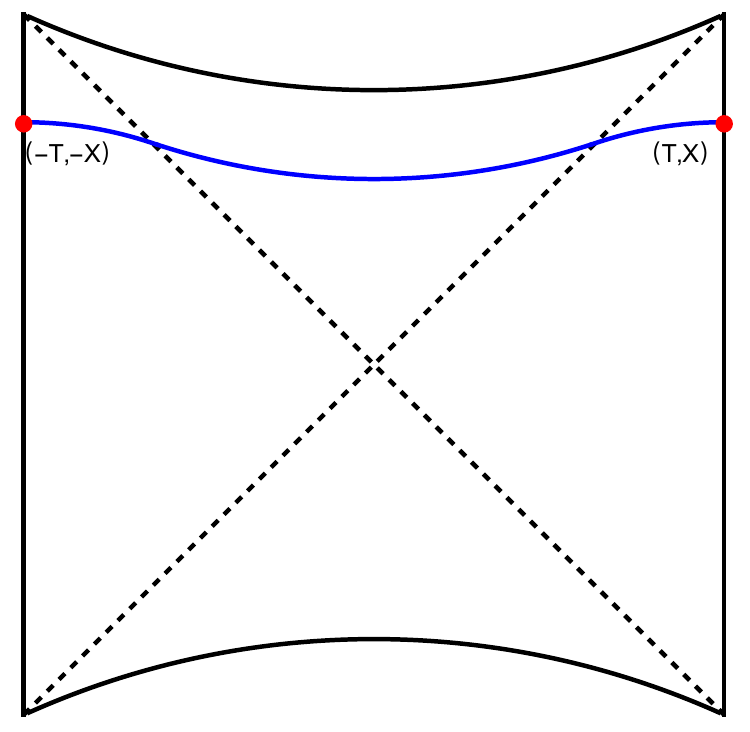}
\qquad
\caption{Penrose diagram of the HRT surface computing the time-dependent entanglement entropy in planar AdS$_{d+1}$-Schwarzschild ($d>2$) with entangling subregion taken to be half the space on both side of the thermofield double. The $\pm X$ displacement in the $x$ direction is perpendicular to the plane of the paper. \label{fig:DispHalfSpaceBTZ}}
\end{figure}

We would like to determine the two-sided entanglement entropy for half spaces extending in the $\vec{y}$ directions at late-time, see Fig.~\ref{fig:DispHalfSpaceBTZ}. We parametrise the HRT surface as $x(u),\, z(u)$, and obtain the area functional for the metric~\eqref{blackBraneInf}:
\begin{equation}
\label{AreaFunct}
\begin{aligned}
S &= s_\text{th} \vol(\p A) \int du  \frac{\sqrt{Q}}{z^{d-1}}\,, \qquad Q\equiv \dot x^2-a(z)-{2\ov b(z)}\dot z\,,
\end{aligned}
\end{equation}
where $\dot f\equiv\frac{df}{du}$. 

As is reviewed in section~\eqref{HMReview}, for $d>2$ black branes the leading behaviour of entanglement entropy in the late time and large entangling subregion limit is determined by the portion of the HRT surfaces that lie inside the horizon; at late time, the HRT surfaces get stuck at some special extremal slice $z=z_*>1$ in the interior instead of moving deeper towards the singularity.\footnote{As we will see in section~\ref{KasnerSec}, these conclusions also hold for $d=2$ hairy BTZ black branes (strings). } Therefore, to extract the leading behaviour of entanglement entropy in the late time and large entangling subregion limit, we can impose the following self-consistent scaling Ansatz~\cite{Mezei:2016zxg,Mezei:2018jco}:
\es{CoordScaling}{
u\rightarrow \Lambda u, \hspace{0.5cm} x \rightarrow \Lambda x, \hspace{0.5cm} z\rightarrow z\,,
}
for $\Lambda\gg1$. In this scaling the $\dot z$ term can be dropped from $Q$ in \eqref{AreaFunct}. Then the $z$ equation of motion becomes algebraic in $z$
\es{projectionz}{
v^2\equiv \dot x^2  = a(z) - \frac{z a'(z)}{2(d-1)} \equiv c(z)\,;
}
e.g.~for neutral black branes $c(z)=1-{d-2\ov 2(d-1) }\, z^d$. The function $c(z)$ is monotonically decreasing and positive in the range $1\leq z \leq z_*$, where $z_*$ is the first zero of $c(z)$. Then $z$ in this range can be solved for in terms of $v^2$ and we have managed to reduce the problem in dimension (from $d+1$ to $d$). The resulting reduced action is the membrane theory introduced around \eqref{MinMemb}:
\begin{equation}
\label{AreaFunct2}
\begin{aligned}
S &= s_\text{th} \vol(\p A) \int du \,  \mathcal{E}(v^2)\,, \qquad
\mathcal{E}(v^2) =\left. \sqrt{\frac{-a'(z)}{2(d-1)z^{2d-3}}}\right|_{z=c^{-1}(v^2)}\,.
\end{aligned}
\end{equation}
The bulk infalling time $u$ assumes the role of boundary theory time. (Recall that at the AdS boundary infalling and boundary times agree.) As the reduction in dimension amounts to integrating out the bulk radial $z$ degree of freedom using~\eqref{AreaFunct2}, we can understand the geometric meaning of the membrane as the $\emph{projection}$ of the codimension-2 bulk HRT surface in the scaling limit~\eqref{CoordScaling} to the AdS boundary along constant infalling time.

We have seen that the RT/HRT surfaces possess universal UV divergences as $z\to 0$~\eqref{areaLawRT},
\begin{align}
\label{AreaLogLaw}
    d> 2:\ \frac{\AreaLaw}{\epsilon^{d-2}} && {\rm or} && d=2:\ \log\frac{2}{\epsilon}
\end{align}
for some UV cutoff $\epsilon\ll 1$. When projecting the HRT surfaces to the boundary along constant infalling time to obtain the membranes, we subtract these UV divergences of the (H)RT surfaces. In the scaling limit, due to the $1/z^2$ factor in~\eqref{blackBraneInf}, the remaining parts of the (H)RT surfaces that compute finite entropy are inside or close to the horizon.\footnote{In the original membrane theory from holographic CFTs~\cite{Mezei:2018jco}, the prescription was to cut the HRT surfaces at the horizon and keep only the interior parts, as it is only these parts that grow linearly with time~\cite{Hartman:2013qma}. In a later work~\cite{Jiang:2024tdj}, it was discovered that for saturated RT surfaces, as well as certain half-space HRT surfaces with displacement in the $x$ direction, portions of the (H)RT surfaces that are outside but near the horizon can also contribute finite entropy in the scaling limit. As such, here we adopt the latter more general prescription to cut out the UV divergences of the (H)RT surfaces.} 

The function $\mathcal{E}(v^2)$ obeys a series of properties. To state them, we further introduce the butterfly velocity $v_B$~\cite{Roberts:2014isa}, related to the out-of-time order correlator (OTOC)
\begin{align}
    v_B=\sqrt{-\frac{a'(1)}{2(d-1)}}\,.\label{vB}
\end{align}
$c(z)$ maps $v_B$ to the horizon at $z=1$. In previous work~\cite{Mezei:2018jco} the near-horizon region around $z=1$ was not treated with enough care. In the later part of this thesis, we fix the resulting imprecisions that involve membrane with regions, where $v=v_B$.

In general, $v_E\leq v_B$ due to the constraint from null energy condition~\cite{Mezei:2016zxg}.\footnote{This inequality also follows from general quantum many-body considerations that are independent from holography~\cite{Mezei:2016wfz}.} We are now ready to state the constraints obeyed by the membrane tension function: it is an even function, monotonically increasing and convex for $0\leq v <1$. It diverges as $v\rightarrow 1$ and certain important values are~\cite{Mezei:2018jco}:\footnote{The behaviour around $v_B$ follows from consistency conditions related to operator spreading~\cite{Jonay:2018yei}. }
\begin{align}
    \mathcal{E}(0)=v_E\,, && \mathcal{E}'(0)=0\,, && \mathcal{E}(v_B)=v_B\,, && \mathcal{E}'(v_B)=1\,, && \mathcal{E}'(v)>0\,, && \mathcal{E}''(v)>0\,. \label{constraints}
\end{align}

In the above discussion, we have focused on strip entangling subregions. Nevertheless, entanglement membranes can be derived from holography with sphere and generic regions, see~\cite{Mezei:2018jco,Mezei:2020knv}.

While entanglement membrane is elegant and profound, the theory is still at the developing stage. An issue of the theory is that when the geometry \eqref{blackBraneInf} is the BTZ black brane, $a(z)=1-z^2,\,b(z)=1$, we get $v_E=v_B= \mathcal{E}(v)=1$. This signals the breakdown of entanglement membrane for two-dimensional CFTs, even though they were some of the first studied examples of entanglement dynamics~\cite{Calabrese:2005in,Calabrese:2007rg} that are generically chaotic for central charge above some threshold value.\footnote{If they only have Virasoro symmetry, the criterion is $c>1$. If they have extended chiral symmetry characterised by an effective central charge $c_\text{currents}$  for the chiral sector of the theory, then the criterion is $c>c_\text{currents}$~\cite{Asplund:2015eha}. Below these threshold values the CFTs are integrable and their entanglement dynamics are described by the quasiparticle effective description~\cite{Calabrese:2005in,Calabrese:2007rg}.} One could dismiss this issue by attributing it to some peculiarity of the dual 3d gravity.
Indeed, the entanglement entropy dynamics in quenches and in the operator entanglement setup obtained using 2d CFT technology is reproduced by the membrane effective description, provided we choose the degenerate membrane tension function ${\cal E}(v)=1$. However, there are two indications that there is a fundamental issue at play: first, the R\'enyi entropy dynamics in 2d CFT for two intervals or more~\cite{Asplund:2015eha} is incompatible with the membrane theory~\cite{Mezei:2018jco}. Second, in quantum circuits the entropy dynamics is translated to the dynamics of a domain wall in a lattice spin model. The diffusion constant of the domain wall is proportional to $1/{\cal E}''(v)$~\cite{Zhou:2019pob}, which diverges for the degenerate membrane tension; indeed, just like 2d CFTs, dual unitary circuits formally give rise to the degenerate membrane tension ${\cal E}(v)=1$~\cite{Zhou:2019pob}, and in the equivalent lattice model the domain wall delocalises. In Chapter~\ref{Chap3}, we will address this issue and propose the effective theory underlying entanglement in 2d CFT to be the $\emph{generalised membrane theory}$ that is equipped with an additional degree of freedom on the membrane. 

\section{Reflected entropy}
\label{ReflectedEntropyReview}
The von Neumann entropy~\eqref{EEdef} that we have been focusing on concerns only systems in bipartite pure states, where we divide the system into two subregions $A$ and $\oA$ and study the entanglement between them. However, in physical systems, it is often of interest to study bipartite mixed states or multipartite states. For these cases, the von Neumann entropy appears to be somewhat insufficient to account for the richer entanglement structures. As such, it would be important to study further information-theoretic quantities that capture aspects of mixed or multipartite state correlations. One such quantity is the mutual information 
\begin{align}
    I(A:B)=S(A)+S(B)-S(AB),\label{MI}
\end{align}
as for a pure state $\ket{\psi}_{AB}$ we have $S(AB)=0$ and $S(A)=S(B)$, so $I(A:B)=2S(A)$.  However, the mutual information is just a linear combination of entanglement entropy, and hence cannot be regarded as a new quantity characterising mixed or multipartite state correlations. In what follows, we will investigate another information-theoretic quantity for mixed or multipartite states known as $\emph{reflected entropy}$~\cite{Dutta:2019gen}. It has the advantage of having a simple holographic dual that can be regarded as natural generalisations of the RT and HRT surface in static and time-dependent cases, respectively. 

For a bipartite system with density matrix $\rho_{AB}$, the reflected entropy is given by
\begin{equation}\label{eq:RefEnt}
S_{R}(A:B) = S(\rho_{AA^*})\,, \quad \rho_{AA^*} = \Tr_{BB^*}\ket{\rho^{1/2}_{AB}}\bra{\rho^{1/2}_{AB}}\,,
\end{equation}
where $\ket{\rho^{1/2}_{AB}}\in \mathcal{H}_{ABA^*B^*}$ is the canonical purification of  $\rho_{AB}$, and $S(\rho)$ is the von Neumann entropy~\cite{Dutta:2019gen}. The extended Hilbert space $\mathcal{H}_{ABA^*B^*}$ is constructed out of mirror copies of the original system~\cite{Engelhardt:2018kcs}.
In applications $A$ and  $B$ are often disjoint regions and are parts of a larger system. Then $\rho_{AB}$ is obtained by reducing the state of this larger system onto the region $A\cup B$.

Given the HRT surfaces $\mathcal{E}_{AB}$ homologous to the entangling region $A\cup B$, the entanglement wedge cross section (EWCS) $\Sigma_{AB}$ is the minimal area surface among the (potentially multiple) extremal surfaces bisecting the entanglement wedge, defined as the causal development of $\mathcal{E}_{AB}\cup A \cup B$~\cite{Takayanagi:2017knl,Akers:2022max}. 

It is believed that the holographic dual of the reflected entropy is the EWCS~\cite{Dutta:2019gen}:
\begin{equation}\label{eq:holoRefEntCov}
S_{R}(A:B) = \min_{\Sigma_{AB}}\frac{1}{2G_N}\text{Area}(\Sigma_{AB})\,,
\end{equation}
where the minimisation is done over all extremal surfaces. See Figure~\ref{fig:EWCSVacConn} for an illustration in vacuum AdS$_3$. 

In the static cases, the HRT surfaces $\mathcal{E}_{AB}$ become RT surfaces lying on a bulk constant time slice, and the EWCS also degenerates to the minimal surface bisecting the entanglement wedge on that slice. The formula~\eqref{eq:holoRefEntCov} has been proved in this case in~\cite{Dutta:2019gen}.

Due to technical difficulties, reflected entropy has been mostly studied in static cases in the literature. In Chapter~\ref{Chap3}, we will employ entanglement membrane to propose effective theories for the time evolution of reflected entropy in the late-time, large-region limit. Most noticeably, we will see that the additional degree of freedom in generalised membrane theory for 2d CFT plays a crucial role; consequently, there are $\emph{two}$ effective theories for 2d CFT and generic chaotic systems, respectively. These effective theories for reflected entropy allows one to further investigate the time evolution of multipartite entanglement signals that we will review in the next section. 

\section{Signals of multiparty entanglement}
\label{multipartySignalReview}
Multipartite states depict a much richer structure than bipartite states. In particular, systems with many degrees of freedom can also display intricate multiparty entanglement patterns. Multiparty entanglement has a intricate structure, and we do not have a good general description of the forms it can take (see the review \cite{Walter:2016lgl}). Some information about multiparty entanglement can be obtained by considering entanglement across different bipartitions of a system; for example the triple information constructed from such data provides a signal of four-party entanglement \cite{Cui:2018dyq}. Further insight can be obtained from the {\it reflected entropy} reviewed in the previous subsection. As we review below, the difference between reflected entropy and mutual information provides a signal of three-party entanglement \cite{Akers:2019gcv,Hayden:2021gno}. By ``signal'' we mean here a quantity whose non-vanishing confirms the presence of multiparty entanglement.  The converse need not be true.  If one of the signals we consider vanishes, there may still be some multiparty entanglement; it is just not detected by the signal in question.  

Recently, the work~\cite{Balasubramanian:2024ysu} initiated a study of multiparty entanglement signals for general numbers of parties.  They used these signals to show that states with holographically dual geometries contain significant amounts of multiparty entanglement.  Using these methods \cite{Balasubramanian:2025hxg} derived an inequality showing that 
purely Greenberger-Horne-Zeilinger-like (GHZ-like) entanglement between any number of parties is forbidden in holography, consistent with arguments for three-party GHZ-like states in \cite{Hayden:2021gno}. 

\subsection{Review of multiparty entanglement signals}
\label{sec:MPEReview}

In this section, we review some multiparty entanglement signals. We will focus primarily on signals of three- and four-party entanglement, but we also include a brief discussion of higher-party entanglement signals. 

We consider  multiparty entanglement in a pure state where the system is divided into $n$ disjoint subsystems. For $n=2$, we have a state $\ket{\psi}_{AB}$, and bipartite entanglement is measured by the von Neumann entropy $S(A)$ of the reduced density matrix $\rho_A = \mbox{tr}_B(\ket{\psi}_{AB} \bra{\psi}_{AB})$.  

For three parties, we can trace the state $\ket{\psi}_{ABC}$ over one party, say $C$, to obtain the reduced density matrix $\rho_{AB}$. The mutual information \eqref{MI} for this reduced density matrix is sensitive to correlation between $A$ and $B$, which includes contributions from both bipartite entanglement between $A$ and $B$ and tripartite entanglement involving $A$, $B$ and $C$. To isolate the tripartite entanglement we need to combine this with another source of information on correlation between $A$ and $B$. The reflected entropy introduced in the previous section is also sensitive to both bipartite entanglement between $A$ and $B$ and tripartite entanglement involving $A$, $B$, and $C$. In the residual information
\begin{align}
    R_3(A:B)=S_R(A:B)-I(A:B),\label{R3}
\end{align}
the contribution from bipartite entanglement $\emph{cancels}$, so the residual information is a signal of three-party entanglement: if $R_3 \neq 0$, the state $\ket{\psi}_{ABC}$ has tripartite entanglement between $A$, $B$, and $C$ \cite{Akers:2019gcv}. To see this more explicitly, let us suppose that $\ket{\psi}_{ABC}$ is a purely bipartite state: 
\begin{align}
    \ket{\psi}_{ABC}=\ket{\psi_1}_{A_1B_1}	\otimes \ket{\psi_2}_{A_2C_1}\otimes\ket{\psi_3}_{B_2C_2}.
\end{align}
One can easily compute the mutual information
\begin{align}
    I(A:B)&=2S(A_1)=2S(B_1),\\
    I(A:C)&=2S(A_2)=2S(C_1),\\
    I(B:C)&=2S(B_2)=2S(C_2).
\end{align}
Now, let us compute $S_R(A:B)$ via canonical purification. Making use of the fact that $S_R$ is additive under tensor product, we can write the reduced density matrix $\rho_{AB}$ as
\begin{align}
    \rho_{AB}&=\Tr_C \ket{\psi}_{ABC} \bra{\psi}_{ABC}\\
    &=\ket{\psi_1}_{A_1B_1} \bra{\psi_1}_{A_1B_1}\otimes\Tr_{C_1}\ket{\psi_2}_{A_2C_1} \bra{\psi_2}_{A_2C_1}\otimes\Tr_{C_2}\ket{\psi_3}_{B_2C_2} \bra{\psi_3}_{B_2C_2}\\
    &=\rho_{A_1B_1}\otimes\rho_{A_2}\otimes\rho_{B_2}.
\end{align}
The canonical purification of $\rho_{A_1B_1}$ is simply a product state of two copies of $\ket{\psi_1}$, hence $S_R(\rho_{A_1B_1})=2S(A_1)=2S(B_1)=I(A:B)$. As there are no entanglements between $A_2$ and $B$ in $\ket{\psi_2}$, we have $S_R(\rho_{A_2})=0$; similarly, $S_R(\rho_{B_2})=0$. Hence, we have $S_R(\rho_{AB})=S_R(\rho_{A_1B_1})=I(A:B)$ for purely bipartite states.

The residual information is provably non-negative \cite{Dutta:2019gen}, but it can nevertheless vanish on states with tripartite entanglement. For example, it vanishes for the GHZ state. 

Note that we chose to trace over $C$; tracing over $B$ or $A$ instead defines other signals of three-party entanglement, $R_3(A:C)$, $R_3(B:C)$. We will see later that these different signals can have qualitatively different behaviour, giving us independent information about aspects of the three-party entanglement. 

For a four party state $\ket{\psi}_{ABCD}$, we can obtain a signal of four-party entanglement by tracing over one of the parties, say $D$, and considering the triple information 
\begin{align}
    I_3(A:B:C)=S_A+S_B+S_C-S_{AB}-S_{AC}-S_{BC}+S_{ABC}\label{I3}
\end{align}
for the resulting reduced density matrix  \cite{Balasubramanian:2014hda}. This is permutation invariant among the four parties, which is made manifest by defining a four-party analogue $I_4(A:B:C:D)$ (see \eqref{ninf} for the general $n$-party definition), and noting that on pure states $I_4(A:B:C:D) = 2 I_3(A:B:C)$. This is analogous to the relation between the mutual information and the von Neumann entropy for two parties. The triple information is not sign-definite in general, although it is always negative on holographic states \cite{Hayden:2011ag}. 

In \cite{Balasubramanian:2024ysu}, another signal of four-party entanglement was defined using canonical purification. If we consider the reduced density matrix $\rho_{AB}$ obtained from $\ket{\psi}_{ABCD}$, using the canonical purification as above we can define the reflected entropy $S_R(A:B)$. Alternatively, if we reduce over $D$ we obtain a reduced density matrix $\rho_{ABC}$, with a canonical purification $\ket{\sqrt{\rho}}_{ABCA^*B^*C^*}$. Reducing this over $CC^*$ defines another reduced density matrix $\bar \rho_{AA^*BB^*}$, with a canonical purification $\ket{\sqrt{\bar{\rho}}}_{AA^*BB^*A_*A^*_*B_*B^*_*}$. Using the von Neumann entropy $S(AA^*A_*A^*_*)$ in this state, we can define the four-party residual entropy 
\begin{equation}
     Q_4=S(AA^*A_*A^*_*)-2S_R(A:B),  \label{Q4}
\end{equation}
which is non-zero only if the state has four-party entanglement. As with the residual information, this is not permutation invariant; we can define different signals by tracing out different subsystems. 

In \cite{Balasubramanian:2024ysu}, the residual information and triple information were generalised to higher numbers of parties. For even numbers of parties we have the $n$-information
\begin{equation} \label{ninf}
    I_n(A_1:\ldots:A_n) = - \sum_{i \leq n} (-1)^n S_i,  
\end{equation}
where $S_i$ is the permutation-invariant combination of all the entropies on sets of $i$ parties chosen from the $n$ parties.\footnote{Note that it has been recently shown that of these quantities, only the mutual information is monotonic until the partial trace \cite{Hernandez-Cuenca:2023iqh}. See also~\cite{Ju:2024hba,Ju:2025tgg} for further studies on $n$-information.} On pure states $I_n = 2 I_{n-1}$, where $I_{n-1}$ is evaluated on the reduced density matrix obtained by tracing out any one party. For $n$ odd $I_n=0$ on pure states, so the $n$-information is not helpful.  For odd $n$ we can define instead the $n$-residual information, by tracing out one party and taking the difference between $I_{n-1}$ evaluated on the canonical purification and $I_{n-1}$ on the density matrix, 
\begin{equation}
    R_n(A_1: \ldots : A_{n-1};A_n) = \frac{1}{2} I_{n-1}(A_1 A_1^*: \ldots :A_{n-1} A_{n-1}^*) - I_{n-1}(A_1: \ldots :A_{n-1}). 
\end{equation}
As with the residual information, this depends on which party we trace out, so there are $n$ independent signals. The $n$-information and the $n$-residual information for $n$ even and odd respectively are all signals of $n$-party entanglement; if they are non-zero the state has $n$-party entanglement. Beyond the cases discussed above nothing is known about their positivity properties.

The discussion in these works, and in the preceding analysis of  \cite{Balasubramanian:2014hda},  was restricted to static or time-symmetric cases, such as vacuum AdS$_3$ and multiboundary wormholes. In Chapter~\ref{Chap4}, we will consider time evolution of multiparty entalgmement signals reviewed in section~\ref{multipartySignalReview} in chaotic systems in the late time, large region limit, in which entanglement dynamics is believed to be universally captured by the membrane effective theory. 

The evolution of the triple information in time-dependent holographic states dual to asymptotically AdS black holes was previously considered in \cite{Balasubramanian:2011at}.\footnote{In the precise setup of~\cite{Balasubramanian:2011at} the membrane theory does not apply because they do not examine a late time, large region limit.} For the analysis there it was sufficient to use the RT/HRT formulas~\cite{Hubeny:2007xt} for entanglement entropy of intervals in holographic theories because the triple information is constructed from such quantities.  To construct the entanglement signals studied in \cite{Balasubramanian:2024ysu} we need to additionally compute reflected entropies.  The time evolution of reflected entropy was first studied in 2d CFT in~\cite{Kudler-Flam:2020url}. More recently, as is detailed in Chapter~\ref{Chap3}, the membrane theory framework was enriched to enable computation of reflected entropy~\cite{Jiang:2024tdj}. We will use this method to study the dynamics of the multiparty signals considered in~\cite{Balasubramanian:2024ysu} in Chapter~\ref{Chap4}. 
 
\chapter{New horizons for inhomogeneous quenches}
\label{Floquet}
\section{Introduction}
In this chapter, we study the SSD CFT briefly reviewed in Chapter~\ref{FloquetIntro} from both holographic and field-theoretic point of view. 

In Section \ref{specquench} we analyse a special quantum quench from $H_0$ to the SSD Hamiltonian $H_1$. In Subsection \ref{SSD: Boundary Horizons} we point out that SSD Hamiltonian leads to a different $\emph{foliation}$ of the 2-dimensional spacetime as is shown in Figure \ref{fig:Triangle}. Furthermore, there is a $\emph{triangle}$-shaped Killing horizon \eqref{triangle} with $\emph{zero}$ temperature. 

Afterwards in Section \ref{Gravity dual of SSD}, we investigate holographic aspects of SSD CFT. In Subsection \ref{4.1}, we show that the gravitational dual of SSD CFT prepared in the ground and thermal states are dual to pure AdS$_3$ and BTZ black hole \cite{Banados:1992gq,Banados:1992wn} with non-trivial $\emph{foliations}$, respectively. Then Subsection \ref{New Horizon on AdS_3} and \ref{New Horizon on BTZ} demonstrate the existence of new $\emph{horizons}$ in both pure AdS$_3$ \eqref{AdSKilling} and BTZ \eqref{BTZNullGeod2}, which are plotted in Figure \ref{fig:AdSHor} and \ref{fig:BTZHor}, respectively. In Subsection \ref{New Horizon on AdS_3} we show that the AdS$_3$ new horizon, which is a Killing horizon, also has $\emph{zero}$ temperature. In Subsection \ref{New Horizon on BTZ} we also compare our new horizon in BTZ geometry with the BTZ black hole horizon and find that the new horizon is $\emph{larger}$ (see Figure \ref{fig:HorCompare}). 

In Subsection \ref{bulk reconstructions}, we comment on the reconstructions of bulk operators in the presence of the aforementioned new horizons and conclude that bulk operators remain $\emph{reconstructable}$ even after falling into these horizons (as long as they remain outside the BTZ black hole horizon).

In Section \ref{SSDq} we generalise the SSD quench results to quenches governed by other $SL(2)$ Virasoro subalgebras. 

Finally, Section \ref{Holographic Entanglement Entropy} is devoted to the study of holographic entanglement entropy \cite{Ryu:2006bv,Ryu:2006ef,Hubeny:2007xt} in the gravity dual of SSD CFT prepared both in the ground and thermal state. We find that due to the non-trivial foliations of the gravity dual of SSD CFT in the bulk, the holographic entanglement entropy is computed by spacelike geodesics anchoring on the boundary that is, in general, at non-equal times, see e.g.~Figure \ref{fig:EEGroundState} for an example. The time evolution of large-radius cutoff surfaces (plotted in Figure \ref{fig:CutoffSurface3d}) plays a crucial role. In cases where CFTs are initially prepared in thermal states, we discover that holographic entanglement entropy shows the entanglement plateau phenomenon \cite{Hubeny:2013gta}. We find precise agreements between our holographic results and field-theoretic calculations in all comparable cases and reproduce holographically the cooling and heating effects \cite{Wen:2022pyj,Goto:2021sqx} discovered earlier in the field theory context.  

\section{A special quantum quench}\label{specquench}
\subsection{Setup}\label{Set up}

We consider CFT$_2$ on a $\emph{cylinder}$ $\mathbb{R}\times S^1$ of circumference $l=2\pi$ with $\emph{periodic}$ boundary conditions,\footnote{The references \cite{Wen:2018vux,Wen:2018agb,Fan:2019upv,Wen:2020wee,Han:2020kwp,Fan:2020orx,Wen:2022pyj} sometimes adopt $\emph{open}$ boundary conditions. We will not use this boundary condition in our work but will point out its appearance when referring to their works.} in either the $\emph{ground}$ state $|0\rangle$ or the $\emph{thermal}$ state $\rho=e^{-\beta H}$ at temperature $\beta^{-1}$. We are interested in the $\emph{Lorentzian}$ time evolution of driven CFTs, although it is technically easier to work in $\emph{Euclidean}$ and then analytically continue to Lorentzian signature. The coordinate on the Euclidean cylinder is 
\begin{equation}
    \begin{aligned}
        w&=\tau+i x,\\
        \overline{w}&=\tau-ix.
    \end{aligned}
    \label{Eucledian_w}
\end{equation}
Here, we use $\tau$ and $t$ to denote imaginary and real time, respectively. The spatial variable $x$ is periodic with periodicity $2\pi$: $x\sim x+2\pi$. The analytic continuation is given by $\tau\to it$, which makes $w$ and $\overline{w}$ (real) light cone coordinates.

We will mostly work in the $\emph{Heisenberg\ picture}$: we take a primary operator $O$ of weights $(h, \overline{h})$ on the $t=0$ time slice with $w=i x$, $\overline{w}=-ix$ and time evolve it using the operator $e^{-\tau H}$ with $H$ made out of the Virasoro generators. Such a time evolution is $\emph{geometric}$, the operator moves $O$ to $\emph{new}$ positions $(w_{{\rm new}}(\tau,x),\overline{w}_{{\rm new}}(\tau,x))$,\footnote{Here $w=i x$, $\overline{w}=-ix$. In later sections we will use the notations $\frac{\partial w_{{\rm new}}}{\partial w}$ and $\frac{\partial w_{{\rm new}}}{\partial (ix)}$, $\frac{\partial \overline{w}_{{\rm new}}}{\partial \overline{w}}$ and $\frac{\partial \overline{w}_{{\rm new}}}{\partial (-ix)}$ interchangeably. }
\begin{align}
    O^{(H)}(\tau,x)\equiv e^{\tau H }O(ix,-ix)e^{-\tau H }=\Big(\frac{\partial w_{{\rm new}}}{\partial w}\Big)^h\Big(\frac{\partial\overline{w}_{{\rm new}}}{\partial\overline{w}}\Big)^{\overline{h}}O(w_{{\rm new}},\overline{w}_{{\rm new}})\,, \label{Heisenberg}
\end{align}
whose explicit expression depends on the Hamiltonian $H$ that drives the system. This is a powerful result that applies irrespective of the state of the system or the boundary conditions (provided they preserve $H$).

 For example, if $H$ is the CFT Hamiltonian 
\begin{align}
    H_0=\int_0^{2\pi} dx\ T_{00}(x)=L_0+\overline{L}_0-\frac{ c}{12}\label{H0},
\end{align}
with the last term being the Casimir energy, then after (imaginary) time $\tau$, $(w_{{\rm new}},\overline{w}_{{\rm new}})$ are simply $(w,\overline{w})$ in \eqref{Eucledian_w}.

\subsection{The sine-squared deformed time evolution} \label{SSDsetup}
The fundamental ingredient of the deformation we would like to study is the sine-squared deformation (SSD) \cite{Wen:2018vux}, which deforms the CFT Hamiltonian \eqref{H0} with the enveloping function
\begin{align}
    v_1(x)=2 \sin^2\Big(\frac{x}{2}\Big)\label{v(x)}.
\end{align}
The resulting deformed Hamiltonian is 
\es{1}{
    H_1&=2\int_0^{2\pi} dx\ \sin ^2\Big(\frac{x}{2}\Big)\,T_{00}(x)\\
    &=L_0-\frac{L_{-1}+L_1}{2}+\overline{L}_0-\frac{\overline{L}_{-1}+\overline{L}_1}{2}-\frac{c}{12}.
}
Note that the SSD enveloping function $v_1(x)$ is $\emph{inhomogeneous}$ in space, in particular, it vanishes at $x=0$ (or $2\pi$). From \eqref{1} one observes $[H_0,H_1]\neq 0$, because $[L_0, L_{\pm 1}]=\mp L_{\pm 1}$; however, $H_0$ and $H_1$ share the same ground state $|0\rangle$, as $L_n|0 \rangle =0$  for $n\ge-1$. 

The use of $H_1$ as a Hamiltonian is unconventional. However, $H_1$ can be conformally mapped to a dilation \cite{Okunishi:2016zat,Wen:2018vux,Fan:2019upv}; mapping back to the cylinder, we obtain $(w_{{\rm new}},\overline{w}_{{\rm new}})$ given by the following logarithms of Mobius transformations \cite{Wen:2018vux, Goto:2021sqx} (see Appendix \ref{SSDreview} for details)
\begin{align}
    w_{{\rm new}}&=\log \Big[\frac{\big(1+\frac{i t}{2}\big)e^{ i x}-\frac{i t}{2}}{\frac{i t}{2}e^{i x}+\big(1-\frac{i t}{2}\big)}\Big]\,,\label{wnew}\\
    \overline{w}_{{\rm new}}&=\log \Big[\frac{\big(1+\frac{i t}{2}\big)e^{-i x}-\frac{i t}{2}}{\frac{i t}{2}e^{-i x}+\big(1-\frac{i t}{2}\big)}\Big]\,,\label{wnewbar}
\end{align}
where we have analytically continued to real-time $\tau\to it$. The logarithm is a multi-valued function; we fix the appropriate sheet by demanding continuity in $t$. The terms inside the logarithm are $SL(2)$ transformations of the variables $z=e^{i x}$ and $\overline{z}=e^{-i x}$ on the unit circle.\footnote{$z$ and $\overline{z}$ are coordinates on the $\emph{complex\ plane}$, related to the cylinder via the exponential maps $z=e^{w}$ and $\overline{z}=e^{\overline{w}}$.} The $SL(2)$ transformation has a $\emph{fixed\ point}$, i.e.~a point that stays invariant under the $SL(2)$ transformation at $x=0$ (identified with $x=2\pi$). In analogy with \eqref{Eucledian_w}, we can define \cite{Goto:2021sqx}
\begin{equation}
\label{tnewxnew}
    \begin{aligned}
        w_{{\rm new}}&= i(t_{{\rm new}}+x_{{\rm new}})\,,\\
        \overline{w}_{{\rm new}}&= i(t_{{\rm new}}-x_{{\rm new}})\,.
    \end{aligned}
\end{equation}
$t_{{\rm new}}$ and $x_{{\rm new}}$ are $\emph{real}$  functions of $t$ and $x$.\footnote{As $SL(2)$ transformations preserve unit circles, $w_{{\rm new}}$ and $\overline{w}_{{\rm new}}$ are completely determined by the argument of the logarithm function and are, therefore, purely imaginary. } We will see that the $SL(2)$ transformation \eqref{wnew} and \eqref{wnewbar} determines (almost) all the physical phenomena of interest: in the $(w_{{\rm new}},\overline{w}_{{\rm new}})$ coordinate, the CFT is in equilibrium, and the time-dependent conformal transformations \eqref{wnew} and \eqref{wnewbar} allow us to deduce non-equilibrium phenomena from known equilibrium physics \cite{Wen:2018vux,moosavi2021inhomogeneous}. 

There is an alternative way of thinking about the SSD CFT: the driving by the Hamiltonian \eqref{1} is equivalent to the natural time evolution of the CFT on the curved spacetime \cite{Wen:2018vux,MacCormack:2018rwq,deBoer:2023lrd}:
\es{SSDMetric3}{
ds^2=-v_1(x)^2dt^2+dx^2\,.
}
For the review of the argument, see Appendix \ref{SSDcurvedreview}. Thus we have three equivalent ways of thinking about correlation functions
\es{3desc}{
\le\langle O^{(H_1)}(\tau,x) \dots \ri\rangle&=\Big(\frac{\partial w_{{\rm new}}}{\partial (i x)}\Big)^h\Big(\frac{\partial\overline{w}_{{\rm new}}}{\partial(-ix))}\Big)^{\overline{h}}\dots \le\langle O(w_{{\rm new}},\overline{w}_{{\rm new}})\dots\ri\rangle_{ds^2=dw_{{\rm new}} d\overline{w}_{{\rm new}}}\\[4pt]
&=\le\langle O(\tau,x) \dots \ri\rangle_{ds'^2=v_1(x)^2d\tau^2+dx^2}\,.
}

Quantum quenches with SSD CFT at $t>0$ have been studied for systems prepared in ground \cite{Wen:2018vux}  and thermal \cite{Goto:2021sqx} states.\footnote{Note that in \cite{Wen:2018vux}, $\emph{open}$ boundary conditions has been imposed to set up a quench problem, which makes the ground state of $H_0$ into an excited state of $H_1$. 
 In this work, we  study the non-trivial time evolution of $H_1$ with periodic boundary conditions. Although this setup cannot strictly be regarded as a quench if the initial state is the ground state, the expression for $w_{{\rm new}}$ \eqref{wnew} and $\overline{w}_{{\rm new}}$ \eqref{wnewbar} remain the same.} Some of the most interesting phenomena that have been found are as follows: In \cite{Goto:2021sqx} it was found that the CFT possesses two energy peaks   that move towards the fixed point $x=0$; the entanglement entropy decreases and approaches the $\emph{vacuum}$ entanglement entropy if the subregion $A$ does not include the fixed point $x=0$ (identified with $2\pi$). In the following, we will provide a geometric perspective on these, predominantly in the context of holography.

\subsection{Boundary triangle} \label{SSD: Boundary Horizons}
\begin{figure}[htbp]
\centering
\includegraphics[width=.68\textwidth]{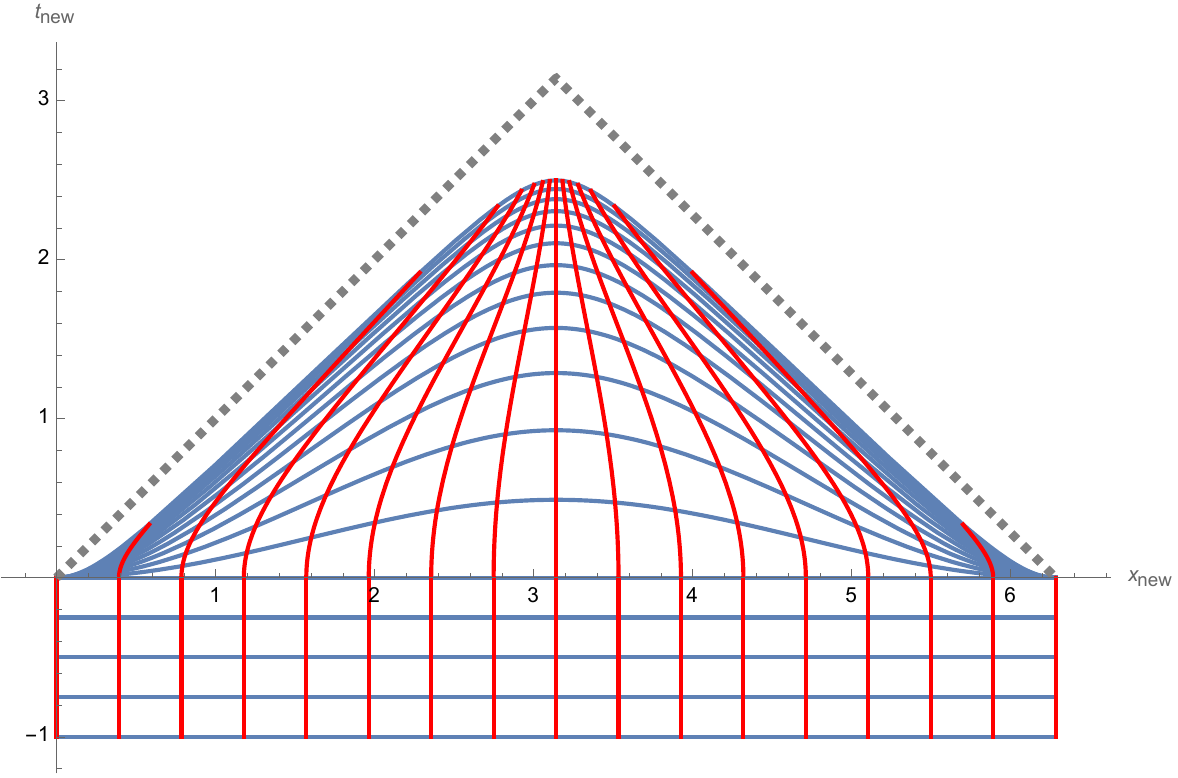}
\qquad
\caption{Constant $t$ lines (blue) and constant-$x$ lines (red) in the $(t_{{\rm new}}, x_{{\rm new}})$-plane. At $t<0$ when the driving Hamiltonian is $H_0$, constant-$t$ lines and constant-$x$ lines are uniformly distributed horizontal and vertical lines, respectively. At $t>0$ when the driving Hamiltonian is suddenly switched to $H_1$, constant-$t$ lines asymptote a triangle-shaped curve \eqref{triangle}, which is plotted as a dashed gray line; constant-$x$ lines converge to the top tip of the triangle, $(\pi,\pi)$. \label{fig:Triangle}} 
\end{figure}

We have seen in Section \ref{SSDsetup} that the non-trivial time evolution of SSD CFTs is completely determined by the coordinate transformations \eqref{wnew} and \eqref{wnewbar}. The spacetime coordinates $t_{{\rm new}}(t,x)$ and $x_{{\rm new}}(t,x)$ then follows from \eqref{tnewxnew}. In Figure \ref{fig:Triangle}, we plot the constant-$t$ lines and constant-$x$ lines on the $(t_{{\rm new}}, x_{{\rm new}})$ plane.\footnote{It is more convenient to plot with the trigonometric form of $t_{{\rm new}}$ \eqref{tnewtrig} and $x_{{\rm new}}$ \eqref{xnewtrig}.}

Figure \ref{fig:Triangle} shows that the action of $H_1$ as Hamiltonian changes the $\emph{foliation}$ of the $(t_{{\rm new}},x_{{\rm new}})$ spacetime in a non-uniform manner.\footnote{This reflects the $\emph{inhomogeneous}$ nature of $H_1$; in particular, the fixed point $x=0$ (which maps to $(t_{{\rm new}},x_{{\rm new}})=(0,0)$) does not time evolve at all. } As time evolves, constant-$t$ lines approach a $\emph{triangle}$-shaped curve 
\begin{equation}
\label{triangle}
    t_{{\rm new}}=
    \left\{
    \begin{aligned}
        &x_{{\rm new}},\ \ 0\leq x_{{\rm new}} < \pi\\
        &2\pi-x_{{\rm new}},\ \ \pi\leq x_{{\rm new}} < 2\pi
    \end{aligned}
    \right.
\end{equation}
instead of moving towards $t_{{\rm new}}\to\infty$ homogeneously as is in the $H_0$ evolution case; constant-$x$ lines converge to the top tip of the triangle, $(\pi,\pi)$. Indeed, from the expression of $t_{{\rm new}}$ \eqref{tnewtrig} and $x_{{\rm new}}$ \eqref{xnewtrig}, one can verify that
\begin{align}
    \lim_{t\to\infty}t_{{\rm new}} = \pi && \lim_{t\to\infty}x_{{\rm new}} = \pi\label{tnewxnewLimit}
\end{align}
regardless of $x$ ($x\neq0,2\pi$). As the triangle \eqref{triangle} separates points that are connected to infinity (outside the triangle) from those that are not (inside the triangle), it can be regarded as a new $\emph{horizon}$ in the $(t_{{\rm new}},x_{{\rm new}})$ plane. Note that the new horizon is ultimately the consequence of the fixed points $x=0,2\pi$ on the circle. If there were no such fixed points, the entire constant $t$ line in $(t_{{\rm new}},x_{{\rm new}})$ would evolve as $t$ grows, and there would be no new horizons.

The SSD Hamiltonian  \eqref{1} evolves the CFT along a conformal Killing vector field, hence the horizon \eqref{triangle} is a $\emph{Killing}$ horizon associated with this vector field. From the expression~\eqref{1}, the conformal Killing vector $k_1$ corresponds to $H_1$ can be written as 
\begin{align}
    k_1&=-z_{{\rm new}}\partial_{z_{{\rm new}}}-\overline{z}_{{\rm new}}\partial_{\overline{z}_{{\rm new}}}+\frac{z_{{\rm new}}^2+1}{2}\partial_{z_{{\rm new}}}+\frac{\overline{z}_{{\rm new}}^2+1}{2}\partial_{\overline{z}_{{\rm new}}}\\
    &= (\cosh  \tau_{{\rm new}}\ \cos x_{{\rm new}}-1)\partial_{\tau_{{\rm new}}}+\sinh \tau_{{\rm new}}\ \sin x_{{\rm new}}\partial_{x_{{\rm new}}}\label{ConfKillingVec},
\end{align}
where $z_{{\rm new}}=e^{\tau_{{\rm new}}+ix_{{\rm new}}}$ and $\overline{z}_{{\rm new}}=e^{\tau_{{\rm new}}-ix_{{\rm new}}}$.\footnote{The conformal Killing vector~\eqref{ConfKillingVec} is a time translation along the $\tau$ direction in the $(\tau_{{\rm new}},x_{{\rm new}})$ coordinate, as can be seen from~\eqref{Heisenberg}. We will only  rarely use $\tau_{{\rm new}}$ as a time coordinate.} One can easily check that $k_1$ obeys the conformal Killing equation
\begin{align}
    \partial_{(\mu}(k_1)_{{\nu})}=\delta_{\mu\nu}\partial_{\rho}(k_1)^{\rho}\,.
\end{align} 
The analytic continuation of the conformal Killing vector \eqref{ConfKillingVec} to Lorentizan signature is
\begin{align}
    k_1=(1-\cos t_{{\rm new}}\ \cos x_{{\rm new}})\partial_{t_{{\rm new}}}+\sin t_{{\rm new}}\ \sin x_{{\rm new}}\partial_{x_{{\rm new}}}\,,\label{ConfKillingVecLor}
\end{align}
where we have removed the overall $i$ so that the vector field implements Lorentzian time evolution. The conformal Killing horizon is located where the conformal Killing vector becomes null, so we compute its norm squared:
\begin{align}
   |k_1|^2=(\cos t_{{\rm new}}-\cos x_{{\rm new}})^2=0\label{3.10} && \quad \implies \quad  t_{{\rm new}}=\pm x_{{\rm new}}+2n\pi\ (n\in\mathbb{Z}). 
\end{align}
As $t_{{\rm new}}\in[0, \pi]$, $t_{{\rm new}}=x_{{\rm new}}$ or $t_{{\rm new}}=-x_{{\rm new}}+2\pi$, which is exactly the expression of the triangle \eqref{triangle}. 
Since \eqref{3.10} has a double zero,
this is an $\emph{extremal}$ conformal Killing horizon, hence it has zero temperature.\footnote{A simple calculation on \eqref{ConfKillingVec} (after analytic continuation) confirms this reasoning, 
\begin{align}
    T=\frac{1}{2\pi}\sqrt{\nabla_{\mu} |k_1|\nabla^{\mu} |k_1|}\Big|_{{\rm hor.}}=\sqrt{-\sin ^2t_{{\rm new}}+\sin ^2x_{{\rm new}}}=0\label{TemperatureBdy}
\end{align}
where $|k_1|$ is given by \eqref{3.10}. } The physical reason for this is that $H_0$ and $H_1$ share the \emph{same} ground state.\footnote{This is in contrast to the Unruh effect, where the Minkowski vacuum is an excited state of and Rindler observer and vice versa.} 

We close this section by comparing the $H_1$ time evolution to those under CFT modular Hamiltonians in the literature, see the comprehensive recent references~\cite{Cardy:2016fqc,deBoer:2021zlm,Das:2024vqe}. The operators $K_\text{full}=s_0L_0+s_1L_1+s_{-1}L_{-1}$ are full/complete modular Hamiltonians associated to an interval $A$ for the vacuum of a 2d CFT on a circle, $K_\text{full}=K_A-K_{\bar A}$. Just like $H_1$, $K_\text{full}$ also annihilates the vacuum.  If we consider the reduced density matrix of the interval, it is a thermal state under modular evolution $\rho\propto e^{-K_A}$. In our study, we do not consider subregions and focus on $H_1$ defined on the entire $S^1$.

\section{Gravity dual for the special quantum quench}\label{Gravity dual of SSD}

\subsection{Bulk geometry}\label{4.1}

The gravity dual of a generic (global) quench involves the changing of boundary conditions for the gravity fields, corresponding to uniform injection of energy into the bulk from the near boundary region. In the simplest scenario, the spacetime reaches equilibrium by settling to a black hole, the dual of thermalisation in the boundary QFT.  

In the special quench we analyse, the Heisenberg evolution on the cylinder can be described by the $t$-depending coordinate transformation $(t_{{\rm new}},x_{{\rm new}})$ \eqref{tnewxnew} expressing non-equilibrium physics from equilibrium results. Correspondingly, the matter stress tensor in the bulk remains zero and the spacetime does not change. What changes is the cutoff surface that the field theory lives on. This among other things leads to the change in energy in the boundary theory.

To construct the holographic dual of SSD CFT, we introduce the emergent bulk radial coordinate $r_{{\rm new}}\in[0,\infty)$. Holographic CFTs in their ground and thermal states are dual to pure AdS$_3$ and BTZ black holes \cite{Banados:1992gq,Banados:1992wn}, respectively. As the CFT in $(t_{{\rm new}},x_{{\rm new}})$ are in equilibrium, the dual bulk metrics in the $(t_{{\rm new}},r_{{\rm new}},x_{{\rm new}})$ coordinate are given by those of the pure AdS$_3$ and BTZ black hole metrics, 
\begin{align}
    &|0\rangle:  && ds^2=-(r_{{\rm new}}^2+1)dt_{{\rm new}}^2+\frac{dr_{{\rm new}}^2}{r_{{\rm new}}^2+1}+r_{{\rm new}}^2dx_{{\rm new}}^2\label{AdSMetric},\\
    &\rho=e^{-\beta H_0}: && ds^2=-(r_{{\rm new}}^2-r_+^2)dt_{{\rm new}}^2+\frac{dr_{{\rm new}}^2}{r_{{\rm new}}^2-r_+^2}+r_{{\rm new}}^2dx_{{\rm new}}^2\label{BTZMetric},
\end{align}
where $r_+$ is the radius of the BTZ black hole, related to the inverse temperature $\beta$ via $r_+=\frac{2\pi}{\beta}$. 

\begin{figure}[htbp]
\centering
\includegraphics[width=.48\textwidth]{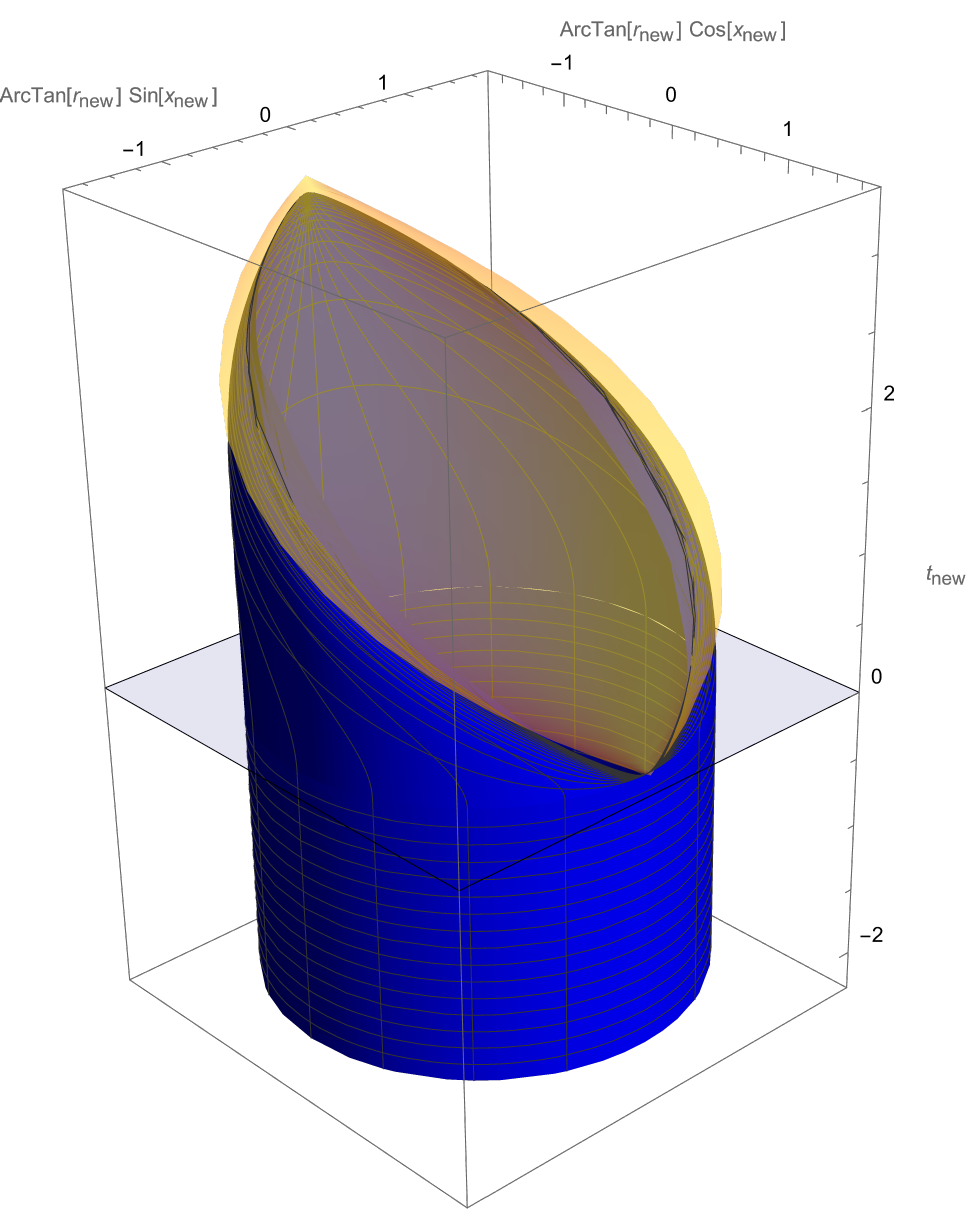}
\includegraphics[width=.37\textwidth]{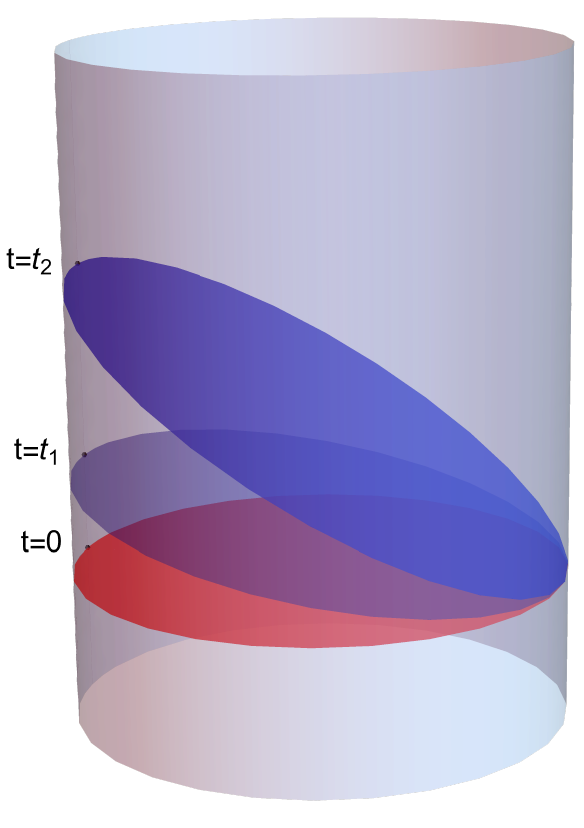}
\qquad
\caption{Left: Plot of the cutoff surface in $(t_{{\rm new}},r_{{\rm new}},x_{{\rm new}})$ spacetime. The light blue surface denotes the $t_{{\rm new}}=t=0$ slice, during which we suddenly switch the Hamiltonian from $H_0$ to $H_1$. At $t<0$ when the driving Hamiltonian is $H_0$, the uniform cutoff $R_\text{new}=\epsilon^{-1}$ is shown in blue. At $t>0$ the driving Hamiltonian becomes $H_1$, and the cutoff surface $R_{{\rm new}}(t_\text{new},x_\text{new})$ is no longer uniform according to \eqref{cutoffpresc}. This significant change leads to the presence of a new horizon, which is depicted above in light yellow. This new horizons is also plotted in Figure \ref{fig:AdSHor} along with its generators. Note that the cutoff surface ends on the horizon (up to finite cutoff corrections). Right: Cartoon of constant $t$ slices in $(t_{{\rm new}},r_{{\rm new}},x_{{\rm new}})$ spacetime in global AdS$_3$. \label{fig:CutoffSurface3d}}
\end{figure}

Besides the movement of the insertion points of operators, there are explicit factors showing up in formulas \eqref{Heisenberg} and \eqref{3desc} multiplying equilibrium correlators. These correspond to taking the holographic cutoff at 
\es{cutoffpresc}{
\epsilon={1\over \sqrt{\frac{\partial w_{{\rm new}}}{\partial w}\frac{\partial\overline{w}_{{\rm new}}}{\partial\overline{w}}}\, R_{{\rm new}}(t_\text{new},x_\text{new})}\,,
}
resulting in a $\emph{nonuniform}$ cutoff surface in the radial coordinate $R_{{\rm new}}(t_\text{new},x_\text{new})$. Note that the formula is applicable for $t>0$, for $t\leq 0$ we have $R_{{\rm new}}=1/\epsilon$. A plot of the cutoff surface \eqref{cutoffpresc} can be found in Figure \ref{fig:CutoffSurface3d}. This formula has appeared in the literature before in~\cite{Goto:2021sqx} based on different arguments; in Appendix~\ref{app:WigglyCutoff} we give a detailed new argument that this  nonuniform cutoff produces the appropriate $\big(\frac{\partial w_{{\rm new}}}{\partial w}\big)^h$ factors in correlators.

We have seen that in the boundary theory a horizon appears, this in the holographic dual leads to a new horizon in the bulk. As constant $t$-slices in the bulk are anchored on the constant $t$-lines on the boundary, this induces a change in the $\emph{foliation}$ of the bulk, see Figure \ref{fig:CutoffSurface3d}. (There is a natural extension of $H_1$ into the bulk in the pure AdS case, but not in BTZ.) As $t\to\infty$, constant-$t$ slices asymptote the bulk null hypersurface anchored on the boundary triangle \eqref{triangle} instead of moving towards $t_{{\rm new}}\to\infty$. As this hypersurface separates bulk points that are connected to the boundary spacetime explored by the CFT from those that are not, it is a new $\emph{horizon}$ in the bulk. 

An alternative approach is provided by the CFT in curved spacetime perspective: we set the boundary metric to agree with \eqref{SSDMetric3}, and use Fefferman-Graham coordinates \cite{Skenderis:1999nb,deHaro:2000vlm} to build a coordinate system adapted to the boundary spacetime. The uniform cutoff surface in the Fefferman-Graham radial coordinate will coincide with the cutoff surface found from \eqref{cutoffpresc}. In 3d pure gravity, the Fefferman-Graham expansion is known to terminate. By solving the Einstein equations order by order in $r$ we obtain the metric~\cite{Skenderis:1999nb}:
\es{FGconcrete}{
ds^2={dr^2\ov r^2}+r^2\big(-v_1(x)^2dt^2+dx^2\big)+g^{(1)}_{ab}(x^c) \,dx^a dx^b+{g^{(2)}_{ab}(x^c) \,dx^a dx^b\ov r^2}\,,
}
where $g^{(j)}_{ab}(x^c)$ is given in Appendix~\ref{app:FGBanados} for the most general solution. There we also give the explicit form of the metric for the case of the vacuum and thermal states. Of course, \eqref{FGconcrete} is just the reslicing of the AdS$_3$ \eqref{AdSMetric} and BTZ \eqref{BTZMetric} spacetimes. (The  reslicing of AdS$_3$ was written down in \cite{MacCormack:2018rwq} before studying the vacuum physics of the SSD CFT.) One advantage of these coordinates is that now the cutoff is at $R=1/\epsilon$ instead of the nonuniform cutoff in $r_{{\rm new}}$ coordinates. 

Note that as discussed in Appendix~\ref{SSDcurvedreview}, the curved spacetime \eqref{SSDMetric3} is obtained by a Weyl transformation from the flat spacetime $ds^2=dw_{{\rm new}} d\overline{w}_{{\rm new}}$. It has been understood from the early days of AdS/CFT~\cite{Witten:1998qj,Henningson:1998gx} that changing the conformal factor of the boundary metric corresponds to choosing a different cutoff prescription. 

\subsection{New horizon in AdS$_3$}\label{New Horizon on AdS_3}

\begin{figure}[htbp]
\centering
\includegraphics[width=.55\textwidth]{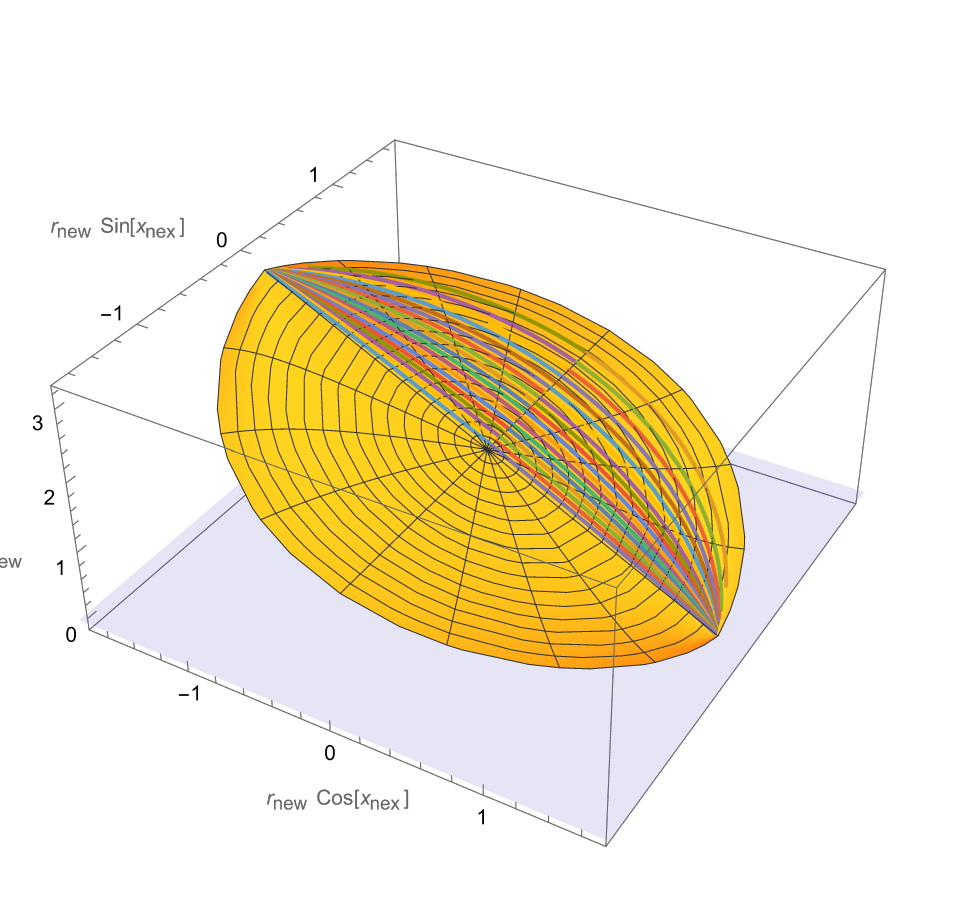}
\qquad
\caption{A plot of the new Killing horizon \eqref{AdSKilling2} in global AdS$_3$. The coloured lines are null geodesics emanating from the top tip of the triangle $(\pi, \pi)$ that generate the hypersurface \eqref{AdSKilling2}; for visualization, we only plotted them on half of the horizon. The light blue surface is the quench surface $t_{{\rm new}}=t=0$. \label{fig:AdSHor}}.
\end{figure}

The isometries of AdS$_3$ form two copies of $SL(2)$ algebras \cite{Brown:1986nw}. The Killing vectors $\zeta_{i}$ and $\overline{\zeta_{i}}$ ($i=-1,0,1$) of AdS$_3$ are given in Appendix \ref{AsymptoticSymmetries}. From these Killing vectors, we can find a linear combination that asymptotes to the conformal Killing vector $k_1^{\mu}$ \eqref{ConfKillingVec}. This provides the natural extension of the $H_1$ evolution into the bulk:
\es{BulkKillingVec}{
K_1&=\zeta_{0}+\overline{\zeta_{0}}-\frac{\zeta_{1}+\overline{\zeta_{1}}}{2}-\frac{\zeta_{-1}+\overline{\zeta_{-1}}}{2}\\
&=\Big(1-\frac{r_{{\rm new}}}{\sqrt{r_{{\rm new}}^2+1}}\cos t_{{\rm new}}\ \cos x_{{\rm new}}\Big)\partial_{t_{{\rm new}}}-\sqrt{r_{{\rm new}}^2+1}\ \sin t_{{\rm new}}\ \cos x_{{\rm new}}\partial_{r_{{\rm new}}}\\
&+\frac{\sqrt{r_{{\rm new}}^2+1}}{r_{{\rm new}}}\sin t_{{\rm new}}\ \sin x_{{\rm new}}\partial_{x_{{\rm new}}}.
}
In the near-boundary $r_{{\rm new}}\to\infty$ limit, the components of $K_1^{\mu}$ \eqref{BulkKillingVec} along the boundary indeed goes to the boundary conformal Killing vector $k_1^{\mu}$ \eqref{ConfKillingVec}, 
\begin{align}
    \lim_{r_{{\rm new}}\to\infty}K_1^{\mu}=(1-\cos t_{{\rm new}}\ \cos x_{{\rm new}}, \sin t_{{\rm new}}\ \sin x_{{\rm new}}\big)=k_1^{\mu}.
\end{align}
The Killing horizon is located where a Killing vector becomes null. From \eqref{BulkKillingVec}, we have for the norm square:
\begin{align}
  |K_1|^2&=\big(\sqrt{r^2+1}\cos t_{{\rm new}}-r\cos x_{{\rm new}}\big)^2=0\label{VNorm}\\
    &\Rightarrow \quad t_{{\rm new}}(r_{{\rm new}},x_{{\rm new}})=\arccos\Big[\frac{r}{\sqrt{r_{{\rm new}}^2+1}}\cos x_{{\rm new}}\Big]\,.\label{AdSKilling}
\end{align}
\eqref{AdSKilling} is the new horizon in pure AdS$_3$. Since $|K_1|^2$ has a double zero, this is an extremal Killing horizon with zero associated temperature.\footnote{A direct calculation shows that this is indeed the case: 
\begin{align}
    T=\frac{1}{2\pi}\sqrt{\nabla_{\mu} |K_1|\nabla^{\mu} |K_1|}\Big|_{{\rm hor.}}=\sqrt{r_{{\rm new}}^2+1}\cos t_{{\rm new}}-r\cos x_{{\rm new}}=0\label{temperature}.
\end{align}
where $|K_1|$ given by \eqref{VNorm}. } Just like on the CFT side, the intuition behind this result is that $H_0$ and $H_1$ share the same vacuum state. A plot of the Killing horizon \eqref{AdSKilling} along with its null generators can be found in Figure \ref{fig:AdSHor}. 

We can find the horizon \eqref{AdSKilling} in an alternative manner, which will be a useful method in the following.
To find the horizon as a null hypersurface, we need to investigate the null geodesics emanating from a point $(t_{{\rm new}},x_{{\rm new}})=(y,y)$ on the boundary triangle. The congruence of null geodesics that maximise $t_{{\rm new}}$ in the bulk then generates the null hypersurface. Here, for simplicity, we study only half of the triangle, i.e.~$y\in[0, \pi]$; the result of the other half follows from reflection. 

We will start with the metric \eqref{AdSMetric} in the $(t_{{\rm new}},r_{{\rm new}},x_{{\rm new}})$ coordinate. The null geodesics anchored on a point $(t_{{\rm new}},x_{{\rm new}})=(y,y)$ on the boundary triangle  (see Appendix \ref{subsubsec:PureAdS} for details) is 
\begin{align}
    t_{{\rm new}}(r_{{\rm new}},x_{{\rm new}},y)={\rm arctan}\frac{-r_{{\rm new}}\ \cot (x_{{\rm new}}-y)}{\sqrt{r_{{\rm new}}^2+\csc ^2(x_{{\rm new}}-y)}}+y-\frac{\pi}{2}\label{AdSNullGeod},
\end{align}
where $y\in[0, \pi]$. \eqref{AdSNullGeod} increases monotonically with $y$ (see \eqref{AdSMono}). Hence, the maximal value of $t$ is at $y=\pi$, which is the top tip of the boundary triangle. After some work detailed in Appendix \ref{subsubsec:PureAdS}, one arrives at 
\begin{align}
    t_{{\rm new}}(r_{{\rm new}},x_{{\rm new}})=\arccos\Big[\frac{r_{{\rm new}}}{\sqrt{r_{{\rm new}}^2+1}}\cos x_{{\rm new}}\Big]\label{AdSKilling2},
\end{align}
agreeing with \eqref{AdSKilling} obtained by the Killing vector method. 

\subsection{New horizon on BTZ}\label{New Horizon on BTZ}
The $SL(2)$ isometry of AdS$_3$ is broken to $\mathbb{R}\times U(1)$ in BTZ geometry \cite{Keski-Vakkuri:1998gmz}. Hence, the new horizon is no longer a Killing horizon, nor can we associate temperature with it. To find the explicit expression of the horizon, one needs to investigate the null geodesic equation using the second method of the previous section. The null geodesics anchored a point on the boundary
triangle $(t_{{\rm new}},x_{{\rm new}})=(y,y)$ (see Appendix \ref{subsubsec:BTZ} for details) is 
\begin{align}
    t_{{\rm new}}(r_{{\rm new}},x_{{\rm new}},y)=\frac{1}{2r_+}\log \frac{r_{{\rm new}}\sqrt{1+r_+^2\sinh ^2(x_{{\rm new}}-y)}-r_+\sqrt{1+r_{{\rm new}}^2\sinh ^2(x_{{\rm new}}-y)}}{r_{{\rm new}}\sqrt{1+r_+^2\sinh ^2(x_{{\rm new}}-y)}+r_+\sqrt{1+r_{{\rm new}}^2\sinh ^2(x_{{\rm new}}-y)}}+y \label{BTZNullGeod1},
\end{align}
where $y\in[0, \pi]$. \eqref{BTZNullGeod1} increases monotonically with $y$ (see \eqref{BTZMono}). Hence, the maximal value of $t$ is at $y=\pi$, which is the top tip of the boundary triangle. After some work detailed in Appendix \ref{subsubsec:BTZ}, one arrives at 
\begin{align}
    t_{{\rm new}}(r,x)=\frac{1}{2r_+}\log \frac{r_{{\rm new}}\sqrt{1+r_+^2\sinh ^2(x_{{\rm new}}-\pi)}-r_+\sqrt{1+r_{{\rm new}}^2\sinh ^2(x_{{\rm new}}-\pi)}}{r_{{\rm new}}\sqrt{1+r_+^2\sinh ^2(x_{{\rm new}}-\pi)}+r_+\sqrt{1+r_{{\rm new}}^2\sinh ^2(x_{{\rm new}}-\pi)}}+\pi\label{BTZNullGeod2}.
\end{align}
\begin{figure}[htbp]
\centering
\includegraphics[width=.55\textwidth]{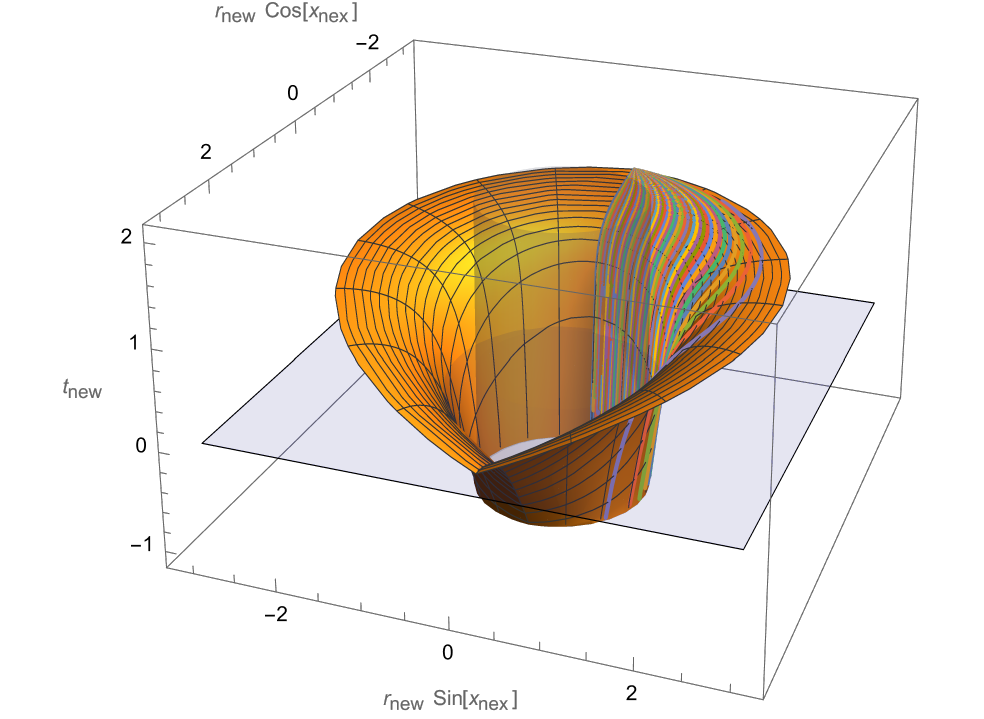}
\qquad
\caption{Plot of the new horizon \eqref{BTZNullGeod2} in the BTZ geometry. The coloured lines are null geodesics emanating from the top tip of the triangle $(\pi, \pi)$ that generate the hypersurface \eqref{BTZNullGeod2}; for visualization, we only plotted them on half of the horizon. The light blue surface is the quench surface $t_{{\rm new}}=t=0$. The BTZ black hole horizon is set to $r_+=1$ and plotted in light gray. The new horizon \eqref{BTZNullGeod2} is larger than the BTZ black hole horizon. \label{fig:BTZHor}}. 
\end{figure}
 A plot of the new horizon \eqref{BTZNullGeod2} along with its null generators can be found in Figure \ref{fig:BTZHor}. 

\begin{figure}[htbp]
\centering
\includegraphics[width=.45\textwidth]{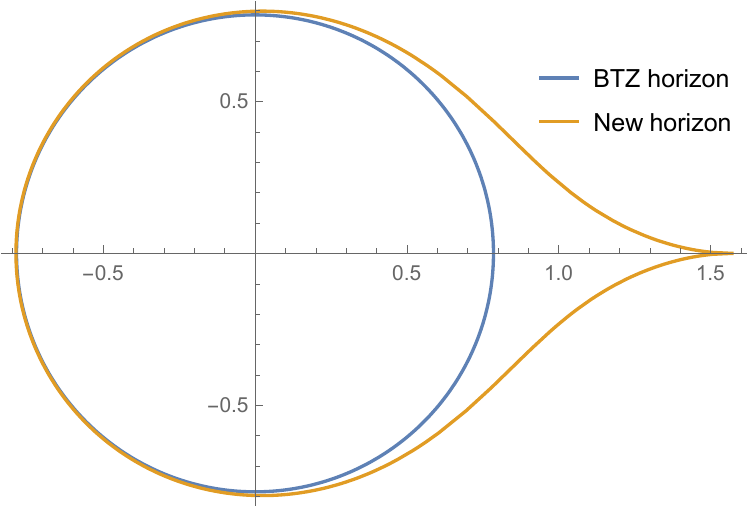}
\qquad
\includegraphics[width=.37\textwidth]{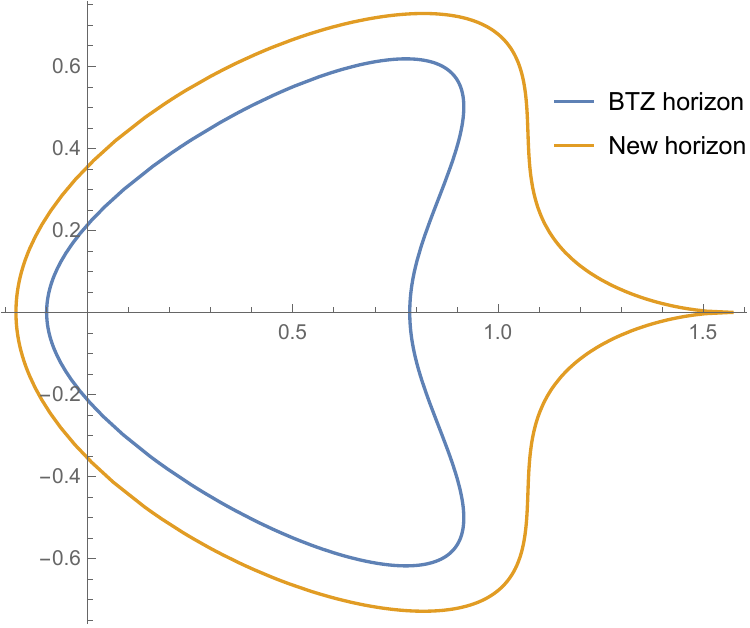}
\caption{Comparison of the new horizon \eqref{BTZNullGeod2} and the BTZ black hole horizon in the coordinates employed by 2112.14388, at $t=0$ (left) and $t=3$ (right), respectively. The blue line denotes the BTZ horizon, whereas the orange line 
 represents the new horizon \eqref{BTZNullGeod2}\label{fig:HorCompare}}
\end{figure}

For comparison with ref.~\cite{Goto:2021sqx},
we use the coordinate system they were working in to plot time slices of both the new horizon \eqref{BTZNullGeod2} and the BTZ horizon.\footnote{The Feffermann-Graham coordinates discussed around \eqref{FGconcrete} and Appendix~\ref{app:FGBanados} generically break down before encountering horizons, as was recently discussed in~\cite{Abajian:2023bqv}. Hence we cannot use them for addressing this question.} (See Appendix \ref{ComparisonRyu} for details.) From Figure~\ref{fig:BTZHor} it is clear that the new horizon will be slightly $\emph{larger}$ than the BTZ black hole horizon, and indeed this is what we find in Figure \ref{fig:HorCompare}. Note also that the new horizon \eqref{BTZNullGeod2} touches the boundary at the fixed point $x=0$.

The horizon \eqref{BTZNullGeod2} in Figure \ref{fig:HorCompare} develops two peaks that move towards the fixed point $x=0$, as is observed for the BTZ horizon in \cite{Goto:2021sqx}. It was also pointed out in \cite{Goto:2021sqx} that the horizon deformation mimics the energy peaks of the boundary stress tensors. We note that the two peaks of the horizons (both the BTZ horizon and the new horizon) originate from the Weyl factor $\frac{\partial w_{{\rm new}}}{\partial w}\frac{\partial\overline{w}_{{\rm new}}}{\partial\overline{w}}$ in \eqref{r'}; in \cite{Goto:2021sqx}, the peaks of the boundary stress tensors also originate from these prefactors $\frac{\partial w_{{\rm new}}}{\partial w}$ and $\frac{\partial\overline{w}_{{\rm new}}}{\partial\overline{w}}$ in $\langle T(w)\rangle$ and $\langle \overline{T}(\overline{w})\rangle$, respectively. This then explains the qualitative\footnote{As the metric \eqref{r'} is not in Fefferman-Graham gauge \cite{Skenderis:1999nb,deHaro:2000vlm}, it is hard to draw a quantitative connection.} correlations between horizon deformations and energy peaks were first observed in~\cite{Goto:2021sqx}. 

\section{Dynamics of entanglement entropy}\label{Holographic Entanglement Entropy}
\begin{figure}[htbp]
\centering
\includegraphics[width=.35\textwidth]{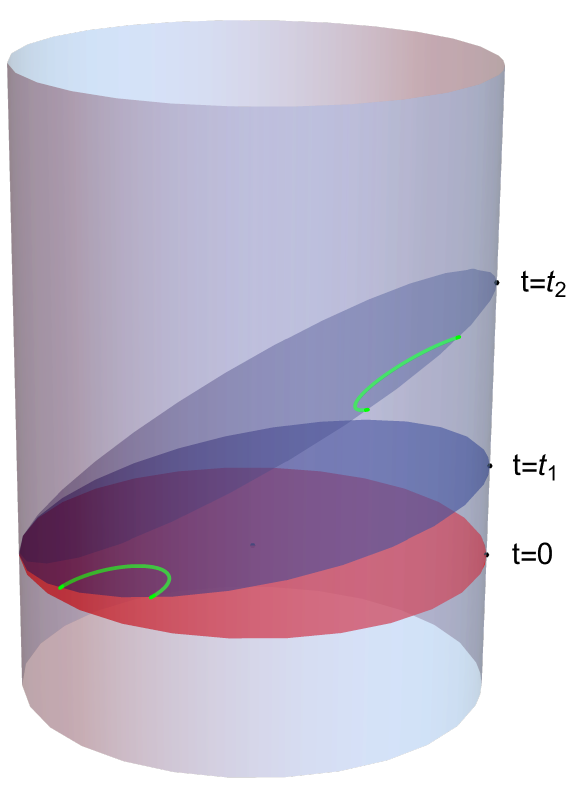}
\qquad
\caption{A demonstration of non-equal-$t_{{\rm new}}$ spacelike geodesics, plotted as green curves in the bulk anchoring on the two endpoints of the subregion $A$ on the boundary constant-$t$ lines.  These spacelike geodesics compute the holographic entanglement entropy. \label{fig:EEGroundState}}
\end{figure}

In this section, we study the holographic entanglement entropy of a single interval $A=[x_1, x_2]$ at a constant $t$-slice. To this end, we again work in the $(t_{{\rm new}},x_{{\rm new}},r_{{\rm new}})$ geometry, and then perform a coordinate transformation back to $(t,x,r)$. At time $t$, the endpoints of $A=[x_1, x_2]$ are obtained from \eqref{wnew}, \eqref{wnewbar}, \eqref{tnewxnew},
\begin{align}
    w_{{{\rm new}},j}=\frac{1}{2i}(t_{{{\rm new}},j}+x_{{{\rm new}},j}), && \overline{w}_{{{\rm new}},j}=\frac{1}{2i}(t_{{{\rm new}},j}-x_{{{\rm new}},j}), && (j=1,2). \label{7.1}
\end{align}
The holographic entanglement entropy of a single interval $A$ is given by the length of the spacelike geodesic anchored on $\partial A=\{(t_{{{\rm new}},1},x_{{{\rm new}},1}), (t_{{{\rm new}},2},x_{{{\rm new}},2})\}$ on the boundary \cite{Ryu:2006bv,Ryu:2006ef}: 
\begin{align}
    S_A=\frac{\mathcal{L}}{4G_N}=\frac{c}{6}\mathcal{L}.
\end{align}
where we have used the Brown-Henneaux central charge, $c=3/2G_N$ \cite{Brown:1986nw}. Therefore, to calculate the holographic entanglement entropy, one needs to obtain the geodesic length $\mathcal{L}$, using the temporal and angular separation $\Delta t_{{{\rm new}}}=t_{{{\rm new}},2}-t_{{{\rm new}},1}$ and $\Delta x_{{{\rm new}}}=x_{{{\rm new}},2}-x_{{{\rm new}},1}$ as inputs, and then transform back to $(t,x,r)$. Note that in general $\Delta t_{{{\rm new}}}\neq0$, so our calculation is $\emph{not}$ of equal time ("time" as $t_{{\rm new}}$). See Figure \ref{fig:EEGroundState} for a demonstration. 

The length of the spacelike geodesic anchoring on the boundary is infinite, therefore, we need to regulate it by imposing a $\emph{cutoff}$ at some large radius $r_{{\rm new}}=R_{{\rm new}}$, and adopt the $\emph{regulated}$ length $\mathcal{L}_{{\rm reg}}$. As is discussed in Subsection \ref{4.1}, the cutoff surface is non-uniform in the radial direction, with $R_{{\rm new}}(t_\text{new},x_\text{new})$ given by \eqref{cutoffpresc}. See Figure \ref{fig:CutoffSurface3d} for a plot of the cutoff surface.

\subsection{Gravity dual of SSD: ground state}
\subsubsection{Holographic computation}\label{sec7.1.1}
The metric \eqref{AdSMetric} leads to two conserved quantities for geodesics, $E=(r_{{\rm new}}^2+1) \frac{dt_{{\rm new}}}{d\lambda}$ and $L=r_{{\rm new}}^2 \frac{dx_{{\rm new}}}{d\lambda}$, where $\lambda$ is the affine parameter along the geodesic. In \cite{Hubeny:2012ry}, it was found that in pure AdS$_3$, the temporal and angular distances are given in terms of the two conserved quantities by 
\begin{align}
    \cos (\Delta t_{{\rm new}})&=\frac{L^2-E^2+1}{\sqrt{(L^2-E^2+1)^2+4E^2}}\label{delta_tAdS},\\
    \cos (\Delta x_{{\rm new}})&=\frac{L^2-E^2-1}{\sqrt{(L^2-E^2-1)^2+4L^2}}\label{delta_xAdS}.
\end{align}
From which one can solve for $E$ and $L$. Using these $E$ and $L$, we compute the regulated length of the spacelike geodesic \cite{Hubeny:2012ry} in Appendix \ref{AdSSpaceLikeGeod}
\begin{align}
    \mathcal{L}_{{\rm reg}}&=-\log \sqrt{(E^2-L^2+1)^2+4L^2}+\log (4R_{{\rm new,1}}R_{{\rm new,2}})\label{geodLenAdS}\\
    &=\log \le[4 \Big|\sinh 
    \big(\frac{1}{2}(w_{{{\rm new}},1}-w_{{{\rm new}},2})\big)\ \sinh \big(\frac{1}{2}(\overline{w}_{{{\rm new}},1}-\overline{w}_{{{\rm new}},2})\big)\Big|\ \ri]+\log (R_{{\rm new,1}}R_{{\rm new,2}})\,,\label{7.7}
\end{align}
where $\log (4R_{{\rm new,1}}R_{{\rm new,2}})$ is the universal diverging piece. From \eqref{cutoffpresc}, we have for the large-radius cutoff
\begin{align}
    R_{{\rm new},i}=\frac{R_i}{\sqrt{\frac{\partial w_{{\rm new},i}}{\partial w_i}\frac{\partial\overline{w}_{{\rm new},i}}{\partial\overline{w}_i}}}\label{regCutoff},
\end{align}
where $i=1,2$ labels the two large-radius cutoffs near the two endpoints of the geodesic. With \eqref{regCutoff}, the holographic entanglement entropy is:\footnote{Note that as $w_{{\rm new}}$ is purely imaginary, the terms inside the logarithm are in fact $\sin $ functions of $t_{{\rm new}}$ and $x_{{\rm new}}$.}
\begin{align}
    S_A=\frac{\mathcal{L}_{{\rm reg}}}{4G_N}=\frac{c}{6} \log \le[4\frac{\Big|\sinh 
    \big(\frac{1}{2}(w_{{{\rm new}},1}-w_{{{\rm new}},2})\big)\ \sinh \big(\frac{1}{2}(\overline{w}_{{{\rm new}},1}-\overline{w}_{{{\rm new}},2})\big)\Big|}{\prod_{i=1}^2\sqrt{\frac{\partial w_{{\rm new},i}}{\partial w_i}\frac{\partial\overline{w}_{{\rm new},i}}{\partial\overline{w}_i}}\epsilon_i}\ri]\label{SSDEEVacAdS},
\end{align}
where we have introduced the regularization parameter $\epsilon_i=R_i^{-1}$. One can easily verify that $S_A=S_{A^c}$, which is a property of pure states. An example of the spacelike geodesic computing the holographic entanglement entropy is in Figure \ref{fig:EEGroundState}.\footnote{One may wonder what general statements we can make about geodesics computing entropies in these coordinates.  Due to the Killing symmetry of AdS$_3$ discussed in Appendix~\ref{AsymptoticSymmetries}, a spacelike geodesic whose endpoints are on a constant-$t$ line on the boundary entirely lies on a constant-$t$ slice in the bulk. As the new horizon~\eqref{AdSKilling} is the $t=\infty$ slice and a Killing horizon, any spacelike geodesic is upper bounded by the new horizon. These statements do not continue to hold in the thermal state.} 

\subsubsection{Match with field theory result}\label{SSDGSEE}
In 1+1 dimensional CFT, the entanglement entropy of a single interval $A=[x_1, x_2]$ can be evaluated from the two-point function of twist operators \cite{Calabrese:2004eu},
\begin{align}
    S_A=-\lim_{n\to 1}\frac{\partial}{\partial n}{\rm Tr}\rho_A^n, && {\rm Tr}\rho_A^n&	\propto \langle \mathcal{T}_{n}(w_1,\overline{w}_1)\mathcal{T}_{-n}(w_2,\overline{w}_2)\rangle.\label{EEfund}
\end{align}
with dimension $h=\overline{h}=\frac{c}{24}(n-\frac{1}{n})$. The time evolution of the two-point function is captured by the $SL(2)$ transformation $(w,\overline{w})\to (w_{{\rm new}}, \overline{w}_{{\rm new}})$: 
\begin{align}
    \langle \mathcal{T}_{n}(w_1,\overline{w}_1)\mathcal{T}_{-n}(w_2,\overline{w}_2)\rangle=\prod_{i=1}^2\Big(\frac{\partial w_{{\rm new},i}}{\partial w}\Big)^h\Big(\frac{\partial\overline{w}_{{\rm new},i}}{\partial\overline{w}}\Big)^h\langle \mathcal{T}_{n}(w_{{{\rm new}},1},\overline{w}_{{{\rm new}},1})\mathcal{T}_{-n}(w_{{{\rm new}},2},\overline{w}_{{{\rm new}},2})\rangle\label{7.12}.
\end{align}
The two-point function $\langle \mathcal{T}_{n}(w_{1,{{\rm new}}},\overline{w}_{1,{{\rm new}}})\mathcal{T}_{-n}(w_{2,{{\rm new}}},\overline{w}_{2,{{\rm new}}})\rangle$ is evaluated following the methods in \cite{Calabrese:2004eu} but paying close attention to the difference between holomorphic and antiholomorphic parts (see Appendix \ref{Calc2pt} for details), 
\begin{align}
    &\langle \mathcal{T}_{n}(w_{{{\rm new}},1},\overline{w}_{{{\rm new}},1})\mathcal{T}_{-n}(w_{{{\rm new}},2},\overline{w}_{{{\rm new}},2})\rangle\nonumber\\
    =&\Big[4\sinh \Big(\frac{1}{2}(w_{{{\rm new}},1}-w_{{{\rm new}},2})\Big)\ \sinh \Big(\frac{1}{2}(\overline{w}_{{{\rm new}},1}-\overline{w}_{{{\rm new}},2})\Big)\Big]^{-2h}\,,\label{2ptFuncNew}
\end{align}
where
\begin{align}
    w_{{\rm new}}=\tau_{{\rm new}}+i x_{{\rm new}}, && \overline{w}_{{\rm new}}=\tau_{{\rm new}}-i x_{{\rm new}}\,.
\end{align}
Substituting \eqref{2ptFuncNew} into \eqref{7.12} and then into \eqref{EEfund} and performing the analytic continuation $\tau_{{\rm new}}\to it_{{\rm new}}$, the CFT entanglement entropy is then
\begin{align}
    S_A&=\frac{c}{6} \log \le[\frac{4 \Big|\sinh \big(\frac{1}{2}(w_{{{\rm new}},1}-w_{{{\rm new}},2})\big)\ \sinh \big(\frac{1}{2}(\overline{w}_{{{\rm new}},1}-\overline{w}_{{{\rm new}},2})\big)\Big|}{\prod_{i=1}^2\sqrt{\frac{\partial w_{{\rm new},i}}{\partial w_i}\frac{\partial\overline{w}_{{\rm new},i}}{\partial\overline{w}_i}}\epsilon_i}\ri]\label{SSDEEVacCFT},
\end{align}
where we have introduced the UV cutoff $\epsilon_i$. The terms multiplying $\epsilon_i$ in the denominator in \eqref{SSDEEVacCFT} come from the Jacobian factors in \eqref{7.12}. The field theory calculation \eqref{SSDEEVacCFT} precisely matches our earlier holographic result \eqref{SSDEEVacAdS}.\footnote{Note, however, that the field theory result is valid for $\emph{generic}$ CFTs, while our holographic result is limited to $\emph{holographic}$, i.e.~large $c$, large gap CFTs only.}

\subsection{Gravity dual of SSD: thermal state}
\subsubsection{Holographic computations}\label{sec7.2.1}
The metric \eqref{BTZMetric} has two conserved quantities, $E=(r_{{\rm new}}^2-r_+^2) \frac{dt_{{\rm new}}}{d\lambda}$ and $L=r_{{\rm new}}^2 \frac{dx_{{\rm new}}}{d\lambda}$, where $\lambda$ is the affine parameter along the geodesic. We will focus on the region outside the BTZ horizon, $r>r_+$. In Appendix \ref{BTZSpaceLikeGeod}, we calculated the temporal and angular separation of the two endpoints of the spacelike geodesic
\begin{align}
    \cosh  (r_+\Delta t_{{\rm new}})&=\frac{L^2-E^2-r_+^2}{\sqrt{(L^2-E^2-r_+^2)^2-4r_+^2E^2}}\label{delta_tBTZ},\\
    \cosh  (r_+\Delta x_{{\rm new}})&=\frac{L^2-E^2+r_+^2}{\sqrt{(L^2-E^2+r_+^2)^2-4r_+^2L^2}}\label{delta_xBTZ},
\end{align}
where $L\ge E+r_+$ so that the spacelike geodesic does not enter the BTZ horizon (see Appendix \ref{BTZSpaceLikeGeod} for details).\footnote{Note that if one sets the BTZ black hole radius $r_+^2=-1$ and hence replaces $\cosh$ with $\cos$  in \eqref{delta_tBTZ} and \eqref{delta_xBTZ}, one recovers \eqref{delta_tAdS} and \eqref{delta_xAdS} in pure AdS$_3$, respectively.} Therefore one can solve for $E$ and $L$. Using these $E$ and $L$, we computed the regulated length of the spacelike geodesic in Appendix \ref{BTZSpaceLikeGeod}.
\begin{align}
    \mathcal{L}_{{\rm reg}}&=-\log \sqrt{(L^2-E^2+r_+^2)^2-4r_+^2L^2}+\log (4R_{{\rm new,1}}R_{{\rm new,2}})\label{geodLenBTZ}\\
    &=\log \le[\Big(\frac{2}{r_+}\Big)^2\Big|\sin 
    \Big(r_+\frac{w_{{{\rm new}},1}-w_{{{\rm new}},2}}{2}\Big)\ \sin \Big(r_+\frac{\overline{w}_{{{\rm new}},1}-\overline{w}_{{{\rm new}},2}}{2}\Big)\Big|\ \ri]+\log (R_{{\rm new,1}}R_{{\rm new,2}})\label{7.18}.
\end{align}
For the BTZ black hole, the radius and inverse temperature are related via $r_+={2\pi/\beta}$. The large-radius cutoff is the same as \eqref{regCutoff}. The holographic entanglement entropy is:\footnote{Note that as $w_{{\rm new}}$ is purely imaginary, the terms inside the logarithm are in fact $\sinh $ functions of $t_{{\rm new}}$ and $x_{{\rm new}}$.}
\begin{align}
    S_A^{(1)}=\frac{\mathcal{L}_{{\rm reg}}}{4G_N}=\frac{c}{6} \log \le[\frac{\big(\frac{\beta}{\pi}\big)^2 \Big|\sin \big(\frac{\pi}{\beta}(w_{{{\rm new}},1}-w_{{{\rm new}},2})\big)\ \sin \big(\frac{\pi}{\beta}(\overline{w}_{{{\rm new}},1}-\overline{w}_{{{\rm new}},2})\big)\Big|}{\prod_{i=1}^2\sqrt{\frac{\partial w_{{\rm new},i}}{\partial w_i}\frac{\partial\overline{w}_{{\rm new},i}}{\partial\overline{w}_i}}\epsilon_i}\ri]\label{HEEfiniteT}.
\end{align}
Where the meaning of the superscript $(1)$ will be clear in later subsections. $S_A^{(1)}$ and $S_{A^c}^{(1)}$ are not equal, as the state is no longer pure but thermal. An example of the spacelike geodesic computing the holographic entanglement entropy is in the left plot of Figure \ref{fig:EEThermalState}. 

\subsubsection{Non-equal-time entanglement plateau}\label{Non-equal-time entanglement plateaux}
As is reviewed in section~\ref{sec:lateauxReview}, an important phenomenon of holographic entanglement entropy in thermal states with finite spatial regions is the entanglement plateaux. Here, the entanglement plateaux take place in a non-equal-time setup, which is a straightforward generalisation of~\cite{Hubeny:2013gta}: in addition to the connected minimal surface (see the red curve in the right plot of Figure \ref{fig:EEThermalState}) that computes \eqref{HEEfiniteT}, there is also a $\emph{disconnected}$ minimal surface, given by the $\emph{union}$ of the minimal surface for $A^c$ (i.e.~$\Delta t_{{\rm new}}$ stays the same while $\Delta x_{{\rm new}}$ becomes the complementary value $2\pi-\Delta x_{{\rm new}}$, see the green curve in the right plot of Figure \ref{fig:EEThermalState}) and the bifurcation surface of the event horizon.\footnote{The bifurcation surface is minimal in all static black hole geometries \cite{Hubeny:2013gta}.} The holographic entanglement entropy is then computed by,
\begin{align}
    S_A={\rm min}\{S_A^{(1)},S_A^{(2)}\},
\end{align}
where $S_A^{(1)}$ is given by \eqref{HEEfiniteT}, and 
\begin{align}
    S_A^{(2)}&=\frac{2\pi^2c}{3\beta}+\frac{c}{6} \log \le[\Big(\frac{\beta}{\pi}\Big)\frac{\big|\sinh \big[\frac{\pi}{\beta}
    ((t_{{{\rm new}},1}-t_{{{\rm new}},2})+(2\pi-(x_{{{\rm new}},1}-x_{{{\rm new}},2}))\big]\Big|}{\prod_{i=1}^2\sqrt{\frac{\partial w_{{\rm new},i}}{\partial w_i}\epsilon_i}}\ri]\nonumber\\[4pt]
    &+\frac{c}{6}\log \le[\Big(\frac{\beta}{\pi}\Big)\frac{\big|\sinh \big[\frac{\pi}{\beta}
    ((t_{{{\rm new}},1}-t_{{{\rm new}},2})-(2\pi-(x_{{{\rm new}},1}-x_{{{\rm new}},2}))\big]\Big|}{\prod_{i=1}^2\sqrt{\frac{\partial \overline{w}_{{\rm new},i}}{\partial \overline{w}_i}\epsilon_i}}\ri]\,.\label{HEEfiniteT2}
\end{align}

The plateau phenomenon in thermal SSD quench was first proposed in \cite{Goto:2021sqx}. We remark that while our results are qualitatively similar, our holographic construction is quantitatively different, as the holographic entanglement entropy is computed by spacelike geodesics anchoring on the boundary at non-equal times. 


\subsubsection{Holographic cooling/heating effect}\label{sec:HoloCoolHeat}
\begin{figure}[htbp]
\centering
\includegraphics[width=.38\textwidth]{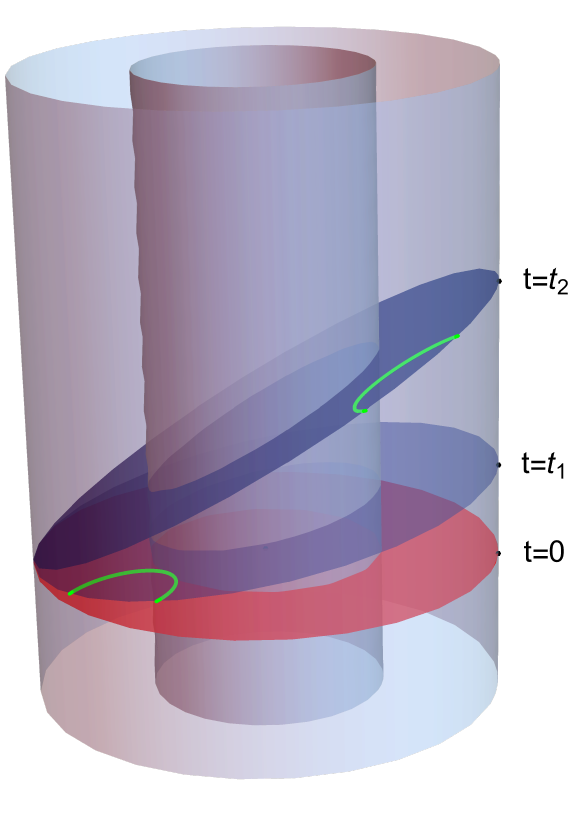}
\includegraphics[width=.38\textwidth]{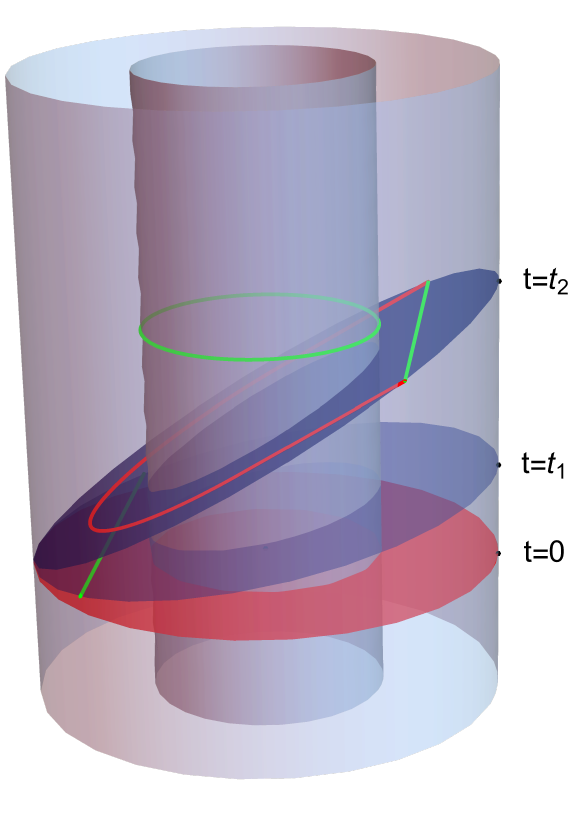}
\qquad
\caption{Non-equal time spacelike geodesics in BTZ geometry. The two geodesic endpoints flow towards the top tip of the triangle $(\pi,\pi)$ as $t$ evolves, from either the same (left) or different (right) side. In the latter case, at late time, the minimal surfaces computing holographic entanglement entropy switches from the red surface homologous to $A$ to the union of the green surface homologous to $A^c$ and the bifurcation surface of the black hole horizon (also coloured in green). \label{fig:EEThermalState}}.
\end{figure}

To further understand the time-evolution of holographic entanglement entropy, let us first note that the endpoints of the interval $t_{{\rm new},i}$ and $x_{{\rm new},i}$ $(i=1,2)$ can flow to the top tip of the triangle \eqref{tnewxnewLimit} from the same or different side, depending on whether $A$ includes the fixed point $x=0$ or not.
\begin{itemize}
    \item If $A$ does not include $x=0$, $(t_{{\rm new},1},x_{{\rm new},1})$ and $(t_{{\rm new},2},x_{{\rm new},2})$ flow to the top tip of the triangle from the $\emph{same}$ side, and $|x_{{{\rm new}},1}-x_{{{\rm new}},2}|\to 0$ in \eqref{HEEfiniteT}. See the left plot of Figure \ref{fig:EEThermalState}. Therefore, $S_A^{(1)}$ is always smaller than $S_A^{(2)}$, so we have $S_A=S_A^{(1)}$ at all $t>0$.  At late-time, one can show \cite{Goto:2021sqx} that $S_A$ is given by the $\emph{vacuum}$ entanglement entropy, 
\begin{align}
        S_A=S_A^{(1)}\to\frac{c}{3}\log \Big[2\Big(\sin \frac{1}{2}(x_2-x_1)\Big)\Big]\label{SSDCoolingEq}.
\end{align}
    \item If $A$ includes $x=0$, $(t_{{\rm new},1},x_{{\rm new},1})$ and $(t_{{\rm new},2},x_{{\rm new},2})$ flow to the top tip of the triangle from $\emph{different}$ sides, making $|x_{{{\rm new}},1}-x_{{{\rm new}},2}|\to 2\pi$ in \eqref{HEEfiniteT}. At some point, therefore, the entanglement entropy switches from $S_A^{(1)}$ to $S_A^{(2)}$. See the right plot of Figure \ref{fig:EEThermalState}. At late time, the second and third terms in \eqref{HEEfiniteT2} become much smaller than the first term,\footnote{For this the time has to compensate the smallness of the cutoff $\epsilon_i$.} so that $S_A$ is dominated by the $\emph{thermal}$ entanglement entropy, 
\begin{align}
    S_A=S_A^{(2)}\to S_{\rho}=\frac{2\pi^2c}{3\beta}\,. \label{SSDHeatingEq}
\end{align}
\end{itemize}

We conclude that whether the entanglement entropy cools down to the vacuum value or heats up and eventually reaches thermal value has a simple geometric interpretation in the bulk, which is whether the spacelike geodesic computing the holographic entanglement entropy winds around the BTZ black hole or not at late time.  Again, \eqref{SSDCoolingEq} and \eqref{SSDHeatingEq} match the results in \cite{Goto:2021sqx} qualitatively, albeit the setup is quantitatively different.  

\subsubsection{Comparison with field theory result on a line}\label{Comparison with field theory result on a line}
Computing the entanglement entropy of 1+1 dimensional CFT on a finite interval at finite temperature $\beta$ is difficult. However, one can compare the holographic result \eqref{HEEfiniteT} with that of the thermal state of CFT$_2$ on a line.  This CFT result can be obtained by Wick rotating $w_{{\rm new}}\to iw_{{\rm new}}$ and rescaling $2 \pi\to\beta$ the cylinder that computes \eqref{SSDEEVacCFT} (see Appendix \ref{Calc2pt} for details), 
\begin{align}
    S_A=\frac{\mathcal{L}_{{\rm reg}}}{4G_N}=\frac{c}{6} \log \le[\frac{\big(\frac{\beta}{\pi}\big)^2 \Big|\sin \big(\frac{\pi}{\beta}(w_{{{\rm new}},1}-w_{{{\rm new}},2})\big)\ \sin \big(\frac{\pi}{\beta}(\overline{w}_{{{\rm new}},1}-\overline{w}_{{{\rm new}},2})\big)\Big|}{\prod_{i=1}^2\sqrt{\frac{\partial w_{{\rm new},i}}{\partial w_i}\frac{\partial\overline{w}_{{\rm new},i}}{\partial\overline{w}_i}}\epsilon_i}\ri]\,.\label{HEEfiniteTCFT}
\end{align}
This result matches with the holographic result $S_A^{(1)}$ before the switch in the dominant extremal surface \eqref{HEEfiniteT2}. In \cite{Asplund:2014coa}, it is explained that the coincidence of \eqref{HEEfiniteT} and \eqref{HEEfiniteTCFT} is due to properties of large-$c$ Virasoro identity blocks arising from a CFT calculation obtaining entanglement entropy from four point functions with two heavy and two light operators of the cyclic orbifold CFT$^n/\mathbb{Z}_n$. In particular, the identity block in the dominant channel corresponds to the spacelike geodesic computing \eqref{HEEfiniteT}. 

\chapter{The entanglement membrane in 2d CFT: reflected entropy and RG flow}
\label{Chap3}
In this chapter, we present holographic studies of the entanglement dynamics of 2d CFTs and their deformations. We derive a generalised membrane with an additional scalar degree of freedom, that we will represent in figures as a colour, on its worldline as an 
effective theory of entanglement dynamics in 2d CFT. We show that upon adding a relevant deformation to the CFT, this additional degree of freedom gets ``gapped out" and we recover the entanglement membrane with non-degenerate membrane tension. This dichotomy is analogous to how hydrodynamics emerges from holomorphic stress tensor dynamics once a 2d CFT is deformed by a relevant operator~\cite{Davison:2024msq}. These observations make us speculate that the additional degree of freedom on the generalised membrane worldline is capturing the infinite-dimensional symmetry charge dynamics of 2d CFT.\footnote{An additional evidence in this direction is the aforementioned failure of the ordinary membrane to capture the R\'enyi entropy for multiple intervals, since R\'enyi entropy should be more sensitive to the conserved charge dynamics. See evidence in this direction in Refs.~\cite{Rakovszky:2019oht,Huang:2019hts,Znidaric:2020czi,Rakovszky:2020idb}.}

We also show that the reflected entropy, that is computed holographically by the entanglement wedge cross section (EWCS), is also captured by the generalised membrane in 2d CFT, and that this quantity, in contrast to entanglement entropy, is not captured by the ordinary membrane with degenerate membrane tension. To complete our investigations, we also study the reflected entropy in higher dimensional CFTs, and demonstrate that the ordinary membrane correctly captures them. 

\section{Entanglement membrane and 2d CFT}\label{Sec:EEMembrane}
From the reviews in section~\ref{sec:membraneReview}, we see that for the $d=2$ BTZ black hole, we get $c(z)=1$, which either enforces  $\dot x^2 =1$ or signals the breakdown of the scaling Ansatz. Indeed, two-sided geodesics reach $e^\Lam$ depths inside the BTZ black hole~\cite{Hartman:2013qma}. Then a different scaling can be applied: let us define $z\equiv e^\xi$ and scale $\xi, u, x$ linearly with $\Lam$. The resulting action in the $d=2$ BTZ black hole is 
\es{AreaFunct3}{
S &= s_\text{th} \int du\,  \sqrt{1+{\dot x^2-1-2\dot z\ov z^2}}\\
&=s_\text{th} \int du \le[1+{d(e^{-\xi})\ov du}+ \frac12\,e^{-2\xi}(\dot x^2-1-\dot \xi^2)+\dots \ri]\,.
}
We note that to leading order the action is simply given by the difference in infalling times $u$. All curves that obey the scaling assumptions and stretch between the same endpoints have the same leading order action. The shape of the curves is determined by the $e^{-2\xi}$ term (since the $e^{-\xi}$ term is a total derivative). We can think of \eqref{AreaFunct3} as a generalised membrane theory, where besides the membrane shape $x(u)$ there is an extra degree of freedom $\xi(u)$ that lives on the membrane. Of course, we have not done much compared to the HRT prescription: $\xi(u)=\log z(u)$ is the logarithm of the bulk depth, and we expanded for large $\xi, u, x$.

Since irrespective of the value of $\xi$ or the membrane shape we get the same value for the entropy, one may question the utility of the above exercise: to obtain the entropy we could just as well use the ordinary membrane theory with ${\cal E}(v)=1$. However, we will show how $\xi$ plays an essential role in computing the reflected entropy in 2d CFT. We will also analyse how the membrane theory arises in an RG flow from a 2d CFT, by gapping out the $\xi$ degree of freedom, which then can be integrated out, leaving us with a local theory for $x(u)$, the membrane shape.

\subsection{The entanglement membrane shape for a half space}\label{ApproxApproxLag}

The membrane shape is determined from the effective Lagrangian 
\es{Leff}{
\mathcal{L}_\text{eff}=\frac12\,e^{-2\xi}(\dot x^2-1-\dot \xi^2)\,.
}
derived from \eqref{AreaFunct3}. The momentum $p=e^{-2\xi} \, \dot x$ is conserved and the $\xi$ equation of motion is
\es{xiEOM}{
\ddot \xi=-1+\dot \xi^2+p^2 e^{4\xi}\,,
}
which motivates us to introduce $p\equiv e^{-2\xi_p}$. We can understand the gross features of the solutions that stay in the large $\xi$ region for long $u$ as follows: For $\xi<\xi_p$ the $p^2 e^{4\xi}$ term is negligible, and the solution is  $\xi(u)= u$ with exponentially small corrections. Once this curve reaches the vicinity of $\xi_p$ it transitions to the solution $\xi(u)= \xi_p$ with exponentially small corrections. The transition between the two behaviours is abrupt and can be neglected in the scaling regime. All of these statements can be verified by straightforward numerical solutions of \eqref{xiEOM}. It turns out that \eqref{xiEOM} can also be solved in closed form, see Appendix~\ref{ApproxLagDispHalf}; the exact solution also confirms the above approximations in the large $T$, large $X$ limit. Finally, since the BTZ black brane is AdS$_3$ with some identifications, we are studying AdS$_3$ geodesics in disguise: this perspective is developed in Appendix~\ref{Displaced half spaces}..

\begin{figure}[htbp]
\centering
\includegraphics[width=.38\textwidth]{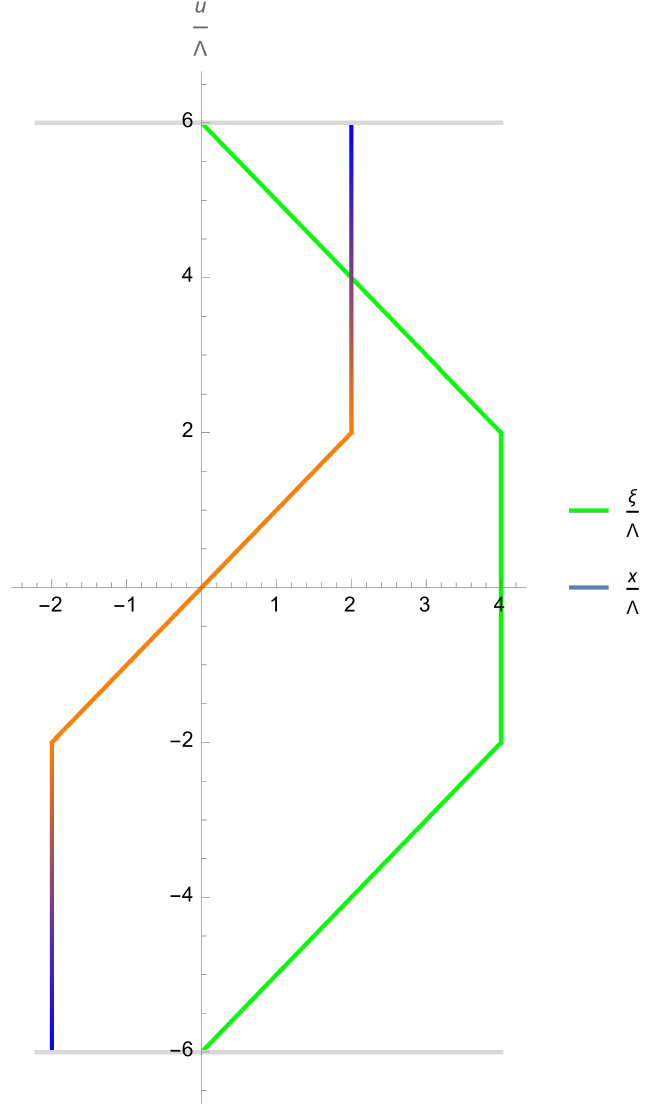}\hspace{2cm}
\includegraphics[width=.22\textwidth]{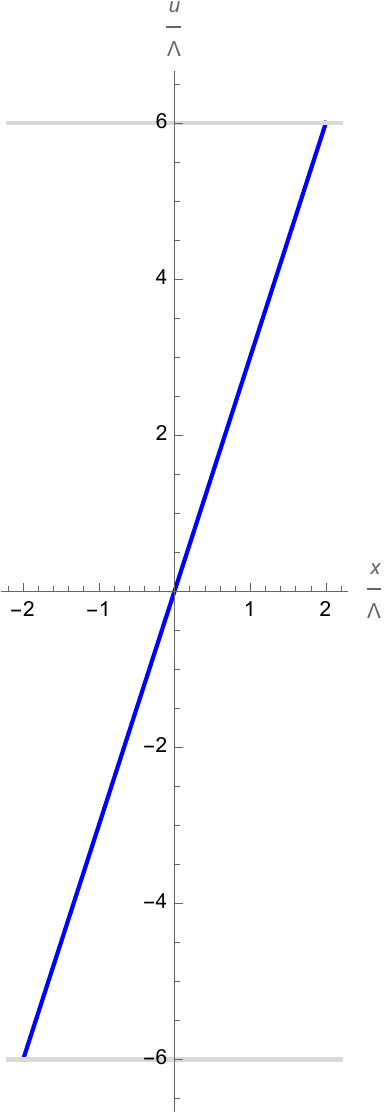}
\qquad
\caption{$\emph{Left}$: A plot of $x(u)$ and $\xi (u)$ (green) in \eqref{FinalXi} in the scaling, i.e. large $\Lambda$ limit in generalised membrane theory. The light gray horizontal lines denote $u=\pm T\Lambda$. As for $x(u)$, we use blue $\to$ orange to denote the increase in the bulk depth in $\xi$. $\emph{Right}$: A plot of the membrane shape function \eqref{MembraneShapeHigherD} in planar AdS$_{d+1}$-Schwarzchild black brane ($d\geq 2$), with $v=\frac{1}{3}$. The membrane is of uniform blue colour as it stays at a constant $z\in O(1)$ at late time. \label{fig:zxApproxLagrangian}}
\end{figure}

In summary we find (when $T>X$)
\es{FinalXi}{
(x(u),\xi(u))=\begin{cases}
\le(-X,T+u\ri)\qquad &-T<u<-X\,,\\
\le(u,T-X\ri)\qquad &-X<u<X\,,\\
\le(X,T-u\ri)\qquad &X<u<T\,,
\end{cases}
} 
where we used the result $\xi_p=T-X$ obtained from matching to boundary conditions. See Fig.~\ref{fig:zxApproxLagrangian} (left) for a plot of \eqref{FinalXi}. In this and subsequent figures we plot the membrane shape $u(x)$ and visualise the extra degree of freedom $\xi$ on the membrane worldline with a colour interpolating between blue ($\xi=0$) to orange ($\xi=\xi_p$). In Fig.~\ref{fig:zxApproxLagrangian} we also plotted $\xi(x)$ with green, which is redundant information. Also in plots we will always take the time and space extents to be $T\Lam$ and $X\Lam$ with $T,X=O(1)$ and $\Lam\to\infty$.

\begin{figure}[htbp]
\centering
\includegraphics[width=.28\textwidth]{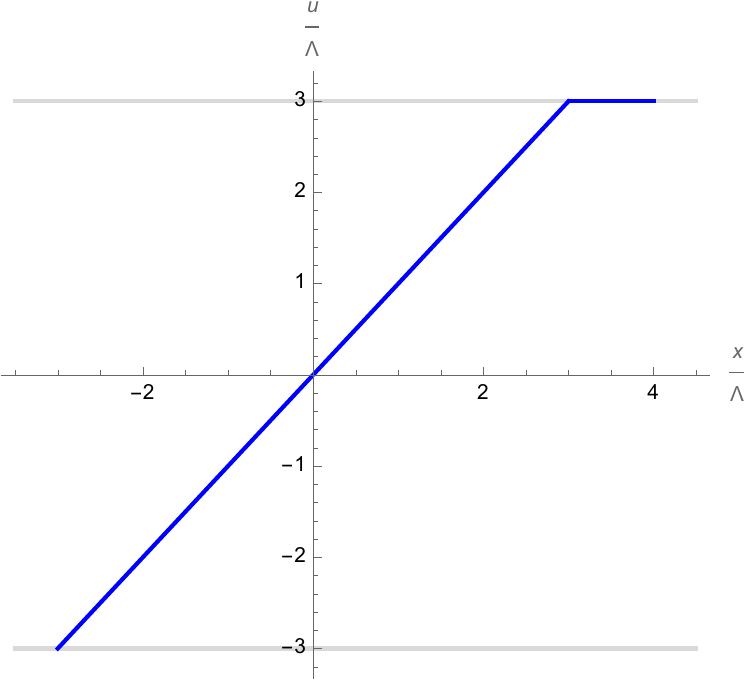}\hspace{1cm}
\includegraphics[width=.28\textwidth]{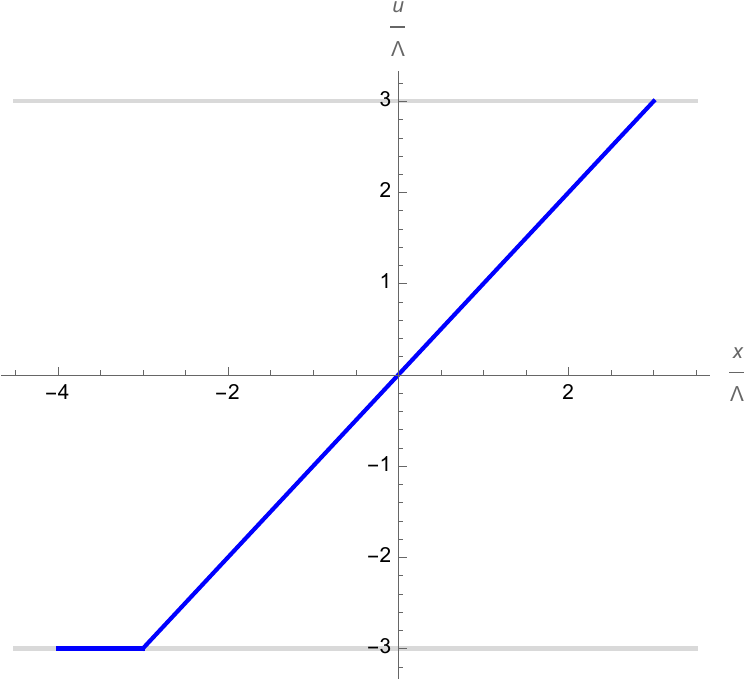}\hspace{1cm}
\includegraphics[width=.28\textwidth]{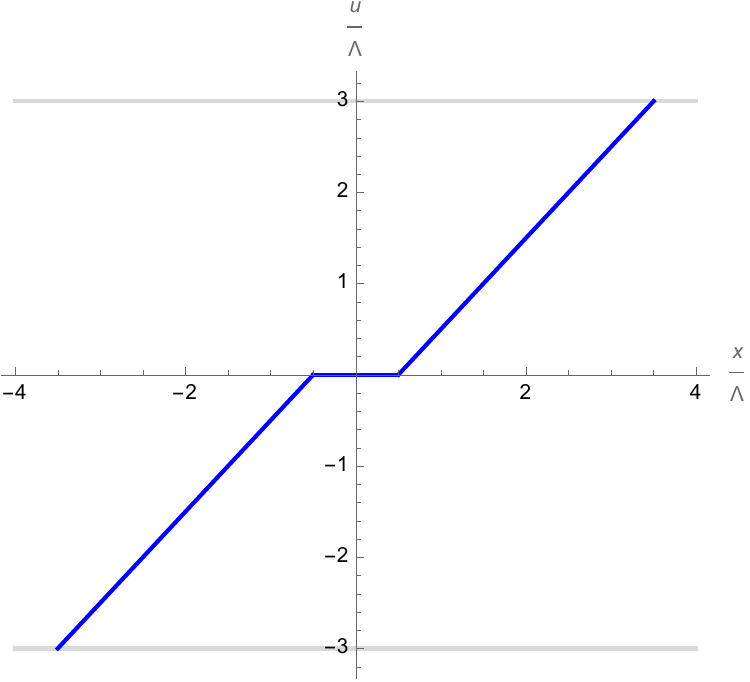}
\caption{Equivalent membranes computing the operator entanglement for $T<X$ in the scaling limit. Here we plotted for $T=6\Lambda$ and $X=7\Lambda$. The membranes are in blue, as the corresponding HRT surface is in the near-horizon region. \label{fig:XgTmem}}
\end{figure}

The corresponding operator entanglement entropy is
\es{OpEntFinal}{
S_{\text{op}}=s_\text{th}\,2T\,,
}
which simply follows from the extent of the generalised membrane in $u$, see the discussion below \eqref{AreaFunct3}. In mapping the HRT surface to the boundary generalised membrane, we have to decide when to switch from the right to the left infalling time. It does not matter at what exact point we switch, as long as the HRT surface is not at or very close to the horizon, see a detailed analysis in~\cite{Mezei:2016zxg}.
The right and left time extents of the membrane then add up to~$2T$.\footnote{A good choice is to switch between the two sides, when the HRT surface reaches the deepest point behind the horizon $z_*$. If we parametrise the membrane in boundary polar coordinates as $r(u,\Om)$, then generically the switch will happen at different locations as we move in polar angle, $u_\text{switch}(\Om)$, so the mapping will be dependent on the subsystem through the membrane shape. A simpler choice would be to switch at $u=0$, but this choice is not time translation invariant, and under large time shifts (corresponding to boosts in the near-horizon region) it leads to the switch happening in the problematic near-horizon region.

There is one extra subtlety: there exist HRT surfaces that live entirely or partly (for some region of $\Om$) in the near-horizon region. Hence they never reach $z_*$ and the above switch procedure is not well-defined for them. These membranes or membrane sections have $v=v_B$, which maps to the near-horizon region $z=1$. In these cases we use the $u_\text{switch}=\text{const}$ switch, with the constant set by continuity in $\Om$ or if the entire membrane has $v=v_B$ we set  $u_\text{switch}=-T,0,T$ depending on what looks most natural, see Fig.~\ref{fig:XgTmem}. This choice however cuts out some portion of the HRT surface, which we need to account for to correctly compute the entropy. We glue in a horizontal membrane for this portion. This rule is consistent, because ${\cal E}(v_B)/v_B=1$. It is also natural from the circuit perspective, where it corresponds to a horizontal cut of legs~\cite{Jonay:2018yei}.  See the first membrane in Fig.~3 in~\cite{Mezei:2018jco} for illustration. Unfortunately, the last membrane on that figure, corresponding to saturation is misleading: we explain below that saturated membranes are more naturally $v_B$ cones. \label{matching_footnote}}    

It is instructive to compare the generalised membrane theory equations of motion \eqref{FinalXi} to that in the ordinary membrane theory~\cite{Mezei:2018jco}, where the HRT surface is described solely by the membrane shape function $x(u)$, since the bulk $z$ degree of freedom can be integrated out. For the two-sided entanglement entropy in half spaces offset by $2X$ $(v_BT>X)$, the membrane shape is a straight line of slope $v$~\cite{Jonay:2018yei,Mezei:2018jco}
\begin{align}
    x(u)=v u\,, && v=\frac{X}{T}\in[0,v_B)\,. \label{MembraneShapeHigherD}
\end{align}
The comparison between ordinary \eqref{MembraneShapeHigherD} and generalised \eqref{FinalXi} membrane theory is illustrated in Fig.~\ref{fig:zxApproxLagrangian}. The corresponding operator entanglement entropy however should not differ between the generalised and ordinary membrane prediction, which is~\cite{Jonay:2018yei,Mezei:2018jco}
\es{OpEntFinal2}{
S_{\text{op}}=s_\text{th}\,2T\,{\cal E}(v)\,.
}
For the degenerate membrane tension ${\cal E}(v)=1$ this indeed agrees with \eqref{OpEntFinal}.


So far we have focused on the $T>X$ case. When $T<X$, one can show that $\xi$ is small, i.e. the geodesic goes just inside the horizon. As $z$ is no longer large, the effective action \eqref{AreaFunct3} is no longer valid. In the discussion above \eqref{AreaFunct3} we encountered the possibility that $z=O(1)$ and $\dot x^2=1$, and this is exactly the HRT surface we need for  $T<X$. 
This conclusion is also supported by the exact BTZ geodesics as shown in Appendix~\ref{Displaced half spaces}. The membrane mapping of this HRT surface is a a bit unnatural, see Fig.~\ref{fig:XgTmem}, as we need to include a horizontal membrane section between sections of $v=v_B=1$ ones. We can add this section at arbitrary positions, and all of these shapes map to the same HRT surface in the bulk. See footnote~\ref{matching_footnote} for an explanation.\footnote{For this setup, it may seem more natural from the bulk point of view to extend the time extent of spacetime from $2T$ to $2X$, and then the membrane would be a straight $v=1$ line. However, such a choice would lead to a very convoluted membrane mapping for more complicated shapes in higher dimensions. To us, the addition of horizontal membrane segments produces the most intuitive membrane theory. }
The corresponding operator entanglement is:
\es{OpEntFinal3}{
S_{\text{op}}=s_\text{th}\,2X\,,
}
in agreement with earlier results in~\cite{Jonay:2018yei,Mezei:2018jco}.

\subsection{Membranes for saturated strips}\label{MembraneStatic2d}

The displaced half space discussion is easily generalised for displaced intervals of width $D$. In this case there are two candidate HRT surfaces or equivalently generalised membranes: the first is the doubling of the HRT surface corresponding to displaced half spaces and it connects the interval endpoints on the two sides, the second is a static RT surface that connects the interval endpoints on the same sides. For early times $T<D$, the first is minimal, and for late times $T>D$ the second becomes dominant, leading to the saturation of operator entanglement entropy. In this section we discuss the static RT surface and its membrane equivalent.

We show the derivation in detail, as the mapping of the near-horizon region to the membrane was not considered with enough care in prior literature. We consider static RT surfaces outside the horizon on a constant Schwarzschild time slice $t=T$.\footnote{The Schwarzschild time $t$ in the geometry \eqref{blackBraneInf} is defined as
\begin{equation}
t = u+\int_0^z\frac{dz'}{a(z')b(z')}.
\end{equation}} We can choose a coordinate system such that our two boundary endpoints are symmetric at $\Big(\pm\frac{D}{2},T\Big)$.\footnote{To study the static geodesic only, it is easiest to set $T=0$ as its geometry does not change with time. }
The static geodesic on the $t=T$ slice in BTZ is depicted in Fig.~\ref{fig:StaticGeod} (see Appendix~\ref{Static geodesic} for the explicit expressions). We observe that the geodesic roughly consists of two parts: a $\emph{plateau}$,\footnote{Throughout this paper, we use the word ``plateau" to refer the portion of the HRT surface that stays at constant $z$ (either inside or outside the horizon) in the scaling limit.} where it lies close to the horizon $z=1$ and moves transversely; and two regions where it $\emph{shoots}$ from the near-horizon region towards the conformal boundary $z=0$. 

\begin{figure}[htbp]
\centering
\includegraphics[width=.55\textwidth]{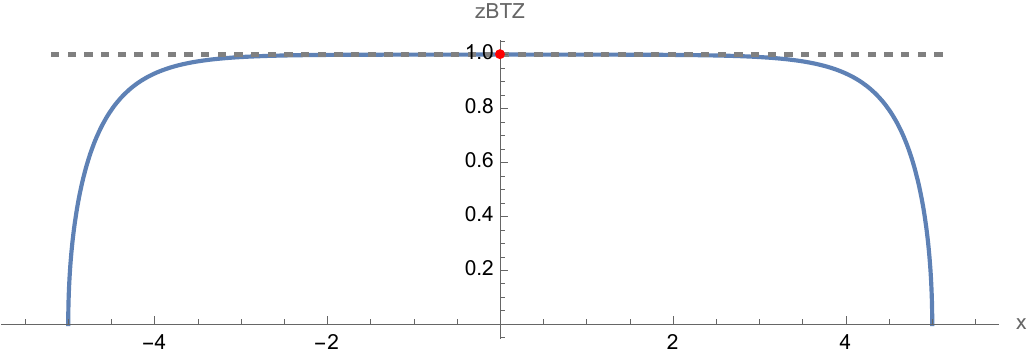}
\qquad
\caption{Plot of the static spacelike geodesic on the $t=T$ slice. The dashed gray line is the black hole horizon. The geodesic consists of a plateau and two parts that shoot towards the conformal boundary. The red dot denotes the deepest point $\tanh \frac{D}{2}$ the geodesic can reach in the bulk. \label{fig:StaticGeod}}
\end{figure}

The static curve shown in Fig.~\ref{fig:StaticGeod} is on the constant Schwarzschild time slice $t=T$. In (generalised) membrane theory, however, we would like to understand the projection of this RT surface to the boundary along constant infalling time. In $d=2$, the infalling time $u$ along the static geodesic on the $t=T$ slice is given by (see Appendix~\ref{SchVSInfallingSec})
\es{u(z)2dExpansion}{
    u(z)&=T-\text{arctanh}z\\
    &\to T+\frac{1}{2}\log\frac{1-z}{2},\ \ (z\to 1^-).
}
Notice that $u(z)\to -\infty$ as $z\to 1^-$. Therefore, close to the horizon, a small shift in $z$ leads to a large shift in $u$. This shift then leads to a finite difference between infalling and Schwarzschild time. See Fig.~\ref{fig:RTInfalling} in Appendix~\ref{Static geodesic} for an illustration of the geodesic in Fig.~\ref{fig:StaticGeod} in infalling coordinates. 

\begin{figure}[htbp]
\centering
\includegraphics[width=.55\textwidth]{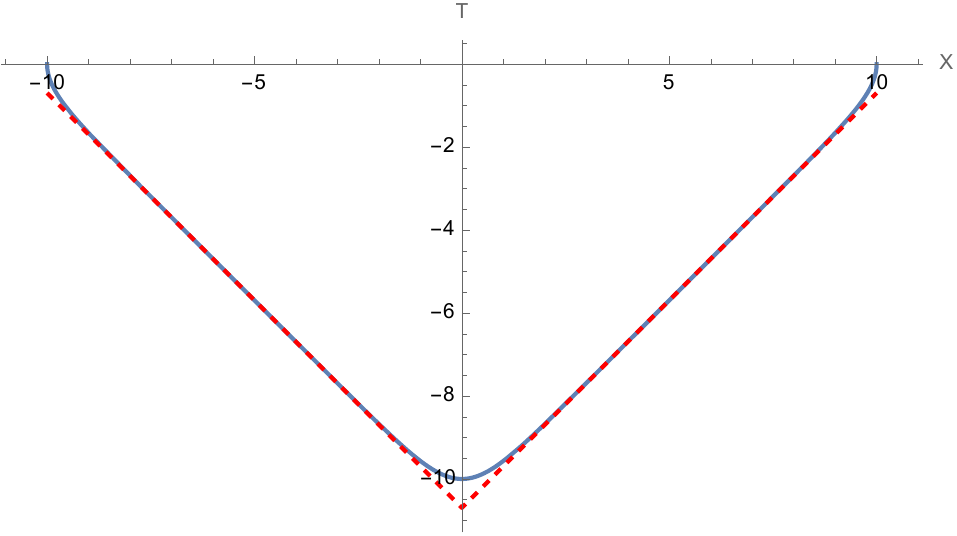}
\qquad
\caption{Projection of the static RT surface (see $\eqref{zLambdaStatic}$ and $\eqref{xLambdaStatic}$ for explicit expressions) to the boundary along constant infalling time. In the scaling limit, the projection is a cone of slope $\pm 1$ \eqref{vBcone2dApprox}, which are depicted here as dashed red lines. Notice that its two corners and tip are all blunt.
 \label{fig:vBcone2d}}
\end{figure}

The explicit projection of the static geodesic shown in Fig.~\ref{fig:StaticGeod} to the boundary along constant infalling time is given by (see Appendix~\ref{Static geodesic} for details)
\begin{equation}
\label{vBcone2dApprox}
    u(x)\approx 
    \left\{
    \begin{aligned}
        &x+T-\frac{D}{2}-\log 2,&&x>0\,,\\
        &-x+T-\frac{D}{2}-\log 2,&&x<0\,,
    \end{aligned}
    \right.
\end{equation}
where we kept the subleading $\log 2$ term to get a more accurate match on Fig.~\ref{fig:vBcone2d}.
We see that the projection is a $\emph{cone}$ with slope $\pm 1$, as is depicted in Fig.~\ref{fig:vBcone2d}. Notice that the majority of the cone comes from the projection of the plateau part of the geodesic; the parts that shoot exponentially towards the boundary are projected to the two blunt corners. The latter is because when $z\to 0$, $\text{arctanh}z$ is small, so $u(z)$ is close to $T$. Notice also that the tip of the cone is blunt in a region vanishingly small in the scaling limit, and hence is not captured by \eqref{vBcone2dApprox}.  

\section{Reflected entropy from the entanglement wedge cross section in 2d CFT}\label{sec:2dEWCS}

The generalised membrane theory and the ordinary membrane theory results with degenerate membrane tension ${\cal E}(v)=1$ predict the same entropy evolution in quenches and for operator entanglement. However, in this and the next section we show that the two give drastically different predictions for reflected entropy: it is the generalised membrane theory computes it correctly for 2d CFTs. 

\subsection{Setup of the problem}
\begin{figure}
    \centering
    \includegraphics[width=0.5\linewidth]{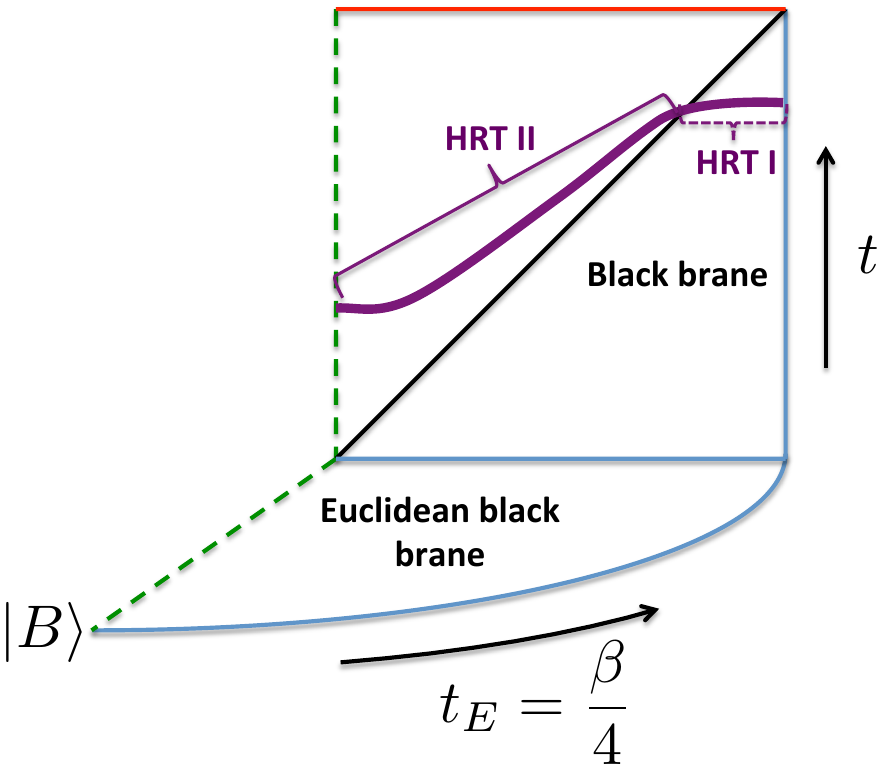}
    \caption{The end-of-the-world (EoW) brane quench setup, with initial state $\ket{\psi_0}=e^{-\frac{\beta}{4}H}\ket{B}$, where $\ket{B}$ is a conformal boundary state.}
    \label{fig:penroseb}
\end{figure}

In order to study the time evolution of the entanglement wedge cross-section, we consider a bipartite system consisting of the union of two strips: $A\cup B = [0,\ell]\cup [\ell+D,\infty)$. As initial state, we consider $\ket{\psi_0}=e^{-\frac{\beta}{4}H}\ket{B}$, where $\ket{B}$ is a conformal boundary state~\cite{Calabrese:2005in}. The time evolution of such a system is described holographically by a one-sided black hole with an end-of-the-world (EoW) brane~\cite{Hartman:2013qma}\footnote{Another well-known set-up is that of a black hole form from the collapse of null dust, described by the Vaidya geometry. The evolution of the EWCS in this scenario has been studied in \cite{BabaeiVelni:2020wfl,BabaeiVelni:2023cge}.}. The location of the EoW brane in infalling time is
\begin{equation}\label{eq:EoWBrane}
u_{\text{EoW}}(z) = \frac{i\pi}{a'(1)b(1)} - \int_0^z \frac{dz'}{a(z')b(z')}\,.
\end{equation}
see Figure~\ref{fig:penroseb} for an illustration. 

The reflected entropy was first studied in this setup in~\cite{Kudler-Flam:2020url}, where three computations were presented: a 2d CFT and the holographic calculation using BTZ geodesics gave agreement, and these were reproduced by a heuristic membrane picture. In what follows, we will reproduce the results of~\cite{Kudler-Flam:2020url}  using the generalised membrane theory, and explain how it is related to their heuristic membrane picture. The main novelty in our treatment is the introduction of a generalised membrane characterising the equilibrium EWCS (section~\ref{EqEWCS}), which describes a spacelike geodesic in BTZ that contains a portion that is null in the scaling limit.\footnote{If extended beyond the HRT surface it ends on, it reaches the BTZ singularity (Appendix~\ref{SatEWCS}).} 
The computations will build on the results on half-space entanglement discussed in section~\ref{Sec:EEMembrane}.

\begin{figure}[ht]
\centering
\begin{tikzpicture}
\node (hrt1) at (0,0)
    {\includegraphics[scale=0.3]{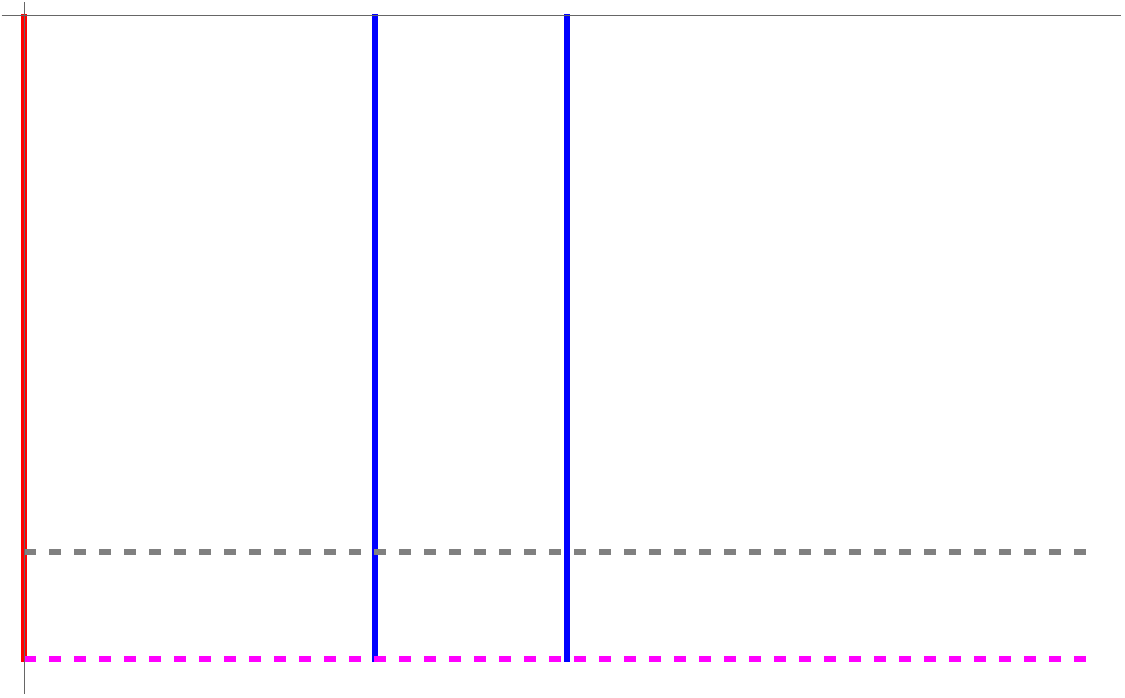}};
\node at (-3,-1.8) {\small $z$};
\node at (4,-1.8) {\small $z$};
\node at (3.2,1.75) {\small $x$};
\node at (10.2,1.75) {\small $x$};
\node at (-2.8,2) {\small $P_1$};
\node at (-1,2) {\small $P_2$};
\node at (0.2,2) {\small $P_3$};
\node at (-1,-1.8) {\small $\ell$};
\node at (0.2,-1.8) {\small $\ell+D$};
\node at (6,2) {\small $\ell$};
\node at (7.2,2) {\small $\ell+D$};
\node (hrt2) at (7,0)
    {\includegraphics[scale=0.3]{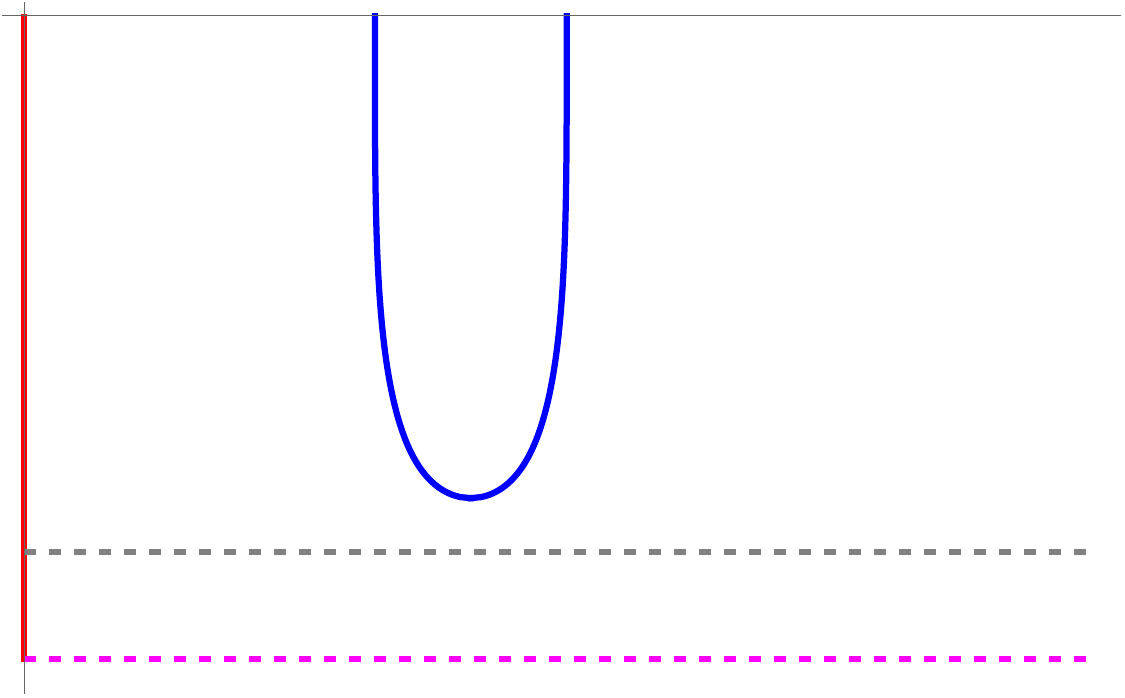}};
\end{tikzpicture}
\caption{HRT and RT surfaces for the interval $[0,\ell]\cup [\ell+D,\infty)$. In the plot, $P_1$, $P_2$ and $P_3$ are on the AdS boundary at some given time $t$, with $x$ coordinates being $0$, $\ell$ and $\ell+D$, respectively. The black dashed line indicates the horizon, while the magenta line corresponds to the end-of-the-world brane, see Figure~\ref{fig:penroseb}. In the right graph, the non-equilibrium HRT surface and equilibrium RT surface are plotted in red and blue, respectively. }
\label{fig:hrts}
\end{figure}

The case $\ell<D$ is uninteresting for our purposes since its entanglement wedge is always disconnected. Hence we will focus on the $\ell>D$ case. The possible extremal surfaces computing the entanglement entropy are shown in Fig.~\ref{fig:hrts}. At early times, the entanglement wedge is disconnected, and so the cross section simply vanishes. At time $t=\frac{D}{2}$, there is a transition to a connected entanglement wedge. Our main goal here will be to characterise the dynamics of the EWCS in this scenario.

\begin{figure}[H]
\centering
\begin{tikzpicture}
\node at (0,0) {\includegraphics[scale=0.4]{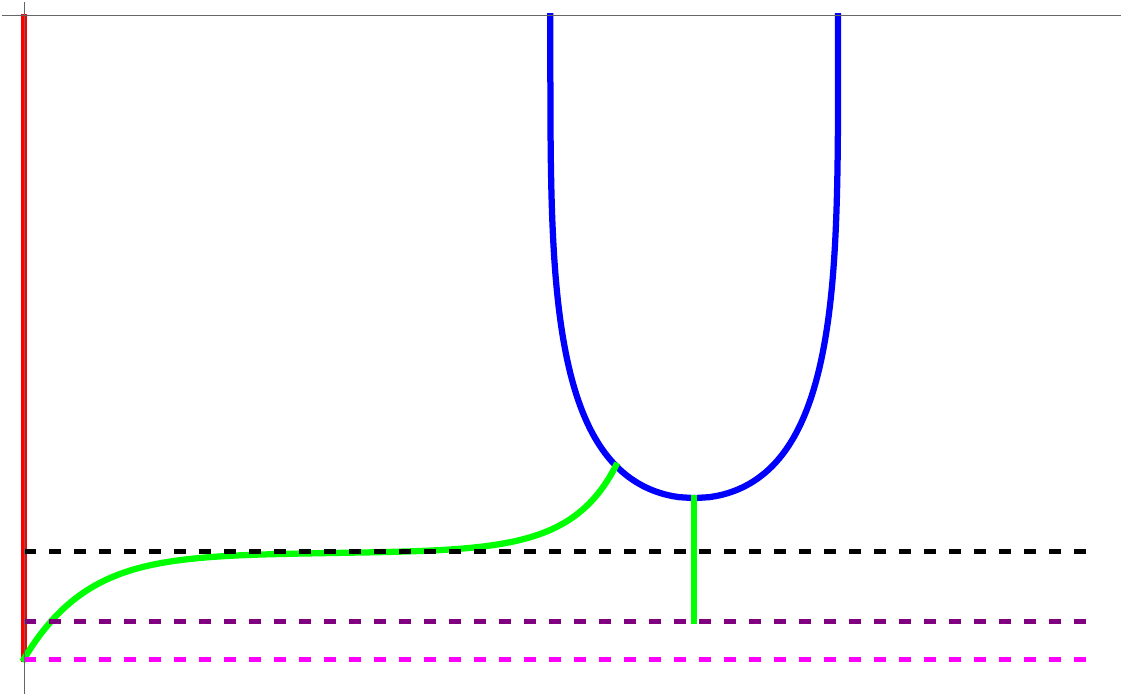}};
\node at (4,2.25) {$x$};
\node at (-3.7,-2.5) {$z$};
\node at (-0.1,2.5) {$\ell$};
\node at (2,2.5) {$\ell+D$};
\node at (4,-1.3) {$z_{h}$};
\node at (4,-1.7) {$z_{0}'$};
\node at (4,-2.1) {$z_{0}$};
\end{tikzpicture}
\caption{Possible non-vanishing entanglement wedge cross sections (green). Since these curves move in time, they may encounter the EoW brane at different $z$'s, which itself is moving according to~\eqref{eq:EoWBrane}.  }
\label{fig:ECSPlot}
\end{figure}

In time-dependent scenarios there are multiple extremal surfaces bisecting the entanglement wedge and, in order to correctly capture the EWCS, one must select the surface with minimal area. For our configuration, the possible surfaces are depicted in Fig.~\ref{fig:ECSPlot}. 

\subsection{Non-equilibrium EWCS}
\begin{figure}[htbp]
\centering
\includegraphics[width=.6\textwidth]{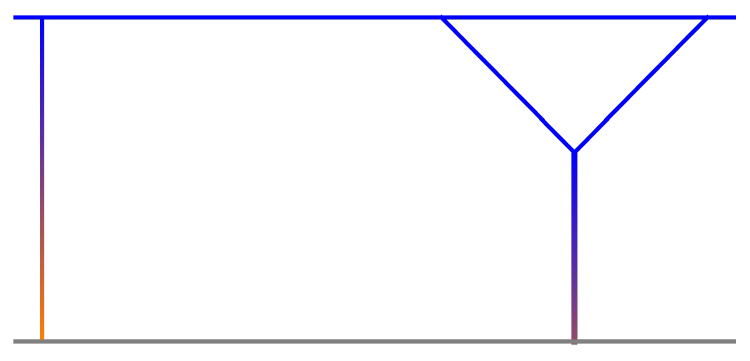}
\qquad
\caption{Projection of non-equilibrium EWCS in $d=2$ along constant infalling time in the scaling limit, for $T=1.2\Lambda$, $\ell=1.5\Lambda$, and $D=\Lambda$. We use blue $\to$ orange to denote the increase in the bulk depth in $\xi$, and plot the EWCS as thick to contrast with HRT. The non-equilibrium EWCS starts from the top tip of the projection of the static RT surface, and then enters the interior where its length grows linearly in time. The non-equilibrium EWCS ends perpendicularly on the EoW brane, whose location is given by $u=0$ for large $z$. We plotted the EoW brane as gray, since HRT and EWCS can end on it at any depth. From this projection, one can easily read off the length of the EWCS to be $T-\frac{D}{2}$. See Fig.~\ref{fig:NeqEWCS} for the full 3d plot. \label{fig:NeqEWCSProj}}
\end{figure}

At $t\geq\frac{D}{2}$, the two HRT surfaces starting from $P_2$ and $P_3$ saturate to an RT surface connecting the two points, see Fig.~\ref{fig:hrts}. The projection of this RT surface is therefore a cone of height $\frac{D}{2}$. The EWCS starts from this RT surface, enters the black hole horizon, and ends on the EoW brane in the interior. The stretch of the spacelike interior time direction~\cite{Hartman:2013qma} then leads the EWCS to grow in time. The EWCS can also be described by the effective Lagrangian method described in section~\ref{ApproxApproxLag}. The symmetry of the problem indicates the EWCS to be a vertical line starting from the tip of the RT surface, where $u=T-\frac{D}{2}$. Thus, we have $X=0$ and $T\to T-\frac{D}{2}$ in \eqref{FinalXi}. The time-dependent EWCS is described by 
\begin{align}
    (x(u),\xi(u))=\left(\ell+\frac{D}{2},T-\frac{D}{2}-u\right).
\end{align}
See Fig.~\ref{fig:NeqEWCSProj}. Therefore, the length of the EWCS is $T-\frac{D}{2}$, which starts from 0 at $T=\frac{D}{2}$ and grows linearly in time with slope $v_E=1$.

\subsection{Equilibrium EWCS}\label{EqEWCS}
To describe the entanglement wedge cross section, we introduce a section of a generalised membrane that moves negligibly in $u$ with a bit of hindsight.\footnote{See Appendix~\ref{SatEWCS} for the exact solution of this problem.} Correspondingly, its contribution to the entropy is zero in the scaling limit. It is describing a portion of an HRT surface that becomes null in the scaling limit. Nevertheless it is important in piecing together the generalised membrane that computes the EWCS. 

\begin{figure}[htbp]
\centering
\includegraphics[width=.5\textwidth]{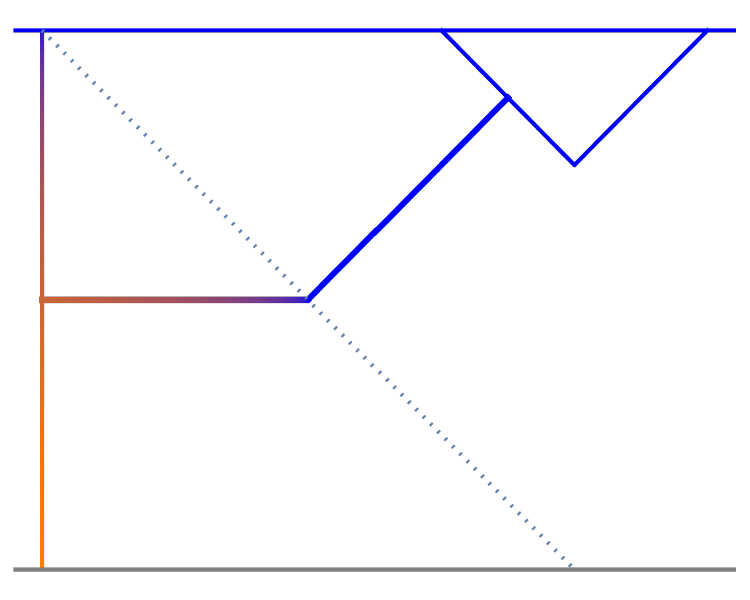}
\qquad
\caption{Projection of equilibrium EWCS in $d=2$ along constant infalling time, for $T=2\Lambda$, $\ell=1.5\Lambda$, and $D=\Lambda$. We use blue $\to$ orange to denote the increase in the bulk depth in $\xi$, and plot the EWCS as thick to contrast it with the HRTs. The equilibrium EWCS starts from the mid-point of the projection of static RT surface, enters the horizon, and arrives at a plateau where it moves in $x$ for a distance $\ell/2$. The EWCS then becomes null and shoots deep into the interior, where it meets the HRT surface. Note that at the connection points the colours of the EWCS matches the colours of the HRT surfaces, as required by the boundary conditions. The dotted 45$^\circ$ denotes an alternative projection of the HRT surface along the null generalised membrane, which allows us to match our results with the heuristic picture of 2001.05501. See Fig.~\ref{fig:SatEWCS} for the full 3d plot. \label{fig:EqEWCSProj}}
\end{figure}

Since this generalised membrane moves a large amount in the $x, \xi$ directions but only a little in the $u$ direction, it is easier to describe it by changing variable from $u$ to $x$. Implementing this change of variables in \eqref{xiEOM} and using momentum conservation $\dot x=e^{2(\xi-\xi_p)}$, we get:
\es{xiEOMx}{
 \xi''&=1-\xi'^2 - e^{4(\xi_p-\xi)}\,,\\
 u'&=e^{-2(\xi-\xi_p)}\,.
}
where $f'=\frac{df}{dx}$. This set of equations has the following solution that will be relevant for us:\footnote{Generically, there is another kink at some $x=x_1$, but this additional feature will not be needed. One can think of the solution presented below as the $x_1\to \infty$ special case of the general solution.}
\begin{equation}
    \label{XiEWCS}
    (u(x),\xi(x))=
    \left\{
    \begin{aligned}
        &(u_p,-x+x_0+\xi_p),&& 0<x<x_0\\
        &(x+u_p-x_0,\xi_p),&& x_0<x
    \end{aligned}
    \right.
\end{equation}
where the constants are chosen for later convenience, see Fig.~\ref{fig:EqEWCSProj}. To match the two HRT surfaces or generalised membranes, we require that
\begin{align}
    &\xi_p+x_0=T-u_p\,,&&\text{to match the non-equilibrium HRT,}\label{matchingConds11}\\
    &(u_p+x_c-x_0, \xi_p)=(T+\ell-x_c, 0)\,,&&\text{to match the equilibrium HRT,}\label{matchingConds22}
\end{align}
where $x_c$ is the connection point between the EWCS and the equilibrium HRT in Fig.~\ref{fig:EqEWCSProj}. In \eqref{matchingConds11}, the non-equilibrium HRT is described by half of \eqref{FinalXi} with $X=0$, hence $x(u)=0$; in \eqref{matchingConds2}, the equilibrium HRT is given by $x=-u+T+\ell$ and $\xi(u)=0$ (as it is outside the horizon). Only three out of four parameters get determined:
\begin{align}
 \xi_p=0\,,&&   u_p=T+{\ell\ov 2}-x_c\,, && x_0=x_c-{\ell\ov 2}\,.\label{matchingConds}
\end{align}
The EWCS is given by the change of $u$ for these generalised membranes
\es{ECWS}{
\ell(\text{EWCS})=x_c-x_0={\ell\ov 2}\,.
}
$\ell(\text{EWCS})$ is independent of $x_c$, so each of the generalised membranes correctly computes the EWCS. This degeneracy does not get resolved even if we solve the approximate equations of motion \eqref{xiEOMx} exactly, as is done in Appendix~\ref{Plateau+Null geodesic}. Once we back off from the large $z$ approximation, subleading terms fix $x_c$ to be $x_c=\ell+{D\ov 4}$, see Appendix~\ref{SatEWCS} for details.

Our generalised membrane description shown in Fig.~\ref{fig:EqEWCSProj} is different from the heuristic membrane picture of~\cite{Kudler-Flam:2020url}. Recall that in getting Fig.~\ref{fig:EqEWCSProj}, we project the bulk HRT surfaces to the boundary along constant infalling time. We can find the connection between our result and that in~\cite{Kudler-Flam:2020url} by projecting instead along the null generalised membrane in~\eqref{XiEWCS}, which amounts to a shift of $\xi$ in the $x$ coordinate compared to projecting along constant infalling time.\footnote{Recall that projecting the HRT surfaces to the boundary along constant infalling time is to plot $(x,u)$ on flat spacetime. When projecting along the null generalised membrane, we plot instead $(x+\xi,u)$.}  Under such a prescription, the horizontal generalised membrane in Fig.~\ref{fig:EqEWCSProj} is projected to a $\emph{point}$, and the non-equilibrium HRT in the left is projected to the dotted 45$^\circ$ line in Fig.~\ref{fig:EqEWCSProj}.\footnote{See Fig.~\ref{fig:SatEWCS} from the exact solution of equilibrium EWCS for a better visualisation.} The resulting projection then matches with the picture of~\cite{Kudler-Flam:2020url}.\footnote{We find it somewhat unsatisfactory that to make the heuristic picture of~\cite{Kudler-Flam:2020url} work for the reflected setup $B'\cup A' =(-\infty,-(\ell+D)]\cup [-\ell,0]$, we would also have to reflect the non-equilibrium membrane, even though it computes the entropy of the complement half-space $(-\infty,0]$ in a pure state. Relatedly, the mapping from the holographic bulk to the membrane picture would need to be dependent on the EWCS we are interested in computing.}

To sum up, in 2d CFT the generalised membrane description of the reflected entropy is given by 
\begin{equation}
\label{GenMembraneEWCS2d}
\frac{S_R}{2} = 
\begin{cases}
0 \ &T < \frac{D}{2}\,, \\ 
T - \frac{D}{2}\ \hspace{0.5cm} &\frac{D}{2} < T < \frac{\ell+D}{2}\,, \\ 
\frac{\ell}{2}  \  &T > \frac{\ell+D}{2}\,.
\end{cases}
\end{equation}
One can compare \eqref{GenMembraneEWCS2d} to the exact solution to this problem detailed in Appendix~\ref{ExactEWCS}, see Figure~\ref{fig:EWCSDyn}. 

\begin{figure}[htbp]
\centering
\includegraphics[width=.55\textwidth]{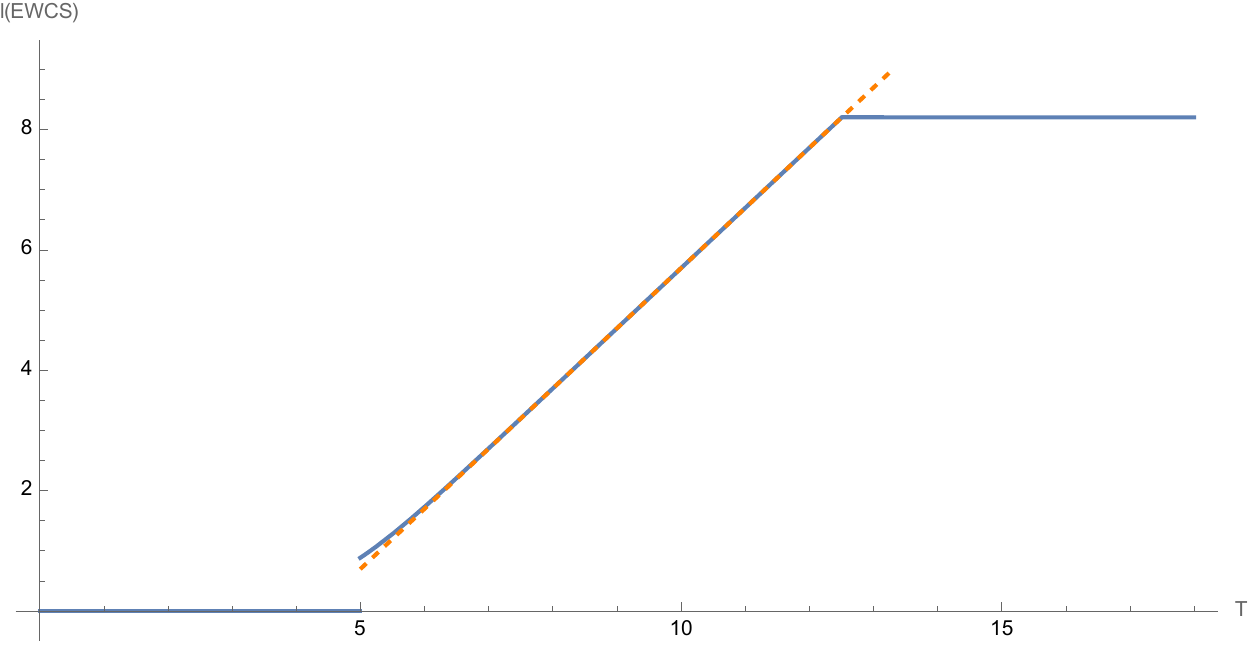}
\qquad
\caption{Time evolution of EWCS, for $\ell=15$, and $D=10$. At early time, the EWCS is 0, as the entanglement wedge is disconnected (Fig.~\ref{fig:hrts} left). After $T=\frac{D}{2}$, the RT surface connecting $P_2$ and $P_3$ saturates (Fig.~\ref{fig:hrts} right), and the EWCS starts to grow linearly with slope 1 in the scaling limit (dashed orange line). Notice that there is a small jump of EWCS length at $T=\frac{D}{2}$; this jump is negligible in the scaling limit. The EWCS then saturates at $\frac{\ell}{2}$ after $T=\frac{\ell+D}{2}$. In the scaling limit, the EWCS dynamics is described by~\eqref{GenMembraneEWCS2d}.\label{fig:EWCSDyn}}
\end{figure}

\section{Reflected entropy in the membrane effective theory}\label{sec:EWCS}
In section~\ref{sec:2dEWCS} we explored the dynamics of the reflected entropy in the context of $2d$ CFTs. Now, we will show how to describe the corresponding dynamics for the case of $d>2$. The setup is the same as that in Fig.~\ref{fig:hrts}, but the equations describing the HRT surfaces and EWCS are different due to the increase in dimensionality. While many of the features explored earlier remain true in higher dimensions, we also observe some important differences. 

While in the preceding sections we only showed computations in the scaling limit that is described by the (generalised) membrane description and relegated finite size computations to appendices, in this section we include finite size numerical computations in the main text, as we believe they may be of methodological interest.

\subsection{Time evolution of entanglement cross-section}
As indicated before, we consider the same configuration as in Fig.~\ref{fig:hrts}. The evolution of entanglement entropy in this setting was first studied in \cite{Hartman:2013qma}, and further analysed in \cite{Liu:2013iza,Liu:2013qca,Mezei:2016zxg,Mezei:2019sla}.

Following the aforementioned references, we parameterise the equilibrium HRT as $(u_1(z),x_1(z) )$. While there is no closed-form solution to the extremisation problem in $d>2$, we can use reparameterisation invariance and translational symmetry to express the solutions in terms of integrals as
\begin{equation}
\label{EqHRTEqn}
\begin{gathered}
u_1(z) = T - \int_0^z \frac{dz'}{a(z')b(z')},\\
x_1(z) = \frac{2\ell+D}{2} + \int_{\Tilde{z}}^{z} \frac{dz'}{b(z')\sqrt{a(z')}}\frac{z'^{d-1}}{\sqrt{\Tilde{z}^{2(d-1)}-z'^{2(d-1)}}}\,.\\
\end{gathered}
\end{equation}
The parameter \(\tilde{z} < 1\) denotes the deepest point reached by the HRT surface in the bulk. This parameter can, in turn, be related to the size of the entangling region by evaluating the integral near the asymptotic boundary, $z=0$. The boundary time at which the entanglement entropy is evaluated is denoted as $T$.

The non-equilibrium HRT surface is simply $(u_2(z),0)$. In this case, reparameterization invariance suffices to solve the extremisation equations as
\begin{equation}
\begin{gathered}
u_2(z) = u_{\text{EoW}}(z_0) - \int_{z_0}^{z} \frac{dz'}{a(z')b(z')}\left(1 - \frac{\sqrt{-a(z_0)}z'^{d-1}}{\sqrt{z_0^{2(d-1)}a(z')-z'^{2(d-1)}a(z_0)}}\right)\,,\\
x_2(z) = 0\,,
\end{gathered}
\end{equation}
where $u_{\text{EoW}}(z)$ is as in eq. \eqref{eq:EoWBrane} and $z_0$ denotes the point at which the surface intersects the end-of-the-world brane. In this case, the boundary time $T$ is related to the parameter $z_0$ by evaluating the integral at $z=0$.

The EWCS that intersects the end-of-the-world brane corresponds to a portion of the non-equilibrium HRT described above. Concretely, it is parameterised as
\begin{equation}\label{intForm}
\begin{gathered}
u_{\text{EWCS}}^{(1)}(z) = u_{\text{EoW}}(z_0') - \int_{z_0'}^z \frac{dz'}{a(z')b(z')}\left(1 - \frac{\sqrt{-a(z_0')}z'^{d-1}}{\sqrt{z_0'^{2(d-1)}a(z')-z'^{2(d-1)}a(z_0')}}\right)\,,\\
x_{\text{EWCS}}^{(1)}(z) = \frac{2\ell+D}{2},
\end{gathered}
\end{equation}
but now the parameter $z_0'$ is determined by the intersection condition
\begin{equation}
\label{Intersection1}
u_{\text{EWCS}}^{(1)}(\Tilde{z}) = u_1(\Tilde{z})\,,
\end{equation}
where $\tilde{z}$ is the same parameter as in \eqref{EqHRTEqn}.

The area of the surface is
\begin{equation}
\label{ECS1}
\mathcal{A}_{\text{EWCS}}^{(1)}(z_0') = \int_{\Tilde{z}}^{z_0'} \frac{dz}{z^{d-1}b(z)}\frac{z_0'^{d-1}}{\sqrt{z_0'^{2(d-1)}a(z)-z^{2(d-1)}a(z_0')}}\,.
\end{equation}
By solving the intersection condition \eqref{Intersection1}, and evaluating the integral \eqref{ECS1} at different values of boundary time $T$, one can obtain the time-dependence of the surface. The result is a monotonic increasing function that is approximately linear for the parameter choices we explored.

The other possible EWCS is a surface connecting the two HRTs, without reaching the EoW brane. Unlike the previous case, this does not correspond to a portion of a known HRT. Fortunately, we can employ the symmetries of the problem to solve the extremisation equations in a similar fashion.

We choose to parameterise this surface as $(u_{\text{EWCS}}^{(2)}(z),x_{\text{EWCS}}^{(2)}(z))$. The area of this surface is given by the functional
\begin{equation}\label{eq:EWCS2Func}
\mathcal{A} = \int_{z_1}^{z_2}\frac{dz}{z^{d-1}}\sqrt{-a(z)\left(\frac{du}{dz}\right)^2- \frac{2}{b(z)}\frac{du}{dz} + \left(\frac{dx}{dz}\right)^2}\,.
\end{equation}
The equations of motion extremising this expression are both reparameterisation and translation invariant, hence we have a couple of conserved quantities denoted as $(j,e)$. In terms of these constants, the solution can be written as
\es{SatEWCSHigherD}{
u_{\text{EWCS}}^{(2)}(z) &= u_2(z_2) - \int_{z_2}^{z} \frac{dz'}{a(z')b(z')}\left(1-\frac{ez'^{d-1}}{\sqrt{H(z')}}\right)\,,\\
x_{\text{EWCS}}^{(2)}(z) &= - \int_{z_2}^{z} dz' \frac{z'^{d-1}}{b(z')\sqrt{H(z')}}\,,\\
H(z) &= \left(e^2 - a(z)\right)z^{2(d-1)} + j^2 a(z)\,,
}
where $(z_1,z_2)$ are the values of $z$ at which the surface intersects the corresponding HRT surface. The integration constants were chosen to guarantee intersection with the non-equilibrium surface.

Up to this point, the calculation is not dissimilar to those of usual HRT surfaces. The main difference arises because this new surface does not obey boundary conditions at $z=0$, instead we demand intersection with the equilibrium surface:
\begin{equation}
\label{Intersection2}
\begin{gathered}
u_{\text{EWCS}}^{(2)}(z_1) = u_1(z_1)\,,\\
x_{\text{EWCS}}^{(2)}(z_1) = x_1(z_1)\,.
\end{gathered}
\end{equation}
The solutions to these conditions provide a family of extremal curves. 

For HRT surfaces, we are able to fix the integration constants, $\tilde{z}$ and $z_0$, by symmetry considerations or by imposing intersection with the end-of-the-world brane. For the new surface associated with the EWCS, we require the intersection with the both HRT surfaces to occur perpendicularly; this guarantees that the surface is not only extremal but also minimal. This perpendicularity condition can be used to determine the integration constants $(j,e)$. We then solve the intersection conditions numerically in order to fix $(z_1,z_2)$.

The area of the surface can be written as
\begin{equation}
\mathcal{A}(z_1,z_2) = \int_{z_1}^{z_2} dz \frac{j}{z^{d-1}b(z)\sqrt{H(z)}}\,,
\end{equation}
which we evaluate numerically after solving the intersection conditions.

\begin{figure}[H]
\centering
\begin{tikzpicture}
\node at (0,0) {\includegraphics[scale=0.4]{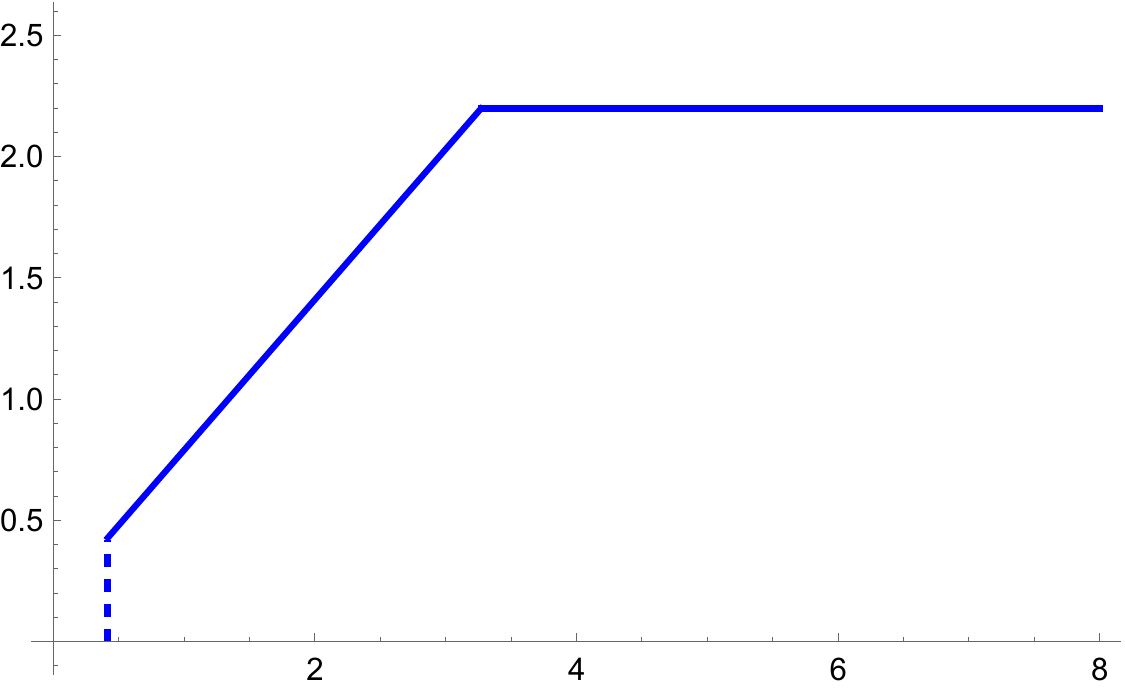}};
\node at (-3.8,2.5) {$\frac{S_R}{2}$};
\node at (4,-2.3) {$T$};
\end{tikzpicture}
\caption{Time evolution of EWCS for $D=1.07$ and $\ell=1.95$ in $d=4$.}
\label{fig:ECSTimeDependence}
\end{figure}

The first candidate surface for the EWCS directly depends on the position of the EoW brane. The second candidate only depends on the position of the EoW brane through the change in shape of the non-equilibrium HRT $u_2(z)$; it approaches a constant as $T\to\infty$. The total time evolution of the EWCS is depicted in Fig.~\ref{fig:ECSTimeDependence}. After the linear growth phase, the time dependence becomes negligible for the choice of parameters considered.

\subsection{Membrane theory for reflected entropy}\label{sec:MemRef}

Let us now develop a description for the general behaviour of the reflected entropy as a function of time, assuming the membrane theory for entanglement entropy is valid. We expect the membrane description to be accurate only in the scaling limit. In order to ease the comparison between the exact numerical results and the membrane prediction, we parameterise the size of the finite interval, $\ell$, as $\ell=\eta D$, with $\eta>1$. We can then take $D\gg1$.

First, we have to determine the scaling limit of the HRT surfaces. The non-equilibrium HRT does not move in $x$, hence it corresponds to the $v=0$ membrane (and most of it sits at $z=z_*$ in the bulk). 
We have seen in section~\ref{MembraneStatic2d} that in $d=2$, the saturated RT surfaces become cones of
of slope $\pm 1$ under the membrane mapping. In $d\geq 2$ dimensions this result generalises as follows:
we can approximate \eqref{EqHRTEqn} by expanding the integrand in $z$ around $\tilde z \to 1^-$, giving
\es{ExpandedInts}{
u(z)&=T-\frac{1}{b(1)a'(1)}\log (1-z)+\dots\,,\\
x(z)&={2\ell+D\over 2}-\frac{\log (1-z)}{b(1)\, \sqrt{-2(d-1)a'(1)}}\,,
}
and eliminating $z$ we get
\es{ExpandedInts2}{
x(u)&=\sqrt{-{a'(1)\over 2(d-1)}}u+\text{const}=v_B u+\text{const}\,,
}
where we used \eqref{vB}.
Therefore, projections of static RT surfaces to the boundary along constant infalling time are $\emph{cones}$ of slope $v_B$. These conclusions remain true for arbitrary shapes as can be inferred from the results from~\cite{Liu:2013una,Mezei:2016wfz,Mezei:2016zxg}. See Appendix~\ref{sec:RTsphereBessel} for a calculation with spherical entangling subregions. 

\begin{figure}[htbp]
\centering
\includegraphics[width=.4\textwidth]{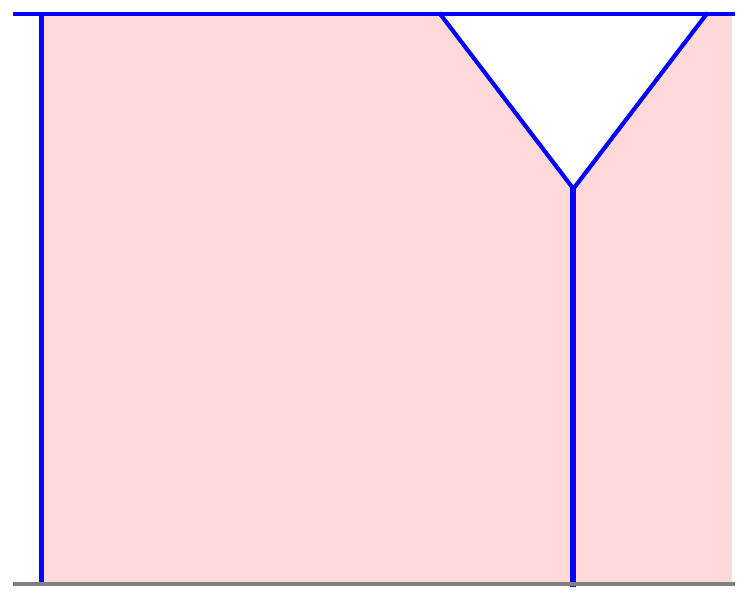}\hspace{0.5cm}
\includegraphics[width=.4\textwidth]{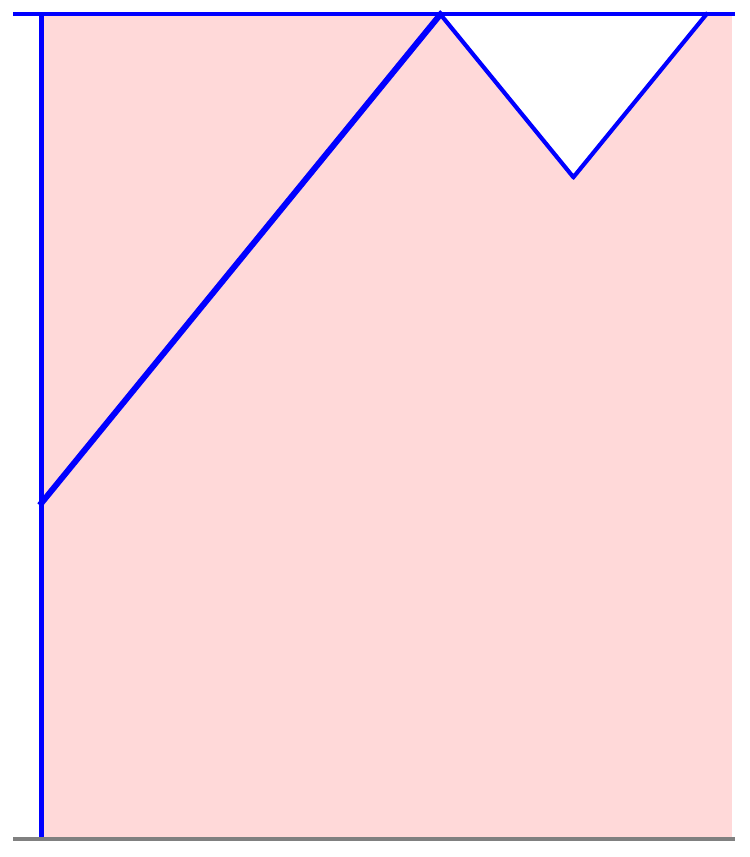}
\qquad
\caption{Projection of EWCS in $d=4$ along constant infalling time in the scaling limit, for $\ell=1.5\Lambda$ and $D=\Lambda$. The entire plots are in blue, as all surfaces are at $z=O(1)$. We plot the EWCS as thick to contrast with HRT. $\emph{Left}:$ Plot at $T=2\Lambda$. The non-equilibrium EWCS starts from the tip of the static RT surface, and then enters the interior and ends on the EoW brane (plotted in gray, since HRT and EWCS can end on it at any depth). Its length grows linearly in time with slope $v_E$. $\emph{Right}:$ Plot at $T=3.1\Lambda$. The equilibrium EWCS skims the horizon from the interior, its projection is therefore a line with slope $v_B$. \label{fig:MembraneEWCS}}
\end{figure}

We can now propose a membrane description for the reflected entropy for the same configuration above. As observed, at early times the entanglement wedge is disconnected and so its cross section vanishes. This continues until the smaller HRT transitions from linear growth to equilibrium, according to the membrane description of entanglement, this occurs at time $t=\frac{D}{2v_E}$. After this time, the cross section is described by a surface connecting the equilibrium RT surface to the end-of-the-world brane. Such a surface is described by the same type of membrane as the linear growth of entanglement entropy. Finally, this linear growth continues until the dominant surface is the one connecting the two HRTs. In the scaling limit it is a straight line with slope $v_B$ and its  contribution is equal to $\ell$. See Fig.~\ref{fig:MembraneEWCS}.  A check of the consistency of the membrane description with the full HRT surface is that the EWCS connecting the two HRTs~\eqref{SatEWCSHigherD} is dominated by the part on the horizon in the large $D$ limit, see Figure~\ref{fig:ECSPlot} which is plotted using~\eqref{SatEWCSHigherD}. A calculation similar to~\eqref{ExpandedInts2} then shows that the projection of this EWCS to the boundary along constant $u$ is indeed a line of $v_B$ slope.\footnote{Note that the membrane theory only characterises this surface under the scaling limit and does not encode its local geometry near the connection points to the two HRTs. To investigate these details, one needs to invoke the full equations of motion~\eqref{SatEWCSHigherD}.}  

In summary, the membrane description of the reflected entropy is
\begin{equation}
\label{MembraneEWCS}
\frac{S_R}{2} = 
\begin{cases}
0 \ &T < \frac{D}{2v_E}\,, \\ 
v_E \left(T - \frac{D}{2v_B}\right) \ \hspace{0.5cm} &\frac{D}{2v_E} < T < \frac{\ell}{v_E}+\frac{D}{2v_B}\,, \\ 
\ell  \  &T > \frac{\ell}{v_E}+\frac{D}{2v_B}\,.
\end{cases}
\end{equation}

\begin{figure}[h!]
\centering
\begin{tikzpicture}
\node at (0,0) {\includegraphics[scale=0.4]{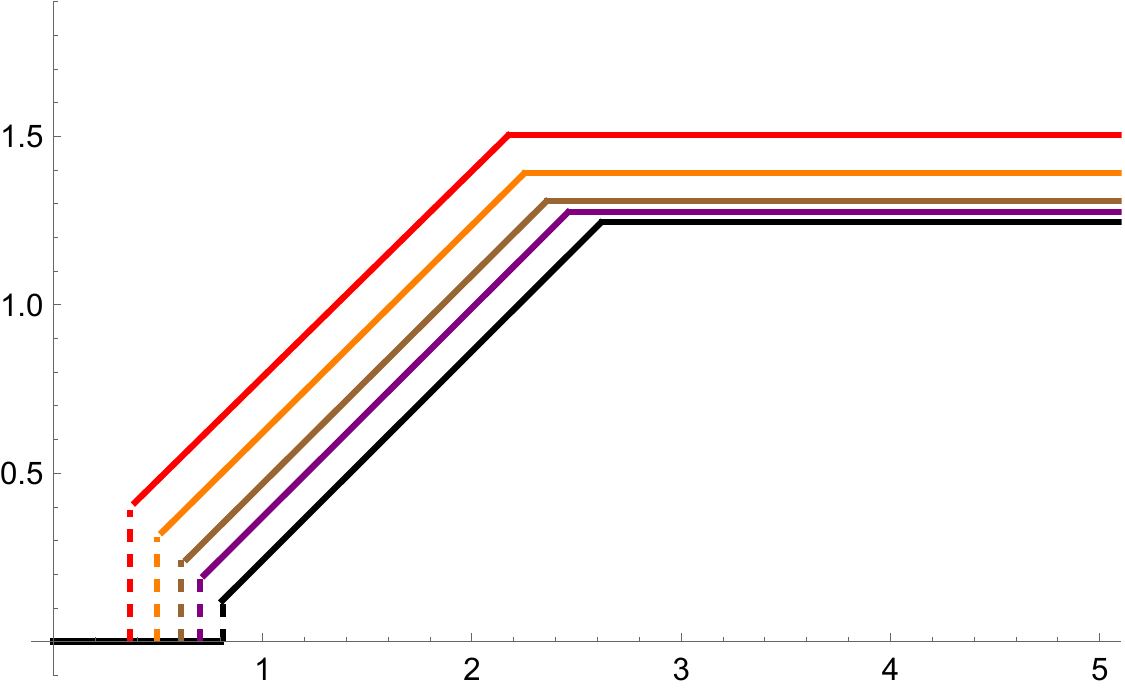}};
\node at (4.5,-2) {$T/D$};
\node at (-3.5,2.5) {$S_R/2D$};
\end{tikzpicture}
\caption{Dynamics of entanglement cross section for $D=1.07$ (red), $D=1.3$ (orange), $D=1.95$ (brown), and $D=3.3$ (purple). This is compared against the membrane prediction (black). As noted in the main text, we emphasise the numerical limitation in solving the intersection problem, which permits us to use only relatively small values of $D$.}
\label{fig:MembraneComp}
\end{figure}

We can compare the membrane theory prediction against the holographic calculation outlined earlier, see Fig.~\ref{fig:MembraneComp}.

As noted earlier, we expect the membrane theory to be accurate in the limit $D\gg 1$. This seems to be supported by the numerical results depicted in Fig.~\ref{fig:MembraneComp}, however we are somewhat constrained from a perfect agreement due to numerical limitations. In Appendix~\ref{App:ScalingECS}, we further explore this point.

\subsection{Comparison between the generalised and ordinary membrane results for reflected entropy}\label{sec:diffs}

\begin{figure}[htbp]
\centering
\includegraphics[width=.45\textwidth]{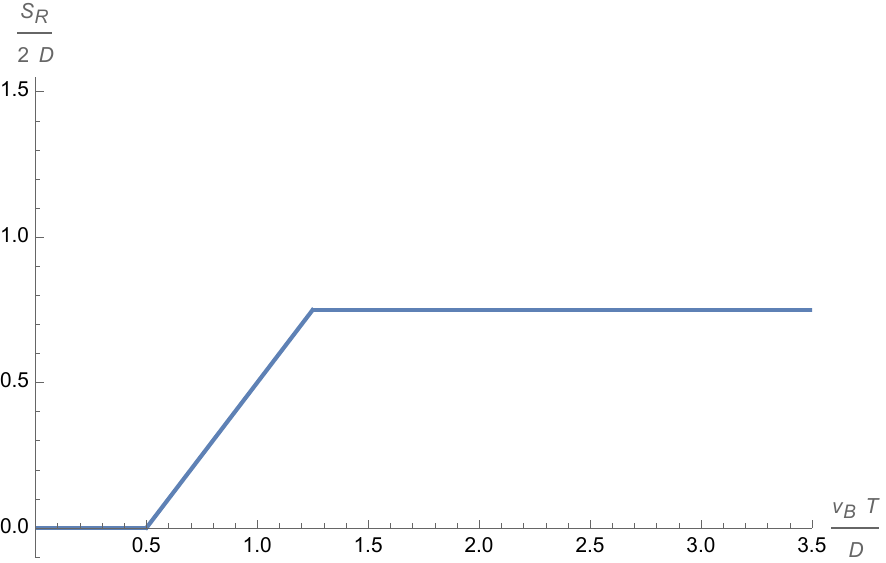}\hspace{0.5cm}
\includegraphics[width=.45\textwidth]{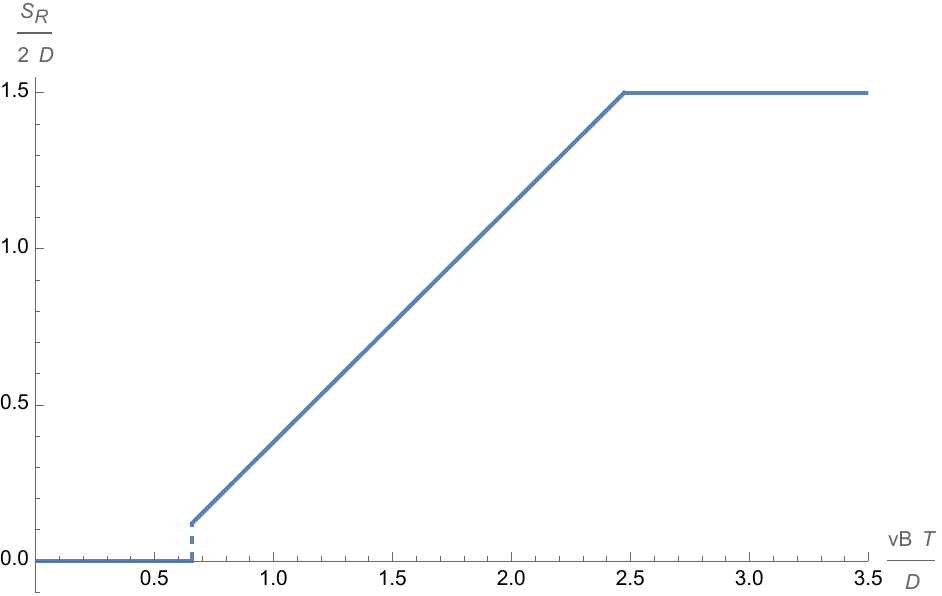}
\qquad
\caption{Time evolution of reflected entropy in $d=2$ (left) and $d=4$ (right) for $\ell/D=3/2$. We use the same ranges on the horizontal and vertical axes to facilitate the comparison. \label{fig:EWCS2d4d}}
\end{figure}

Let us compare the results for reflected entropy from the generalised \eqref{GenMembraneEWCS2d} and ordinary membrane results  \eqref{MembraneEWCS} applicable to 2d and higher-dimensional CFTs (and other generic chaotic systems). There are key differences that we illustrate on Fig.~\ref{fig:EWCS2d4d}. On the horizontal axis we put $v_B T/D$ to make them look as similar as possible. The first difference that we notice that the membrane theory result is not continuous, it starts with a jump of size 
\es{JumpSize}{
{\Delta S_R\ov 2D}=\frac{1}{2}\left(1-\frac{v_E}{v_B}\right)D\,, \qquad \text{at ${v_B T\ov D}={v_B\ov 2 v_E}$.}
}
Note however that this difference would vanish if $v_E=v_B$ were to hold, i.e.~in the limit of the degenerate membrane tension ${\cal E}(v)=1$. The slope of the two curves are also different, by a factor $v_E/v_B$, which again would go away for $v_E=v_B$. However, the plateau values of the curves $\ell/2D$ and $\ell/D$ do not match even for  the degenerate membrane tension: \emph{this is the prediction that genuinely distinguishes the two effective theories}.

\section{Entanglement membrane and 2d RG flows}\label{KasnerSec}
In section~\ref{Sec:EEMembrane}, we have explored the generalised entanglement membrane in 2d CFT with a degenerate membrane tension, and compared the resulting equation of motion with that of ordinary membrane theory~\cite{Mezei:2018jco}. In this section, we show that deforming a 2d CFT with a relevant operator leads to a
a renormalisation group (RG) flow between the generalised and ordinary membrane theories and the emergence of a  
 non-degenerate membrane tension.  In the small deformation limit we maintain full analytic control over this RG interpolation. 

\subsection{Holographic RG flow from BTZ to AdS$_3$-Kasner}
\label{sec:Basic Setup}
We consider a two-sided planar BTZ black hole with a relevant deformation. Such a deformation breaks the infinite-dimensional conformal symmetry of the dual (1+1)-dimensional field theory. The action is given by
\begin{align}
    \mathcal{L}=\frac{1}{2}(R+2)-\frac{1}{2}(g^{ab}\partial_a\phi \partial_b\phi+m^2\phi^2)\,.\label{action}
\end{align}
Here we choose $m^2=-\frac{3}{4}$, which is above the Breitenlohner-Freedman bound $m^2_\text{BF}=-1$~\cite{Breitenlohner:1982jf}. It is convenient to rewrite the metric \eqref{blackBraneInf} in the following form: 
\begin{align}
    ds^2=\frac{1}{z^2}\Big(-f(z)e^{-\chi(z)}du^2-2e^{-\chi(z)/2}dudz+dx^2\Big)\,,\label{metricInfalling}
\end{align}
where the horizon is at $z_+=1$, hence $f(1)=0$. Without the relevant deformation, the metric \eqref{metricInfalling} becomes the planar BTZ metric, where $f(z)=1-z^2$ and $\chi(z)=0$. Unlike BTZ, the metric \eqref{metricInfalling} does contain a curvature singularity at $z=\infty$. 
The Einstein-scalar equations of motions can be obtained by varying the action \eqref{action}:\footnote{See~\cite{Caceres:2021fuw} for the Einstein-scalar equations of motions in general $d\geq 2$ dimensions with arbitrary $\Delta$. }
\begin{align}
    \phi''+\Big(\frac{f'}{f}-\frac{1}{z}-\frac{\chi'}{2}\Big)\phi'+\frac{3}{4}\frac{\phi}{z^2f}&=0\label{EoM1},\\
    \chi'-2\frac{f'}{f}-\frac{3}{4}\frac{\phi^2}{zf}-\frac{4}{zf}+\frac{4}{z}&=0\label{EoM2},\\
    \chi'-z\phi'^2&=0\label{EoM3}.
\end{align}

The scalar field and the geometry behave asymptotically ($z\to 0$) as~\cite{Witten:1998qj,Gubser:1998bc,deHaro:2000vlm}: 
\begin{align}
    \phi=\phi_0 z^{\frac{1}{2}}+\langle O\rangle z^{\frac{3}{2}}+..., && \chi=\frac{\phi_0^2}{4}z+\frac{3}{4}\phi_0\langle O\rangle z^2+..., && e^{-\chi}f=1-\langle T_{tt}\rangle z^2+...\label{falloffs},
\end{align}
where $\phi_0$ serves as the source for the dual boundary operator whose expectation value is $\langle O\rangle$. As $-\frac{d^2}{4}<m^2<-\frac{d^2}{4}+1$, both modes are normalizable~\cite{Klebanov:1999tb}. Here we implement standard quantisation, with the dimension of $O$ given by $\Delta=\frac{d}{2}+\sqrt{\frac{d^2}{4}+m^2}=\frac{3}{2}$.
The temperature of the black hole is: 
\begin{align}
    T=\frac{|f_+'|e^{-\frac{\chi_+}{2}}}{4\pi}\label{temperature2},
\end{align}
where $\chi_+$ and $f_+'$ are the horizon value of $\chi$ and $f'$, respectively. 

From the equations of motion \eqref{EoM1}-\eqref{EoM3}, one can read off the following near-singularity ($z\to\infty$) behaviours:  
\begin{align}
    \phi=\sqrt{2}c\log{z}+..., && \chi=2c^2\log{ z}+\chi_{\infty}+...,&& f=-f_{\infty}z^{2+c^2}+...\,,\label{Kasner}
\end{align}
where the constant $c=0$ for BTZ. Let $z^{2+c^2}=\frac{1}{\tau^2}$, the metrics \eqref{metricInfalling} then takes the Kasner form~\cite{Frenkel:2020ysx,Kasner:1921zz,Belinski:1973zz} in the interior 
\begin{align}
    ds^2\sim -d\tau^2+\tau^{2p_{t}}dt^2+\tau^{2p_{x}}dx^2\,, && \phi\sim -\sqrt{2}p_{\phi}\log{\tau}\,,
\end{align}
where the Kasner exponents are given by 
\begin{align}
    p_t=\frac{c^2}{2+c^2}\,, && p_x=\frac{2}{2+c^2}\,, && p_{\phi}=\frac{2c}{2+c^2}\,.
\end{align}
We have $p_t+p_x=1$, $p_t^2+p_x^2+p_{\phi}^2=1$.  For the BTZ black hole, $p_t=p_{\phi}=0$ and $p_x=1$. One can think of $p_t$ as measuring the deformation of the geometry away from BTZ and therefore characterising the relevant deformation. 

\begin{figure}[htbp]
\centering
\includegraphics[width=.48\textwidth]{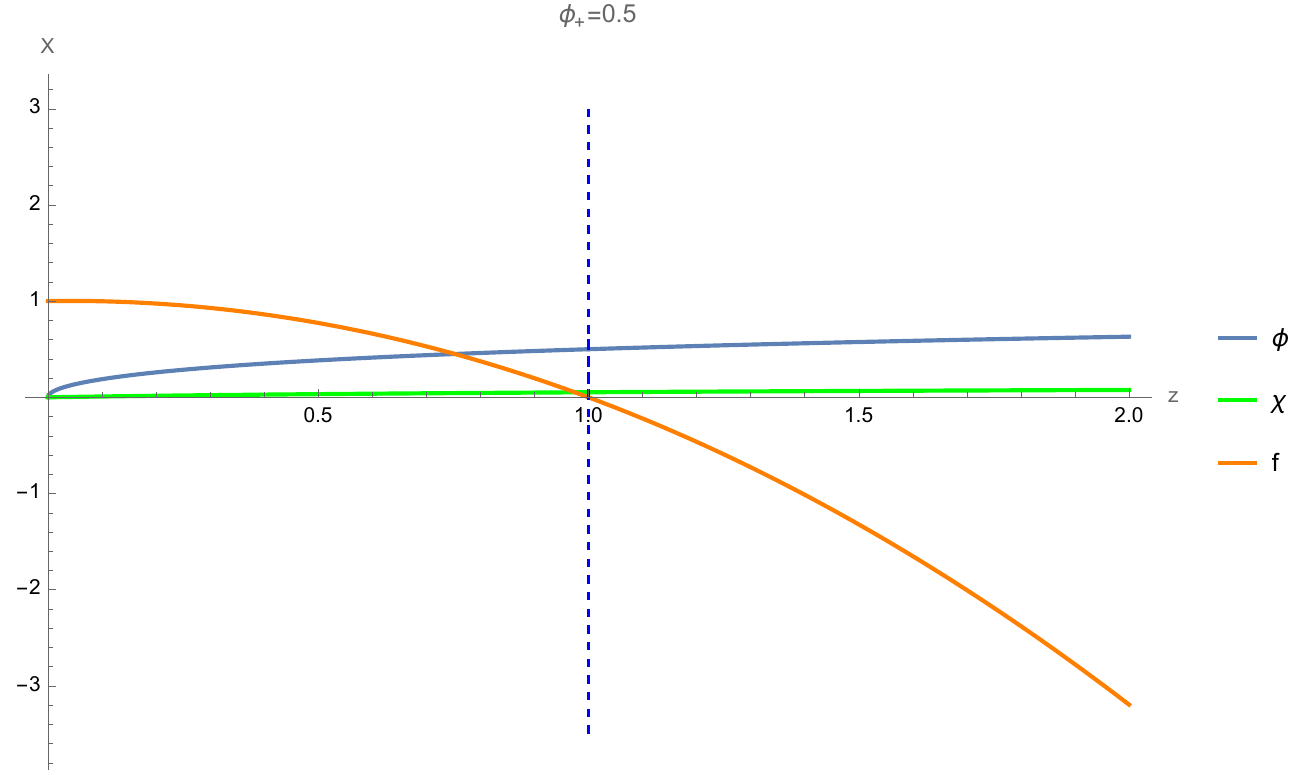}
\includegraphics[width=.45\textwidth]{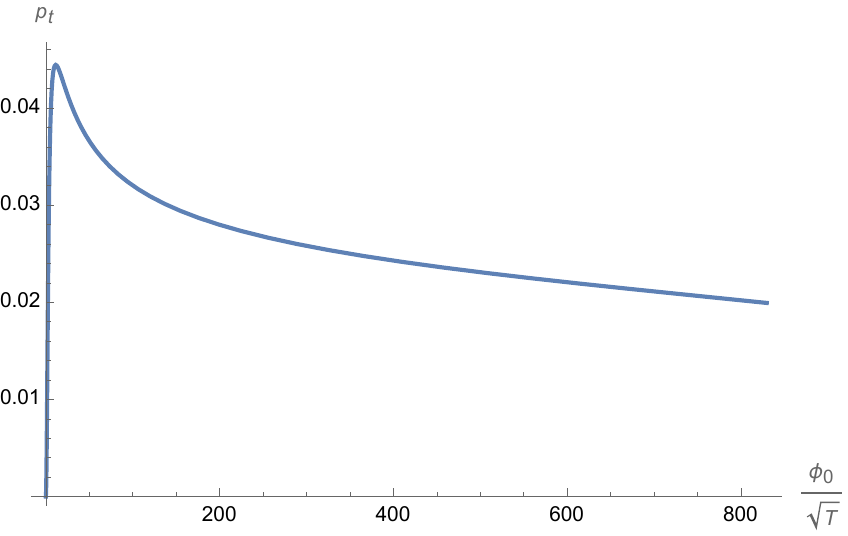}
\qquad
\caption{A demonstration of the shooting solutions. $\emph{Left}$: Plot of $\phi$ (blue), $\chi$ (green) and $f$ (orange) when $\phi_+=0.5$. The vertical dashed blue line is the black hole horizon $z=1$. $\emph{Right}$: Plot of the Kasner exponent $p_t$ with respect to the dimensionless parameter $\frac{\phi_0}{\sqrt{T}}$. Notice that for larger values of $\frac{\phi_0}{\sqrt{T}}$, $p_t$ returns to its BTZ value $p_t=0$. \label{fig:ShootingEgl}}
\end{figure}

\begin{figure}[htbp]
\centering
\includegraphics[width=.65\textwidth]{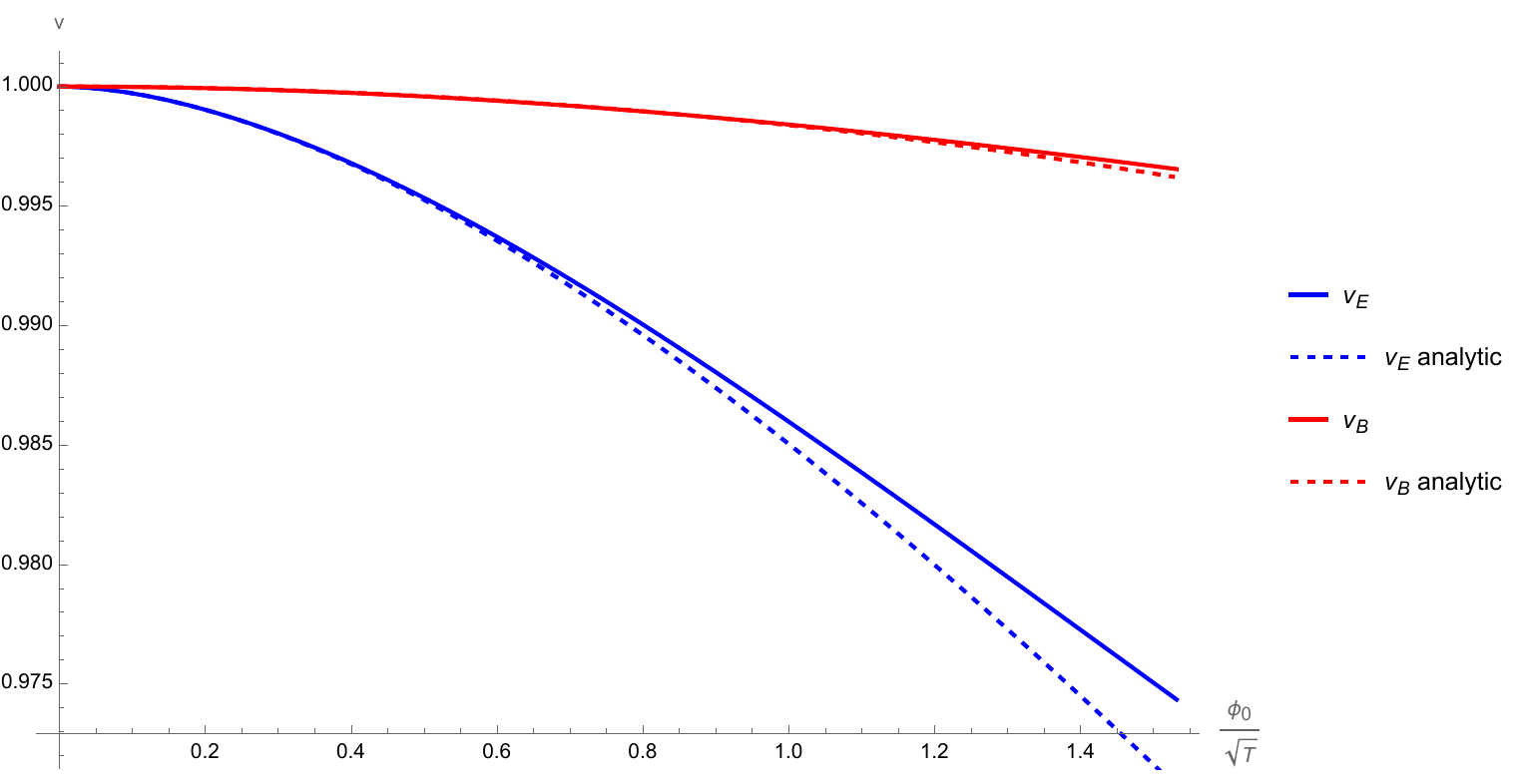}
\qquad
\caption{Plot of $v_E$ (blue) and $v_B$ (red) with respect to the dimensionless parameter $\frac{\phi_0}{\sqrt{T}}$. The solid lines are numerical solutions, whereas the dashed lines are analytic expansions in the small-$\phi$ limit. At $\phi_0=0$, i.e. when the geometry is planar BTZ black hole, $v_E=v_B=1$. When $\phi_0>0$, we have $v_E<v_B$ due to the constraint from null energy condition. When $\frac{\phi_0}{\sqrt{T}}$ is small, the numerical solutions of both $v_E$ and $v_B$ are well approximated by small-$\phi$ perturbation theory. \label{fig:vEvB}}
\end{figure}

\begin{figure}[htbp]
\centering
\includegraphics[width=.5\textwidth]{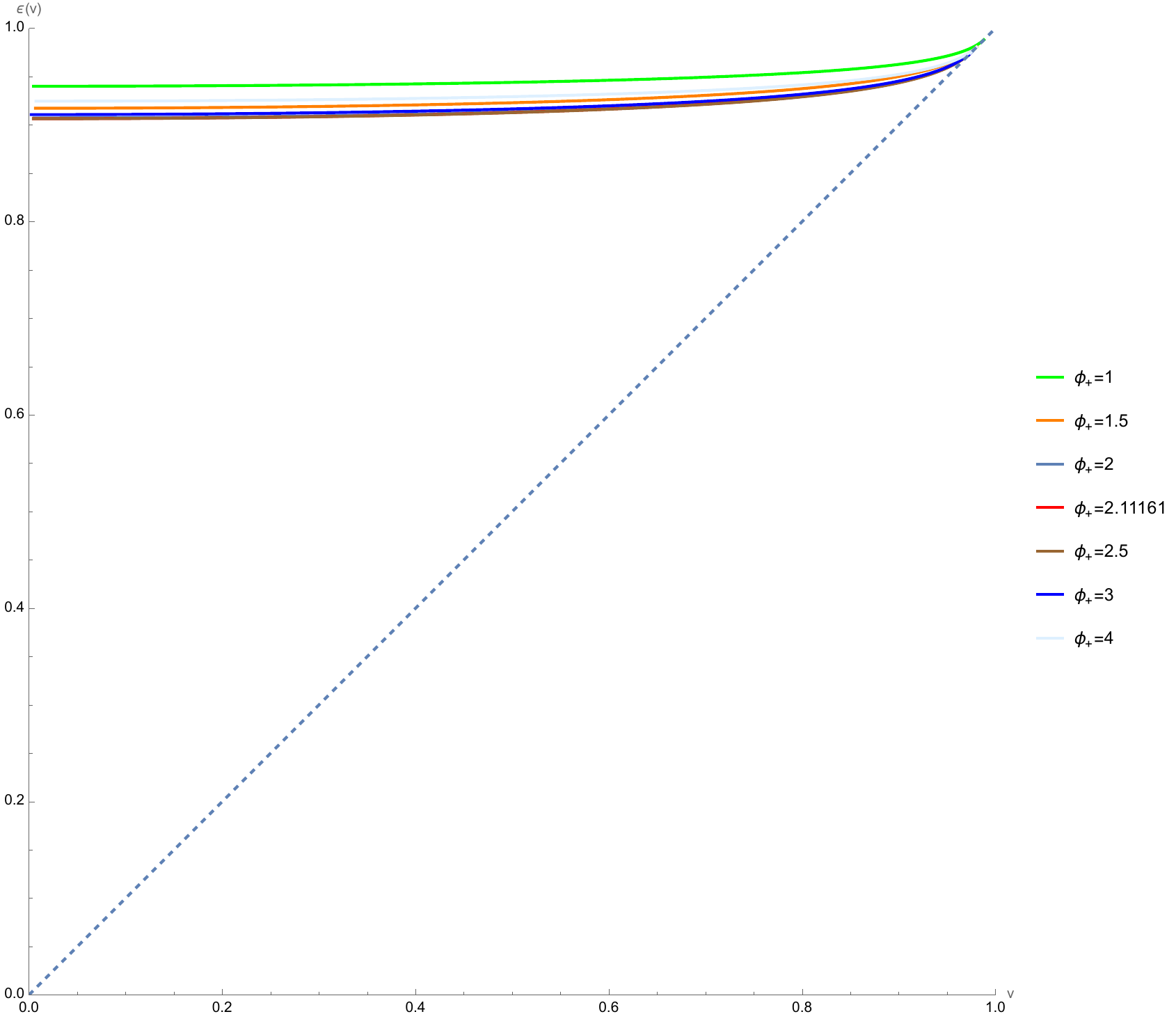}
\qquad
\caption{Plot of $\mathcal{E}(v)$ in AdS$_3$-Kasner geometry, for $\phi_+=1,1.5,2,,2.11,2.5,3,4$. The dashed line is at $45^{\circ}$. As $\phi_+$ gets large, the membrane tension returns to its undeformed degenerate BTZ value $\mathcal{E}(v)=1$, see Fig.~\ref{fig:ShootingEgl} (right). \label{fig:Ev}}
\end{figure}

The equations of motion \eqref{EoM1}-\eqref{EoM3} can be solved numerically using the so-called ``shooting" method~\cite{Hartnoll:2008vx,Hartnoll:2008kx}. First, expand $\phi$, $\chi$ and $f$ around the horizon $z_+=1$:
\begin{align}
    \phi(z)=\sum_{n=0}^N\phi_{(n)}(z-1)^n\,, && \chi(z)=\sum_{n=0}^N\chi_{(n)}(z-1)^n\,, && f(z)=\sum_{n=1}^N f_{(n)}(z-1)^n \,\label{shootingExpansion}
\end{align}
where $N$ is some positive integer,  and we have used $f(1)=0$. Then, plugging \eqref{shootingExpansion} into the equations of motion \eqref{EoM1}-\eqref{EoM3} allows one to express all the expansion parameters in \eqref{shootingExpansion} in terms of the horizon value of the scalar field $\phi_{(0)}=\phi_+$. One can therefore integrate the field equations numerically in terms of the dimensionless parameter $\frac{\phi_0}{\sqrt{T}}$ from an $\epsilon$-ball centred at the horizon $z=1$. Here, in addition to integrating to the boundary $z=0$~\cite{Hartnoll:2008vx,Hartnoll:2008kx}, we also integrate towards the singularity $z=\infty$~\cite{Frenkel:2020ysx}. A demonstration of the shooting solution is presented in Fig.~\ref{fig:ShootingEgl}, where we show the plots of $\phi$, $\chi$, and $f$, as well as the Kasner exponent $p_t$.\footnote{Notice that $p_t>0$ in $d=2$, whereas for $d\geq 3$, $p_t<0$~\cite{Frenkel:2020ysx}. } Notice that as $\frac{\phi_0}{\sqrt{T}}$ increases from 0 to $\infty$, $p_t$ first increases and then decreases after reaching its maximal value, indicating that the geometry returns to BTZ in the large $\phi_+$ limit. The translation between the metric functions featuring in \eqref{blackBraneInf} and \eqref{metricInfalling} is $a(z)=f(z)e^{-\chi(z)}$ and $b(z)=e^{\chi(z)/2}$, hence $a(z),\, b(z)$ move away from their BTZ values under the relevant deformation, and $v_E$ \eqref{vE} and $v_B$ \eqref{vB} also become non-trivial functions of $\frac{\phi_0}{\sqrt{T}}$, see Fig.~\ref{fig:vEvB}. For the same reason, the membrane tension $\mathcal{E}(v)$ \eqref{AreaFunct2} is no longer degenerate. See Fig.~\ref{fig:Ev} for a plot of numerical membrane tensions in AdS$_3$-Kasner geometry. It is found that as $\phi_+$ increases from 0, $\mathcal{E}(v)$ starts to deflect from $\mathcal{E}(v)=1$ and becomes non-trivial; for large $\phi_+$, however, we observe that $\mathcal{E}(v)$ asymptotes to $\mathcal{E}(v)=1$. This is likely because as $\phi_+\to\infty$ the Kasner exponent $p_t\to 0$, the BTZ value.

\subsection{Small $\phi$ perturbation theory}
\label{smallPhiPertTh}
In the previous subsection, we have solved the equations of motion \eqref{EoM1}--\eqref{EoM3} numerically for generic perturbations. For small values of $\phi$, however, one can establish a perturbation theory and find analytic solutions order by order. Let us consider the leading order perturbation in $\phi$
\begin{align}
    \phi(z)=\lambda \phi_1(z)\label{phiLambdaExp},
\end{align}
where  $\lambda$ is a small constant. From the equations of motion \eqref{EoM3} and the action \eqref{action}, both $\chi(z)$ and $f(z)$ are second order in $\lambda$, 
\begin{align}
    \chi(z)&=\lambda^2 \chi_1(z)\label{chiLambdaExp},\\
    f(z)&=1-z^2+\lambda^2 f_1(z)\label{fLambdaExp}.
\end{align}
Plugging \eqref{phiLambdaExp}-\eqref{fLambdaExp} into the equations of motions \eqref{EoM1}-\eqref{EoM3}, one can obtain   $\phi_1(z)$ in closed form, and obtain integral expressions for $\chi_1(z)$ and $f_1(z)$,
\begin{align}
    \phi_1(z)&=\frac{2}{\pi}\sqrt{z}K\Big(\frac{1-z}{2}\Big)\,,\label{phi1}\\
    \chi_1(z)&=\int_0^{z}dz' \frac{\left((z'+1) K\left(\frac{1-z'}{2}\right)-2 z'
   E\left(\frac{1-z'}{2}\right)\right)^2}{\pi ^2 \left(z'^2-1\right)^2}\,, \label{chi1IntExpr}\\
   f_1(z)&=z^2 \int_1^{z}dz' \frac{-(2 z'^2+z'-1) K\left(\frac{1-z'}{2}\right)^2+2 z' (z'+1)
   K\left(\frac{1-z'}{2}\right) E\left(\frac{1-z'}{2}\right)-2 z'^2
   E\left(\frac{1-z'}{2}\right)^2}{\pi ^2 z'^2 \left(z'^2-1\right)}\,,  \label{f1IntExpr}
\end{align}
where $K(z)$ and $E(z)$ denote the complete elliptic integral of the first and second kind, respectively. Integrating \eqref{EoM3} from 0 to $z$ in \eqref{chi1IntExpr} ensures $\chi(0)=0$. When $\phi_+$ is small, one can check that~\eqref{phi1} matches well with numerical solution. According to \eqref{phi1}, $\phi(1)=\lambda$ at the horizon $z=1$. From \eqref{phi1}, one learns that $\phi(z)$  behaves near the boundary $z\to 0$ as 
\begin{align}
    \phi(z)\to\lambda\frac{ \Gamma \left(1/4\right)}{\sqrt{2 \pi }\  \Gamma
   \left(3/4\right)}z^{\frac{1}{2}}+\lambda\frac{ \Gamma \left(-1/4\right)}{8 \sqrt{2
   \pi }\ \Gamma \left(5/4\right)}z^{\frac{3}{2}}+...,\label{phi1bdy}
\end{align}
and near the singularity $z\to \infty$ as
\begin{align}
    \phi(z)\to\lambda\frac{\sqrt{2}}{\pi }\log{z}+\dots. \label{phi1sing}
\end{align}
Comparing \eqref{phi1bdy} to \eqref{falloffs}, and \eqref{phi1sing} to \eqref{Kasner} then yields a relation among the small perturbation parameter $\lambda$, the value of $\phi$ at the horizon $\phi_+$, the falloff $\phi_0$, as well as the Kasner parameter $c$, 
\begin{align}
    \lambda=\phi_+=\pi c= \sqrt{2 \pi }\ \frac{\Gamma
   \left(3/4\right)}{\Gamma \left(1/4\right)} \phi_0\,.
\end{align} 


The integral expressions \eqref{chi1IntExpr} and \eqref{f1IntExpr} make it hard to derive a closed form expression for $\mathcal{E}(v)$ \eqref{AreaFunct2} even in perturbation theory. To get more insight into its behaviour, in the next subsection we expand $\mathcal{E}(v)$ around $v=0$ and $v=v_B$ to leading non-trivial orders in $\lambda$. As $v^2=c(z)$ \eqref{projectionz} is monotonically decreasing in $z$, these expansions take place around $z=z_*$ and $z=1$, respectively. 

\subsubsection{Expansion near the singularity}\label{Expansion near the singularity}
Let us first consider the expansion of $\mathcal{E}(v)$ at $v=0$ or equivalently $z=z_*$, where $\mathcal{E}(0)=v_E$. Intuitively, $z_*$ is large for small $\lambda$, as $z_*=\infty$ in the undeformed BTZ black hole. Therefore, to find the expansion of $\mathcal{E}(v)$ at $v=0$, one can start by studying the behaviours of $\chi$ and $f$ at large $z$:
\begin{align}
    \chi_1(z)&\approx \frac{2}{\pi^2}\log{z}+\frac{-2-\pi+12\log{2}}{2\pi^2}\,,\label{chi1C}\\
    f_1(z)&\approx -\frac{z^2}{\pi^2}(\log{z}+3\log{2})\,.\label{f1C}
\end{align}
\eqref{chi1C} and \eqref{f1C} then allow us to find the small $\lambda$ expansion of $v^2(z)$ \eqref{projectionz} to $O(\lambda^2)$ 
\begin{align}
    v^2(z)\approx 1-\frac{1}{2\pi^2}\lambda^2 z^2 &&
    \Rightarrow && z(v^2)=\frac{\sqrt{2}\pi}{\lambda}\sqrt{1-v^2}\,,\label{zvhor}
\end{align}
which is a monotonically decreasing function in $z$ deviating from $v^2=1$ in BTZ. $z_*$ is given by the solution of $v^2(z)=0$, 
\begin{align}
    z_*=\frac{\sqrt{2}\pi}{\lambda}\,.\label{zStar}
\end{align}
We see that the relevant deformation \eqref{phiLambdaExp} of BTZ pulls $z_*$ from $\infty$ to a finite value. The entanglement dynamics in AdS$_3$-Kasner geometry is thus in reminiscent of those in planar AdS$_{d+1}$-Schwarzschild black branes in $d\geq 3$ dimensions, in the sense that at late time, the HRT surface computing entanglement entropy stays at a finite $z_*$ instead of moving towards $z=\infty$. Unlike $z_{*}\in O(1)$ in AdS$_{d+1}$-Schwarzschild, however, $z_*$ in AdS$_3$-Kasner~\eqref{zStar} is large for small perturbations. When $\lambda\to 0$, $z_*\to\infty$, and the entanglement dynamics returns to that in BTZ. As $v_E$ \eqref{vE} is determined by the interior geometry at $z=z_*$, one can then proceed to obtain the small $\lambda$ expansion of $v_E$ 
\begin{align}
    v_E\approx 1+\frac{1}{2\pi^2}\lambda^2\log{\lambda}+\frac{1+\pi -7 \log 2 -2 \log \pi }{8 \pi ^3}\lambda^2\,.\label{vEexpansion}
\end{align}
Notice that the leading non-trivial term in $\lambda$ is at $O(\lambda^2\log{\lambda})$. A plot of \eqref{vEexpansion} with respect to the dimensionless parameter $\frac{\phi_0}{\sqrt{T}}$ can be found in Fig.~\ref{fig:vEvB}.\footnote{In plotting $v_E$ as a function of the dimensionless quantity $\frac{\phi_0}{\sqrt{T}}$, we also need to include the correction of temperature $T$ \eqref{temperature} due to the deformation \eqref{phiLambdaExp}: 
\begin{align}
    T\approx\frac{1}{2\pi}\Big(1+\frac{\lambda^2}{4\pi}\Big)\label{TwithLambda}.
\end{align}
Notice that the $O(\lambda^2)$ correction drops out if we only keep the expansion of $v_E$ to $O(\lambda^2)$. The same applies to $v_B$. } For small $\phi_0$, \eqref{vEexpansion} matches well with numerical result. 

As we have already seen from the numerical solutions, the membrane tension is no longer degenerate in AdS$_3$-Kasner geometry. By plugging the above results into~\eqref{AreaFunct2}, we obtain the small $\lambda$ expansion of $\mathcal{E}(v)$:
\begin{align}
    \mathcal{E}(v)\approx v_E-\frac{\lambda^2}{4 \pi ^2}\log{(1-v^2)}\label{EvNearSing}.
\end{align}
This formula was obtained
from the large $z$ expansion of the metric functions, and this limits its regime of validity in $v$. From \eqref{zvhor}, we see that unless $1-v=O(\lam^2)$, we stay in the large $z$ regime, hence  \eqref{EvNearSing} is widely applicable.
Indeed, this formula obeys the constraints $\mathcal{E}(0)=v_E$, $\mathcal{E}'(0)=0$ $\mathcal{E}'(v)\geq 0$, $\mathcal{E}''(v)\geq 0$ from \eqref{constraints}. A plot of the expansion \eqref{EvNearSing} can be found in Fig.~\ref{fig:EvAnalytic}. For small $\lambda$, \eqref{EvNearSing} matches well with numerical results. 

\subsubsection{Expansion near the horizon}
We expect that the butterfly velocity will move from its 2d CFT value $v_B=1$ to  $v_B=1-O(\lam^2)$. Plugging this into \eqref{EvNearSing} generates a $O(\lam^2\log(\lam))$ term that threatens the validity of the expansion. Indeed $v_B$ is determined by the geometry at the horizon $z=1$, as in  \eqref{vB}, whereas \eqref{EvNearSing} was obtained from a large $z$ expansion. Thus we examine the behaviour of $\mathcal{E}(v)$ near the butterfly velocity more carefully.

\begin{figure}[htbp]
\centering
\includegraphics[width=.66\textwidth]{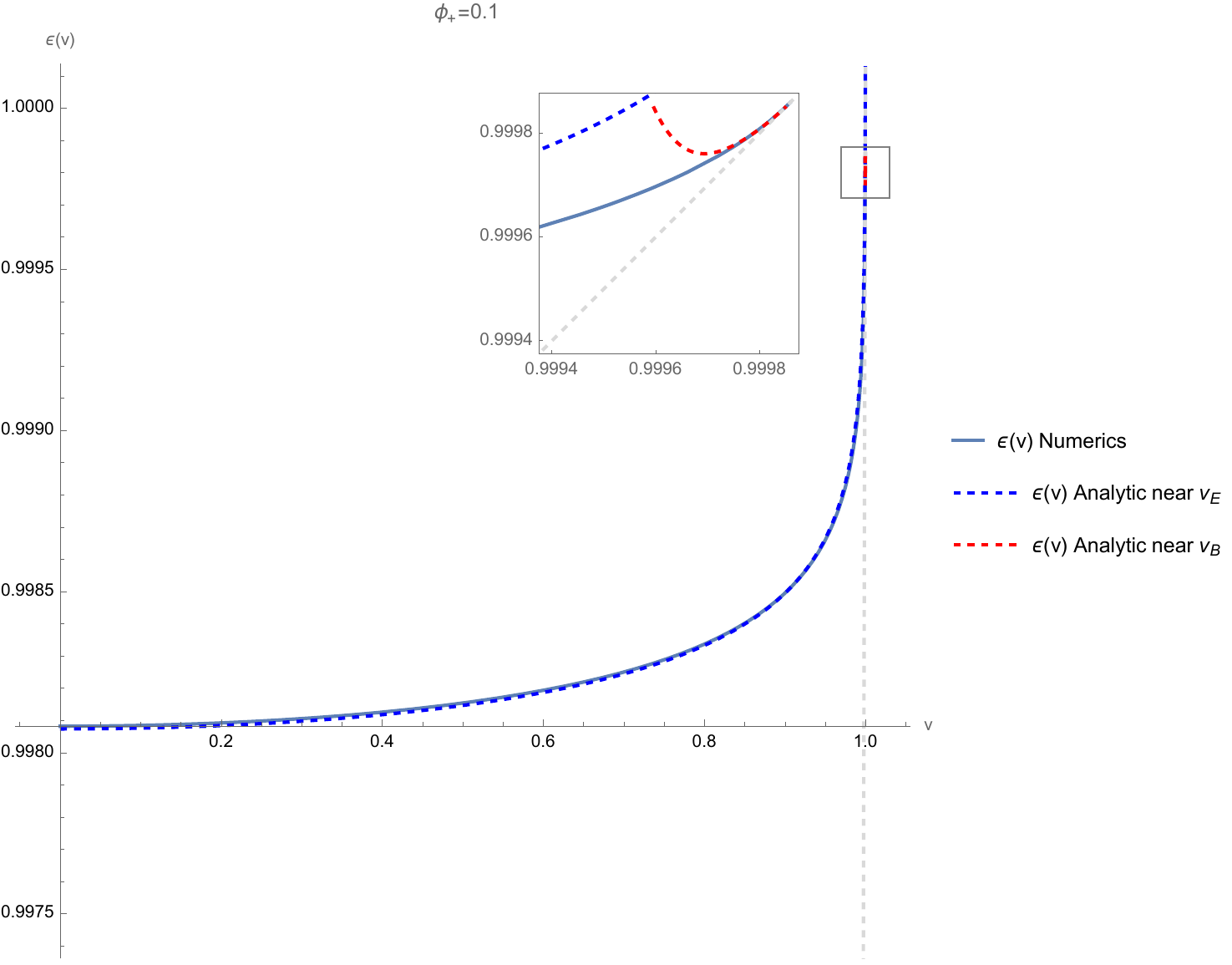}
\qquad
\caption{Plot of the membrane tension $\mathcal{E}(v)$ when $\phi_+=0.1$. For small $\phi_+$, $\mathcal{E}(v)$ is close to 1. The solid line is obtained numerically from shooting solutions, whereas the blue and red dashed lines are perturbative expansions around the deepest point $z_*$ \eqref{EvNearSing} and the horizon $z=1$ \eqref{vzEz2}, respectively. The dashed gray line has unit slope, greatly exaggerated here by the scale of the vertical axis.  Notice that the near-$v_B$ expansion of $\mathcal{E}(v)$ is valid only in a small interval near $v=v_B$, which is difficult to visualise and hence is plotted in the inset. \label{fig:EvAnalytic}}
\end{figure}

It is a straightforward exercise to obtain a series representation in $\de z\equiv z-1$ of the different metric components near the horizon. From these using \eqref{vB}, \eqref{projectionz}, and \eqref{AreaFunct2}  we can form:\footnote{To obtain these expansions we need to evaluate the following definite integral:
\begin{align}
    \chi_1(1)&=\int_0^{1}dz' \frac{\left((z'+1) K\left(\frac{1-z'}{2}\right)-2 z'
   E\left(\frac{1-z'}{2}\right)\right)^2}{\pi ^2 \left(z'^2-1\right)^2}=\frac{3}{8}-\frac{1}{2\pi}\label{chi1(1)}\,.
\end{align}}
\es{vzEz}{
{v_B-1\ov \lam^2}&=-{3\ov 32}+{1\ov 4\pi}<0\,,\\
{v^2-1\ov \lam^2}&={c(z)-1\ov \lam^2}=\le(-{3\ov 16}+{1\ov 2\pi}\ri)-{9\ov 128}\de z-{99\ov 2048}\de z^2+\dots\,,\\
{\mathcal{E}(v)-1\ov \lam^2}&={\sqrt{\frac{-a'(z)}{2z}}-1\ov \lam^2}=\le(-{3\ov 32}+{1\ov 4\pi}\ri)-{9\ov 256}\de z+{45\ov 4096}\de z^2+\dots\,,
}
where we have explicitly computed terms up to $O(\de z^4)$. By expressing $\de z$ as a function of 
\es{nudef}{
\nu\equiv {v^2-v_B^2\ov \lam^2}\ll1\,,
}
we can express $\mathcal{E}(v)$ in a $v$ expansion near the butterfly velocity (with a radius of validity in $v$ very small, $O(\lam^2)$) as follows:
\es{vzEz2}{
\mathcal{E}(v)=1+\lam^2\le(\le(-{3\ov 32}+{1\ov 4\pi}\ri)+{\nu\ov 2}+{64\nu^2\ov 9}+\frac{35840 \nu ^3}{243}+\frac{7813120 \nu ^4}{2187}+\dots\ri)\,.
}
This expression satisfies the constraints $\mathcal{E}(v_B)=v_B,\, \mathcal{E}'(v_B)=1$, $\mathcal{E}'(v)\geq 0$, $\mathcal{E}''(v)\geq 0$ from  \eqref{constraints}, as required. In Fig.~\ref{fig:EvAnalytic}, we plot the expansion \eqref{vzEz2}. It is found that \eqref{vzEz2} agrees with numerical results when $v$ is close to $v_B$.

It is instructive to convert our holographic results~\eqref{vzEz} to those in conformal perturbation theory~\cite{Davison:2024msq} 
\begin{align}
\label{eq:CPTaction}
    S=S_{{\rm CFT}}+\sqrt{c}\,\mu \int d^2 x O,
\end{align}
where $\mu$ is the relevant coupling that deforms the CFT, and the scalar primary operator $O$ is normalised such that its vacuum two-point function is $|x-x'|^{-2\Delta}$. This normalisation is related to our convention via $\phi_0=-4\sqrt{3}\pi\mu$.\footnote{The conversion between holographic and CFT normalisations   is~\cite{Davison:2024msq}
\begin{align}
    \phi=\frac{2\sqrt{3}}{1-\Delta}\mu z^{2-\Delta}+\dots.\label{CFTNorm}
\end{align}
Comparing~\eqref{CFTNorm} with our falloff~\eqref{falloffs}, we have $\phi_0=\frac{2\sqrt{3}}{1-\Delta}\mu$. } The butterfly velocity in~\eqref{vzEz} can then be rewritten in terms of the dimensionless coupling $\frac{\mu}{\sqrt{T}}$ as 
\begin{align}
\label{eq:vBHairyBTZ}
    v_B=1-\frac{3 \pi  (3 \pi -8) \Gamma \left(\frac{3}{4}\right)^2}{2 \Gamma
   \left(\frac{1}{4}\right)^2}\Big(\frac{\mu}{\sqrt{T}}\Big)^2\,.
\end{align}
The butterfly velocity~\eqref{eq:vBHairyBTZ}, and more generally that with general $\Delta\in(1,2)$ 
    \begin{align}
        v_B=1-\frac{3 (2 \pi )^{2 (\Delta -1)} \Gamma (2-\Delta ) \Gamma \left(\frac{\Delta }{2}\right)^2}{\Gamma \left(1-\frac{\Delta}{2}\right)^2 \Gamma (\Delta )} \left(\frac{\pi \Delta  (\Delta -2)   \cot
        \left(\frac{\pi  \Delta }{2}\right)}{4 (\Delta -1)}-1\right) \left(\frac{\mu}{T^{2-\Delta}}\right)^2, 
    \end{align}
calculated using the same method in obtaining~\eqref{vzEz}~\cite{Davison:2025xdj}, is in exact agreement with the prediction from universal hydrodynamics in two-dimensional QFT at high temperature~\cite{Davison:2024msq,Davison:2025xdj}. Such a prediction is made through locations of pole skipping points of the stress tensor thermal two-point functions in momentum space. See~\cite{Davison:2025xdj} for studies of holographic pole skipping in hairy BTZ black hole backgrounds dual to~\eqref{eq:CPTaction}.

\chapter{Time Evolution of Multiparty Entanglement Signals}
\label{Chap4}
In this chapter, we study the real-time ddynamics of multipartite entanglement signals reviewed in section~\ref{multipartySignalReview} using the membrane effective theories for reflected entropy proposed in Chapter~\ref{Chap3}.  

We present the dynamics of three-party and four-party entanglement signals  in Secs.~\ref{sec:r3} and  \ref{sec:I3}. In Sec.~\ref{sec:higher}, we consider extensions to more parties. In Sec.~\ref{app:gen} we discuss multiparty entanglement signals in a generalised membrane theory describing two-dimensional conformal field theories.   

Overall, we find that scrambling dynamics generally produces  multiparty entanglement with rich structure that varies over time. Notably:
\begin{itemize}
\item {\it Different entanglement signals show different dynamical behaviours:} the fact that there are many distinct multiparty entanglement signals is useful, and perhaps necessary, because they likely capture different {\it kinds} of entanglement patterns.\footnote{See \cite{Verstraete:2002gqj} for a classification of four qubit entanglement patterns.} Indeed, we see in our examples that the different signals we consider have qualitatively different dynamical behaviour, capturing different aspects of the dynamics of multiparty entanglement. 
\item {\it Discontinuities:} signals based on bipartitions, such as the triple information, are always continuous, but signals based on the reflected entropy can have discontinuities. We will see that there can be dynamical discontinuities in the three-party entanglement in the membrane description. This suggests phase transitions in the entanglement structure of the state.
\item {\it Non-monotonicity:} we  consider entanglement between large regions, while the initial state has entanglement at small scales. So our multiparty entanglement signals always vanish in the initial state.\footnote{When the state is dual to Vaidya-BTZ spacetime \cite{Balasubramanian:2010ce}, the results in~\cite{Balasubramanian:2024ysu} indicate that there are non-zero initial multiparty entanglement signals. But these initial signals are negligible compared to entropy in the late time and large region limit~\cite{Mezei:2018jco}.} Surprisingly, the subsequent evolution is not monotonic: after an initial increase, in some cases the entanglement signal decreases again.  We will argue that this behaviour reflects the absence of some kinds of  multiparty entanglement in generic holographic states.
\item {\it Saturation of bounds:} the authors of  \cite{Balasubramanian:2024ysu} constructed bounds on  multiparty entanglement. We will see that in some cases, there are periods where the entanglement captured by the membrane description saturates one of these bounds, indicating that the extensive part of the entanglement of one of the regions is {\it entirely} multipartite. 
\end{itemize}

\section{Dynamics of $R_3$}
\label{sec:r3}

In this section we use the membrane effective description for the reflected entropy obtained in \cite{Jiang:2024tdj} to calculate the residual information in the simplest cases. We consider an infinite system with a division including one or two finite regions. We discuss here the ordinary membrane theory; the generalised membrane theory which describes the holographic calculations for $d=2$ has some quantitative differences, and is discussed in Section \ref{app:gen}.

\subsection{A single finite region}\label{SingleFiniteReg}

The simplest case to illustrate the dynamics of $R_3$ is where we divide the boundary into three regions along $x$: $A = (-\infty, 0)$,  $B = (0, \ell)$, and  $C= (\ell, \infty)$. 

\begin{figure}[htbp]
\centering
\includegraphics[width=.35\textwidth]{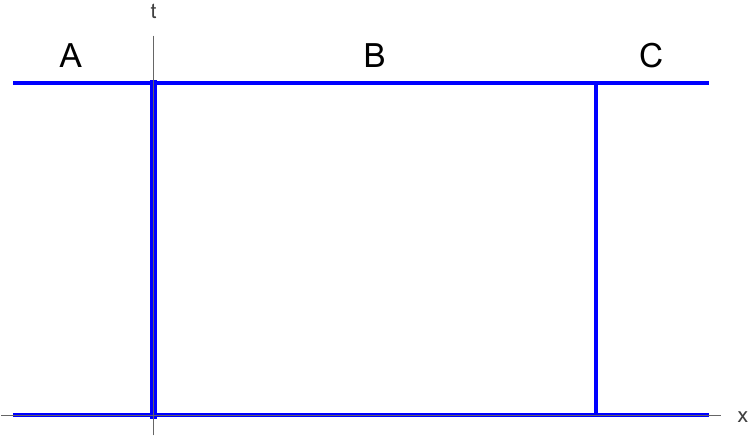}\hspace{5mm}
\includegraphics[width=.35\textwidth]{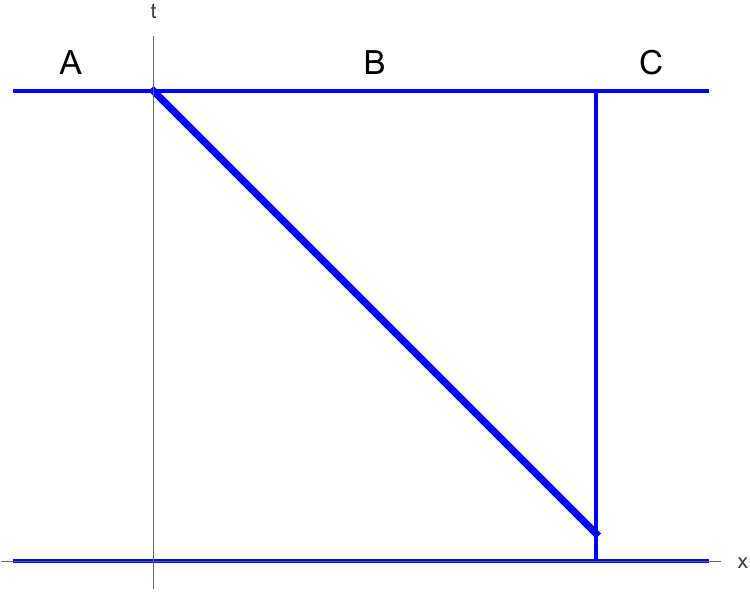}
\qquad
\caption{Membranes computing $S_R(A:B)$ (thick) in the setup of a single finite region $A = (-\infty, 0)$,  $B = (0, \ell)$, and  $C= (\ell, \infty)$.\label{fig:SRABC}}
\end{figure}

In this case, $S(A)$ and $S(C)$ grow linearly for all times, while $S(B)$ has linear growth followed by saturation. In this  case $S(A) = S(C)$ as they are both vertical surfaces. This gives

%
\begin{equation}
I(A:B) = S(A) + S(B) - S(C) = S(B) = s_{\text{th}}\label{IABSingleFiniteReg}
\begin{cases}
2  t &  t < \frac{\ell}{2} \,, \\
\ell &  t > \frac{\ell}{2} \,,  
\end{cases}
\end{equation}
where we have used the fact that for a pure state, $S(AB)=S(C)$. We also have
\begin{equation}
\label{IACSingleFiniteReg}
I(A:C) = S(A) + S(C) - S(B) = s_{\text{th}} 
\begin{cases}
0 &  t < \frac{\ell}{2} \,, \\
2 t - \ell &  t > \frac{\ell}{2} \,.
\end{cases}
\end{equation}
Since $A$, $B$ are adjoining regions, the $AB$ entanglement wedge cross-section is always non-empty. The cross-section is initially a vertical membrane hanging from the boundary, growing linearly in time. At $t=\ell$, this changes to a diagonal membrane extending across the $B$ region, giving a constant extensive answer for the entanglement wedge cross-section, See Figure~\ref{fig:SRABC}.\footnote{In the case of holographic CFTs, the saturated membrane on the right of Figure~\ref{fig:SRABC} is obtained from an Entanglement Wedge Cross-Section (EWCS) that skims through the planar black brane horizon and then ends on the HRT surface corresponding to the vertical membrane for S(C). This has been checked numerically in certain cases in~\cite{Jiang:2024tdj}. Notice that the diagonal membrane only captures the scaling behaviour of the EWCS and does not reflect its intersection with the HRT. } 
So we have\footnote{In the case of adjacent intervals, $S_R$ is not UV finite~\cite{Dutta:2019gen,Kudler-Flam:2020url}. In this case, we can  regularize the EWCS the same way as (H)RT, by subtracting UV divergences of the form~\eqref{AreaLogLaw}. These UV divergences cancel in $R_3$. \label{footnoteSR}}
\begin{equation}
\label{SRABSingReg}
S_R(A:B) = s_{\text{th}} 
\begin{cases}
 2 t &  t < \ell \,, \\
2\ell &  t > \ell \,.  
\end{cases}
\end{equation}
Hence 
\begin{equation}
\label{R3ABSingInt}
R_3(A:B) = S_R - I= s_{\text{th}} 
\begin{cases}
0 &  t < \frac{\ell}{2} \,, \\
2 t - \ell & \frac{\ell}{2} <  t < \ell \,, \\
\ell &  t > \ell \,.  
\end{cases}
\end{equation}
This exhibits the same qualitative dynamics as in the semi-infinite case considered in \cite{Jiang:2024tdj}; it is initially zero until the initial local entanglement has spread sufficiently to contribute to multiparty entanglement between the three regions, then it grows linearly for a period of time and then saturates. 

\begin{figure}[htbp]
\centering
\includegraphics[width=.35\textwidth]{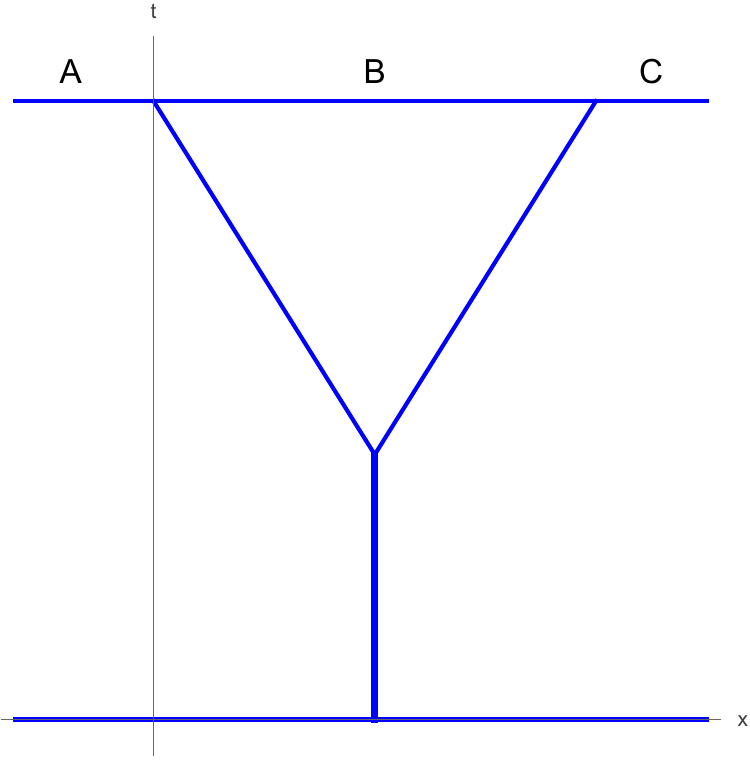}
\qquad
\caption{Membrane computing $S_R(A:C)$ (thick) in the setup of a single finite region.\label{fig:SRABCac}}
\end{figure}

By contrast, the $AC$ entanglement wedge is initially disconnected, only becoming connected when $S(B)$ saturates. Subsequently, the only possible cross-section is a linearly growing vertical surface, which hangs from the tip of the membrane describing $S(B)$, see Figure~\ref{fig:SRABCac}.

\begin{figure}[htbp]
\centering
\includegraphics[width=.6\textwidth]{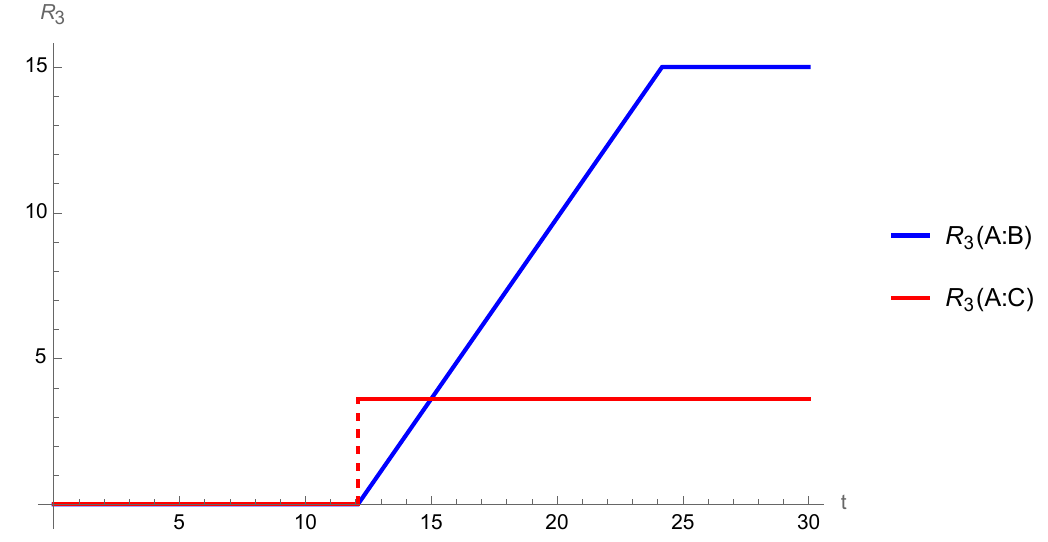}
\qquad
\caption{Time evolution of $R_3(A:B)$~\eqref{R3ABSingInt} (blue) and $R_3(A:C)$~\eqref{R3ACSingInt} (red) when $\ell=15$. \label{fig:R3SingInt}}
\end{figure}

This gives 
\begin{equation}
\label{SRACSingleFiniteReg}
S_R(A:C) = s_{\text{th}} 
\begin{cases}
0 &  t < \frac{\ell}{2} \,, \\
2 t-  \frac{v_E}{v_B} \ell &  t > \frac{\ell}{2} \, .
\end{cases}
\end{equation}
Note there is a discontinuity in $S_R$ at this transition, and we have an offset in time, as in the case discussed in \cite{Jiang:2024tdj}. As a result there is also a discontinuity in the residual information,
\begin{equation}
\label{R3ACSingInt}
R_3(A:C) = s_{\text{th}} 
\begin{cases}
0 &  t < \frac{\ell}{2} \,, \\
\ell \left( 1- \frac{v_E}{v_B} \right)  &  t > \frac{\ell}{2} \,.  
\end{cases}
\end{equation}
This illustrates the general point that different signals can have qualitatively different behaviour, see Figure~\ref{fig:R3SingInt}.

\subsection{Two finite regions}
\label{sec:r3twor}

Consider now a case where we divide the boundary into four pieces, that is with two finite intervals. To calculate $R_3$, we first need to combine two of the pieces to form a single region. There are two qualitatively different choices: to combine one of the semi-infinite pieces with a finite piece, or to combine the two semi-infinite pieces. The former is essentially what was considered in \cite{Jiang:2024tdj}, so we focus on the latter. That is, we consider dividing the boundary into three regions along $x$ which are now $A = (0,\ell_A)$, $B= (\ell_A, \ell_A + \ell_B)$, and $C = (-\infty,0) \cup (\ell_A + \ell_B,\infty)$. Let us suppose without loss of generality that $\ell_B < \ell_A$. We will consider both $R_3(A:B)$ and $R_3(A:C)$ ($R_3(B:C)$ is similar to $R_3(A:C)$). 

\begin{figure}[htbp]
\centering
\includegraphics[width=.35\textwidth]{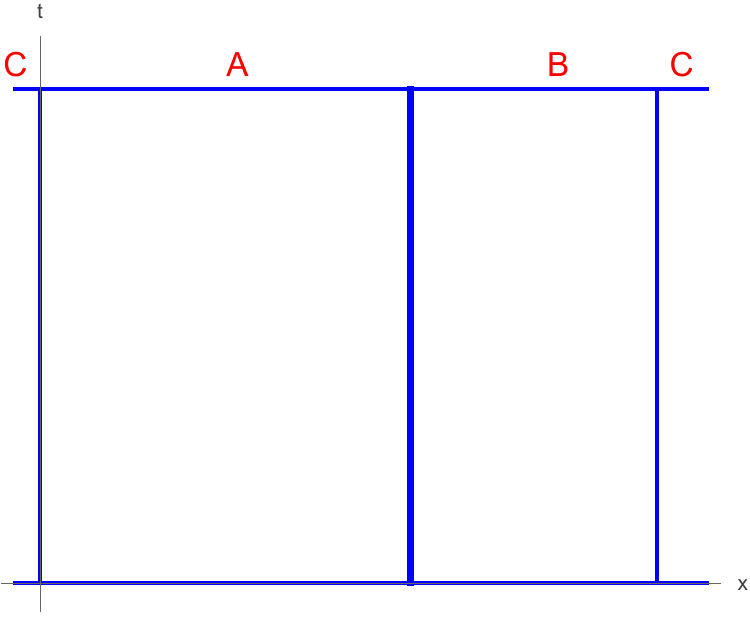}\hspace{5mm}
\includegraphics[width=.35\textwidth]{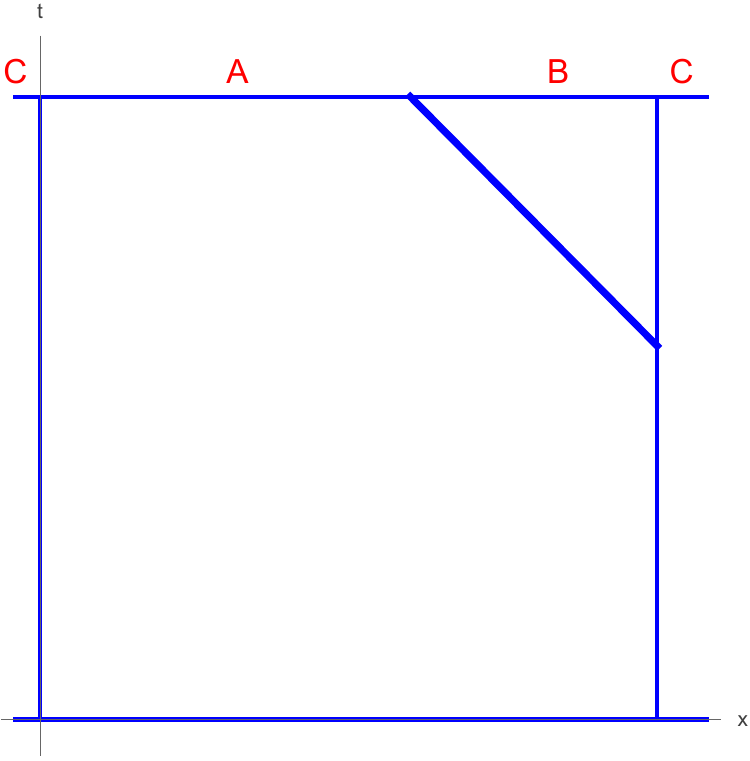}
\qquad
\caption{Membranes computing $S_R(A:B)$ (thick) in the setup of two finite regions $A = (0,\ell_A)$, $B= (\ell_A, \ell_A + \ell_B)$, and $C = (-\infty,0) \cup (\ell_A + \ell_B,\infty)$ ($\ell_A>\ell_B$). After the saturation of the $S(AB)$ to a $v_B$ cone, we have $S_R(A:B)=0$, see Figure~\ref{fig:SatEWCSAdj}. \label{fig:SRCABC}}
\end{figure}

\begin{figure}[htbp]
\centering
\includegraphics[width=.5\textwidth]{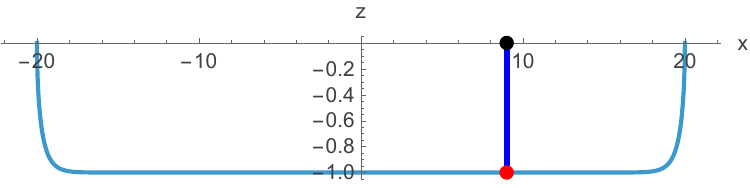}\vspace{0.5cm}
\includegraphics[width=.5\textwidth]{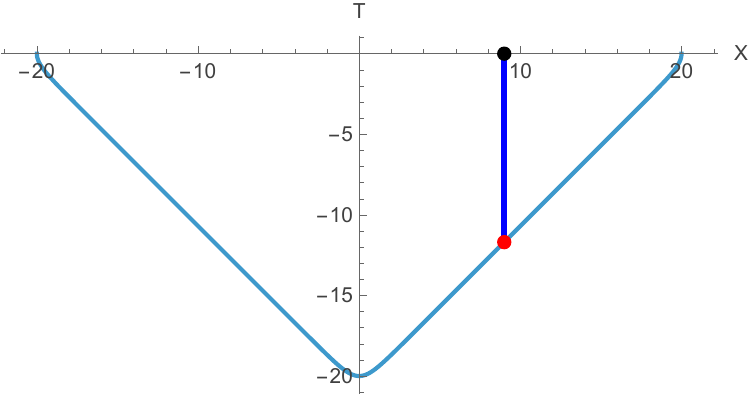}
\qquad
\caption{$\emph{Top}$: The minimal cross-section (blue) in finite adjacent intervals after saturation. It is almost radial as it needs to end perpendicularly  on the plateau of the RT surface perpendicularly, see Appendix~\ref{SatSRAdj}. $\emph{Bottom}:$ 
Projections of the minimal cross-section (blue) to the AdS boundary along constant infalling time. As discussed in appendix~\ref{SatSRAdj}, this vertical surface does not have a membrane description, and we have $S_R=0$ in this case contrary to what one might expect from the figure.  \label{fig:SatEWCSAdj}}
\end{figure}

The entropies $S(A)$, $S(B)$ and $S(AB) = S(C)$ exhibit the general linear growth followed by saturation behaviour of finite intervals. We thus have 
\begin{equation}
I(A:B) = s_{\text{th}} \label{IABTwoFiniteReg}
\begin{cases}
2  t &  t < \frac{\ell_B}{2} \,, \\
\ell_B & \frac{\ell_B}{2} <  t < \frac{\ell_A}{2} \,,   \\
\ell_A + \ell_B - 2  t & \frac{\ell_A}{2} <  t < \frac{\ell_A+\ell_B}{2} \,,  \\
0 &   t > \frac{\ell_A+ \ell_B}{2} \,.
\end{cases}
\end{equation}
The $AB$ entanglement wedge is always connected. The cross-section is initially a vertical surface hanging down from the boundary. It transitions to a surface extending across the $B$ region when $ t = \ell_B$, see Figure~\ref{fig:SRCABC}. When $S(AB)$ saturates at $ t = \ell_A + \ell_B$, the cross-section goes back to hanging down from the boundary but just to the $v_B$ cone membrane corresponding to $S(AB)$ (see Figure~\ref{fig:SatEWCSAdj}). However, this cross-section is no longer described by membrane theory. Intuitively, this is because it does not grow with $t$ or $x$ in the scaling limit. In Appendix~\ref{SatSRAdj}, we show using holography that for this cross-section $S_R=0$ in the scaling limit (after subtracting the UV divergences 
as discussed in footnote~\ref{footnoteSR}). So 

%
\begin{equation} \label{2srab}
S_R(A:B) = s_{\text{th}} 
\begin{cases}
2  t &  t < \ell_B \,, \\
2\ell_B & \ell_B <  t < \frac{\ell_A+\ell_B}{2} \,,  \\
0  &   t > \frac{\ell_A+ \ell_B}{2} \,.
\end{cases}
\end{equation}
The order of transitions in the residual information depends on whether $2 \ell_B$ is bigger or smaller than $\ell_A$. Suppose for definiteness $\ell_A < 2 \ell_B$,\footnote{The other case with $\ell_A>2\ell_B$ shows the same qualitative behaviour.} which includes the case where the two finite intervals are equal size. Then taking the difference between~\eqref{2srab} and~\eqref{IABTwoFiniteReg} we have 
\begin{equation}
\label{R3ABTwoReg1}
R_3(A:B) = s_{\text{th}} 
\begin{cases}
0 &  t < \frac{\ell_B}{2} \,, \\
2 t - \ell_B & \frac{\ell_B}{2} <  t < \frac{\ell_A}{2} \,,   \\
4 t - \ell_A - \ell_B & \frac{\ell_A}{2} <  t < \ell_B \,,   \\
2  t + \ell_B - \ell_A & \ell_B <  t < \frac{\ell_A+\ell_B}{2} \,,  \\
0  &   t > \frac{\ell_A+ \ell_B}{2} \,.
\end{cases}
\end{equation}
There is a single discontinuity in the residual information, at the final transition. At the transition point, the residual information reaches its maximum, $R_3(A:B) = 2\ell_B$, saturating a bound from \cite{Balasubramanian:2024ysu}, indicating that the extensive part of the entanglement of $B$ is entirely multipartite. 

\begin{figure}[htbp]
\centering
\includegraphics[width=.7\textwidth]{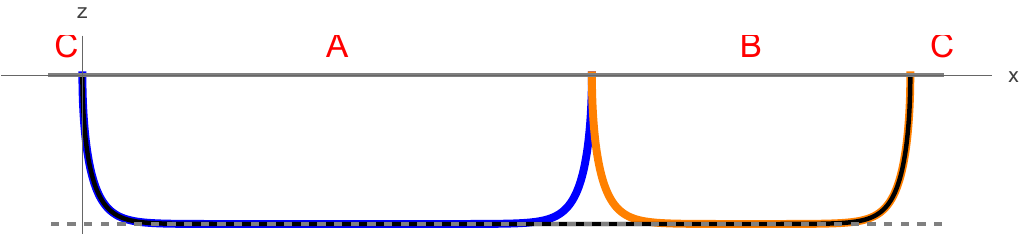}
\qquad
\caption{In the large subregion limit, RT surfaces have large plateau regions where they skim along the black hole horizon (dashed gray line), so the A RT surface (blue) and B RT surface (orange) lie close to the AB RT surface (black). \label{fig:RTClose}}
\end{figure}

Also, the jump is downward, so the residual information is not monotonically increasing. The residual information is only a signal of multiparty entanglement, so in general this does not mean that there is no multiparty entanglement in the final equilibrium state. But in holographic cases, in the final equilibrium here the $A$ and $B$ RT surfaces lie close to the $AB$ RT surface (see Figure~\ref{fig:RTClose}),  so it is reasonable to believe that the extensive entanglement here {\it is} purely bipartite, between $A$ and $C$ and $B$ and $C$.\footnote{The intuition here is that the entanglement can be modelled by a tensor network with locally entangled EPR pairs across the RT surface. This is a good model for the high-temperature thermofield double state, and was also shown to model hot multi-boundary wormholes~\cite{Marolf:2015vma}, which have a similar structure where the RT surfaces for different boundaries lie close to each other along most of their length. When the RT surfaces lie close to each other, the entangled EPR pairs across each surface simply join up to give local bipartite entanglement between the two sides.}

\begin{figure}[htbp]
\centering
\includegraphics[width=.3\textwidth]{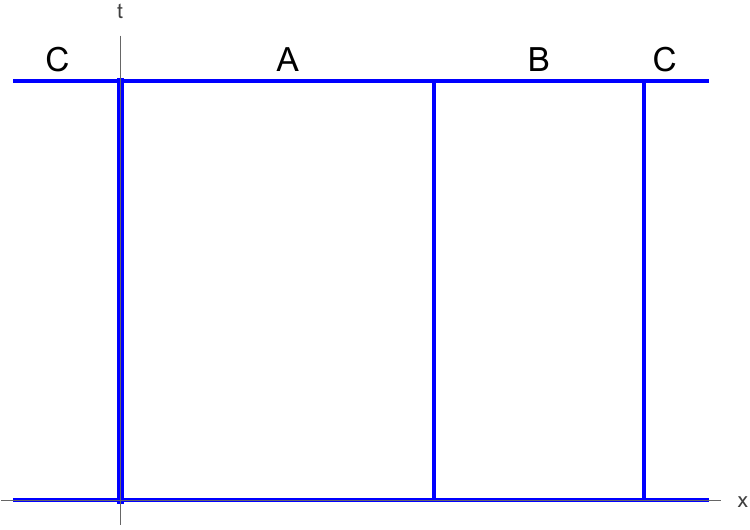}\hspace{3mm}
\includegraphics[width=.3\textwidth]{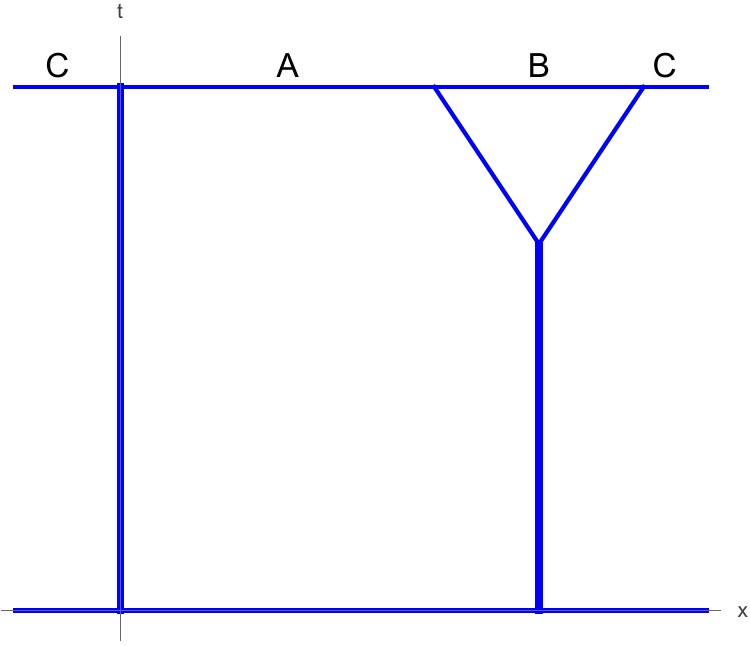}\hspace{3mm}
\includegraphics[width=.3\textwidth]{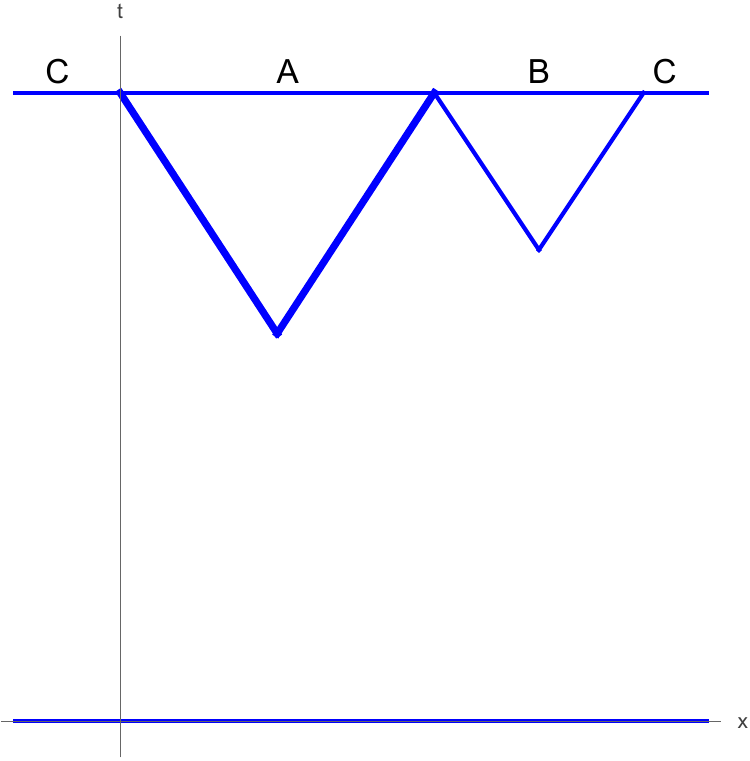}
\qquad
\caption{Membranes computing $S_R(A:C)$ (thick) in the setup of two finite regions. \label{fig:SRCABCac}}
\end{figure}

Let us also consider $R_3(A:C)$. We have
\begin{equation}
\label{IACTwoFiniteReg}
I(A:C) = s_{\text{th}} 
\begin{cases}
2 t &  t < \frac{\ell_B}{2} \,, \\
4 t - \ell_B & \frac{\ell_B}{2} <  t < \frac{\ell_A}{2} \,,   \\
2 t + \ell_A - \ell_B & \frac{\ell_A}{2} <  t < \frac{\ell_A+\ell_B}{2} \,,  \\
2 \ell_A &   t > \frac{\ell_A+ \ell_B}{2} \,.
\end{cases}
\end{equation}
The $AC$ entanglement wedge is always connected between $A$ and $C$. Initially the cross-section is just the vertical surface at $x=0$. When $B$ saturates, it acquires an additional component, a vertical surface hanging down from the $B$ membrane, then the two components join up. See Figure~\ref{fig:SRCABCac}.

Thus
\begin{equation} \label{2srac}
S_R(A:C) = s_{\text{th}} 
\begin{cases}
2 t &  t < \frac{\ell_B}{2} \,, \\
4 t - \frac{v_E}{v_B} \ell_B  & \frac{\ell_B}{2} < t < \frac{\ell_A}{2} + \frac{v_E}{v_B} \frac{\ell_B}{4} \,,   \\
2\ell_A &  t > \frac{\ell_A}{2} + \frac{v_E}{v_B} \frac{\ell_B}{4}  \,.
\end{cases}
\end{equation}
This gives
\begin{equation}
\label{R3ACTwoReg}
R_3(A:C) = s_{\text{th}} 
\begin{cases}
0 &  t < \frac{\ell_B}{2} \,, \\
\ell_B \left( 1 - \frac{v_E}{v_B} \right) & \frac{\ell_B}{2} <  t < \frac{\ell_A}{2} \,,   \\
2  t - \ell_A + \ell_B \left( 1 - \frac{v_E}{v_B} \right)  & \frac{\ell_A}{2} <  t < \frac{\ell_A}{2} + \frac{v_E}{v_B} \frac{\ell_B}{4} \,,  \\
\ell_A + \ell_B - 2 t & \frac{\ell_A}{2} + \frac{v_E}{v_B} \frac{\ell_B}{4}<  t < \frac{\ell_A+\ell_B}{2} \,,  \\
0 &   t > \frac{\ell_A+ \ell_B}{2} \,.
\end{cases}
\end{equation}
We see that the two residual information are non-zero in the same region, but their behaviour is otherwise quite different. $R_3(A:C)$ has an initial discontinuity, and then goes smoothly to zero at the end, so we see that $R_3$ can also decrease smoothly. It is also interesting to notice that $R_3(A:C)$ does not saturate the $2\ell_B$ bound. 


\begin{figure}[htbp]
\centering
\includegraphics[width=.6\textwidth]{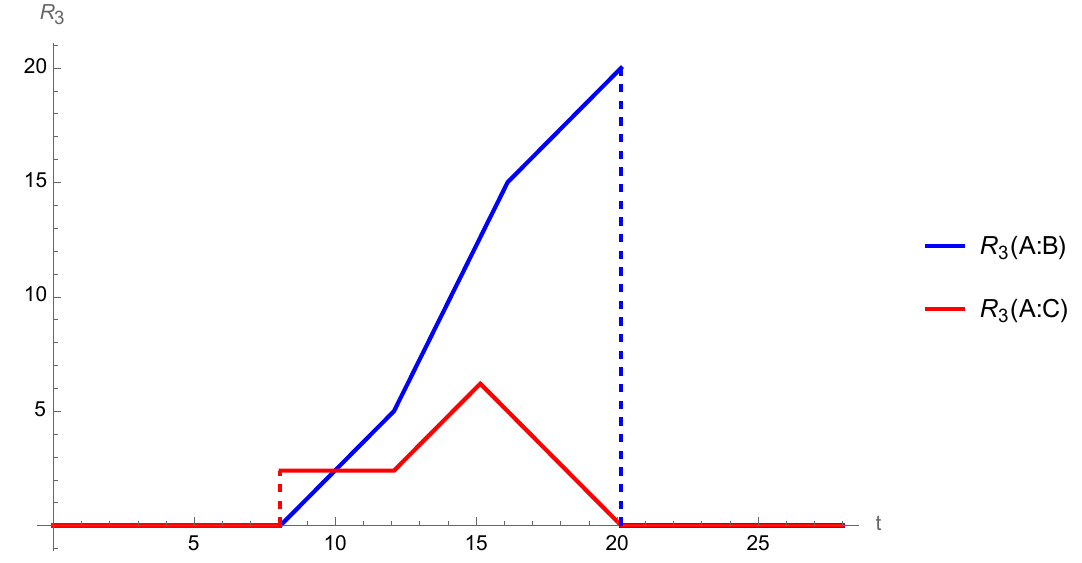}
\qquad
\caption{Time evolution of $R_3(A:B)$ (blue) and $R_3(A:C)$ (red) when $\ell_A=15$ and $\ell_B=10$. \label{fig:R3TwoReg}}
\end{figure}


\section{Dynamics of $I_3$ and $Q_4$}
\label{sec:I3}

We now turn to the dynamics of $I_3$ and $Q_4$. We will consider a single example, with regions $A = (0, \ell_A)$, $B = (\ell_A, \ell_A+\ell_B)$, $C= (\ell_A+\ell_B, \ell_A + \ell_B+\ell_C)$ and $D = (-\infty,0) \cup (\ell_A + \ell_B + \ell_C, \infty)$. For simplicity we restrict to $\ell_A = \ell_C$. The time-dependence of $I_3$ in this situation at finite time was previously considered in \cite{Balasubramanian:2011at}. We consider just the late-time, large region behaviour, in the context of the membrane theory. This captures the main qualitative features of the previous discussion. Our main interest here is to compare the behaviour of $I_3$ and the novel multiparty signal $Q_4$ introduced in \cite{Balasubramanian:2024ysu}. As we will see below, we find non-monotonicity\footnote{As already noted for $I_3$ in \cite{Balasubramanian:2011at}.} and saturation of bounds for both signals. For $I_3$ we cannot have discontinuities as it is just a combination of von Neumann entropies, but we do find discontinuities for $Q_4$.

If $\ell_A=\ell_C<\ell_B$, we find
\begin{equation}
\label{I3AdjacentNeqLenCase1}
I_3(A:B:C) = s_{\text{th}} 
\begin{cases}
0  &0<t < \frac{\ell_B}{2}\,, \\ 
\ell_B-2 t &\frac{\ell_B}{2}<t < \frac{\ell_A+\ell_B}{2}\,, \\ 
2 t-2\ell_A-\ell_B &\frac{\ell_A+\ell_B}{2} < t < \frac{2\ell_A+\ell_B}{2}\,, \\ 
0 &t>\frac{2\ell_A+\ell_B}{2}\,.
\end{cases}
\end{equation}
This gives a simple triangular profile for $-I_3$, initially increasing and then decreasing smoothly to zero.  


If  $\ell_B<\ell_A=\ell_C<2\ell_B$, we have a slightly more complex behaviour, 
\begin{equation}
\label{I3AdjacentNeqLenCase2}
I_3(A:B:C) = s_{\text{th}} 
\begin{cases}
0 \ &0<t < \frac{\ell_A}{2}\,, \\ 
2\ell_A-4 t\ &\frac{\ell_A}{2}<t < \frac{2\ell_A-\ell_B}{2}\,, \\ 
\ell_B-2 t\ &\frac{2\ell_A-\ell_B}{2}<t < \frac{\ell_A+\ell_B}{2}\,, \\ 
2 t-2\ell_A-\ell_B &\frac{\ell_A+\ell_B}{2} < t < \frac{2\ell_A+\ell_B}{2}\,, \\ 
0\ &t>\frac{2\ell_A+\ell_B}{2}\,.
\end{cases}
\end{equation}


The most interesting case is when $\ell_A=\ell_C>2\ell_B$:
\begin{equation}
\label{I3AdjacentNeqLenCase3}
I_3(A:B:C) = s_{\text{th}} 
\begin{cases}
0 \ &0<t < \frac{\ell_A}{2}\,, \\ 
2\ell_A-4 t\ &\frac{\ell_A}{2}<t < \frac{\ell_A+\ell_B}{2}\,, \\ 
-2\ell_B\ &\frac{\ell_A+\ell_B}{2}<t < \frac{2\ell_A-\ell_B}{2}\,, \\ 
2 t-2\ell_A-\ell_B &\frac{2\ell_A-\ell_B}{2} < t < \frac{2\ell_A+\ell_B}{2}\,, \\ 
0\ &t>\frac{2\ell_A+\ell_B}{2}\,.
\end{cases}
\end{equation}
In this case there is a plateau region, where $I_3 = -2\ell_B$. This saturates a bound on $I_3$, indicating that the entanglement of $B$ is entirely four-partite. These results are plotted in in figures \ref{fig:I3Q41},
\ref{fig:I3Q42}, \ref{fig:I3Q43}.

\begin{figure}[htbp]
\centering
\includegraphics[width=.35\textwidth]{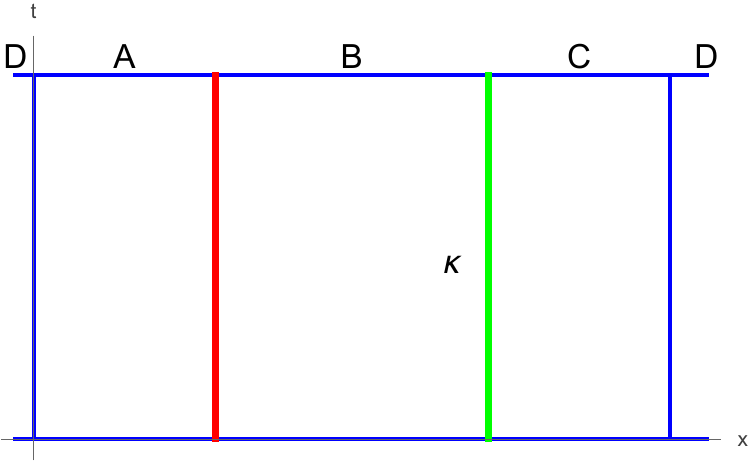}\hspace{3mm}
\includegraphics[width=.35\textwidth]{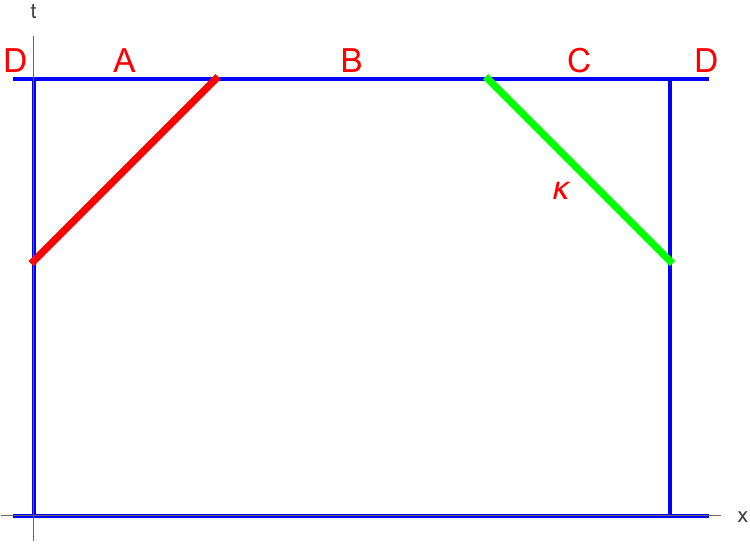}\vspace{6mm}
\includegraphics[width=.31\textwidth]{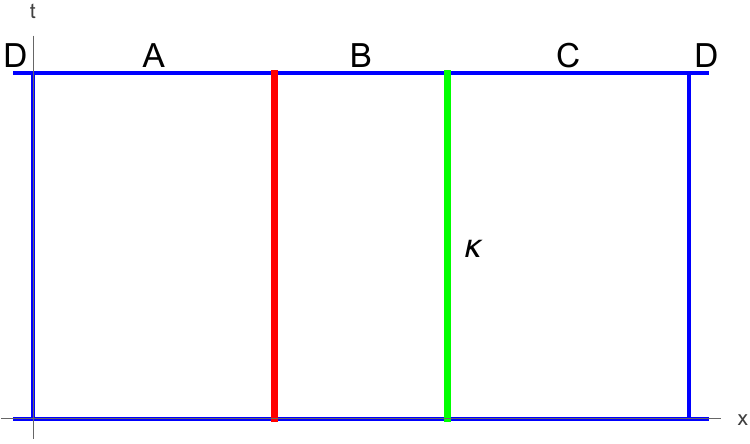}\hspace{2mm}
\includegraphics[width=.31\textwidth]{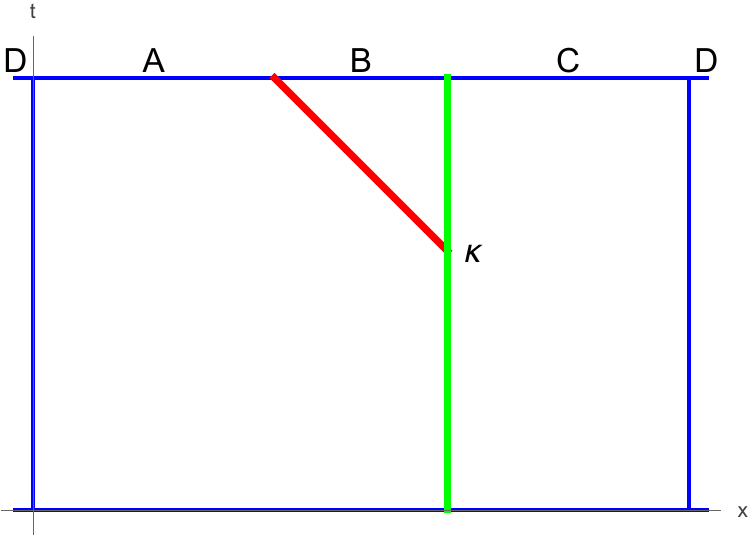}\hspace{2mm}
\includegraphics[width=.31\textwidth]{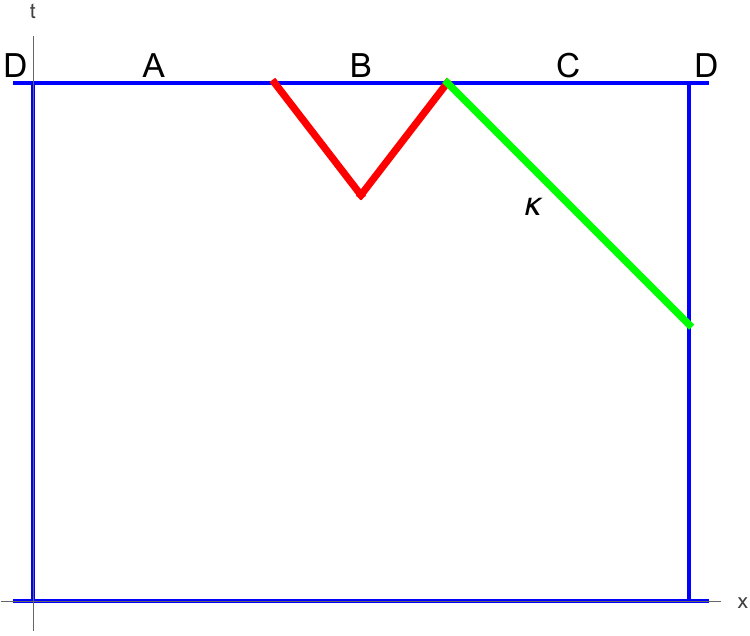}
\qquad
\caption{Minimal cross-sections computing $S_{AA^*A_*A^*_*}$ (red, thick) in the entanglement membrane description when $\ell_A<\ell_B$ (top row), and $\ell_A>\ell_B$ (bottom row). Notice that in the latter case, although the saturation of $\kappa$ (green, thick) changes the shape of the membrane computing $S_{AA^*A_*A^*_*}$, it does not change the value of $S_{AA^*A_*A^*_*}$ in the scaling limit. We only plotted membranes before the saturation of the $ABC$ entanglement wedge, after which all the entanglement wedge cross-sections have vanishing area in the membrane description. \label{fig:kappa}}
\end{figure}

\begin{figure}[htbp]
\centering
\includegraphics[width=.6\textwidth]{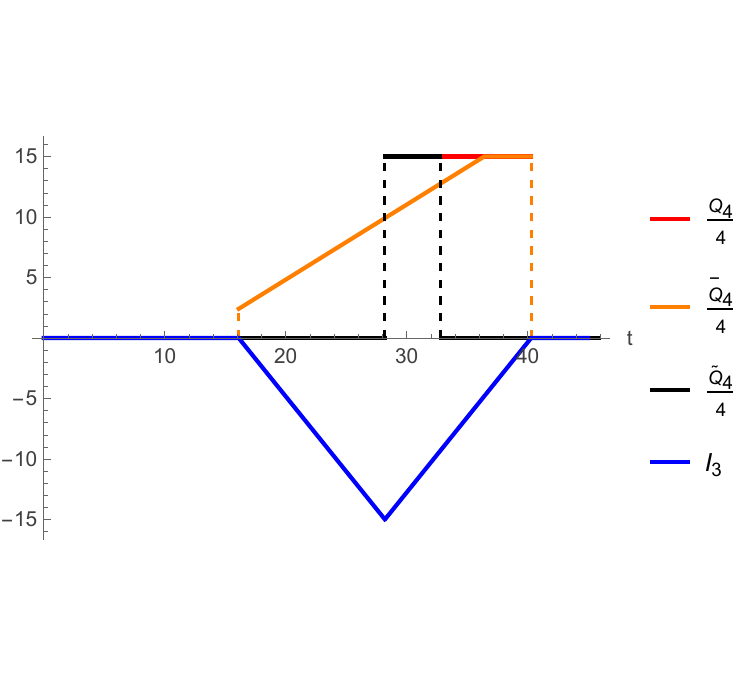}
\caption{Time evolution of $I_3(A:B:C)$~\eqref{I3AdjacentNeqLenCase1} (blue), $Q_4$~\eqref{Q4DABCD1} (red), $\Tilde{Q}_4$~\eqref{Q4T} (black) and $\bar{Q}_4$~\eqref{Q4B} (orange) for $\ell_A = \ell_C < \ell_B$. Here, we plot for $\ell_A=15$ and $\ell_B=20$. In this plot, the time at which $Q_4$ and $\Tilde{Q}_4$ start to be non-zero (the first black dashed vertical line) is the same; the time at which $Q_4$ and $\bar{Q}_4$ drop to 0 (orange dashed vertical line) is the same. \label{fig:I3Q41}}
\end{figure}

\begin{figure}[htbp]
\centering
\includegraphics[width=.6\textwidth]{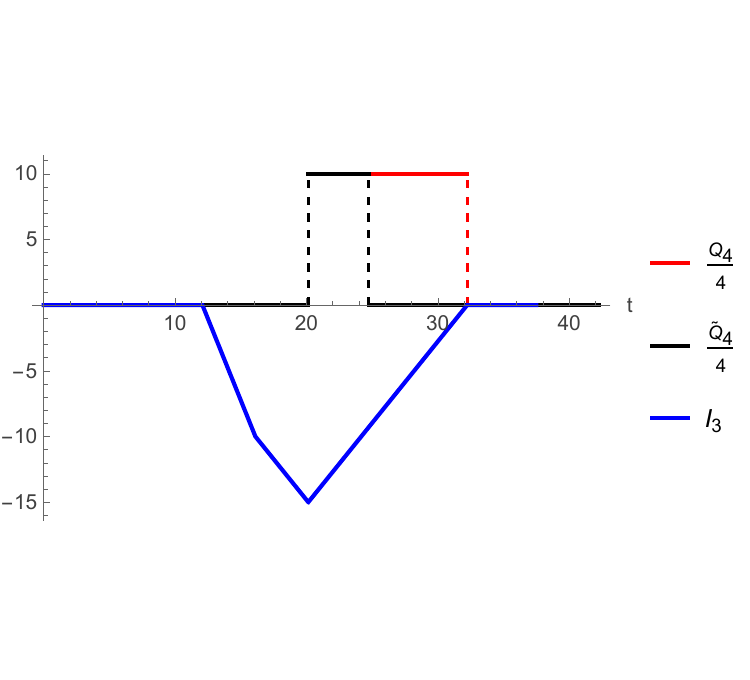}
\caption{Time evolution of $I_3(A:B:C)$~\eqref{I3AdjacentNeqLenCase1} (blue), $Q_4$~\eqref{Q4DABCD1} (red), and $\Tilde{Q}_4$~\eqref{Q4T} (black), for $\ell_B < \ell_A = \ell_C < 2 \ell_B$. Here, we plot for $\ell_A=15$ and $\ell_B=10$. In this plot, the time at which $Q_4$ and $\Tilde{Q}_4$ start to be non-zero (the first black dashed vertical line) is the same. \label{fig:I3Q42}}
\end{figure}

\begin{figure}[htbp]
\centering
\includegraphics[width=.6\textwidth]{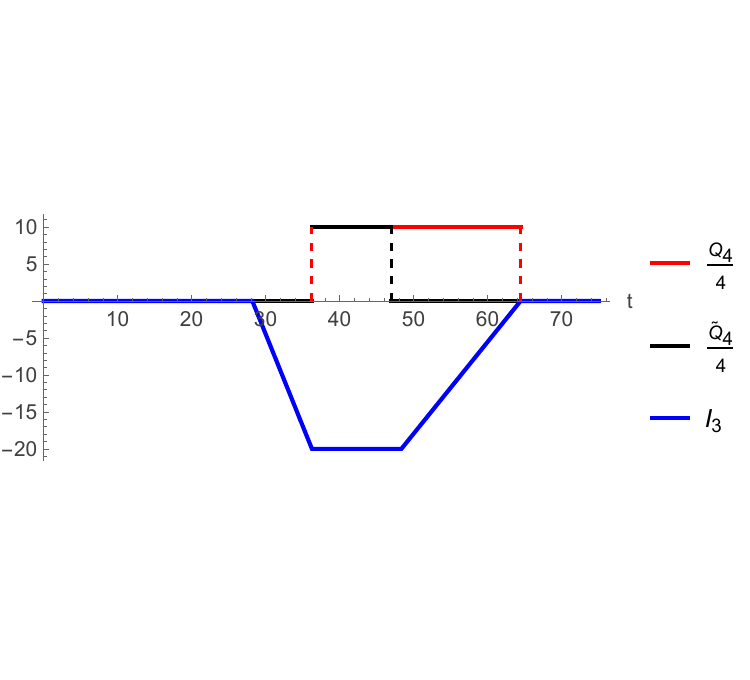}
\caption{Time evolution of $I_3(A:B:C)$~\eqref{I3AdjacentNeqLenCase1} (blue), $Q_4$~\eqref{Q4DABCD1} (red), and $\Tilde{Q}_4$~\eqref{Q4T} (black), for $2\ell_B < \ell_A = \ell_C$. Here, we plot for $\ell_A=35$ and $\ell_B=10$. \label{fig:I3Q43}}
\end{figure}

Next, we consider the calculation of $Q_4$ as in \cite{Balasubramanian:2024ysu}, tracing over $D$ first and then $CC^*$. 
The calculation of $S_R(A:B)$ in~\eqref{Q4} is similar to that in~\eqref{SRABSingReg}: we regard $CD$ as a single region, and the $AB$ entanglement wedge cross-section is then the same as those in Figure~\ref{fig:SRCABC}.\footnote{In Figure~\ref{fig:SRCABC} we considered only the case when $\ell_A>\ell_B$. Here we also allow $\ell_A<\ell_B$ and the $AB$ entanglement wedge cross-section saturates to the smaller value of $\ell_A$ and $\ell_B$.} To calculate $S_{AA^*A_*A^*_*}$, we consider the $AA^*BB^*$ entanglement wedge cross-section in the canonical purification of $\rho_{ABC}$. The first purification results in $EW(AA^*BB^*)$ bounded by a membrane $\kappa$ bisecting the EW of $AB$  (regarded as a single region) and $C$ (Figure~\ref{fig:kappa}). Notice that $\kappa$ itself possesses a non-trivial dynamic. After the second purification, $S_{AA^*A_*A^*_*}$ is computed by the minimal membrane bisecting $EW(AA^*BB^*)$ (Figure~\ref{fig:kappa}). 

By comparing the membranes computing $S_{AA^*A_*A^*_*}$ in Figure~\ref{fig:kappa} with those calculating $S_R(A:B)$ in Figure~\ref{fig:SRCABC}, we see that the EWCS for $S_{AA^*A_*A^*_*}$ and $S_R(A:B)$ at early times are the same.\footnote{In the case with $\ell_A > \ell_B$, the EWCS for $S_{AA^*A_*A^*_*}$ in the last panel in figure~\ref{fig:kappa} changes, but it has the same area.} Thus, the contributions cancel and $Q_4=0$. This ends when $S_{AB}$ saturates and we get figure~\ref{fig:SatEWCSAdj}. Then $S_R(A:B)=0$ in the membrane description, and $Q_4\neq 0$. This continues until $S_{ABC}$ similarly saturates, after which $S_{AA^*A_*A^*_*}$ also vanishes so we have $Q_4=0$ again. So $Q_4$ follows a unit-pulse-like time evolution profile

\begin{equation}
\label{Q4DABCD1}
Q_4 = s_{\text{th}} 
\begin{cases}
0 \ &0<t < \frac{\ell_A+\ell_B}{2}\,, \\ 
4 \text{min}\{\ell_A,\ell_B\} &\frac{\ell_A+\ell_B}{2}<t < \frac{\ell_A+\ell_B+\ell_C}{2}\,, \\ 
0\ &t>\frac{\ell_A+\ell_B+\ell_C}{2}\,.
\end{cases}
\end{equation}
where the minimisation in $Q_4$ between $\ell_A$ and $\ell_B$ captures how the saturated cross-section computing $S_{AA^*A_*A^*_*}$ exits the $AA^*BB^*$ entanglement wedge, see Figure~\ref{fig:kappa}. 

We plot the results when $\ell_A = \ell_C$ in Figures \ref{fig:I3Q41},
\ref{fig:I3Q42}, \ref{fig:I3Q43}. We see that the behaviour of $I_3$ and $Q_4$ are quite different; $I_3$ changes smoothly, while $Q_4$ is piecewise constant, and non-zero only in a subregion of the region where $I_3$ is varying. The discontinuities in $Q_4$ suggest phase transitions in the entanglement structure of the state, which are not evident from just considering $I_3$. In particular, $I_3$ goes smoothly to zero, but $Q_4$ shows that multiparty entanglement is {\it not} smoothly decreasing. It is still large at least up until $t = \frac{1}{2} (\ell_A + \ell_B + \ell_C)$, when it drops suddenly to zero, suggesting a sudden change of entanglement phase.

In Figure \ref{fig:I3Q43}, a bound from \cite{Balasubramanian:2024ysu} is saturated at the  plateau for $I_3$, indicating that the extensive entanglement of $B$ is purely four-party  at least in this period. In $Q_4$, the plateau also saturates a bound, which in this case indicates that the extensive entanglement of either $B$ (if $\ell_B < \ell_A$) or $A$ and $C$ (if $\ell_A = \ell_C < \ell_B$) is purely four-party  in that period. 

As in the calculation of $R_3$ in the previous section, the vanishing of $Q_4$ and $I_3$ at late times does not necessarily indicate the absence of four-party entanglement, but in the holographic cases, once all the finite regions have saturated the $ABC$ RT surface lies close to the individual $A$, $B$, $C$ RT surfaces, suggesting that the extensive part of the entanglement is entirely bipartite entanglement between each of the finite regions and $D$.

\begin{figure}[htbp]
\centering
\includegraphics[width=.4\textwidth]{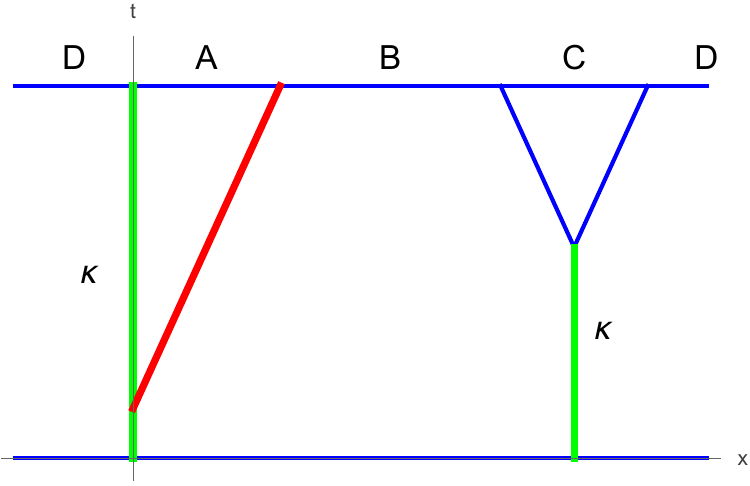}\hspace{4mm}
\includegraphics[width=.4\textwidth]{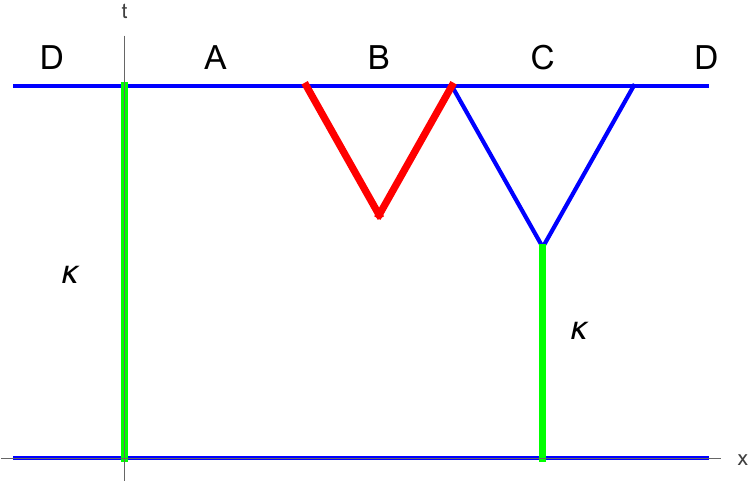}
\qquad
\caption{Examples of membranes computing $S_{AA^*A_*A^*_*}$ (red, thick) in $\tilde{Q}_4$. Notice that $\kappa$ (green, thick) behaves as $S_R(A:C)$ by regarding $AB$ as a single interval; $S_{AA^*A_*A^*_*}=0$ after $\kappa$ saturates. \label{fig:Q4T}}
\end{figure}

In obtaining~\eqref{Q4DABCD1}, we have traced out $D$ first and then $CC^*$. We can also consider tracing out $C$ first and the $DD^*$. This gives an independent signal, which we will denote as $\Tilde{Q}_4$. When tracing out $C$ in the first purification, the entanglement wedge of $AB$ (regarded as a single interval) and $D$ is bipartitioned by minimal membranes $\kappa$ that behave as $S_R(A:C)$ in Figure~\ref{fig:SRCABCac}. $S_{AA^*A_*A^*_*}$ in the second purification is then computed by the minimal cross-section that bisects the $AA^*BB^*$ entanglement wedge bounded by $\kappa$, see Figure~\ref{fig:Q4T}. As such, when fixing $\ell_A=\ell_C$, we have 
\begin{equation}
\label{Q4T}
\Tilde{Q}_4 = s_{\text{th}} 
\begin{cases}
0 \ &0<t < \frac{\ell_A+\ell_B}{2}\,, \\ 
4 \text{min}\{\ell_A,\ell_B\} &\frac{\ell_A+\ell_B}{2}<t < \frac{\ell_A+\ell_B}{2}+\frac{v_E}{4v_B}\ell_A\,, \\ 
0\ &t>\frac{\ell_A+\ell_B}{2}+\frac{v_E}{4v_B}\ell_A\,.
\end{cases}
\end{equation}
This is also plotted in figures~\ref{fig:I3Q41},
\ref{fig:I3Q42}, \ref{fig:I3Q43}. Note that compared to $Q_4$, the duration for $\Tilde{Q}_4$ to be non-zero is shorter. This illustrates the advantage of having multiple signals, as $Q_4$ and $I_3$ capture the existence of 4-party entanglement that $\Tilde{Q}_4$ fails to detect after its disappearance. 

\begin{figure}[htbp]
\centering
\includegraphics[width=.45\textwidth]{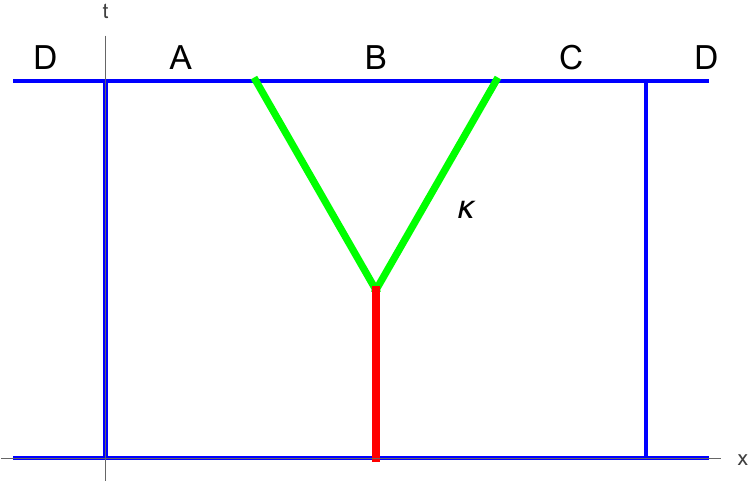}
\qquad
\caption{An example of the membrane computing $S_R(A:C)$ or $S_{AA^*A_*A^*_*}$ in $\bar{Q}_4$, depicted in red. In the former case, $\kappa$ along with the two outermost vertical membranes together are the membranes for $S(AC)$. \label{fig:Q4B}}
\end{figure}

One can also consider tracing over different regions in the definition of $Q_4$. For example, we can trace out $D$ and $B$ to obtain another version of $Q_4$, which we denote as $\bar{Q}_4$. When $\ell_A>\ell_B$, the membranes for $S_R(A:C)$ consist of a vertical membrane dangling from the tip of the $v_B$ cone anchored to $B$ at early time as shown in Figure~\ref{fig:Q4B}, followed by a diagonal membrane of slope $|v_B|$ exiting the $AC$ entangling wedge through either of its boundary. When $\ell_A<\ell_B$, we simply have $S_R(A:C)=0$ at all time, as the $AC$ entanglement wedge is always disconnected. 

To calculate $S_{AA^*A_*A^*_*}$, let us first consider the minimal cross-section $\kappa$ that bisects the entanglement wedge of $AC$ (regarded as one disjoint subregion) and $B$ in the first purification. At early time, $\kappa$ consists of two vertical membranes. These will later be replaced by  a single triangle anchored to $B$ (see Figure~\ref{fig:Q4B}) or two diagonal membranes with slope $\pm v_B$ exiting the $ABC$ entanglement wedge, depending on whether $\ell_B<2\ell_A$ or $\ell_B>2\ell_A$. In the latter case, the $AA^*CC^*$ entanglement wedge is always disconnected, and $S_{AA^*A_*A^*_*}=0$ at all time. As for the former, $S_{AA^*A_*A^*_*}$ is computed by the minimal cross-section bi-partitioning the $AA^*CC^*$ entanglement wedge bounded by $\kappa$, which is a vertical membrane dangling from
the tip of the $v_B$ cone at early time (as  shown in Figure~\ref{fig:Q4B}) and a diagonal membrane of slope $|v_B|$ exiting the $AA^*CC^*$ EW at late time. Notice that these are the $\emph{same}$ membranes as those computing $S_R(A:C)$. Hence, $\bar{Q}_4\neq 0$ only when $\ell_A<\ell_B<2\ell_A$: for this choice of $\ell_A$ and $\ell_B$, $S_R(A:C)=0$ at all times, but $S_{AA^*A_*A^*_*}$ undergoes a non-trivial dynamic. When $\ell_B>\ell_A$, both $S_R(A:C)$ and $S_{AA^*A_*A^*_*}$ are $0$ all the time; when $\ell_A>\ell_B$, both $S_R(A:C)$ and $S_{AA^*A_*A^*_*}$ have non-trivial dynamics, but they are computed by the same membranes at any time, so their contributions cancel. As such, we have that for $\ell_A<\ell_B<2\ell_A$,
\begin{equation} \label{Q4B}
\bar{Q}_4 = s_{\text{th}} 
\begin{cases}
0 &  t < \frac{\ell_B}{2} \,, \\
4 \big(t - \frac{v_E}{2 v_B} \ell_B\big)  &  \frac{\ell_B}{2} < t < \ell_A + \frac{v_E}{2v_B} \ell_B \,, \\
4 \ell_A & \ell_A + \frac{v_E}{2v_B} \ell_B < t < \ell_A + \frac{\ell_B}{2}  \,,   \\
0 &  t > \ell_A + \frac{\ell_B}{2} \,.
\end{cases}
\end{equation}
This is plotted in Figure~\ref{fig:I3Q41}. In the cases in Figures~\ref{fig:I3Q42} and ~\ref{fig:I3Q43}, $\bar Q_4$ is always zero.

Unlike $Q_4$ or $\Tilde{Q}_4$, $\bar{Q}_4$ does not display the dynamics of a unit pulse. Moreover, $\bar{Q}_4$ is non-zero in the whole interval where $I_3$ is nonvanishing. We also note (see Figure~\ref{fig:I3Q41}) that the times at which $Q_4$, $\Tilde{Q}_4$ and $\bar{Q}_4$ saturate the $4\ell_A$ bound are different. We believe that the time at which the extensive part of the entanglement of $A$ is entirely multipartite is when $Q_4$ stays at its bound $Q_4= 4\ell_A$, i.e. when $\frac{\ell_A+\ell_B}{2}<t < \frac{\ell_A+\ell_B+\ell_C}{2}$. During this time $\Tilde{Q}_4$ and $\bar{Q}_4$ could be 0 or non-maximal, which means that they fail to reflect the fact that the entanglement of $A$ is entirely multipartite. The above different behaviours among $Q_4$, $\Tilde{Q}_4$, and $\bar{Q}_4$ again exemplify the advantage of having multiple signals, as each of them only captures certain aspects of 4-partite entanglement.


There are a number of other versions of $Q_4$: these can be either positive or negative between $\pm 4{\rm min}\{\ell_A,\ell_B\}$, and might saturate these bounds or not, depending on the subregion sizes. 
%



\chapter{Conclusion}
In this thesis, we studied several aspects of the real-time dynamics of entanglement entropy in various systems that admit holographic gravity duals. 

In chapter~\ref{Floquet}, we studied the gravitational dual of SSD CFT~\cite{Wen:2018vux} prepared in both vacuum and thermal states with periodic boundary conditions. We found that the holographic dual of the non-trivial time evolution of SSD CFT in the ground and thermal states can be understood as pure AdS$_3$ and BTZ black hole spacetime with novel foliations. Remarkably, this leads to the presence of $\emph{new horizons}$ in both cases. Subsequently, we computed the holographic entanglement entropy in the gravity dual of SSD CFT, found agreement with CFT$_2$ results in all cases that are comparable, and reproduced the cooling/heating effects holographically. 

In chapter~\ref{Chap3}, we addressed the degeneracy of entanglement membrane~\cite{Jonay:2018yei,Mezei:2018jco} in 2d CFT. Using holography, we found that one needs to introduce an additional degree of freedom on the membrane, which we depicted in colour in our figures. Beyond EE, this $\emph{generalised membrane theory}$ also captures the reflected entropy computed holographically by the entanglement wedge cross sections. To study the interpolation between generalised and ordinary membrane theory, we introduced a relevant deformation to the 2d CFT, which is holographically dual to a hairy planar BTZ black hole with an interior Kasner universe~\cite{Frenkel:2020ysx}. Under such deformation, the conventional non-degenerate membrane tension emerges, which we computed both numerically and from conformal perturbation theory. We also studied reflected entropy in higher dimensional CFTs, and showed that their dynamics in the scaling limit are described by ordinary membrane theory. We found that due to the additional degree of freedom in generalised membrane theory for 2d CFT, the dynamics of reflected entropy show different behaviours in $d=2$ and $d\geq3$. 

In chapter~\ref{Chap4}, we applied the (generalised) membrane theory for reflected entropy introduced in chapter~\ref{Chap3} to address the dynamics of multipartite entanglement signals~\cite{Balasubramanian:2024ysu} in chaotic systems including but not necessarily restricted to holographic theories. We found that scrambling dynamics generates multiparty entanglement with rich structure including: (a) qualitatively different dynamical behaviours for different signals, likely reflecting different dynamics for different kinds of entanglement patterns, (b) discontinuities possibly indicating dynamical phase transitions in the entanglement structure, (c) transient and non-monotonic multiparty entanglement signals,  and (d) periods during which the extensive entanglement of some regions is entirely multipartite.


\renewcommand{\bibname}{Bibliography}
\bibliographystyle{apacite}
\bibliography{Bibliography.bib}

\begin{thebibliography}{}

\bibitem [\protect \citeauthoryear {%
Abajian%
, Aprile%
, Myers%
\BCBL {}\ \BBA {} Vieira%
}{%
Abajian%
\ \protect \BOthers {.}}{%
{\protect \APACyear {2024}}%
}]{%
Abajian:2023bqv}
\APACinsertmetastar {%
Abajian:2023bqv}%
\begin{APACrefauthors}%
Abajian, J.%
, Aprile, F.%
, Myers, R\BPBI C.%
\BCBL {}\ \BBA {} Vieira, P.%
\end{APACrefauthors}%
\unskip\
\newblock
\APACrefYearMonthDay{2024}{}{}.
\newblock
{\BBOQ}\APACrefatitle {{Correlation functions of huge operators in AdS$_{3}$/CFT$_{2}$: domes, doors and book pages}} {{Correlation functions of huge operators in AdS$_{3}$/CFT$_{2}$: domes, doors and book pages}}.{\BBCQ}
\newblock
\APACjournalVolNumPages{JHEP}{03}{}{118}.
\newblock
\begin{APACrefDOI} \doi{10.1007/JHEP03(2024)118} \end{APACrefDOI}
\PrintBackRefs{\CurrentBib}

\bibitem [\protect \citeauthoryear {%
Akers%
, Faulkner%
, Lin%
\BCBL {}\ \BBA {} Rath%
}{%
Akers%
\ \protect \BOthers {.}}{%
{\protect \APACyear {2022}}%
}]{%
Akers:2022max}
\APACinsertmetastar {%
Akers:2022max}%
\begin{APACrefauthors}%
Akers, C.%
, Faulkner, T.%
, Lin, S.%
\BCBL {}\ \BBA {} Rath, P.%
\end{APACrefauthors}%
\unskip\
\newblock
\APACrefYearMonthDay{2022}{}{}.
\newblock
{\BBOQ}\APACrefatitle {{The Page curve for reflected entropy}} {{The Page curve for reflected entropy}}.{\BBCQ}
\newblock
\APACjournalVolNumPages{JHEP}{06}{}{089}.
\newblock
\begin{APACrefDOI} \doi{10.1007/JHEP06(2022)089} \end{APACrefDOI}
\PrintBackRefs{\CurrentBib}

\bibitem [\protect \citeauthoryear {%
Akers%
\ \BBA {} Rath%
}{%
Akers%
\ \BBA {} Rath%
}{%
{\protect \APACyear {2020}}%
}]{%
Akers:2019gcv}
\APACinsertmetastar {%
Akers:2019gcv}%
\begin{APACrefauthors}%
Akers, C.%
\BCBT {}\ \BBA {} Rath, P.%
\end{APACrefauthors}%
\unskip\
\newblock
\APACrefYearMonthDay{2020}{}{}.
\newblock
{\BBOQ}\APACrefatitle {{Entanglement Wedge Cross Sections Require Tripartite Entanglement}} {{Entanglement Wedge Cross Sections Require Tripartite Entanglement}}.{\BBCQ}
\newblock
\APACjournalVolNumPages{JHEP}{04}{}{208}.
\newblock
\begin{APACrefDOI} \doi{10.1007/JHEP04(2020)208} \end{APACrefDOI}
\PrintBackRefs{\CurrentBib}

\bibitem [\protect \citeauthoryear {%
Araki%
\ \BBA {} Lieb%
}{%
Araki%
\ \BBA {} Lieb%
}{%
{\protect \APACyear {1970}}%
}]{%
Araki:1970ba}
\APACinsertmetastar {%
Araki:1970ba}%
\begin{APACrefauthors}%
Araki, H.%
\BCBT {}\ \BBA {} Lieb, E\BPBI H.%
\end{APACrefauthors}%
\unskip\
\newblock
\APACrefYearMonthDay{1970}{}{}.
\newblock
{\BBOQ}\APACrefatitle {{Entropy inequalities}} {{Entropy inequalities}}.{\BBCQ}
\newblock
\APACjournalVolNumPages{Commun. Math. Phys.}{18}{}{160--170}.
\newblock
\begin{APACrefDOI} \doi{10.1007/BF01646092} \end{APACrefDOI}
\PrintBackRefs{\CurrentBib}

\bibitem [\protect \citeauthoryear {%
Asplund%
, Bernamonti%
, Galli%
\BCBL {}\ \BBA {} Hartman%
}{%
Asplund%
\ \protect \BOthers {.}}{%
{\protect \APACyear {2015}}%
{\protect \APACexlab {{\protect \BCnt {1}}}}}]{%
Asplund:2015eha}
\APACinsertmetastar {%
Asplund:2015eha}%
\begin{APACrefauthors}%
Asplund, C\BPBI T.%
, Bernamonti, A.%
, Galli, F.%
\BCBL {}\ \BBA {} Hartman, T.%
\end{APACrefauthors}%
\unskip\
\newblock
\APACrefYearMonthDay{2015{\protect \BCnt {1}}}{}{}.
\newblock
{\BBOQ}\APACrefatitle {{Entanglement Scrambling in 2d Conformal Field Theory}} {{Entanglement Scrambling in 2d Conformal Field Theory}}.{\BBCQ}
\newblock
\APACjournalVolNumPages{JHEP}{09}{}{110}.
\newblock
\begin{APACrefDOI} \doi{10.1007/JHEP09(2015)110} \end{APACrefDOI}
\PrintBackRefs{\CurrentBib}

\bibitem [\protect \citeauthoryear {%
Asplund%
, Bernamonti%
, Galli%
\BCBL {}\ \BBA {} Hartman%
}{%
Asplund%
\ \protect \BOthers {.}}{%
{\protect \APACyear {2015}}%
{\protect \APACexlab {{\protect \BCnt {2}}}}}]{%
Asplund:2014coa}
\APACinsertmetastar {%
Asplund:2014coa}%
\begin{APACrefauthors}%
Asplund, C\BPBI T.%
, Bernamonti, A.%
, Galli, F.%
\BCBL {}\ \BBA {} Hartman, T.%
\end{APACrefauthors}%
\unskip\
\newblock
\APACrefYearMonthDay{2015{\protect \BCnt {2}}}{}{}.
\newblock
{\BBOQ}\APACrefatitle {{Holographic Entanglement Entropy from 2d CFT: Heavy States and Local Quenches}} {{Holographic Entanglement Entropy from 2d CFT: Heavy States and Local Quenches}}.{\BBCQ}
\newblock
\APACjournalVolNumPages{JHEP}{02}{}{171}.
\newblock
\begin{APACrefDOI} \doi{10.1007/JHEP02(2015)171} \end{APACrefDOI}
\PrintBackRefs{\CurrentBib}

\bibitem [\protect \citeauthoryear {%
Babaei~Velni%
, Mohammadi~Mozaffar%
\BCBL {}\ \BBA {} Vahidinia%
}{%
Babaei~Velni%
\ \protect \BOthers {.}}{%
{\protect \APACyear {2020}}%
}]{%
BabaeiVelni:2020wfl}
\APACinsertmetastar {%
BabaeiVelni:2020wfl}%
\begin{APACrefauthors}%
Babaei~Velni, K.%
, Mohammadi~Mozaffar, M\BPBI R.%
\BCBL {}\ \BBA {} Vahidinia, M\BPBI H.%
\end{APACrefauthors}%
\unskip\
\newblock
\APACrefYearMonthDay{2020}{}{}.
\newblock
{\BBOQ}\APACrefatitle {{Evolution of entanglement wedge cross section following a global quench}} {{Evolution of entanglement wedge cross section following a global quench}}.{\BBCQ}
\newblock
\APACjournalVolNumPages{JHEP}{08}{}{129}.
\newblock
\begin{APACrefDOI} \doi{10.1007/JHEP08(2020)129} \end{APACrefDOI}
\PrintBackRefs{\CurrentBib}

\bibitem [\protect \citeauthoryear {%
Babaei~Velni%
, Mohammadi~Mozaffar%
\BCBL {}\ \BBA {} Vahidinia%
}{%
Babaei~Velni%
\ \protect \BOthers {.}}{%
{\protect \APACyear {2023}}%
}]{%
BabaeiVelni:2023cge}
\APACinsertmetastar {%
BabaeiVelni:2023cge}%
\begin{APACrefauthors}%
Babaei~Velni, K.%
, Mohammadi~Mozaffar, M\BPBI R.%
\BCBL {}\ \BBA {} Vahidinia, M\BPBI H.%
\end{APACrefauthors}%
\unskip\
\newblock
\APACrefYearMonthDay{2023}{}{}.
\newblock
{\BBOQ}\APACrefatitle {{Entanglement wedge cross section growth during thermalization}} {{Entanglement wedge cross section growth during thermalization}}.{\BBCQ}
\newblock
\APACjournalVolNumPages{Phys. Rev. D}{107}{10}{106014}.
\newblock
\begin{APACrefDOI} \doi{10.1103/PhysRevD.107.106014} \end{APACrefDOI}
\PrintBackRefs{\CurrentBib}

\bibitem [\protect \citeauthoryear {%
Balasubramanian%
, Bernamonti%
, Copland%
, Craps%
\BCBL {}\ \BBA {} Galli%
}{%
Balasubramanian%
, Bernamonti%
, Copland%
\BCBL {}\ \protect \BOthers {.}}{%
{\protect \APACyear {2011}}%
}]{%
Balasubramanian:2011at}
\APACinsertmetastar {%
Balasubramanian:2011at}%
\begin{APACrefauthors}%
Balasubramanian, V.%
, Bernamonti, A.%
, Copland, N.%
, Craps, B.%
\BCBL {}\ \BBA {} Galli, F.%
\end{APACrefauthors}%
\unskip\
\newblock
\APACrefYearMonthDay{2011}{}{}.
\newblock
{\BBOQ}\APACrefatitle {{Thermalization of mutual and tripartite information in strongly coupled two dimensional conformal field theories}} {{Thermalization of mutual and tripartite information in strongly coupled two dimensional conformal field theories}}.{\BBCQ}
\newblock
\APACjournalVolNumPages{Phys. Rev. D}{84}{}{105017}.
\newblock
\begin{APACrefDOI} \doi{10.1103/PhysRevD.84.105017} \end{APACrefDOI}
\PrintBackRefs{\CurrentBib}

\bibitem [\protect \citeauthoryear {%
Balasubramanian%
, Bernamonti%
, de Boer%
\BCBL {}\ \protect \BOthers {.}}{%
Balasubramanian%
, Bernamonti%
, de Boer%
\BCBL {}\ \protect \BOthers {.}}{%
{\protect \APACyear {2011}}%
{\protect \APACexlab {{\protect \BCnt {1}}}}}]{%
Balasubramanian:2011ur}
\APACinsertmetastar {%
Balasubramanian:2011ur}%
\begin{APACrefauthors}%
Balasubramanian, V.%
, Bernamonti, A.%
, de Boer, J.%
, Copland, N.%
, Craps, B.%
, Keski-Vakkuri, E.%
\BDBL {}Staessens, W.%
\end{APACrefauthors}%
\unskip\
\newblock
\APACrefYearMonthDay{2011{\protect \BCnt {1}}}{}{}.
\newblock
{\BBOQ}\APACrefatitle {{Holographic Thermalization}} {{Holographic Thermalization}}.{\BBCQ}
\newblock
\APACjournalVolNumPages{Phys. Rev. D}{84}{}{026010}.
\newblock
\begin{APACrefDOI} \doi{10.1103/PhysRevD.84.026010} \end{APACrefDOI}
\PrintBackRefs{\CurrentBib}

\bibitem [\protect \citeauthoryear {%
Balasubramanian%
, Bernamonti%
, de Boer%
\BCBL {}\ \protect \BOthers {.}}{%
Balasubramanian%
, Bernamonti%
, de Boer%
\BCBL {}\ \protect \BOthers {.}}{%
{\protect \APACyear {2011}}%
{\protect \APACexlab {{\protect \BCnt {2}}}}}]{%
Balasubramanian:2010ce}
\APACinsertmetastar {%
Balasubramanian:2010ce}%
\begin{APACrefauthors}%
Balasubramanian, V.%
, Bernamonti, A.%
, de Boer, J.%
, Copland, N.%
, Craps, B.%
, Keski-Vakkuri, E.%
\BDBL {}Staessens, W.%
\end{APACrefauthors}%
\unskip\
\newblock
\APACrefYearMonthDay{2011{\protect \BCnt {2}}}{}{}.
\newblock
{\BBOQ}\APACrefatitle {{Thermalization of Strongly Coupled Field Theories}} {{Thermalization of Strongly Coupled Field Theories}}.{\BBCQ}
\newblock
\APACjournalVolNumPages{Phys. Rev. Lett.}{106}{}{191601}.
\newblock
\begin{APACrefDOI} \doi{10.1103/PhysRevLett.106.191601} \end{APACrefDOI}
\PrintBackRefs{\CurrentBib}

\bibitem [\protect \citeauthoryear {%
Balasubramanian%
, Hayden%
, Maloney%
, Marolf%
\BCBL {}\ \BBA {} Ross%
}{%
Balasubramanian%
\ \protect \BOthers {.}}{%
{\protect \APACyear {2014}}%
}]{%
Balasubramanian:2014hda}
\APACinsertmetastar {%
Balasubramanian:2014hda}%
\begin{APACrefauthors}%
Balasubramanian, V.%
, Hayden, P.%
, Maloney, A.%
, Marolf, D.%
\BCBL {}\ \BBA {} Ross, S\BPBI F.%
\end{APACrefauthors}%
\unskip\
\newblock
\APACrefYearMonthDay{2014}{}{}.
\newblock
{\BBOQ}\APACrefatitle {{Multiboundary Wormholes and Holographic Entanglement}} {{Multiboundary Wormholes and Holographic Entanglement}}.{\BBCQ}
\newblock
\APACjournalVolNumPages{Class. Quant. Grav.}{31}{}{185015}.
\newblock
\begin{APACrefDOI} \doi{10.1088/0264-9381/31/18/185015} \end{APACrefDOI}
\PrintBackRefs{\CurrentBib}

\bibitem [\protect \citeauthoryear {%
Balasubramanian%
, Kang%
, Cummings%
, Murdia%
\BCBL {}\ \BBA {} Ross%
}{%
Balasubramanian%
, Kang%
, Cummings%
\BCBL {}\ \protect \BOthers {.}}{%
{\protect \APACyear {2025}}%
}]{%
Balasubramanian:2025hxg}
\APACinsertmetastar {%
Balasubramanian:2025hxg}%
\begin{APACrefauthors}%
Balasubramanian, V.%
, Kang, M\BPBI J.%
, Cummings, C.%
, Murdia, C.%
\BCBL {}\ \BBA {} Ross, S\BPBI F.%
\end{APACrefauthors}%
\unskip\
\newblock
\APACrefYearMonthDay{2025}{9}{}.
\newblock
{\BBOQ}\APACrefatitle {{Purely GHZ-like entanglement is forbidden in holography}} {{Purely GHZ-like entanglement is forbidden in holography}}.{\BBCQ}
\newblock

\PrintBackRefs{\CurrentBib}

\bibitem [\protect \citeauthoryear {%
Balasubramanian%
, Kang%
, Murdia%
\BCBL {}\ \BBA {} Ross%
}{%
Balasubramanian%
, Kang%
, Murdia%
\BCBL {}\ \BBA {} Ross%
}{%
{\protect \APACyear {2025}}%
}]{%
Balasubramanian:2024ysu}
\APACinsertmetastar {%
Balasubramanian:2024ysu}%
\begin{APACrefauthors}%
Balasubramanian, V.%
, Kang, M\BPBI J.%
, Murdia, C.%
\BCBL {}\ \BBA {} Ross, S\BPBI F.%
\end{APACrefauthors}%
\unskip\
\newblock
\APACrefYearMonthDay{2025}{}{}.
\newblock
{\BBOQ}\APACrefatitle {{Signals of multiparty entanglement and holography}} {{Signals of multiparty entanglement and holography}}.{\BBCQ}
\newblock
\APACjournalVolNumPages{JHEP}{06}{}{068}.
\newblock
\begin{APACrefDOI} \doi{10.1007/JHEP06(2025)068} \end{APACrefDOI}
\PrintBackRefs{\CurrentBib}

\bibitem [\protect \citeauthoryear {%
Banados%
}{%
Banados%
}{%
{\protect \APACyear {1999}}%
}]{%
Banados:1998gg}
\APACinsertmetastar {%
Banados:1998gg}%
\begin{APACrefauthors}%
Banados, M.%
\end{APACrefauthors}%
\unskip\
\newblock
\APACrefYearMonthDay{1999}{}{}.
\newblock
{\BBOQ}\APACrefatitle {{Three-dimensional quantum geometry and black holes}} {{Three-dimensional quantum geometry and black holes}}.{\BBCQ}
\newblock
\APACjournalVolNumPages{AIP Conf. Proc.}{484}{1}{147--169}.
\newblock
\begin{APACrefDOI} \doi{10.1063/1.59661} \end{APACrefDOI}
\PrintBackRefs{\CurrentBib}

\bibitem [\protect \citeauthoryear {%
Banados%
, Henneaux%
, Teitelboim%
\BCBL {}\ \BBA {} Zanelli%
}{%
Banados%
\ \protect \BOthers {.}}{%
{\protect \APACyear {1993}}%
}]{%
Banados:1992gq}
\APACinsertmetastar {%
Banados:1992gq}%
\begin{APACrefauthors}%
Banados, M.%
, Henneaux, M.%
, Teitelboim, C.%
\BCBL {}\ \BBA {} Zanelli, J.%
\end{APACrefauthors}%
\unskip\
\newblock
\APACrefYearMonthDay{1993}{}{}.
\newblock
{\BBOQ}\APACrefatitle {{Geometry of the (2+1) black hole}} {{Geometry of the (2+1) black hole}}.{\BBCQ}
\newblock
\APACjournalVolNumPages{Phys. Rev. D}{48}{}{1506--1525}.
\newblock
\APACrefnote{[Erratum: Phys.Rev.D 88, 069902 (2013)]}
\newblock
\begin{APACrefDOI} \doi{10.1103/PhysRevD.48.1506} \end{APACrefDOI}
\PrintBackRefs{\CurrentBib}

\bibitem [\protect \citeauthoryear {%
Banados%
, Teitelboim%
\BCBL {}\ \BBA {} Zanelli%
}{%
Banados%
\ \protect \BOthers {.}}{%
{\protect \APACyear {1992}}%
}]{%
Banados:1992wn}
\APACinsertmetastar {%
Banados:1992wn}%
\begin{APACrefauthors}%
Banados, M.%
, Teitelboim, C.%
\BCBL {}\ \BBA {} Zanelli, J.%
\end{APACrefauthors}%
\unskip\
\newblock
\APACrefYearMonthDay{1992}{}{}.
\newblock
{\BBOQ}\APACrefatitle {{The Black hole in three-dimensional space-time}} {{The Black hole in three-dimensional space-time}}.{\BBCQ}
\newblock
\APACjournalVolNumPages{Phys. Rev. Lett.}{69}{}{1849--1851}.
\newblock
\begin{APACrefDOI} \doi{10.1103/PhysRevLett.69.1849} \end{APACrefDOI}
\PrintBackRefs{\CurrentBib}

\bibitem [\protect \citeauthoryear {%
Banks%
, Douglas%
, Horowitz%
\BCBL {}\ \BBA {} Martinec%
}{%
Banks%
\ \protect \BOthers {.}}{%
{\protect \APACyear {1998}}%
}]{%
Banks:1998dd}
\APACinsertmetastar {%
Banks:1998dd}%
\begin{APACrefauthors}%
Banks, T.%
, Douglas, M\BPBI R.%
, Horowitz, G\BPBI T.%
\BCBL {}\ \BBA {} Martinec, E\BPBI J.%
\end{APACrefauthors}%
\unskip\
\newblock
\APACrefYearMonthDay{1998}{8}{}.
\newblock
{\BBOQ}\APACrefatitle {{AdS dynamics from conformal field theory}} {{AdS dynamics from conformal field theory}}.{\BBCQ}
\newblock

\PrintBackRefs{\CurrentBib}

\bibitem [\protect \citeauthoryear {%
Belinski%
\ \BBA {} Khalatnikov%
}{%
Belinski%
\ \BBA {} Khalatnikov%
}{%
{\protect \APACyear {1973}}%
}]{%
Belinski:1973zz}
\APACinsertmetastar {%
Belinski:1973zz}%
\begin{APACrefauthors}%
Belinski, V\BPBI A.%
\BCBT {}\ \BBA {} Khalatnikov, I\BPBI M.%
\end{APACrefauthors}%
\unskip\
\newblock
\APACrefYearMonthDay{1973}{}{}.
\newblock
{\BBOQ}\APACrefatitle {{Effect of Scalar and Vector Fields on the Nature of the Cosmological Singularity}} {{Effect of Scalar and Vector Fields on the Nature of the Cosmological Singularity}}.{\BBCQ}
\newblock
\APACjournalVolNumPages{Sov. Phys. JETP}{36}{}{591}.
\PrintBackRefs{\CurrentBib}

\bibitem [\protect \citeauthoryear {%
Bernamonti%
, Galli%
\BCBL {}\ \BBA {} Ge%
}{%
Bernamonti%
\ \protect \BOthers {.}}{%
{\protect \APACyear {2024}}%
}]{%
Bernamonti:2024fgx}
\APACinsertmetastar {%
Bernamonti:2024fgx}%
\begin{APACrefauthors}%
Bernamonti, A.%
, Galli, F.%
\BCBL {}\ \BBA {} Ge, D.%
\end{APACrefauthors}%
\unskip\
\newblock
\APACrefYearMonthDay{2024}{2}{}.
\newblock
{\BBOQ}\APACrefatitle {{Boundary-induced transitions in M\"obius quenches of holographic BCFT}} {{Boundary-induced transitions in M\"obius quenches of holographic BCFT}}.{\BBCQ}
\newblock

\PrintBackRefs{\CurrentBib}

\bibitem [\protect \citeauthoryear {%
Blake%
, Davison%
, Grozdanov%
\BCBL {}\ \BBA {} Liu%
}{%
Blake%
, Davison%
\BCBL {}\ \protect \BOthers {.}}{%
{\protect \APACyear {2018}}%
}]{%
Blake:2018leo}
\APACinsertmetastar {%
Blake:2018leo}%
\begin{APACrefauthors}%
Blake, M.%
, Davison, R\BPBI A.%
, Grozdanov, S.%
\BCBL {}\ \BBA {} Liu, H.%
\end{APACrefauthors}%
\unskip\
\newblock
\APACrefYearMonthDay{2018}{}{}.
\newblock
{\BBOQ}\APACrefatitle {{Many-body chaos and energy dynamics in holography}} {{Many-body chaos and energy dynamics in holography}}.{\BBCQ}
\newblock
\APACjournalVolNumPages{JHEP}{10}{}{035}.
\newblock
\begin{APACrefDOI} \doi{10.1007/JHEP10(2018)035} \end{APACrefDOI}
\PrintBackRefs{\CurrentBib}

\bibitem [\protect \citeauthoryear {%
Blake%
, Lee%
\BCBL {}\ \BBA {} Liu%
}{%
Blake%
, Lee%
\BCBL {}\ \BBA {} Liu%
}{%
{\protect \APACyear {2018}}%
}]{%
Blake:2017ris}
\APACinsertmetastar {%
Blake:2017ris}%
\begin{APACrefauthors}%
Blake, M.%
, Lee, H.%
\BCBL {}\ \BBA {} Liu, H.%
\end{APACrefauthors}%
\unskip\
\newblock
\APACrefYearMonthDay{2018}{}{}.
\newblock
{\BBOQ}\APACrefatitle {{A quantum hydrodynamical description for scrambling and many-body chaos}} {{A quantum hydrodynamical description for scrambling and many-body chaos}}.{\BBCQ}
\newblock
\APACjournalVolNumPages{JHEP}{10}{}{127}.
\newblock
\begin{APACrefDOI} \doi{10.1007/JHEP10(2018)127} \end{APACrefDOI}
\PrintBackRefs{\CurrentBib}

\bibitem [\protect \citeauthoryear {%
Bombelli%
, Koul%
, Lee%
\BCBL {}\ \BBA {} Sorkin%
}{%
Bombelli%
\ \protect \BOthers {.}}{%
{\protect \APACyear {1986}}%
}]{%
Bombelli:1986rw}
\APACinsertmetastar {%
Bombelli:1986rw}%
\begin{APACrefauthors}%
Bombelli, L.%
, Koul, R\BPBI K.%
, Lee, J.%
\BCBL {}\ \BBA {} Sorkin, R\BPBI D.%
\end{APACrefauthors}%
\unskip\
\newblock
\APACrefYearMonthDay{1986}{}{}.
\newblock
{\BBOQ}\APACrefatitle {{A Quantum Source of Entropy for Black Holes}} {{A Quantum Source of Entropy for Black Holes}}.{\BBCQ}
\newblock
\APACjournalVolNumPages{Phys. Rev. D}{34}{}{373--383}.
\newblock
\begin{APACrefDOI} \doi{10.1103/PhysRevD.34.373} \end{APACrefDOI}
\PrintBackRefs{\CurrentBib}

\bibitem [\protect \citeauthoryear {%
Breitenlohner%
\ \BBA {} Freedman%
}{%
Breitenlohner%
\ \BBA {} Freedman%
}{%
{\protect \APACyear {1982}}%
}]{%
Breitenlohner:1982jf}
\APACinsertmetastar {%
Breitenlohner:1982jf}%
\begin{APACrefauthors}%
Breitenlohner, P.%
\BCBT {}\ \BBA {} Freedman, D\BPBI Z.%
\end{APACrefauthors}%
\unskip\
\newblock
\APACrefYearMonthDay{1982}{}{}.
\newblock
{\BBOQ}\APACrefatitle {{Stability in Gauged Extended Supergravity}} {{Stability in Gauged Extended Supergravity}}.{\BBCQ}
\newblock
\APACjournalVolNumPages{Annals Phys.}{144}{}{249}.
\newblock
\begin{APACrefDOI} \doi{10.1016/0003-4916(82)90116-6} \end{APACrefDOI}
\PrintBackRefs{\CurrentBib}

\bibitem [\protect \citeauthoryear {%
Brown%
\ \BBA {} Henneaux%
}{%
Brown%
\ \BBA {} Henneaux%
}{%
{\protect \APACyear {1986}}%
}]{%
Brown:1986nw}
\APACinsertmetastar {%
Brown:1986nw}%
\begin{APACrefauthors}%
Brown, J\BPBI D.%
\BCBT {}\ \BBA {} Henneaux, M.%
\end{APACrefauthors}%
\unskip\
\newblock
\APACrefYearMonthDay{1986}{}{}.
\newblock
{\BBOQ}\APACrefatitle {{Central Charges in the Canonical Realization of Asymptotic Symmetries: An Example from Three-Dimensional Gravity}} {{Central Charges in the Canonical Realization of Asymptotic Symmetries: An Example from Three-Dimensional Gravity}}.{\BBCQ}
\newblock
\APACjournalVolNumPages{Commun. Math. Phys.}{104}{}{207--226}.
\newblock
\begin{APACrefDOI} \doi{10.1007/BF01211590} \end{APACrefDOI}
\PrintBackRefs{\CurrentBib}

\bibitem [\protect \citeauthoryear {%
Caceres%
, Kundu%
, Patra%
\BCBL {}\ \BBA {} Shashi%
}{%
Caceres%
\ \protect \BOthers {.}}{%
{\protect \APACyear {2022}}%
}]{%
Caceres:2021fuw}
\APACinsertmetastar {%
Caceres:2021fuw}%
\begin{APACrefauthors}%
Caceres, E.%
, Kundu, A.%
, Patra, A\BPBI K.%
\BCBL {}\ \BBA {} Shashi, S.%
\end{APACrefauthors}%
\unskip\
\newblock
\APACrefYearMonthDay{2022}{}{}.
\newblock
{\BBOQ}\APACrefatitle {{Page curves and bath deformations}} {{Page curves and bath deformations}}.{\BBCQ}
\newblock
\APACjournalVolNumPages{SciPost Phys. Core}{5}{}{033}.
\newblock
\begin{APACrefDOI} \doi{10.21468/SciPostPhysCore.5.2.033} \end{APACrefDOI}
\PrintBackRefs{\CurrentBib}

\bibitem [\protect \citeauthoryear {%
Calabrese%
\ \BBA {} Cardy%
}{%
Calabrese%
\ \BBA {} Cardy%
}{%
{\protect \APACyear {2007}}%
}]{%
Calabrese:2007rg}
\APACinsertmetastar {%
Calabrese:2007rg}%
\begin{APACrefauthors}%
Calabrese, P.%
\BCBT {}\ \BBA {} Cardy, J.%
\end{APACrefauthors}%
\unskip\
\newblock
\APACrefYearMonthDay{2007}{}{}.
\newblock
{\BBOQ}\APACrefatitle {{Quantum Quenches in Extended Systems}} {{Quantum Quenches in Extended Systems}}.{\BBCQ}
\newblock
\APACjournalVolNumPages{J. Stat. Mech.}{0706}{}{P06008}.
\newblock
\begin{APACrefDOI} \doi{10.1088/1742-5468/2007/06/P06008} \end{APACrefDOI}
\PrintBackRefs{\CurrentBib}

\bibitem [\protect \citeauthoryear {%
Calabrese%
\ \BBA {} Cardy%
}{%
Calabrese%
\ \BBA {} Cardy%
}{%
{\protect \APACyear {2004}}%
}]{%
Calabrese:2004eu}
\APACinsertmetastar {%
Calabrese:2004eu}%
\begin{APACrefauthors}%
Calabrese, P.%
\BCBT {}\ \BBA {} Cardy, J\BPBI L.%
\end{APACrefauthors}%
\unskip\
\newblock
\APACrefYearMonthDay{2004}{}{}.
\newblock
{\BBOQ}\APACrefatitle {{Entanglement entropy and quantum field theory}} {{Entanglement entropy and quantum field theory}}.{\BBCQ}
\newblock
\APACjournalVolNumPages{J. Stat. Mech.}{0406}{}{P06002}.
\newblock
\begin{APACrefDOI} \doi{10.1088/1742-5468/2004/06/P06002} \end{APACrefDOI}
\PrintBackRefs{\CurrentBib}

\bibitem [\protect \citeauthoryear {%
Calabrese%
\ \BBA {} Cardy%
}{%
Calabrese%
\ \BBA {} Cardy%
}{%
{\protect \APACyear {2005}}%
}]{%
Calabrese:2005in}
\APACinsertmetastar {%
Calabrese:2005in}%
\begin{APACrefauthors}%
Calabrese, P.%
\BCBT {}\ \BBA {} Cardy, J\BPBI L.%
\end{APACrefauthors}%
\unskip\
\newblock
\APACrefYearMonthDay{2005}{}{}.
\newblock
{\BBOQ}\APACrefatitle {{Evolution of entanglement entropy in one-dimensional systems}} {{Evolution of entanglement entropy in one-dimensional systems}}.{\BBCQ}
\newblock
\APACjournalVolNumPages{J. Stat. Mech.}{0504}{}{P04010}.
\newblock
\begin{APACrefDOI} \doi{10.1088/1742-5468/2005/04/P04010} \end{APACrefDOI}
\PrintBackRefs{\CurrentBib}

\bibitem [\protect \citeauthoryear {%
Cardy%
\ \BBA {} Tonni%
}{%
Cardy%
\ \BBA {} Tonni%
}{%
{\protect \APACyear {2016}}%
}]{%
Cardy:2016fqc}
\APACinsertmetastar {%
Cardy:2016fqc}%
\begin{APACrefauthors}%
Cardy, J.%
\BCBT {}\ \BBA {} Tonni, E.%
\end{APACrefauthors}%
\unskip\
\newblock
\APACrefYearMonthDay{2016}{}{}.
\newblock
{\BBOQ}\APACrefatitle {{Entanglement hamiltonians in two-dimensional conformal field theory}} {{Entanglement hamiltonians in two-dimensional conformal field theory}}.{\BBCQ}
\newblock
\APACjournalVolNumPages{J. Stat. Mech.}{1612}{12}{123103}.
\newblock
\begin{APACrefDOI} \doi{10.1088/1742-5468/2016/12/123103} \end{APACrefDOI}
\PrintBackRefs{\CurrentBib}

\bibitem [\protect \citeauthoryear {%
Casini%
, Liu%
\BCBL {}\ \BBA {} Mezei%
}{%
Casini%
\ \protect \BOthers {.}}{%
{\protect \APACyear {2016}}%
}]{%
Casini:2015zua}
\APACinsertmetastar {%
Casini:2015zua}%
\begin{APACrefauthors}%
Casini, H.%
, Liu, H.%
\BCBL {}\ \BBA {} Mezei, M.%
\end{APACrefauthors}%
\unskip\
\newblock
\APACrefYearMonthDay{2016}{}{}.
\newblock
{\BBOQ}\APACrefatitle {{Spread of entanglement and causality}} {{Spread of entanglement and causality}}.{\BBCQ}
\newblock
\APACjournalVolNumPages{JHEP}{07}{}{077}.
\newblock
\begin{APACrefDOI} \doi{10.1007/JHEP07(2016)077} \end{APACrefDOI}
\PrintBackRefs{\CurrentBib}

\bibitem [\protect \citeauthoryear {%
Chowdhury%
, Georges%
, Parcollet%
\BCBL {}\ \BBA {} Sachdev%
}{%
Chowdhury%
\ \protect \BOthers {.}}{%
{\protect \APACyear {2022}}%
}]{%
Chowdhury:2021qpy}
\APACinsertmetastar {%
Chowdhury:2021qpy}%
\begin{APACrefauthors}%
Chowdhury, D.%
, Georges, A.%
, Parcollet, O.%
\BCBL {}\ \BBA {} Sachdev, S.%
\end{APACrefauthors}%
\unskip\
\newblock
\APACrefYearMonthDay{2022}{}{}.
\newblock
{\BBOQ}\APACrefatitle {{Sachdev-Ye-Kitaev models and beyond: Window into non-Fermi liquids}} {{Sachdev-Ye-Kitaev models and beyond: Window into non-Fermi liquids}}.{\BBCQ}
\newblock
\APACjournalVolNumPages{Rev. Mod. Phys.}{94}{3}{035004}.
\newblock
\begin{APACrefDOI} \doi{10.1103/RevModPhys.94.035004} \end{APACrefDOI}
\PrintBackRefs{\CurrentBib}

\bibitem [\protect \citeauthoryear {%
Cui%
\ \protect \BOthers {.}}{%
Cui%
\ \protect \BOthers {.}}{%
{\protect \APACyear {2019}}%
}]{%
Cui:2018dyq}
\APACinsertmetastar {%
Cui:2018dyq}%
\begin{APACrefauthors}%
Cui, S\BPBI X.%
, Hayden, P.%
, He, T.%
, Headrick, M.%
, Stoica, B.%
\BCBL {}\ \BBA {} Walter, M.%
\end{APACrefauthors}%
\unskip\
\newblock
\APACrefYearMonthDay{2019}{}{}.
\newblock
{\BBOQ}\APACrefatitle {{Bit Threads and Holographic Monogamy}} {{Bit Threads and Holographic Monogamy}}.{\BBCQ}
\newblock
\APACjournalVolNumPages{Commun. Math. Phys.}{376}{1}{609--648}.
\newblock
\begin{APACrefDOI} \doi{10.1007/s00220-019-03510-8} \end{APACrefDOI}
\PrintBackRefs{\CurrentBib}

\bibitem [\protect \citeauthoryear {%
Das%
, Ezhuthachan%
, Porey%
\BCBL {}\ \BBA {} Roy%
}{%
Das%
\ \protect \BOthers {.}}{%
{\protect \APACyear {2024}}%
}]{%
Das:2024vqe}
\APACinsertmetastar {%
Das:2024vqe}%
\begin{APACrefauthors}%
Das, S.%
, Ezhuthachan, B.%
, Porey, S.%
\BCBL {}\ \BBA {} Roy, B.%
\end{APACrefauthors}%
\unskip\
\newblock
\APACrefYearMonthDay{2024}{6}{}.
\newblock
{\BBOQ}\APACrefatitle {{Notes on heating phase dynamics in Floquet CFTs and Modular quantization}} {{Notes on heating phase dynamics in Floquet CFTs and Modular quantization}}.{\BBCQ}
\newblock

\PrintBackRefs{\CurrentBib}

\bibitem [\protect \citeauthoryear {%
Davison%
\ \BBA {} Delacretaz%
}{%
Davison%
\ \BBA {} Delacretaz%
}{%
{\protect \APACyear {2024}}%
}]{%
Davison:2024msq}
\APACinsertmetastar {%
Davison:2024msq}%
\begin{APACrefauthors}%
Davison, R\BPBI A.%
\BCBT {}\ \BBA {} Delacretaz, L\BPBI V.%
\end{APACrefauthors}%
\unskip\
\newblock
\APACrefYearMonthDay{2024}{9}{}.
\newblock
{\BBOQ}\APACrefatitle {{Universal thermalization dynamics in (1+1)d QFTs}} {{Universal thermalization dynamics in (1+1)d QFTs}}.{\BBCQ}
\newblock

\PrintBackRefs{\CurrentBib}

\bibitem [\protect \citeauthoryear {%
Davison%
\ \BBA {} Jiang%
}{%
Davison%
\ \BBA {} Jiang%
}{%
{\protect \APACyear {2026}}%
}]{%
Davison:2025xdj}
\APACinsertmetastar {%
Davison:2025xdj}%
\begin{APACrefauthors}%
Davison, R\BPBI A.%
\BCBT {}\ \BBA {} Jiang, H.%
\end{APACrefauthors}%
\unskip\
\newblock
\APACrefYearMonthDay{2026}{}{}.
\newblock
{\BBOQ}\APACrefatitle {{Pole skipping from universal hydrodynamics of (1+1)d QFTs}} {{Pole skipping from universal hydrodynamics of (1+1)d QFTs}}.{\BBCQ}
\newblock
\APACjournalVolNumPages{JHEP}{04}{}{162}.
\newblock
\begin{APACrefDOI} \doi{10.1007/JHEP04(2026)162} \end{APACrefDOI}
\PrintBackRefs{\CurrentBib}

\bibitem [\protect \citeauthoryear {%
de Boer%
\ \protect \BOthers {.}}{%
de Boer%
\ \protect \BOthers {.}}{%
{\protect \APACyear {2022}}%
}]{%
deBoer:2021zlm}
\APACinsertmetastar {%
deBoer:2021zlm}%
\begin{APACrefauthors}%
de Boer, J.%
, Esp\'\i{}ndola, R.%
, Najian, B.%
, Patramanis, D.%
, van~der Heijden, J.%
\BCBL {}\ \BBA {} Zukowski, C.%
\end{APACrefauthors}%
\unskip\
\newblock
\APACrefYearMonthDay{2022}{}{}.
\newblock
{\BBOQ}\APACrefatitle {{Virasoro entanglement Berry phases}} {{Virasoro entanglement Berry phases}}.{\BBCQ}
\newblock
\APACjournalVolNumPages{JHEP}{03}{}{179}.
\newblock
\begin{APACrefDOI} \doi{10.1007/JHEP03(2022)179} \end{APACrefDOI}
\PrintBackRefs{\CurrentBib}

\bibitem [\protect \citeauthoryear {%
de Boer%
, Godet%
, Kastikainen%
\BCBL {}\ \BBA {} Keski-Vakkuri%
}{%
de Boer%
\ \protect \BOthers {.}}{%
{\protect \APACyear {2023}}%
}]{%
deBoer:2023lrd}
\APACinsertmetastar {%
deBoer:2023lrd}%
\begin{APACrefauthors}%
de Boer, J.%
, Godet, V.%
, Kastikainen, J.%
\BCBL {}\ \BBA {} Keski-Vakkuri, E.%
\end{APACrefauthors}%
\unskip\
\newblock
\APACrefYearMonthDay{2023}{}{}.
\newblock
{\BBOQ}\APACrefatitle {{Quantum information geometry of driven CFTs}} {{Quantum information geometry of driven CFTs}}.{\BBCQ}
\newblock
\APACjournalVolNumPages{JHEP}{09}{}{087}.
\newblock
\begin{APACrefDOI} \doi{10.1007/JHEP09(2023)087} \end{APACrefDOI}
\PrintBackRefs{\CurrentBib}

\bibitem [\protect \citeauthoryear {%
de Haro%
, Solodukhin%
\BCBL {}\ \BBA {} Skenderis%
}{%
de Haro%
\ \protect \BOthers {.}}{%
{\protect \APACyear {2001}}%
}]{%
deHaro:2000vlm}
\APACinsertmetastar {%
deHaro:2000vlm}%
\begin{APACrefauthors}%
de Haro, S.%
, Solodukhin, S\BPBI N.%
\BCBL {}\ \BBA {} Skenderis, K.%
\end{APACrefauthors}%
\unskip\
\newblock
\APACrefYearMonthDay{2001}{}{}.
\newblock
{\BBOQ}\APACrefatitle {{Holographic reconstruction of space-time and renormalization in the AdS / CFT correspondence}} {{Holographic reconstruction of space-time and renormalization in the AdS / CFT correspondence}}.{\BBCQ}
\newblock
\APACjournalVolNumPages{Commun. Math. Phys.}{217}{}{595--622}.
\newblock
\begin{APACrefDOI} \doi{10.1007/s002200100381} \end{APACrefDOI}
\PrintBackRefs{\CurrentBib}

\bibitem [\protect \citeauthoryear {%
Deutsch%
}{%
Deutsch%
}{%
{\protect \APACyear {1991}}%
}]{%
Deutsch:1991msp}
\APACinsertmetastar {%
Deutsch:1991msp}%
\begin{APACrefauthors}%
Deutsch, J\BPBI M.%
\end{APACrefauthors}%
\unskip\
\newblock
\APACrefYearMonthDay{1991}{}{}.
\newblock
{\BBOQ}\APACrefatitle {{Quantum statistical mechanics in a closed system}} {{Quantum statistical mechanics in a closed system}}.{\BBCQ}
\newblock
\APACjournalVolNumPages{Phys. Rev. A}{43}{4}{2046}.
\newblock
\begin{APACrefDOI} \doi{10.1103/PhysRevA.43.2046} \end{APACrefDOI}
\PrintBackRefs{\CurrentBib}

\bibitem [\protect \citeauthoryear {%
Dong%
, Harlow%
\BCBL {}\ \BBA {} Wall%
}{%
Dong%
, Harlow%
\BCBL {}\ \BBA {} Wall%
}{%
{\protect \APACyear {2016}}%
}]{%
Dong:2016eik}
\APACinsertmetastar {%
Dong:2016eik}%
\begin{APACrefauthors}%
Dong, X.%
, Harlow, D.%
\BCBL {}\ \BBA {} Wall, A\BPBI C.%
\end{APACrefauthors}%
\unskip\
\newblock
\APACrefYearMonthDay{2016}{}{}.
\newblock
{\BBOQ}\APACrefatitle {{Reconstruction of Bulk Operators within the Entanglement Wedge in Gauge-Gravity Duality}} {{Reconstruction of Bulk Operators within the Entanglement Wedge in Gauge-Gravity Duality}}.{\BBCQ}
\newblock
\APACjournalVolNumPages{Phys. Rev. Lett.}{117}{2}{021601}.
\newblock
\begin{APACrefDOI} \doi{10.1103/PhysRevLett.117.021601} \end{APACrefDOI}
\PrintBackRefs{\CurrentBib}

\bibitem [\protect \citeauthoryear {%
Dong%
, Lewkowycz%
\BCBL {}\ \BBA {} Rangamani%
}{%
Dong%
, Lewkowycz%
\BCBL {}\ \BBA {} Rangamani%
}{%
{\protect \APACyear {2016}}%
}]{%
Dong:2016hjy}
\APACinsertmetastar {%
Dong:2016hjy}%
\begin{APACrefauthors}%
Dong, X.%
, Lewkowycz, A.%
\BCBL {}\ \BBA {} Rangamani, M.%
\end{APACrefauthors}%
\unskip\
\newblock
\APACrefYearMonthDay{2016}{}{}.
\newblock
{\BBOQ}\APACrefatitle {{Deriving covariant holographic entanglement}} {{Deriving covariant holographic entanglement}}.{\BBCQ}
\newblock
\APACjournalVolNumPages{JHEP}{11}{}{028}.
\newblock
\begin{APACrefDOI} \doi{10.1007/JHEP11(2016)028} \end{APACrefDOI}
\PrintBackRefs{\CurrentBib}

\bibitem [\protect \citeauthoryear {%
Dutta%
\ \BBA {} Faulkner%
}{%
Dutta%
\ \BBA {} Faulkner%
}{%
{\protect \APACyear {2019}}%
}]{%
Dutta:2019gen}
\APACinsertmetastar {%
Dutta:2019gen}%
\begin{APACrefauthors}%
Dutta, S.%
\BCBT {}\ \BBA {} Faulkner, T.%
\end{APACrefauthors}%
\unskip\
\newblock
\APACrefYearMonthDay{2019}{}{}.
\newblock
{\BBOQ}\APACrefatitle {{A canonical purification for the entanglement wedge cross-section}} {{A canonical purification for the entanglement wedge cross-section}}.{\BBCQ}
\newblock

\PrintBackRefs{\CurrentBib}

\bibitem [\protect \citeauthoryear {%
Engelhardt%
\ \BBA {} Wall%
}{%
Engelhardt%
\ \BBA {} Wall%
}{%
{\protect \APACyear {2019}}%
}]{%
Engelhardt:2018kcs}
\APACinsertmetastar {%
Engelhardt:2018kcs}%
\begin{APACrefauthors}%
Engelhardt, N.%
\BCBT {}\ \BBA {} Wall, A\BPBI C.%
\end{APACrefauthors}%
\unskip\
\newblock
\APACrefYearMonthDay{2019}{}{}.
\newblock
{\BBOQ}\APACrefatitle {{Coarse Graining Holographic Black Holes}} {{Coarse Graining Holographic Black Holes}}.{\BBCQ}
\newblock
\APACjournalVolNumPages{JHEP}{05}{}{160}.
\newblock
\begin{APACrefDOI} \doi{10.1007/JHEP05(2019)160} \end{APACrefDOI}
\PrintBackRefs{\CurrentBib}

\bibitem [\protect \citeauthoryear {%
Fan%
, Gu%
, Vishwanath%
\BCBL {}\ \BBA {} Wen%
}{%
Fan%
\ \protect \BOthers {.}}{%
{\protect \APACyear {2020}}%
}]{%
Fan:2019upv}
\APACinsertmetastar {%
Fan:2019upv}%
\begin{APACrefauthors}%
Fan, R.%
, Gu, Y.%
, Vishwanath, A.%
\BCBL {}\ \BBA {} Wen, X.%
\end{APACrefauthors}%
\unskip\
\newblock
\APACrefYearMonthDay{2020}{}{}.
\newblock
{\BBOQ}\APACrefatitle {{Emergent Spatial Structure and Entanglement Localization in Floquet Conformal Field Theory}} {{Emergent Spatial Structure and Entanglement Localization in Floquet Conformal Field Theory}}.{\BBCQ}
\newblock
\APACjournalVolNumPages{Phys. Rev. X}{10}{3}{031036}.
\newblock
\begin{APACrefDOI} \doi{10.1103/PhysRevX.10.031036} \end{APACrefDOI}
\PrintBackRefs{\CurrentBib}

\bibitem [\protect \citeauthoryear {%
Fan%
, Gu%
, Vishwanath%
\BCBL {}\ \BBA {} Wen%
}{%
Fan%
\ \protect \BOthers {.}}{%
{\protect \APACyear {2021}}%
}]{%
Fan:2020orx}
\APACinsertmetastar {%
Fan:2020orx}%
\begin{APACrefauthors}%
Fan, R.%
, Gu, Y.%
, Vishwanath, A.%
\BCBL {}\ \BBA {} Wen, X.%
\end{APACrefauthors}%
\unskip\
\newblock
\APACrefYearMonthDay{2021}{}{}.
\newblock
{\BBOQ}\APACrefatitle {{Floquet conformal field theories with generally deformed Hamiltonians}} {{Floquet conformal field theories with generally deformed Hamiltonians}}.{\BBCQ}
\newblock
\APACjournalVolNumPages{SciPost Phys.}{10}{2}{049}.
\newblock
\begin{APACrefDOI} \doi{10.21468/SciPostPhys.10.2.049} \end{APACrefDOI}
\PrintBackRefs{\CurrentBib}

\bibitem [\protect \citeauthoryear {%
Fidkowski%
, Hubeny%
, Kleban%
\BCBL {}\ \BBA {} Shenker%
}{%
Fidkowski%
\ \protect \BOthers {.}}{%
{\protect \APACyear {2004}}%
}]{%
Fidkowski:2003nf}
\APACinsertmetastar {%
Fidkowski:2003nf}%
\begin{APACrefauthors}%
Fidkowski, L.%
, Hubeny, V.%
, Kleban, M.%
\BCBL {}\ \BBA {} Shenker, S.%
\end{APACrefauthors}%
\unskip\
\newblock
\APACrefYearMonthDay{2004}{}{}.
\newblock
{\BBOQ}\APACrefatitle {{The Black hole singularity in AdS / CFT}} {{The Black hole singularity in AdS / CFT}}.{\BBCQ}
\newblock
\APACjournalVolNumPages{JHEP}{02}{}{014}.
\newblock
\begin{APACrefDOI} \doi{10.1088/1126-6708/2004/02/014} \end{APACrefDOI}
\PrintBackRefs{\CurrentBib}

\bibitem [\protect \citeauthoryear {%
Fisher%
, Khemani%
, Nahum%
\BCBL {}\ \BBA {} Vijay%
}{%
Fisher%
\ \protect \BOthers {.}}{%
{\protect \APACyear {2023}}%
}]{%
Fisher:2022qey}
\APACinsertmetastar {%
Fisher:2022qey}%
\begin{APACrefauthors}%
Fisher, M\BPBI P\BPBI A.%
, Khemani, V.%
, Nahum, A.%
\BCBL {}\ \BBA {} Vijay, S.%
\end{APACrefauthors}%
\unskip\
\newblock
\APACrefYearMonthDay{2023}{}{}.
\newblock
{\BBOQ}\APACrefatitle {{Random Quantum Circuits}} {{Random Quantum Circuits}}.{\BBCQ}
\newblock
\APACjournalVolNumPages{Ann. Rev. Condensed Matter Phys.}{14}{}{335--379}.
\newblock
\begin{APACrefDOI} \doi{10.1146/annurev-conmatphys-031720-030658} \end{APACrefDOI}
\PrintBackRefs{\CurrentBib}

\bibitem [\protect \citeauthoryear {%
Freedman%
, Gubser%
, Pilch%
\BCBL {}\ \BBA {} Warner%
}{%
Freedman%
\ \protect \BOthers {.}}{%
{\protect \APACyear {1999}}%
}]{%
Freedman:1999gp}
\APACinsertmetastar {%
Freedman:1999gp}%
\begin{APACrefauthors}%
Freedman, D\BPBI Z.%
, Gubser, S\BPBI S.%
, Pilch, K.%
\BCBL {}\ \BBA {} Warner, N\BPBI P.%
\end{APACrefauthors}%
\unskip\
\newblock
\APACrefYearMonthDay{1999}{}{}.
\newblock
{\BBOQ}\APACrefatitle {{Renormalization group flows from holography supersymmetry and a c theorem}} {{Renormalization group flows from holography supersymmetry and a c theorem}}.{\BBCQ}
\newblock
\APACjournalVolNumPages{Adv. Theor. Math. Phys.}{3}{}{363--417}.
\newblock
\begin{APACrefDOI} \doi{10.4310/ATMP.1999.v3.n2.a7} \end{APACrefDOI}
\PrintBackRefs{\CurrentBib}

\bibitem [\protect \citeauthoryear {%
Frenkel%
, Hartnoll%
, Kruthoff%
\BCBL {}\ \BBA {} Shi%
}{%
Frenkel%
\ \protect \BOthers {.}}{%
{\protect \APACyear {2020}}%
}]{%
Frenkel:2020ysx}
\APACinsertmetastar {%
Frenkel:2020ysx}%
\begin{APACrefauthors}%
Frenkel, A.%
, Hartnoll, S\BPBI A.%
, Kruthoff, J.%
\BCBL {}\ \BBA {} Shi, Z\BPBI D.%
\end{APACrefauthors}%
\unskip\
\newblock
\APACrefYearMonthDay{2020}{}{}.
\newblock
{\BBOQ}\APACrefatitle {{Holographic flows from CFT to the Kasner universe}} {{Holographic flows from CFT to the Kasner universe}}.{\BBCQ}
\newblock
\APACjournalVolNumPages{JHEP}{08}{}{003}.
\newblock
\begin{APACrefDOI} \doi{10.1007/JHEP08(2020)003} \end{APACrefDOI}
\PrintBackRefs{\CurrentBib}

\bibitem [\protect \citeauthoryear {%
Gendiar%
, Dani{\v{s}}ka%
, Lee%
\BCBL {}\ \BBA {} Nishino%
}{%
Gendiar%
\ \protect \BOthers {.}}{%
{\protect \APACyear {2011}}%
}]{%
gendiar2011suppression}
\APACinsertmetastar {%
gendiar2011suppression}%
\begin{APACrefauthors}%
Gendiar, A.%
, Dani{\v{s}}ka, M.%
, Lee, Y.%
\BCBL {}\ \BBA {} Nishino, T.%
\end{APACrefauthors}%
\unskip\
\newblock
\APACrefYearMonthDay{2011}{}{}.
\newblock
{\BBOQ}\APACrefatitle {Suppression of finite-size effects in one-dimensional correlated systems} {Suppression of finite-size effects in one-dimensional correlated systems}.{\BBCQ}
\newblock
\APACjournalVolNumPages{Physical Review A}{83}{5}{052118}.
\PrintBackRefs{\CurrentBib}

\bibitem [\protect \citeauthoryear {%
Gendiar%
, Krcmar%
\BCBL {}\ \BBA {} Nishino%
}{%
Gendiar%
\ \protect \BOthers {.}}{%
{\protect \APACyear {2009}}%
}]{%
Gendiar:2008udd}
\APACinsertmetastar {%
Gendiar:2008udd}%
\begin{APACrefauthors}%
Gendiar, A.%
, Krcmar, R.%
\BCBL {}\ \BBA {} Nishino, T.%
\end{APACrefauthors}%
\unskip\
\newblock
\APACrefYearMonthDay{2009}{}{}.
\newblock
{\BBOQ}\APACrefatitle {{Spherical Deformation for One-Dimensional Quantum Systems}} {{Spherical Deformation for One-Dimensional Quantum Systems}}.{\BBCQ}
\newblock
\APACjournalVolNumPages{Prog. Theor. Phys.}{122}{4}{953--967}.
\newblock
\APACrefnote{[Erratum: Prog.Theor.Phys. 123, 393 (2010)]}
\newblock
\begin{APACrefDOI} \doi{10.1143/PTP.122.953} \end{APACrefDOI}
\PrintBackRefs{\CurrentBib}

\bibitem [\protect \citeauthoryear {%
Girardello%
, Petrini%
, Porrati%
\BCBL {}\ \BBA {} Zaffaroni%
}{%
Girardello%
\ \protect \BOthers {.}}{%
{\protect \APACyear {1998}}%
}]{%
Girardello:1998pd}
\APACinsertmetastar {%
Girardello:1998pd}%
\begin{APACrefauthors}%
Girardello, L.%
, Petrini, M.%
, Porrati, M.%
\BCBL {}\ \BBA {} Zaffaroni, A.%
\end{APACrefauthors}%
\unskip\
\newblock
\APACrefYearMonthDay{1998}{}{}.
\newblock
{\BBOQ}\APACrefatitle {{Novel local CFT and exact results on perturbations of N=4 superYang Mills from AdS dynamics}} {{Novel local CFT and exact results on perturbations of N=4 superYang Mills from AdS dynamics}}.{\BBCQ}
\newblock
\APACjournalVolNumPages{JHEP}{12}{}{022}.
\newblock
\begin{APACrefDOI} \doi{10.1088/1126-6708/1998/12/022} \end{APACrefDOI}
\PrintBackRefs{\CurrentBib}

\bibitem [\protect \citeauthoryear {%
Girardello%
, Petrini%
, Porrati%
\BCBL {}\ \BBA {} Zaffaroni%
}{%
Girardello%
\ \protect \BOthers {.}}{%
{\protect \APACyear {2000}}%
}]{%
Girardello:1999bd}
\APACinsertmetastar {%
Girardello:1999bd}%
\begin{APACrefauthors}%
Girardello, L.%
, Petrini, M.%
, Porrati, M.%
\BCBL {}\ \BBA {} Zaffaroni, A.%
\end{APACrefauthors}%
\unskip\
\newblock
\APACrefYearMonthDay{2000}{}{}.
\newblock
{\BBOQ}\APACrefatitle {{The Supergravity dual of N=1 superYang-Mills theory}} {{The Supergravity dual of N=1 superYang-Mills theory}}.{\BBCQ}
\newblock
\APACjournalVolNumPages{Nucl. Phys. B}{569}{}{451--469}.
\newblock
\begin{APACrefDOI} \doi{10.1016/S0550-3213(99)00764-6} \end{APACrefDOI}
\PrintBackRefs{\CurrentBib}

\bibitem [\protect \citeauthoryear {%
Goto%
, Nozaki%
, Ryu%
, Tamaoka%
\BCBL {}\ \BBA {} Tan%
}{%
Goto%
\ \protect \BOthers {.}}{%
{\protect \APACyear {2023}}%
}]{%
Goto:2023wai}
\APACinsertmetastar {%
Goto:2023wai}%
\begin{APACrefauthors}%
Goto, K.%
, Nozaki, M.%
, Ryu, S.%
, Tamaoka, K.%
\BCBL {}\ \BBA {} Tan, M\BPBI T.%
\end{APACrefauthors}%
\unskip\
\newblock
\APACrefYearMonthDay{2023}{2}{}.
\newblock
{\BBOQ}\APACrefatitle {{Scrambling and Recovery of Quantum Information in Inhomogeneous Quenches in Two-dimensional Conformal Field Theories}} {{Scrambling and Recovery of Quantum Information in Inhomogeneous Quenches in Two-dimensional Conformal Field Theories}}.{\BBCQ}
\newblock

\PrintBackRefs{\CurrentBib}

\bibitem [\protect \citeauthoryear {%
Goto%
, Nozaki%
, Tamaoka%
, Tan%
\BCBL {}\ \BBA {} Ryu%
}{%
Goto%
\ \protect \BOthers {.}}{%
{\protect \APACyear {2021}}%
}]{%
Goto:2021sqx}
\APACinsertmetastar {%
Goto:2021sqx}%
\begin{APACrefauthors}%
Goto, K.%
, Nozaki, M.%
, Tamaoka, K.%
, Tan, M\BPBI T.%
\BCBL {}\ \BBA {} Ryu, S.%
\end{APACrefauthors}%
\unskip\
\newblock
\APACrefYearMonthDay{2021}{12}{}.
\newblock
{\BBOQ}\APACrefatitle {{Non-Equilibrating a Black Hole with Inhomogeneous Quantum Quench}} {{Non-Equilibrating a Black Hole with Inhomogeneous Quantum Quench}}.{\BBCQ}
\newblock

\PrintBackRefs{\CurrentBib}

\bibitem [\protect \citeauthoryear {%
Grozdanov%
, Schalm%
\BCBL {}\ \BBA {} Scopelliti%
}{%
Grozdanov%
\ \protect \BOthers {.}}{%
{\protect \APACyear {2018}}%
}]{%
Grozdanov:2017ajz}
\APACinsertmetastar {%
Grozdanov:2017ajz}%
\begin{APACrefauthors}%
Grozdanov, S.%
, Schalm, K.%
\BCBL {}\ \BBA {} Scopelliti, V.%
\end{APACrefauthors}%
\unskip\
\newblock
\APACrefYearMonthDay{2018}{}{}.
\newblock
{\BBOQ}\APACrefatitle {{Black hole scrambling from hydrodynamics}} {{Black hole scrambling from hydrodynamics}}.{\BBCQ}
\newblock
\APACjournalVolNumPages{Phys. Rev. Lett.}{120}{23}{231601}.
\newblock
\begin{APACrefDOI} \doi{10.1103/PhysRevLett.120.231601} \end{APACrefDOI}
\PrintBackRefs{\CurrentBib}

\bibitem [\protect \citeauthoryear {%
Gubser%
, Klebanov%
\BCBL {}\ \BBA {} Polyakov%
}{%
Gubser%
\ \protect \BOthers {.}}{%
{\protect \APACyear {1998}}%
}]{%
Gubser:1998bc}
\APACinsertmetastar {%
Gubser:1998bc}%
\begin{APACrefauthors}%
Gubser, S\BPBI S.%
, Klebanov, I\BPBI R.%
\BCBL {}\ \BBA {} Polyakov, A\BPBI M.%
\end{APACrefauthors}%
\unskip\
\newblock
\APACrefYearMonthDay{1998}{}{}.
\newblock
{\BBOQ}\APACrefatitle {{Gauge theory correlators from noncritical string theory}} {{Gauge theory correlators from noncritical string theory}}.{\BBCQ}
\newblock
\APACjournalVolNumPages{Phys. Lett. B}{428}{}{105--114}.
\newblock
\begin{APACrefDOI} \doi{10.1016/S0370-2693(98)00377-3} \end{APACrefDOI}
\PrintBackRefs{\CurrentBib}

\bibitem [\protect \citeauthoryear {%
Hamilton%
, Kabat%
, Lifschytz%
\BCBL {}\ \BBA {} Lowe%
}{%
Hamilton%
\ \protect \BOthers {.}}{%
{\protect \APACyear {2006}}%
}]{%
Hamilton:2006az}
\APACinsertmetastar {%
Hamilton:2006az}%
\begin{APACrefauthors}%
Hamilton, A.%
, Kabat, D\BPBI N.%
, Lifschytz, G.%
\BCBL {}\ \BBA {} Lowe, D\BPBI A.%
\end{APACrefauthors}%
\unskip\
\newblock
\APACrefYearMonthDay{2006}{}{}.
\newblock
{\BBOQ}\APACrefatitle {{Holographic representation of local bulk operators}} {{Holographic representation of local bulk operators}}.{\BBCQ}
\newblock
\APACjournalVolNumPages{Phys. Rev. D}{74}{}{066009}.
\newblock
\begin{APACrefDOI} \doi{10.1103/PhysRevD.74.066009} \end{APACrefDOI}
\PrintBackRefs{\CurrentBib}

\bibitem [\protect \citeauthoryear {%
Han%
\ \BBA {} Wen%
}{%
Han%
\ \BBA {} Wen%
}{%
{\protect \APACyear {2020}}%
}]{%
Han:2020kwp}
\APACinsertmetastar {%
Han:2020kwp}%
\begin{APACrefauthors}%
Han, B.%
\BCBT {}\ \BBA {} Wen, X.%
\end{APACrefauthors}%
\unskip\
\newblock
\APACrefYearMonthDay{2020}{}{}.
\newblock
{\BBOQ}\APACrefatitle {{Classification of $SL_2$ deformed Floquet conformal field theories}} {{Classification of $SL_2$ deformed Floquet conformal field theories}}.{\BBCQ}
\newblock
\APACjournalVolNumPages{Phys. Rev. B}{102}{20}{205125}.
\newblock
\begin{APACrefDOI} \doi{10.1103/PhysRevB.102.205125} \end{APACrefDOI}
\PrintBackRefs{\CurrentBib}

\bibitem [\protect \citeauthoryear {%
Hartman%
}{%
Hartman%
}{%
{\protect \APACyear {{\protect \bibnodate {}}}}%
}]{%
Hartman}
\APACinsertmetastar {%
Hartman}%
\begin{APACrefauthors}%
Hartman, T.%
\end{APACrefauthors}%
\unskip\
\newblock
\APACrefYearMonthDay{{\protect \bibnodate {}}}{}{}.
\newblock
{\BBOQ}\APACrefatitle {{Lectures on Quantum Gravity and Black Holes}} {{Lectures on Quantum Gravity and Black Holes}}.{\BBCQ}
\newblock

\PrintBackRefs{\CurrentBib}

\bibitem [\protect \citeauthoryear {%
Hartman%
\ \BBA {} Maldacena%
}{%
Hartman%
\ \BBA {} Maldacena%
}{%
{\protect \APACyear {2013}}%
}]{%
Hartman:2013qma}
\APACinsertmetastar {%
Hartman:2013qma}%
\begin{APACrefauthors}%
Hartman, T.%
\BCBT {}\ \BBA {} Maldacena, J.%
\end{APACrefauthors}%
\unskip\
\newblock
\APACrefYearMonthDay{2013}{}{}.
\newblock
{\BBOQ}\APACrefatitle {{Time Evolution of Entanglement Entropy from Black Hole Interiors}} {{Time Evolution of Entanglement Entropy from Black Hole Interiors}}.{\BBCQ}
\newblock
\APACjournalVolNumPages{JHEP}{05}{}{014}.
\newblock
\begin{APACrefDOI} \doi{10.1007/JHEP05(2013)014} \end{APACrefDOI}
\PrintBackRefs{\CurrentBib}

\bibitem [\protect \citeauthoryear {%
Hartnoll%
, Herzog%
\BCBL {}\ \BBA {} Horowitz%
}{%
Hartnoll%
\ \protect \BOthers {.}}{%
{\protect \APACyear {2008}}%
{\protect \APACexlab {{\protect \BCnt {1}}}}}]{%
Hartnoll:2008vx}
\APACinsertmetastar {%
Hartnoll:2008vx}%
\begin{APACrefauthors}%
Hartnoll, S\BPBI A.%
, Herzog, C\BPBI P.%
\BCBL {}\ \BBA {} Horowitz, G\BPBI T.%
\end{APACrefauthors}%
\unskip\
\newblock
\APACrefYearMonthDay{2008{\protect \BCnt {1}}}{}{}.
\newblock
{\BBOQ}\APACrefatitle {{Building a Holographic Superconductor}} {{Building a Holographic Superconductor}}.{\BBCQ}
\newblock
\APACjournalVolNumPages{Phys. Rev. Lett.}{101}{}{031601}.
\newblock
\begin{APACrefDOI} \doi{10.1103/PhysRevLett.101.031601} \end{APACrefDOI}
\PrintBackRefs{\CurrentBib}

\bibitem [\protect \citeauthoryear {%
Hartnoll%
, Herzog%
\BCBL {}\ \BBA {} Horowitz%
}{%
Hartnoll%
\ \protect \BOthers {.}}{%
{\protect \APACyear {2008}}%
{\protect \APACexlab {{\protect \BCnt {2}}}}}]{%
Hartnoll:2008kx}
\APACinsertmetastar {%
Hartnoll:2008kx}%
\begin{APACrefauthors}%
Hartnoll, S\BPBI A.%
, Herzog, C\BPBI P.%
\BCBL {}\ \BBA {} Horowitz, G\BPBI T.%
\end{APACrefauthors}%
\unskip\
\newblock
\APACrefYearMonthDay{2008{\protect \BCnt {2}}}{}{}.
\newblock
{\BBOQ}\APACrefatitle {{Holographic Superconductors}} {{Holographic Superconductors}}.{\BBCQ}
\newblock
\APACjournalVolNumPages{JHEP}{12}{}{015}.
\newblock
\begin{APACrefDOI} \doi{10.1088/1126-6708/2008/12/015} \end{APACrefDOI}
\PrintBackRefs{\CurrentBib}

\bibitem [\protect \citeauthoryear {%
Hartnoll%
\ \BBA {} Neogi%
}{%
Hartnoll%
\ \BBA {} Neogi%
}{%
{\protect \APACyear {2023}}%
}]{%
Hartnoll:2022rdv}
\APACinsertmetastar {%
Hartnoll:2022rdv}%
\begin{APACrefauthors}%
Hartnoll, S\BPBI A.%
\BCBT {}\ \BBA {} Neogi, N.%
\end{APACrefauthors}%
\unskip\
\newblock
\APACrefYearMonthDay{2023}{}{}.
\newblock
{\BBOQ}\APACrefatitle {{AdS black holes with a bouncing interior}} {{AdS black holes with a bouncing interior}}.{\BBCQ}
\newblock
\APACjournalVolNumPages{SciPost Phys.}{14}{4}{074}.
\newblock
\begin{APACrefDOI} \doi{10.21468/SciPostPhys.14.4.074} \end{APACrefDOI}
\PrintBackRefs{\CurrentBib}

\bibitem [\protect \citeauthoryear {%
Hayden%
, Headrick%
\BCBL {}\ \BBA {} Maloney%
}{%
Hayden%
\ \protect \BOthers {.}}{%
{\protect \APACyear {2013}}%
}]{%
Hayden:2011ag}
\APACinsertmetastar {%
Hayden:2011ag}%
\begin{APACrefauthors}%
Hayden, P.%
, Headrick, M.%
\BCBL {}\ \BBA {} Maloney, A.%
\end{APACrefauthors}%
\unskip\
\newblock
\APACrefYearMonthDay{2013}{}{}.
\newblock
{\BBOQ}\APACrefatitle {{Holographic Mutual Information is Monogamous}} {{Holographic Mutual Information is Monogamous}}.{\BBCQ}
\newblock
\APACjournalVolNumPages{Phys. Rev. D}{87}{4}{046003}.
\newblock
\begin{APACrefDOI} \doi{10.1103/PhysRevD.87.046003} \end{APACrefDOI}
\PrintBackRefs{\CurrentBib}

\bibitem [\protect \citeauthoryear {%
Hayden%
, Parrikar%
\BCBL {}\ \BBA {} Sorce%
}{%
Hayden%
\ \protect \BOthers {.}}{%
{\protect \APACyear {2021}}%
}]{%
Hayden:2021gno}
\APACinsertmetastar {%
Hayden:2021gno}%
\begin{APACrefauthors}%
Hayden, P.%
, Parrikar, O.%
\BCBL {}\ \BBA {} Sorce, J.%
\end{APACrefauthors}%
\unskip\
\newblock
\APACrefYearMonthDay{2021}{}{}.
\newblock
{\BBOQ}\APACrefatitle {{The Markov gap for geometric reflected entropy}} {{The Markov gap for geometric reflected entropy}}.{\BBCQ}
\newblock
\APACjournalVolNumPages{JHEP}{10}{}{047}.
\newblock
\begin{APACrefDOI} \doi{10.1007/JHEP10(2021)047} \end{APACrefDOI}
\PrintBackRefs{\CurrentBib}

\bibitem [\protect \citeauthoryear {%
Henningson%
\ \BBA {} Skenderis%
}{%
Henningson%
\ \BBA {} Skenderis%
}{%
{\protect \APACyear {1998}}%
}]{%
Henningson:1998gx}
\APACinsertmetastar {%
Henningson:1998gx}%
\begin{APACrefauthors}%
Henningson, M.%
\BCBT {}\ \BBA {} Skenderis, K.%
\end{APACrefauthors}%
\unskip\
\newblock
\APACrefYearMonthDay{1998}{}{}.
\newblock
{\BBOQ}\APACrefatitle {{The Holographic Weyl anomaly}} {{The Holographic Weyl anomaly}}.{\BBCQ}
\newblock
\APACjournalVolNumPages{JHEP}{07}{}{023}.
\newblock
\begin{APACrefDOI} \doi{10.1088/1126-6708/1998/07/023} \end{APACrefDOI}
\PrintBackRefs{\CurrentBib}

\bibitem [\protect \citeauthoryear {%
Hern{\'a}ndez-Cuenca%
, Hubeny%
\BCBL {}\ \BBA {} Jia%
}{%
Hern{\'a}ndez-Cuenca%
\ \protect \BOthers {.}}{%
{\protect \APACyear {2024}}%
}]{%
Hernandez-Cuenca:2023iqh}
\APACinsertmetastar {%
Hernandez-Cuenca:2023iqh}%
\begin{APACrefauthors}%
Hern{\'a}ndez-Cuenca, S.%
, Hubeny, V\BPBI E.%
\BCBL {}\ \BBA {} Jia, H\BPBI F.%
\end{APACrefauthors}%
\unskip\
\newblock
\APACrefYearMonthDay{2024}{}{}.
\newblock
{\BBOQ}\APACrefatitle {{Holographic entropy inequalities and multipartite entanglement}} {{Holographic entropy inequalities and multipartite entanglement}}.{\BBCQ}
\newblock
\APACjournalVolNumPages{JHEP}{08}{}{238}.
\newblock
\begin{APACrefDOI} \doi{10.1007/JHEP08(2024)238} \end{APACrefDOI}
\PrintBackRefs{\CurrentBib}

\bibitem [\protect \citeauthoryear {%
Hikihara%
\ \BBA {} Nishino%
}{%
Hikihara%
\ \BBA {} Nishino%
}{%
{\protect \APACyear {2011}}%
}]{%
Hikihara:2011mtb}
\APACinsertmetastar {%
Hikihara:2011mtb}%
\begin{APACrefauthors}%
Hikihara, T.%
\BCBT {}\ \BBA {} Nishino, T.%
\end{APACrefauthors}%
\unskip\
\newblock
\APACrefYearMonthDay{2011}{}{}.
\newblock
{\BBOQ}\APACrefatitle {{Connecting distant ends of one-dimensional critical systems by a sine-square deformation}} {{Connecting distant ends of one-dimensional critical systems by a sine-square deformation}}.{\BBCQ}
\newblock
\APACjournalVolNumPages{Phys. Rev. B}{83}{6}{060414}.
\newblock
\begin{APACrefDOI} \doi{10.1103/physrevb.83.060414} \end{APACrefDOI}
\PrintBackRefs{\CurrentBib}

\bibitem [\protect \citeauthoryear {%
Holzhey%
, Larsen%
\BCBL {}\ \BBA {} Wilczek%
}{%
Holzhey%
\ \protect \BOthers {.}}{%
{\protect \APACyear {1994}}%
}]{%
Holzhey:1994we}
\APACinsertmetastar {%
Holzhey:1994we}%
\begin{APACrefauthors}%
Holzhey, C.%
, Larsen, F.%
\BCBL {}\ \BBA {} Wilczek, F.%
\end{APACrefauthors}%
\unskip\
\newblock
\APACrefYearMonthDay{1994}{}{}.
\newblock
{\BBOQ}\APACrefatitle {{Geometric and renormalized entropy in conformal field theory}} {{Geometric and renormalized entropy in conformal field theory}}.{\BBCQ}
\newblock
\APACjournalVolNumPages{Nucl. Phys. B}{424}{}{443--467}.
\newblock
\begin{APACrefDOI} \doi{10.1016/0550-3213(94)90402-2} \end{APACrefDOI}
\PrintBackRefs{\CurrentBib}

\bibitem [\protect \citeauthoryear {%
Huang%
}{%
Huang%
}{%
{\protect \APACyear {2019}}%
}]{%
Huang:2019hts}
\APACinsertmetastar {%
Huang:2019hts}%
\begin{APACrefauthors}%
Huang, Y.%
\end{APACrefauthors}%
\unskip\
\newblock
\APACrefYearMonthDay{2019}{2}{}.
\newblock
{\BBOQ}\APACrefatitle {{Dynamics of R\'enyi entanglement entropy in diffusive qudit systems}} {{Dynamics of R\'enyi entanglement entropy in diffusive qudit systems}}.{\BBCQ}
\newblock

\newblock
\begin{APACrefDOI} \doi{10.1088/2633-1357/abd1e2} \end{APACrefDOI}
\PrintBackRefs{\CurrentBib}

\bibitem [\protect \citeauthoryear {%
Hubeny%
}{%
Hubeny%
}{%
{\protect \APACyear {2012}}%
}]{%
Hubeny:2012ry}
\APACinsertmetastar {%
Hubeny:2012ry}%
\begin{APACrefauthors}%
Hubeny, V\BPBI E.%
\end{APACrefauthors}%
\unskip\
\newblock
\APACrefYearMonthDay{2012}{}{}.
\newblock
{\BBOQ}\APACrefatitle {{Extremal surfaces as bulk probes in AdS/CFT}} {{Extremal surfaces as bulk probes in AdS/CFT}}.{\BBCQ}
\newblock
\APACjournalVolNumPages{JHEP}{07}{}{093}.
\newblock
\begin{APACrefDOI} \doi{10.1007/JHEP07(2012)093} \end{APACrefDOI}
\PrintBackRefs{\CurrentBib}

\bibitem [\protect \citeauthoryear {%
Hubeny%
, Maxfield%
, Rangamani%
\BCBL {}\ \BBA {} Tonni%
}{%
Hubeny%
\ \protect \BOthers {.}}{%
{\protect \APACyear {2013}}%
}]{%
Hubeny:2013gta}
\APACinsertmetastar {%
Hubeny:2013gta}%
\begin{APACrefauthors}%
Hubeny, V\BPBI E.%
, Maxfield, H.%
, Rangamani, M.%
\BCBL {}\ \BBA {} Tonni, E.%
\end{APACrefauthors}%
\unskip\
\newblock
\APACrefYearMonthDay{2013}{}{}.
\newblock
{\BBOQ}\APACrefatitle {{Holographic entanglement plateaux}} {{Holographic entanglement plateaux}}.{\BBCQ}
\newblock
\APACjournalVolNumPages{JHEP}{08}{}{092}.
\newblock
\begin{APACrefDOI} \doi{10.1007/JHEP08(2013)092} \end{APACrefDOI}
\PrintBackRefs{\CurrentBib}

\bibitem [\protect \citeauthoryear {%
Hubeny%
, Rangamani%
\BCBL {}\ \BBA {} Takayanagi%
}{%
Hubeny%
\ \protect \BOthers {.}}{%
{\protect \APACyear {2007}}%
}]{%
Hubeny:2007xt}
\APACinsertmetastar {%
Hubeny:2007xt}%
\begin{APACrefauthors}%
Hubeny, V\BPBI E.%
, Rangamani, M.%
\BCBL {}\ \BBA {} Takayanagi, T.%
\end{APACrefauthors}%
\unskip\
\newblock
\APACrefYearMonthDay{2007}{}{}.
\newblock
{\BBOQ}\APACrefatitle {{A Covariant holographic entanglement entropy proposal}} {{A Covariant holographic entanglement entropy proposal}}.{\BBCQ}
\newblock
\APACjournalVolNumPages{JHEP}{07}{}{062}.
\newblock
\begin{APACrefDOI} \doi{10.1088/1126-6708/2007/07/062} \end{APACrefDOI}
\PrintBackRefs{\CurrentBib}

\bibitem [\protect \citeauthoryear {%
Ishibashi%
\ \BBA {} Tada%
}{%
Ishibashi%
\ \BBA {} Tada%
}{%
{\protect \APACyear {2015}}%
}]{%
Ishibashi:2015jba}
\APACinsertmetastar {%
Ishibashi:2015jba}%
\begin{APACrefauthors}%
Ishibashi, N.%
\BCBT {}\ \BBA {} Tada, T.%
\end{APACrefauthors}%
\unskip\
\newblock
\APACrefYearMonthDay{2015}{}{}.
\newblock
{\BBOQ}\APACrefatitle {{Infinite circumference limit of conformal field theory}} {{Infinite circumference limit of conformal field theory}}.{\BBCQ}
\newblock
\APACjournalVolNumPages{J. Phys. A}{48}{31}{315402}.
\newblock
\begin{APACrefDOI} \doi{10.1088/1751-8113/48/31/315402} \end{APACrefDOI}
\PrintBackRefs{\CurrentBib}

\bibitem [\protect \citeauthoryear {%
Ishibashi%
\ \BBA {} Tada%
}{%
Ishibashi%
\ \BBA {} Tada%
}{%
{\protect \APACyear {2016}}%
}]{%
Ishibashi:2016bey}
\APACinsertmetastar {%
Ishibashi:2016bey}%
\begin{APACrefauthors}%
Ishibashi, N.%
\BCBT {}\ \BBA {} Tada, T.%
\end{APACrefauthors}%
\unskip\
\newblock
\APACrefYearMonthDay{2016}{}{}.
\newblock
{\BBOQ}\APACrefatitle {{Dipolar quantization and the infinite circumference limit of two-dimensional conformal field theories}} {{Dipolar quantization and the infinite circumference limit of two-dimensional conformal field theories}}.{\BBCQ}
\newblock
\APACjournalVolNumPages{Int. J. Mod. Phys. A}{31}{32}{1650170}.
\newblock
\begin{APACrefDOI} \doi{10.1142/S0217751X16501700} \end{APACrefDOI}
\PrintBackRefs{\CurrentBib}

\bibitem [\protect \citeauthoryear {%
Jafferis%
, Lewkowycz%
, Maldacena%
\BCBL {}\ \BBA {} Suh%
}{%
Jafferis%
\ \protect \BOthers {.}}{%
{\protect \APACyear {2016}}%
}]{%
Jafferis:2015del}
\APACinsertmetastar {%
Jafferis:2015del}%
\begin{APACrefauthors}%
Jafferis, D\BPBI L.%
, Lewkowycz, A.%
, Maldacena, J.%
\BCBL {}\ \BBA {} Suh, S\BPBI J.%
\end{APACrefauthors}%
\unskip\
\newblock
\APACrefYearMonthDay{2016}{}{}.
\newblock
{\BBOQ}\APACrefatitle {{Relative entropy equals bulk relative entropy}} {{Relative entropy equals bulk relative entropy}}.{\BBCQ}
\newblock
\APACjournalVolNumPages{JHEP}{06}{}{004}.
\newblock
\begin{APACrefDOI} \doi{10.1007/JHEP06(2016)004} \end{APACrefDOI}
\PrintBackRefs{\CurrentBib}

\bibitem [\protect \citeauthoryear {%
Jiang%
\ \BBA {} Mezei%
}{%
Jiang%
\ \BBA {} Mezei%
}{%
{\protect \APACyear {2025}}%
}]{%
Jiang:2024hgt}
\APACinsertmetastar {%
Jiang:2024hgt}%
\begin{APACrefauthors}%
Jiang, H.%
\BCBT {}\ \BBA {} Mezei, M.%
\end{APACrefauthors}%
\unskip\
\newblock
\APACrefYearMonthDay{2025}{}{}.
\newblock
{\BBOQ}\APACrefatitle {{New horizons for inhomogeneous quenches and Floquet CFT}} {{New horizons for inhomogeneous quenches and Floquet CFT}}.{\BBCQ}
\newblock
\APACjournalVolNumPages{JHEP}{04}{}{025}.
\newblock
\begin{APACrefDOI} \doi{10.1007/JHEP04(2025)025} \end{APACrefDOI}
\PrintBackRefs{\CurrentBib}

\bibitem [\protect \citeauthoryear {%
Jiang%
, Mezei%
\BCBL {}\ \BBA {} Virrueta%
}{%
Jiang%
\ \protect \BOthers {.}}{%
{\protect \APACyear {2025}}%
}]{%
Jiang:2024tdj}
\APACinsertmetastar {%
Jiang:2024tdj}%
\begin{APACrefauthors}%
Jiang, H.%
, Mezei, M.%
\BCBL {}\ \BBA {} Virrueta, J.%
\end{APACrefauthors}%
\unskip\
\newblock
\APACrefYearMonthDay{2025}{}{}.
\newblock
{\BBOQ}\APACrefatitle {{The entanglement membrane in 2d CFT: reflected entropy, RG flow, and information velocity}} {{The entanglement membrane in 2d CFT: reflected entropy, RG flow, and information velocity}}.{\BBCQ}
\newblock
\APACjournalVolNumPages{JHEP}{06}{}{114}.
\newblock
\begin{APACrefDOI} \doi{10.1007/JHEP06(2025)114} \end{APACrefDOI}
\PrintBackRefs{\CurrentBib}

\bibitem [\protect \citeauthoryear {%
Jonay%
, Huse%
\BCBL {}\ \BBA {} Nahum%
}{%
Jonay%
\ \protect \BOthers {.}}{%
{\protect \APACyear {2018}}%
}]{%
Jonay:2018yei}
\APACinsertmetastar {%
Jonay:2018yei}%
\begin{APACrefauthors}%
Jonay, C.%
, Huse, D\BPBI A.%
\BCBL {}\ \BBA {} Nahum, A.%
\end{APACrefauthors}%
\unskip\
\newblock
\APACrefYearMonthDay{2018}{2}{}.
\newblock
{\BBOQ}\APACrefatitle {{Coarse-grained dynamics of operator and state entanglement}} {{Coarse-grained dynamics of operator and state entanglement}}.{\BBCQ}
\newblock

\PrintBackRefs{\CurrentBib}

\bibitem [\protect \citeauthoryear {%
Ju%
, Pan%
, Sun%
\BCBL {}\ \BBA {} Zhao%
}{%
Ju%
\ \protect \BOthers {.}}{%
{\protect \APACyear {2024}}%
}]{%
Ju:2024hba}
\APACinsertmetastar {%
Ju:2024hba}%
\begin{APACrefauthors}%
Ju, X\BHBI X.%
, Pan, W\BHBI B.%
, Sun, Y\BHBI W.%
\BCBL {}\ \BBA {} Zhao, Y.%
\end{APACrefauthors}%
\unskip\
\newblock
\APACrefYearMonthDay{2024}{11}{}.
\newblock
{\BBOQ}\APACrefatitle {{Holographic multipartite entanglement from the upper bound of $n$-partite information}} {{Holographic multipartite entanglement from the upper bound of $n$-partite information}}.{\BBCQ}
\newblock

\PrintBackRefs{\CurrentBib}

\bibitem [\protect \citeauthoryear {%
Ju%
, Sun%
\BCBL {}\ \BBA {} Zhao%
}{%
Ju%
\ \protect \BOthers {.}}{%
{\protect \APACyear {2025}}%
}]{%
Ju:2025tgg}
\APACinsertmetastar {%
Ju:2025tgg}%
\begin{APACrefauthors}%
Ju, X\BHBI X.%
, Sun, Y\BHBI W.%
\BCBL {}\ \BBA {} Zhao, Y.%
\end{APACrefauthors}%
\unskip\
\newblock
\APACrefYearMonthDay{2025}{}{}.
\newblock
{\BBOQ}\APACrefatitle {{Upper bound of holographic entanglement entropy combinations}} {{Upper bound of holographic entanglement entropy combinations}}.{\BBCQ}
\newblock
\APACjournalVolNumPages{JHEP}{09}{}{085}.
\newblock
\begin{APACrefDOI} \doi{10.1007/JHEP09(2025)085} \end{APACrefDOI}
\PrintBackRefs{\CurrentBib}

\bibitem [\protect \citeauthoryear {%
Kasner%
}{%
Kasner%
}{%
{\protect \APACyear {1921}}%
}]{%
Kasner:1921zz}
\APACinsertmetastar {%
Kasner:1921zz}%
\begin{APACrefauthors}%
Kasner, E.%
\end{APACrefauthors}%
\unskip\
\newblock
\APACrefYearMonthDay{1921}{}{}.
\newblock
{\BBOQ}\APACrefatitle {{Geometrical theorems on Einstein's cosmological equations}} {{Geometrical theorems on Einstein's cosmological equations}}.{\BBCQ}
\newblock
\APACjournalVolNumPages{Am. J. Math.}{43}{}{217--221}.
\newblock
\begin{APACrefDOI} \doi{10.2307/2370192} \end{APACrefDOI}
\PrintBackRefs{\CurrentBib}

\bibitem [\protect \citeauthoryear {%
Katsura%
}{%
Katsura%
}{%
{\protect \APACyear {2012}}%
}]{%
Katsura:2011ss}
\APACinsertmetastar {%
Katsura:2011ss}%
\begin{APACrefauthors}%
Katsura, H.%
\end{APACrefauthors}%
\unskip\
\newblock
\APACrefYearMonthDay{2012}{}{}.
\newblock
{\BBOQ}\APACrefatitle {{Sine-square deformation of solvable spin chains and conformal field theories}} {{Sine-square deformation of solvable spin chains and conformal field theories}}.{\BBCQ}
\newblock
\APACjournalVolNumPages{J. Phys. A}{45}{}{115003}.
\newblock
\begin{APACrefDOI} \doi{10.1088/1751-8113/45/11/115003} \end{APACrefDOI}
\PrintBackRefs{\CurrentBib}

\bibitem [\protect \citeauthoryear {%
Kaufman%
\ \protect \BOthers {.}}{%
Kaufman%
\ \protect \BOthers {.}}{%
{\protect \APACyear {2016}}%
}]{%
Kaufman:2016mif}
\APACinsertmetastar {%
Kaufman:2016mif}%
\begin{APACrefauthors}%
Kaufman, A\BPBI M.%
, Tai, M\BPBI E.%
, Lukin, A.%
, Rispoli, M.%
, Schittko, R.%
, Preiss, P\BPBI M.%
\BCBL {}\ \BBA {} Greiner, M.%
\end{APACrefauthors}%
\unskip\
\newblock
\APACrefYearMonthDay{2016}{}{}.
\newblock
{\BBOQ}\APACrefatitle {{Quantum thermalization through entanglement in an isolated many-body system}} {{Quantum thermalization through entanglement in an isolated many-body system}}.{\BBCQ}
\newblock
\APACjournalVolNumPages{Science}{353}{6301}{aaf6725}.
\newblock
\begin{APACrefDOI} \doi{10.1126/science.aaf6725} \end{APACrefDOI}
\PrintBackRefs{\CurrentBib}

\bibitem [\protect \citeauthoryear {%
Keski-Vakkuri%
}{%
Keski-Vakkuri%
}{%
{\protect \APACyear {1999}}%
}]{%
Keski-Vakkuri:1998gmz}
\APACinsertmetastar {%
Keski-Vakkuri:1998gmz}%
\begin{APACrefauthors}%
Keski-Vakkuri, E.%
\end{APACrefauthors}%
\unskip\
\newblock
\APACrefYearMonthDay{1999}{}{}.
\newblock
{\BBOQ}\APACrefatitle {{Bulk and boundary dynamics in BTZ black holes}} {{Bulk and boundary dynamics in BTZ black holes}}.{\BBCQ}
\newblock
\APACjournalVolNumPages{Phys. Rev. D}{59}{}{104001}.
\newblock
\begin{APACrefDOI} \doi{10.1103/PhysRevD.59.104001} \end{APACrefDOI}
\PrintBackRefs{\CurrentBib}

\bibitem [\protect \citeauthoryear {%
Kitaev%
\ \BBA {} Preskill%
}{%
Kitaev%
\ \BBA {} Preskill%
}{%
{\protect \APACyear {2006}}%
}]{%
Kitaev:2005dm}
\APACinsertmetastar {%
Kitaev:2005dm}%
\begin{APACrefauthors}%
Kitaev, A.%
\BCBT {}\ \BBA {} Preskill, J.%
\end{APACrefauthors}%
\unskip\
\newblock
\APACrefYearMonthDay{2006}{}{}.
\newblock
{\BBOQ}\APACrefatitle {{Topological entanglement entropy}} {{Topological entanglement entropy}}.{\BBCQ}
\newblock
\APACjournalVolNumPages{Phys. Rev. Lett.}{96}{}{110404}.
\newblock
\begin{APACrefDOI} \doi{10.1103/PhysRevLett.96.110404} \end{APACrefDOI}
\PrintBackRefs{\CurrentBib}

\bibitem [\protect \citeauthoryear {%
Klebanov%
\ \BBA {} Witten%
}{%
Klebanov%
\ \BBA {} Witten%
}{%
{\protect \APACyear {1999}}%
}]{%
Klebanov:1999tb}
\APACinsertmetastar {%
Klebanov:1999tb}%
\begin{APACrefauthors}%
Klebanov, I\BPBI R.%
\BCBT {}\ \BBA {} Witten, E.%
\end{APACrefauthors}%
\unskip\
\newblock
\APACrefYearMonthDay{1999}{}{}.
\newblock
{\BBOQ}\APACrefatitle {{AdS / CFT correspondence and symmetry breaking}} {{AdS / CFT correspondence and symmetry breaking}}.{\BBCQ}
\newblock
\APACjournalVolNumPages{Nucl. Phys. B}{556}{}{89--114}.
\newblock
\begin{APACrefDOI} \doi{10.1016/S0550-3213(99)00387-9} \end{APACrefDOI}
\PrintBackRefs{\CurrentBib}

\bibitem [\protect \citeauthoryear {%
Kudler-Flam%
, Kusuki%
\BCBL {}\ \BBA {} Ryu%
}{%
Kudler-Flam%
\ \protect \BOthers {.}}{%
{\protect \APACyear {2020}}%
}]{%
Kudler-Flam:2020url}
\APACinsertmetastar {%
Kudler-Flam:2020url}%
\begin{APACrefauthors}%
Kudler-Flam, J.%
, Kusuki, Y.%
\BCBL {}\ \BBA {} Ryu, S.%
\end{APACrefauthors}%
\unskip\
\newblock
\APACrefYearMonthDay{2020}{}{}.
\newblock
{\BBOQ}\APACrefatitle {{Correlation measures and the entanglement wedge cross-section after quantum quenches in two-dimensional conformal field theories}} {{Correlation measures and the entanglement wedge cross-section after quantum quenches in two-dimensional conformal field theories}}.{\BBCQ}
\newblock
\APACjournalVolNumPages{JHEP}{04}{}{074}.
\newblock
\begin{APACrefDOI} \doi{10.1007/JHEP04(2020)074} \end{APACrefDOI}
\PrintBackRefs{\CurrentBib}

\bibitem [\protect \citeauthoryear {%
Kudler-Flam%
, Nozaki%
, Numasawa%
, Ryu%
\BCBL {}\ \BBA {} Tan%
}{%
Kudler-Flam%
\ \protect \BOthers {.}}{%
{\protect \APACyear {2023}}%
}]{%
Kudler-Flam:2023ahk}
\APACinsertmetastar {%
Kudler-Flam:2023ahk}%
\begin{APACrefauthors}%
Kudler-Flam, J.%
, Nozaki, M.%
, Numasawa, T.%
, Ryu, S.%
\BCBL {}\ \BBA {} Tan, M\BPBI T.%
\end{APACrefauthors}%
\unskip\
\newblock
\APACrefYearMonthDay{2023}{9}{}.
\newblock
{\BBOQ}\APACrefatitle {{Bridging two quantum quench problems - local joining quantum quench and M\"obius quench - and their holographic dual descriptions}} {{Bridging two quantum quench problems - local joining quantum quench and M\"obius quench - and their holographic dual descriptions}}.{\BBCQ}
\newblock

\PrintBackRefs{\CurrentBib}

\bibitem [\protect \citeauthoryear {%
Levin%
\ \BBA {} Wen%
}{%
Levin%
\ \BBA {} Wen%
}{%
{\protect \APACyear {2006}}%
}]{%
Levin:2006zz}
\APACinsertmetastar {%
Levin:2006zz}%
\begin{APACrefauthors}%
Levin, M.%
\BCBT {}\ \BBA {} Wen, X\BHBI G.%
\end{APACrefauthors}%
\unskip\
\newblock
\APACrefYearMonthDay{2006}{}{}.
\newblock
{\BBOQ}\APACrefatitle {{Detecting Topological Order in a Ground State Wave Function}} {{Detecting Topological Order in a Ground State Wave Function}}.{\BBCQ}
\newblock
\APACjournalVolNumPages{Phys. Rev. Lett.}{96}{}{110405}.
\newblock
\begin{APACrefDOI} \doi{10.1103/PhysRevLett.96.110405} \end{APACrefDOI}
\PrintBackRefs{\CurrentBib}

\bibitem [\protect \citeauthoryear {%
Lewkowycz%
\ \BBA {} Maldacena%
}{%
Lewkowycz%
\ \BBA {} Maldacena%
}{%
{\protect \APACyear {2013}}%
}]{%
Lewkowycz:2013nqa}
\APACinsertmetastar {%
Lewkowycz:2013nqa}%
\begin{APACrefauthors}%
Lewkowycz, A.%
\BCBT {}\ \BBA {} Maldacena, J.%
\end{APACrefauthors}%
\unskip\
\newblock
\APACrefYearMonthDay{2013}{}{}.
\newblock
{\BBOQ}\APACrefatitle {{Generalized gravitational entropy}} {{Generalized gravitational entropy}}.{\BBCQ}
\newblock
\APACjournalVolNumPages{JHEP}{08}{}{090}.
\newblock
\begin{APACrefDOI} \doi{10.1007/JHEP08(2013)090} \end{APACrefDOI}
\PrintBackRefs{\CurrentBib}

\bibitem [\protect \citeauthoryear {%
Lin%
}{%
Lin%
}{%
{\protect \APACyear {2017}}%
}]{%
Lin:2017uzr}
\APACinsertmetastar {%
Lin:2017uzr}%
\begin{APACrefauthors}%
Lin, J.%
\end{APACrefauthors}%
\unskip\
\newblock
\APACrefYearMonthDay{2017}{4}{}.
\newblock
{\BBOQ}\APACrefatitle {{Ryu-Takayanagi Area as an Entanglement Edge Term}} {{Ryu-Takayanagi Area as an Entanglement Edge Term}}.{\BBCQ}
\newblock

\PrintBackRefs{\CurrentBib}

\bibitem [\protect \citeauthoryear {%
Lin%
\ \BBA {} Radi{\v{c}}evi{\'c}%
}{%
Lin%
\ \BBA {} Radi{\v{c}}evi{\'c}%
}{%
{\protect \APACyear {2020}}%
}]{%
Lin:2018bud}
\APACinsertmetastar {%
Lin:2018bud}%
\begin{APACrefauthors}%
Lin, J.%
\BCBT {}\ \BBA {} Radi{\v{c}}evi{\'c}, {\DJ}.%
\end{APACrefauthors}%
\unskip\
\newblock
\APACrefYearMonthDay{2020}{}{}.
\newblock
{\BBOQ}\APACrefatitle {{Comments on defining entanglement entropy}} {{Comments on defining entanglement entropy}}.{\BBCQ}
\newblock
\APACjournalVolNumPages{Nucl. Phys. B}{958}{}{115118}.
\newblock
\begin{APACrefDOI} \doi{10.1016/j.nuclphysb.2020.115118} \end{APACrefDOI}
\PrintBackRefs{\CurrentBib}

\bibitem [\protect \citeauthoryear {%
Liu%
\ \BBA {} Mezei%
}{%
Liu%
\ \BBA {} Mezei%
}{%
{\protect \APACyear {2014}}%
}]{%
Liu:2013una}
\APACinsertmetastar {%
Liu:2013una}%
\begin{APACrefauthors}%
Liu, H.%
\BCBT {}\ \BBA {} Mezei, M.%
\end{APACrefauthors}%
\unskip\
\newblock
\APACrefYearMonthDay{2014}{}{}.
\newblock
{\BBOQ}\APACrefatitle {{Probing renormalization group flows using entanglement entropy}} {{Probing renormalization group flows using entanglement entropy}}.{\BBCQ}
\newblock
\APACjournalVolNumPages{JHEP}{01}{}{098}.
\newblock
\begin{APACrefDOI} \doi{10.1007/JHEP01(2014)098} \end{APACrefDOI}
\PrintBackRefs{\CurrentBib}

\bibitem [\protect \citeauthoryear {%
Liu%
\ \BBA {} Suh%
}{%
Liu%
\ \BBA {} Suh%
}{%
{\protect \APACyear {2014}}%
{\protect \APACexlab {{\protect \BCnt {1}}}}}]{%
Liu:2013qca}
\APACinsertmetastar {%
Liu:2013qca}%
\begin{APACrefauthors}%
Liu, H.%
\BCBT {}\ \BBA {} Suh, S\BPBI J.%
\end{APACrefauthors}%
\unskip\
\newblock
\APACrefYearMonthDay{2014{\protect \BCnt {1}}}{}{}.
\newblock
{\BBOQ}\APACrefatitle {{Entanglement growth during thermalization in holographic systems}} {{Entanglement growth during thermalization in holographic systems}}.{\BBCQ}
\newblock
\APACjournalVolNumPages{Phys. Rev. D}{89}{6}{066012}.
\newblock
\begin{APACrefDOI} \doi{10.1103/PhysRevD.89.066012} \end{APACrefDOI}
\PrintBackRefs{\CurrentBib}

\bibitem [\protect \citeauthoryear {%
Liu%
\ \BBA {} Suh%
}{%
Liu%
\ \BBA {} Suh%
}{%
{\protect \APACyear {2014}}%
{\protect \APACexlab {{\protect \BCnt {2}}}}}]{%
Liu:2013iza}
\APACinsertmetastar {%
Liu:2013iza}%
\begin{APACrefauthors}%
Liu, H.%
\BCBT {}\ \BBA {} Suh, S\BPBI J.%
\end{APACrefauthors}%
\unskip\
\newblock
\APACrefYearMonthDay{2014{\protect \BCnt {2}}}{}{}.
\newblock
{\BBOQ}\APACrefatitle {{Entanglement Tsunami: Universal Scaling in Holographic Thermalization}} {{Entanglement Tsunami: Universal Scaling in Holographic Thermalization}}.{\BBCQ}
\newblock
\APACjournalVolNumPages{Phys. Rev. Lett.}{112}{}{011601}.
\newblock
\begin{APACrefDOI} \doi{10.1103/PhysRevLett.112.011601} \end{APACrefDOI}
\PrintBackRefs{\CurrentBib}

\bibitem [\protect \citeauthoryear {%
MacCormack%
, Liu%
, Nozaki%
\BCBL {}\ \BBA {} Ryu%
}{%
MacCormack%
\ \protect \BOthers {.}}{%
{\protect \APACyear {2019}}%
}]{%
MacCormack:2018rwq}
\APACinsertmetastar {%
MacCormack:2018rwq}%
\begin{APACrefauthors}%
MacCormack, I.%
, Liu, A.%
, Nozaki, M.%
\BCBL {}\ \BBA {} Ryu, S.%
\end{APACrefauthors}%
\unskip\
\newblock
\APACrefYearMonthDay{2019}{}{}.
\newblock
{\BBOQ}\APACrefatitle {{Holographic Duals of Inhomogeneous Systems: The Rainbow Chain and the Sine-Square Deformation Model}} {{Holographic Duals of Inhomogeneous Systems: The Rainbow Chain and the Sine-Square Deformation Model}}.{\BBCQ}
\newblock
\APACjournalVolNumPages{J. Phys. A}{52}{50}{505401}.
\newblock
\begin{APACrefDOI} \doi{10.1088/1751-8121/ab3944} \end{APACrefDOI}
\PrintBackRefs{\CurrentBib}

\bibitem [\protect \citeauthoryear {%
Maldacena%
}{%
Maldacena%
}{%
{\protect \APACyear {1998}}%
}]{%
Maldacena:1997re}
\APACinsertmetastar {%
Maldacena:1997re}%
\begin{APACrefauthors}%
Maldacena, J\BPBI M.%
\end{APACrefauthors}%
\unskip\
\newblock
\APACrefYearMonthDay{1998}{}{}.
\newblock
{\BBOQ}\APACrefatitle {{The Large N limit of superconformal field theories and supergravity}} {{The Large N limit of superconformal field theories and supergravity}}.{\BBCQ}
\newblock
\APACjournalVolNumPages{Adv. Theor. Math. Phys.}{2}{}{231--252}.
\newblock
\begin{APACrefDOI} \doi{10.4310/ATMP.1998.v2.n2.a1} \end{APACrefDOI}
\PrintBackRefs{\CurrentBib}

\bibitem [\protect \citeauthoryear {%
Marolf%
, Maxfield%
, Peach%
\BCBL {}\ \BBA {} Ross%
}{%
Marolf%
\ \protect \BOthers {.}}{%
{\protect \APACyear {2015}}%
}]{%
Marolf:2015vma}
\APACinsertmetastar {%
Marolf:2015vma}%
\begin{APACrefauthors}%
Marolf, D.%
, Maxfield, H.%
, Peach, A.%
\BCBL {}\ \BBA {} Ross, S\BPBI F.%
\end{APACrefauthors}%
\unskip\
\newblock
\APACrefYearMonthDay{2015}{}{}.
\newblock
{\BBOQ}\APACrefatitle {{Hot multiboundary wormholes from bipartite entanglement}} {{Hot multiboundary wormholes from bipartite entanglement}}.{\BBCQ}
\newblock
\APACjournalVolNumPages{Class. Quant. Grav.}{32}{21}{215006}.
\newblock
\begin{APACrefDOI} \doi{10.1088/0264-9381/32/21/215006} \end{APACrefDOI}
\PrintBackRefs{\CurrentBib}

\bibitem [\protect \citeauthoryear {%
Marolf%
\ \BBA {} Wall%
}{%
Marolf%
\ \BBA {} Wall%
}{%
{\protect \APACyear {2016}}%
}]{%
Marolf:2016dob}
\APACinsertmetastar {%
Marolf:2016dob}%
\begin{APACrefauthors}%
Marolf, D.%
\BCBT {}\ \BBA {} Wall, A\BPBI C.%
\end{APACrefauthors}%
\unskip\
\newblock
\APACrefYearMonthDay{2016}{}{}.
\newblock
{\BBOQ}\APACrefatitle {{State-Dependent Divergences in the Entanglement Entropy}} {{State-Dependent Divergences in the Entanglement Entropy}}.{\BBCQ}
\newblock
\APACjournalVolNumPages{JHEP}{10}{}{109}.
\newblock
\begin{APACrefDOI} \doi{10.1007/JHEP10(2016)109} \end{APACrefDOI}
\PrintBackRefs{\CurrentBib}

\bibitem [\protect \citeauthoryear {%
Mezei%
}{%
Mezei%
}{%
{\protect \APACyear {2017}}%
}]{%
Mezei:2016zxg}
\APACinsertmetastar {%
Mezei:2016zxg}%
\begin{APACrefauthors}%
Mezei, M.%
\end{APACrefauthors}%
\unskip\
\newblock
\APACrefYearMonthDay{2017}{}{}.
\newblock
{\BBOQ}\APACrefatitle {{On entanglement spreading from holography}} {{On entanglement spreading from holography}}.{\BBCQ}
\newblock
\APACjournalVolNumPages{JHEP}{05}{}{064}.
\newblock
\begin{APACrefDOI} \doi{10.1007/JHEP05(2017)064} \end{APACrefDOI}
\PrintBackRefs{\CurrentBib}

\bibitem [\protect \citeauthoryear {%
Mezei%
}{%
Mezei%
}{%
{\protect \APACyear {2018}}%
}]{%
Mezei:2018jco}
\APACinsertmetastar {%
Mezei:2018jco}%
\begin{APACrefauthors}%
Mezei, M.%
\end{APACrefauthors}%
\unskip\
\newblock
\APACrefYearMonthDay{2018}{}{}.
\newblock
{\BBOQ}\APACrefatitle {{Membrane theory of entanglement dynamics from holography}} {{Membrane theory of entanglement dynamics from holography}}.{\BBCQ}
\newblock
\APACjournalVolNumPages{Phys. Rev. D}{98}{10}{106025}.
\newblock
\begin{APACrefDOI} \doi{10.1103/PhysRevD.98.106025} \end{APACrefDOI}
\PrintBackRefs{\CurrentBib}

\bibitem [\protect \citeauthoryear {%
Mezei%
\ \BBA {} Rajgadia%
}{%
Mezei%
\ \BBA {} Rajgadia%
}{%
{\protect \APACyear {2025}}%
}]{%
Mezei:2025fcd}
\APACinsertmetastar {%
Mezei:2025fcd}%
\begin{APACrefauthors}%
Mezei, M.%
\BCBT {}\ \BBA {} Rajgadia, H.%
\end{APACrefauthors}%
\unskip\
\newblock
\APACrefYearMonthDay{2025}{12}{}.
\newblock
{\BBOQ}\APACrefatitle {{Entanglement membrane in the Brownian SYK chain}} {{Entanglement membrane in the Brownian SYK chain}}.{\BBCQ}
\newblock

\PrintBackRefs{\CurrentBib}

\bibitem [\protect \citeauthoryear {%
Mezei%
\ \BBA {} Stanford%
}{%
Mezei%
\ \BBA {} Stanford%
}{%
{\protect \APACyear {2017}}%
}]{%
Mezei:2016wfz}
\APACinsertmetastar {%
Mezei:2016wfz}%
\begin{APACrefauthors}%
Mezei, M.%
\BCBT {}\ \BBA {} Stanford, D.%
\end{APACrefauthors}%
\unskip\
\newblock
\APACrefYearMonthDay{2017}{}{}.
\newblock
{\BBOQ}\APACrefatitle {{On entanglement spreading in chaotic systems}} {{On entanglement spreading in chaotic systems}}.{\BBCQ}
\newblock
\APACjournalVolNumPages{JHEP}{05}{}{065}.
\newblock
\begin{APACrefDOI} \doi{10.1007/JHEP05(2017)065} \end{APACrefDOI}
\PrintBackRefs{\CurrentBib}

\bibitem [\protect \citeauthoryear {%
Mezei%
\ \BBA {} van~der Schee%
}{%
Mezei%
\ \BBA {} van~der Schee%
}{%
{\protect \APACyear {2020}}%
}]{%
Mezei:2020knv}
\APACinsertmetastar {%
Mezei:2020knv}%
\begin{APACrefauthors}%
Mezei, M.%
\BCBT {}\ \BBA {} van~der Schee, W.%
\end{APACrefauthors}%
\unskip\
\newblock
\APACrefYearMonthDay{2020}{}{}.
\newblock
{\BBOQ}\APACrefatitle {{Black holes often saturate entanglement entropy the fastest}} {{Black holes often saturate entanglement entropy the fastest}}.{\BBCQ}
\newblock
\APACjournalVolNumPages{Phys. Rev. Lett.}{124}{20}{201601}.
\newblock
\begin{APACrefDOI} \doi{10.1103/PhysRevLett.124.201601} \end{APACrefDOI}
\PrintBackRefs{\CurrentBib}

\bibitem [\protect \citeauthoryear {%
Mezei%
\ \BBA {} Virrueta%
}{%
Mezei%
\ \BBA {} Virrueta%
}{%
{\protect \APACyear {2019}}%
}]{%
Mezei:2019sla}
\APACinsertmetastar {%
Mezei:2019sla}%
\begin{APACrefauthors}%
Mezei, M.%
\BCBT {}\ \BBA {} Virrueta, J.%
\end{APACrefauthors}%
\unskip\
\newblock
\APACrefYearMonthDay{2019}{9}{}.
\newblock
{\BBOQ}\APACrefatitle {{The Quantum Null Energy Condition and Entanglement Entropy in Quenches}} {{The Quantum Null Energy Condition and Entanglement Entropy in Quenches}}.{\BBCQ}
\newblock

\PrintBackRefs{\CurrentBib}

\bibitem [\protect \citeauthoryear {%
Mezei%
\ \BBA {} Virrueta%
}{%
Mezei%
\ \BBA {} Virrueta%
}{%
{\protect \APACyear {2020}}%
}]{%
Mezei:2019zyt}
\APACinsertmetastar {%
Mezei:2019zyt}%
\begin{APACrefauthors}%
Mezei, M.%
\BCBT {}\ \BBA {} Virrueta, J.%
\end{APACrefauthors}%
\unskip\
\newblock
\APACrefYearMonthDay{2020}{}{}.
\newblock
{\BBOQ}\APACrefatitle {{Exploring the Membrane Theory of Entanglement Dynamics}} {{Exploring the Membrane Theory of Entanglement Dynamics}}.{\BBCQ}
\newblock
\APACjournalVolNumPages{JHEP}{02}{}{013}.
\newblock
\begin{APACrefDOI} \doi{10.1007/JHEP02(2020)013} \end{APACrefDOI}
\PrintBackRefs{\CurrentBib}

\bibitem [\protect \citeauthoryear {%
Moosavi%
}{%
Moosavi%
}{%
{\protect \APACyear {2021}}%
}]{%
moosavi2021inhomogeneous}
\APACinsertmetastar {%
moosavi2021inhomogeneous}%
\begin{APACrefauthors}%
Moosavi, P.%
\end{APACrefauthors}%
\unskip\
\newblock
\APACrefYearMonthDay{2021}{}{}.
\newblock
{\BBOQ}\APACrefatitle {Inhomogeneous conformal field theory out of equilibrium} {Inhomogeneous conformal field theory out of equilibrium}.{\BBCQ}
\newblock
\BIn{} \APACrefbtitle {Annales Henri Poincar{\'e}} {Annales henri poincar{\'e}}\ (\BPGS\ 1--40).
\PrintBackRefs{\CurrentBib}

\bibitem [\protect \citeauthoryear {%
Nahum%
, Ruhman%
, Vijay%
\BCBL {}\ \BBA {} Haah%
}{%
Nahum%
\ \protect \BOthers {.}}{%
{\protect \APACyear {2017}}%
}]{%
Nahum:2016muy}
\APACinsertmetastar {%
Nahum:2016muy}%
\begin{APACrefauthors}%
Nahum, A.%
, Ruhman, J.%
, Vijay, S.%
\BCBL {}\ \BBA {} Haah, J.%
\end{APACrefauthors}%
\unskip\
\newblock
\APACrefYearMonthDay{2017}{}{}.
\newblock
{\BBOQ}\APACrefatitle {{Quantum Entanglement Growth Under Random Unitary Dynamics}} {{Quantum Entanglement Growth Under Random Unitary Dynamics}}.{\BBCQ}
\newblock
\APACjournalVolNumPages{Phys. Rev. X}{7}{3}{031016}.
\newblock
\begin{APACrefDOI} \doi{10.1103/PhysRevX.7.031016} \end{APACrefDOI}
\PrintBackRefs{\CurrentBib}

\bibitem [\protect \citeauthoryear {%
Okunishi%
}{%
Okunishi%
}{%
{\protect \APACyear {2016}}%
}]{%
Okunishi:2016zat}
\APACinsertmetastar {%
Okunishi:2016zat}%
\begin{APACrefauthors}%
Okunishi, K.%
\end{APACrefauthors}%
\unskip\
\newblock
\APACrefYearMonthDay{2016}{}{}.
\newblock
{\BBOQ}\APACrefatitle {{Sine-square deformation and M\"obius quantization of 2D conformal field theory}} {{Sine-square deformation and M\"obius quantization of 2D conformal field theory}}.{\BBCQ}
\newblock
\APACjournalVolNumPages{PTEP}{2016}{6}{063A02}.
\newblock
\begin{APACrefDOI} \doi{10.1093/ptep/ptw060} \end{APACrefDOI}
\PrintBackRefs{\CurrentBib}

\bibitem [\protect \citeauthoryear {%
Page%
}{%
Page%
}{%
{\protect \APACyear {1993}}%
}]{%
Page:1993df}
\APACinsertmetastar {%
Page:1993df}%
\begin{APACrefauthors}%
Page, D\BPBI N.%
\end{APACrefauthors}%
\unskip\
\newblock
\APACrefYearMonthDay{1993}{}{}.
\newblock
{\BBOQ}\APACrefatitle {{Average entropy of a subsystem}} {{Average entropy of a subsystem}}.{\BBCQ}
\newblock
\APACjournalVolNumPages{Phys. Rev. Lett.}{71}{}{1291--1294}.
\newblock
\begin{APACrefDOI} \doi{10.1103/PhysRevLett.71.1291} \end{APACrefDOI}
\PrintBackRefs{\CurrentBib}

\bibitem [\protect \citeauthoryear {%
Rakovszky%
, Pollmann%
\BCBL {}\ \BBA {} von Keyserlingk%
}{%
Rakovszky%
\ \protect \BOthers {.}}{%
{\protect \APACyear {2019}}%
}]{%
Rakovszky:2019oht}
\APACinsertmetastar {%
Rakovszky:2019oht}%
\begin{APACrefauthors}%
Rakovszky, T.%
, Pollmann, F.%
\BCBL {}\ \BBA {} von Keyserlingk, C\BPBI W.%
\end{APACrefauthors}%
\unskip\
\newblock
\APACrefYearMonthDay{2019}{}{}.
\newblock
{\BBOQ}\APACrefatitle {{Sub-ballistic Growth of R\'enyi Entropies due to Diffusion}} {{Sub-ballistic Growth of R\'enyi Entropies due to Diffusion}}.{\BBCQ}
\newblock
\APACjournalVolNumPages{Phys. Rev. Lett.}{122}{25}{250602}.
\newblock
\begin{APACrefDOI} \doi{10.1103/PhysRevLett.122.250602} \end{APACrefDOI}
\PrintBackRefs{\CurrentBib}

\bibitem [\protect \citeauthoryear {%
Rakovszky%
, Pollmann%
\BCBL {}\ \BBA {} von Keyserlingk%
}{%
Rakovszky%
\ \protect \BOthers {.}}{%
{\protect \APACyear {2020}}%
}]{%
Rakovszky:2020idb}
\APACinsertmetastar {%
Rakovszky:2020idb}%
\begin{APACrefauthors}%
Rakovszky, T.%
, Pollmann, F.%
\BCBL {}\ \BBA {} von Keyserlingk, C\BPBI W.%
\end{APACrefauthors}%
\unskip\
\newblock
\APACrefYearMonthDay{2020}{10}{}.
\newblock
{\BBOQ}\APACrefatitle {{Comment on ''Entanglement growth in diffusive systems''}} {{Comment on ''Entanglement growth in diffusive systems''}}.{\BBCQ}
\newblock

\PrintBackRefs{\CurrentBib}

\bibitem [\protect \citeauthoryear {%
Rampp%
, Rather%
\BCBL {}\ \BBA {} Claeys%
}{%
Rampp%
\ \protect \BOthers {.}}{%
{\protect \APACyear {2024}}%
}]{%
Rampp:2023vah}
\APACinsertmetastar {%
Rampp:2023vah}%
\begin{APACrefauthors}%
Rampp, M\BPBI A.%
, Rather, S\BPBI A.%
\BCBL {}\ \BBA {} Claeys, P\BPBI W.%
\end{APACrefauthors}%
\unskip\
\newblock
\APACrefYearMonthDay{2024}{}{}.
\newblock
{\BBOQ}\APACrefatitle {{Entanglement membrane in exactly solvable lattice models}} {{Entanglement membrane in exactly solvable lattice models}}.{\BBCQ}
\newblock
\APACjournalVolNumPages{Phys. Rev. Res.}{6}{3}{033271}.
\newblock
\begin{APACrefDOI} \doi{10.1103/PhysRevResearch.6.033271} \end{APACrefDOI}
\PrintBackRefs{\CurrentBib}

\bibitem [\protect \citeauthoryear {%
Rangamani%
\ \BBA {} Takayanagi%
}{%
Rangamani%
\ \BBA {} Takayanagi%
}{%
{\protect \APACyear {2017}}%
}]{%
Rangamani:2016dms}
\APACinsertmetastar {%
Rangamani:2016dms}%
\begin{APACrefauthors}%
Rangamani, M.%
\BCBT {}\ \BBA {} Takayanagi, T.%
\end{APACrefauthors}%
\unskip\
\newblock
\APACrefYear{2017}.
\newblock
\APACrefbtitle {{Holographic Entanglement Entropy}} {{Holographic Entanglement Entropy}}\ (\BVOL~931).
\newblock
\APACaddressPublisher{}{Springer}.
\newblock
\begin{APACrefDOI} \doi{10.1007/978-3-319-52573-0} \end{APACrefDOI}
\PrintBackRefs{\CurrentBib}

\bibitem [\protect \citeauthoryear {%
Rigol%
, Dunjko%
\BCBL {}\ \BBA {} Olshanii%
}{%
Rigol%
\ \protect \BOthers {.}}{%
{\protect \APACyear {2008}}%
}]{%
Rigol:2007juv}
\APACinsertmetastar {%
Rigol:2007juv}%
\begin{APACrefauthors}%
Rigol, M.%
, Dunjko, V.%
\BCBL {}\ \BBA {} Olshanii, M.%
\end{APACrefauthors}%
\unskip\
\newblock
\APACrefYearMonthDay{2008}{}{}.
\newblock
{\BBOQ}\APACrefatitle {{Thermalization and its mechanism for generic isolated quantum systems}} {{Thermalization and its mechanism for generic isolated quantum systems}}.{\BBCQ}
\newblock
\APACjournalVolNumPages{Nature}{452}{7189}{854--858}.
\newblock
\begin{APACrefDOI} \doi{10.1038/nature06838} \end{APACrefDOI}
\PrintBackRefs{\CurrentBib}

\bibitem [\protect \citeauthoryear {%
Roberts%
, Stanford%
\BCBL {}\ \BBA {} Susskind%
}{%
Roberts%
\ \protect \BOthers {.}}{%
{\protect \APACyear {2015}}%
}]{%
Roberts:2014isa}
\APACinsertmetastar {%
Roberts:2014isa}%
\begin{APACrefauthors}%
Roberts, D\BPBI A.%
, Stanford, D.%
\BCBL {}\ \BBA {} Susskind, L.%
\end{APACrefauthors}%
\unskip\
\newblock
\APACrefYearMonthDay{2015}{}{}.
\newblock
{\BBOQ}\APACrefatitle {{Localized shocks}} {{Localized shocks}}.{\BBCQ}
\newblock
\APACjournalVolNumPages{JHEP}{03}{}{051}.
\newblock
\begin{APACrefDOI} \doi{10.1007/JHEP03(2015)051} \end{APACrefDOI}
\PrintBackRefs{\CurrentBib}

\bibitem [\protect \citeauthoryear {%
Rosenhaus%
}{%
Rosenhaus%
}{%
{\protect \APACyear {2019}}%
}]{%
Rosenhaus:2018dtp}
\APACinsertmetastar {%
Rosenhaus:2018dtp}%
\begin{APACrefauthors}%
Rosenhaus, V.%
\end{APACrefauthors}%
\unskip\
\newblock
\APACrefYearMonthDay{2019}{}{}.
\newblock
{\BBOQ}\APACrefatitle {{An introduction to the SYK model}} {{An introduction to the SYK model}}.{\BBCQ}
\newblock
\APACjournalVolNumPages{J. Phys. A}{52}{}{323001}.
\newblock
\begin{APACrefDOI} \doi{10.1088/1751-8121/ab2ce1} \end{APACrefDOI}
\PrintBackRefs{\CurrentBib}

\bibitem [\protect \citeauthoryear {%
Ryu%
\ \BBA {} Takayanagi%
}{%
Ryu%
\ \BBA {} Takayanagi%
}{%
{\protect \APACyear {2006}}%
{\protect \APACexlab {{\protect \BCnt {1}}}}}]{%
Ryu:2006ef}
\APACinsertmetastar {%
Ryu:2006ef}%
\begin{APACrefauthors}%
Ryu, S.%
\BCBT {}\ \BBA {} Takayanagi, T.%
\end{APACrefauthors}%
\unskip\
\newblock
\APACrefYearMonthDay{2006{\protect \BCnt {1}}}{}{}.
\newblock
{\BBOQ}\APACrefatitle {{Aspects of Holographic Entanglement Entropy}} {{Aspects of Holographic Entanglement Entropy}}.{\BBCQ}
\newblock
\APACjournalVolNumPages{JHEP}{08}{}{045}.
\newblock
\begin{APACrefDOI} \doi{10.1088/1126-6708/2006/08/045} \end{APACrefDOI}
\PrintBackRefs{\CurrentBib}

\bibitem [\protect \citeauthoryear {%
Ryu%
\ \BBA {} Takayanagi%
}{%
Ryu%
\ \BBA {} Takayanagi%
}{%
{\protect \APACyear {2006}}%
{\protect \APACexlab {{\protect \BCnt {2}}}}}]{%
Ryu:2006bv}
\APACinsertmetastar {%
Ryu:2006bv}%
\begin{APACrefauthors}%
Ryu, S.%
\BCBT {}\ \BBA {} Takayanagi, T.%
\end{APACrefauthors}%
\unskip\
\newblock
\APACrefYearMonthDay{2006{\protect \BCnt {2}}}{}{}.
\newblock
{\BBOQ}\APACrefatitle {{Holographic derivation of entanglement entropy from AdS/CFT}} {{Holographic derivation of entanglement entropy from AdS/CFT}}.{\BBCQ}
\newblock
\APACjournalVolNumPages{Phys. Rev. Lett.}{96}{}{181602}.
\newblock
\begin{APACrefDOI} \doi{10.1103/PhysRevLett.96.181602} \end{APACrefDOI}
\PrintBackRefs{\CurrentBib}

\bibitem [\protect \citeauthoryear {%
Shenker%
\ \BBA {} Stanford%
}{%
Shenker%
\ \BBA {} Stanford%
}{%
{\protect \APACyear {2014}}%
}]{%
Shenker:2013pqa}
\APACinsertmetastar {%
Shenker:2013pqa}%
\begin{APACrefauthors}%
Shenker, S\BPBI H.%
\BCBT {}\ \BBA {} Stanford, D.%
\end{APACrefauthors}%
\unskip\
\newblock
\APACrefYearMonthDay{2014}{}{}.
\newblock
{\BBOQ}\APACrefatitle {{Black holes and the butterfly effect}} {{Black holes and the butterfly effect}}.{\BBCQ}
\newblock
\APACjournalVolNumPages{JHEP}{03}{}{067}.
\newblock
\begin{APACrefDOI} \doi{10.1007/JHEP03(2014)067} \end{APACrefDOI}
\PrintBackRefs{\CurrentBib}

\bibitem [\protect \citeauthoryear {%
Skenderis%
\ \BBA {} Solodukhin%
}{%
Skenderis%
\ \BBA {} Solodukhin%
}{%
{\protect \APACyear {2000}}%
}]{%
Skenderis:1999nb}
\APACinsertmetastar {%
Skenderis:1999nb}%
\begin{APACrefauthors}%
Skenderis, K.%
\BCBT {}\ \BBA {} Solodukhin, S\BPBI N.%
\end{APACrefauthors}%
\unskip\
\newblock
\APACrefYearMonthDay{2000}{}{}.
\newblock
{\BBOQ}\APACrefatitle {{Quantum effective action from the AdS / CFT correspondence}} {{Quantum effective action from the AdS / CFT correspondence}}.{\BBCQ}
\newblock
\APACjournalVolNumPages{Phys. Lett. B}{472}{}{316--322}.
\newblock
\begin{APACrefDOI} \doi{10.1016/S0370-2693(99)01467-7} \end{APACrefDOI}
\PrintBackRefs{\CurrentBib}

\bibitem [\protect \citeauthoryear {%
Son%
\ \BBA {} Starinets%
}{%
Son%
\ \BBA {} Starinets%
}{%
{\protect \APACyear {2002}}%
}]{%
Son:2002sd}
\APACinsertmetastar {%
Son:2002sd}%
\begin{APACrefauthors}%
Son, D\BPBI T.%
\BCBT {}\ \BBA {} Starinets, A\BPBI O.%
\end{APACrefauthors}%
\unskip\
\newblock
\APACrefYearMonthDay{2002}{}{}.
\newblock
{\BBOQ}\APACrefatitle {{Minkowski space correlators in AdS / CFT correspondence: Recipe and applications}} {{Minkowski space correlators in AdS / CFT correspondence: Recipe and applications}}.{\BBCQ}
\newblock
\APACjournalVolNumPages{JHEP}{09}{}{042}.
\newblock
\begin{APACrefDOI} \doi{10.1088/1126-6708/2002/09/042} \end{APACrefDOI}
\PrintBackRefs{\CurrentBib}

\bibitem [\protect \citeauthoryear {%
Srednicki%
}{%
Srednicki%
}{%
{\protect \APACyear {1993}}%
}]{%
Srednicki:1993im}
\APACinsertmetastar {%
Srednicki:1993im}%
\begin{APACrefauthors}%
Srednicki, M.%
\end{APACrefauthors}%
\unskip\
\newblock
\APACrefYearMonthDay{1993}{}{}.
\newblock
{\BBOQ}\APACrefatitle {{Entropy and area}} {{Entropy and area}}.{\BBCQ}
\newblock
\APACjournalVolNumPages{Phys. Rev. Lett.}{71}{}{666--669}.
\newblock
\begin{APACrefDOI} \doi{10.1103/PhysRevLett.71.666} \end{APACrefDOI}
\PrintBackRefs{\CurrentBib}

\bibitem [\protect \citeauthoryear {%
Srednicki%
}{%
Srednicki%
}{%
{\protect \APACyear {1994}}%
}]{%
Srednicki:1994mfb}
\APACinsertmetastar {%
Srednicki:1994mfb}%
\begin{APACrefauthors}%
Srednicki, M.%
\end{APACrefauthors}%
\unskip\
\newblock
\APACrefYearMonthDay{1994}{3}{}.
\newblock
{\BBOQ}\APACrefatitle {{Chaos and Quantum Thermalization}} {{Chaos and Quantum Thermalization}}.{\BBCQ}
\newblock
\APACjournalVolNumPages{Phys. Rev. E}{50}{}{}.
\newblock
\begin{APACrefDOI} \doi{10.1103/PhysRevE.50.888} \end{APACrefDOI}
\PrintBackRefs{\CurrentBib}

\bibitem [\protect \citeauthoryear {%
Susskind%
\ \BBA {} Witten%
}{%
Susskind%
\ \BBA {} Witten%
}{%
{\protect \APACyear {1998}}%
}]{%
Susskind:1998dq}
\APACinsertmetastar {%
Susskind:1998dq}%
\begin{APACrefauthors}%
Susskind, L.%
\BCBT {}\ \BBA {} Witten, E.%
\end{APACrefauthors}%
\unskip\
\newblock
\APACrefYearMonthDay{1998}{5}{}.
\newblock
{\BBOQ}\APACrefatitle {{The Holographic bound in anti-de Sitter space}} {{The Holographic bound in anti-de Sitter space}}.{\BBCQ}
\newblock

\PrintBackRefs{\CurrentBib}

\bibitem [\protect \citeauthoryear {%
Takayanagi%
\ \BBA {} Umemoto%
}{%
Takayanagi%
\ \BBA {} Umemoto%
}{%
{\protect \APACyear {2018}}%
}]{%
Takayanagi:2017knl}
\APACinsertmetastar {%
Takayanagi:2017knl}%
\begin{APACrefauthors}%
Takayanagi, T.%
\BCBT {}\ \BBA {} Umemoto, K.%
\end{APACrefauthors}%
\unskip\
\newblock
\APACrefYearMonthDay{2018}{}{}.
\newblock
{\BBOQ}\APACrefatitle {{Entanglement of purification through holographic duality}} {{Entanglement of purification through holographic duality}}.{\BBCQ}
\newblock
\APACjournalVolNumPages{Nature Phys.}{14}{6}{573--577}.
\newblock
\begin{APACrefDOI} \doi{10.1038/s41567-018-0075-2} \end{APACrefDOI}
\PrintBackRefs{\CurrentBib}

\bibitem [\protect \citeauthoryear {%
Tamura%
\ \BBA {} Katsura%
}{%
Tamura%
\ \BBA {} Katsura%
}{%
{\protect \APACyear {2017}}%
}]{%
Tamura:2017vbx}
\APACinsertmetastar {%
Tamura:2017vbx}%
\begin{APACrefauthors}%
Tamura, S.%
\BCBT {}\ \BBA {} Katsura, H.%
\end{APACrefauthors}%
\unskip\
\newblock
\APACrefYearMonthDay{2017}{}{}.
\newblock
{\BBOQ}\APACrefatitle {{Zero-energy states in conformal field theory with sine-square deformation}} {{Zero-energy states in conformal field theory with sine-square deformation}}.{\BBCQ}
\newblock
\APACjournalVolNumPages{PTEP}{2017}{11}{113A01}.
\newblock
\begin{APACrefDOI} \doi{10.1093/ptep/ptx147} \end{APACrefDOI}
\PrintBackRefs{\CurrentBib}

\bibitem [\protect \citeauthoryear {%
Vardhan%
\ \BBA {} Moudgalya%
}{%
Vardhan%
\ \BBA {} Moudgalya%
}{%
{\protect \APACyear {2024}}%
}]{%
Vardhan:2024wxb}
\APACinsertmetastar {%
Vardhan:2024wxb}%
\begin{APACrefauthors}%
Vardhan, S.%
\BCBT {}\ \BBA {} Moudgalya, S.%
\end{APACrefauthors}%
\unskip\
\newblock
\APACrefYearMonthDay{2024}{7}{}.
\newblock
{\BBOQ}\APACrefatitle {{Entanglement dynamics from universal low-lying modes}} {{Entanglement dynamics from universal low-lying modes}}.{\BBCQ}
\newblock

\PrintBackRefs{\CurrentBib}

\bibitem [\protect \citeauthoryear {%
Verstraete%
, Dehaene%
, De~Moor%
\BCBL {}\ \BBA {} Verschelde%
}{%
Verstraete%
\ \protect \BOthers {.}}{%
{\protect \APACyear {2002}}%
}]{%
Verstraete:2002gqj}
\APACinsertmetastar {%
Verstraete:2002gqj}%
\begin{APACrefauthors}%
Verstraete, F.%
, Dehaene, J.%
, De~Moor, B.%
\BCBL {}\ \BBA {} Verschelde, H.%
\end{APACrefauthors}%
\unskip\
\newblock
\APACrefYearMonthDay{2002}{}{}.
\newblock
{\BBOQ}\APACrefatitle {{Four qubits can be entangled in nine different ways}} {{Four qubits can be entangled in nine different ways}}.{\BBCQ}
\newblock
\APACjournalVolNumPages{Phys. Rev. A}{65}{5}{052112}.
\newblock
\begin{APACrefDOI} \doi{10.1103/PhysRevA.65.052112} \end{APACrefDOI}
\PrintBackRefs{\CurrentBib}

\bibitem [\protect \citeauthoryear {%
\v{Z}nidari\v{c}%
}{%
\v{Z}nidari\v{c}%
}{%
{\protect \APACyear {2020}}%
}]{%
Znidaric:2020czi}
\APACinsertmetastar {%
Znidaric:2020czi}%
\begin{APACrefauthors}%
\v{Z}nidari\v{c}, M.%
\end{APACrefauthors}%
\unskip\
\newblock
\APACrefYearMonthDay{2020}{}{}.
\newblock
{\BBOQ}\APACrefatitle {{Entanglement growth in diffusive systems}} {{Entanglement growth in diffusive systems}}.{\BBCQ}
\newblock
\APACjournalVolNumPages{Commun. Phys.}{3}{1}{100}.
\newblock
\begin{APACrefDOI} \doi{10.1038/s42005-020-0366-7} \end{APACrefDOI}
\PrintBackRefs{\CurrentBib}

\bibitem [\protect \citeauthoryear {%
Walter%
, Gross%
\BCBL {}\ \BBA {} Eisert%
}{%
Walter%
\ \protect \BOthers {.}}{%
{\protect \APACyear {2016}}%
}]{%
Walter:2016lgl}
\APACinsertmetastar {%
Walter:2016lgl}%
\begin{APACrefauthors}%
Walter, M.%
, Gross, D.%
\BCBL {}\ \BBA {} Eisert, J.%
\end{APACrefauthors}%
\unskip\
\newblock
\APACrefYearMonthDay{2016}{12}{}.
\newblock
{\BBOQ}\APACrefatitle {{Multi-partite entanglement}} {{Multi-partite entanglement}}.{\BBCQ}
\newblock

\PrintBackRefs{\CurrentBib}

\bibitem [\protect \citeauthoryear {%
Wen%
, Fan%
\BCBL {}\ \BBA {} Vishwanath%
}{%
Wen%
\ \protect \BOthers {.}}{%
{\protect \APACyear {2022}}%
}]{%
Wen:2022pyj}
\APACinsertmetastar {%
Wen:2022pyj}%
\begin{APACrefauthors}%
Wen, X.%
, Fan, R.%
\BCBL {}\ \BBA {} Vishwanath, A.%
\end{APACrefauthors}%
\unskip\
\newblock
\APACrefYearMonthDay{2022}{10}{}.
\newblock
{\BBOQ}\APACrefatitle {{Floquet's Refrigerator: Conformal Cooling in Driven Quantum Critical Systems}} {{Floquet's Refrigerator: Conformal Cooling in Driven Quantum Critical Systems}}.{\BBCQ}
\newblock

\PrintBackRefs{\CurrentBib}

\bibitem [\protect \citeauthoryear {%
Wen%
, Fan%
, Vishwanath%
\BCBL {}\ \BBA {} Gu%
}{%
Wen%
\ \protect \BOthers {.}}{%
{\protect \APACyear {2021}}%
}]{%
Wen:2020wee}
\APACinsertmetastar {%
Wen:2020wee}%
\begin{APACrefauthors}%
Wen, X.%
, Fan, R.%
, Vishwanath, A.%
\BCBL {}\ \BBA {} Gu, Y.%
\end{APACrefauthors}%
\unskip\
\newblock
\APACrefYearMonthDay{2021}{}{}.
\newblock
{\BBOQ}\APACrefatitle {{Periodically, quasiperiodically, and randomly driven conformal field theories}} {{Periodically, quasiperiodically, and randomly driven conformal field theories}}.{\BBCQ}
\newblock
\APACjournalVolNumPages{Phys. Rev. Res.}{3}{2}{023044}.
\newblock
\begin{APACrefDOI} \doi{10.1103/PhysRevResearch.3.023044} \end{APACrefDOI}
\PrintBackRefs{\CurrentBib}

\bibitem [\protect \citeauthoryear {%
Wen%
, Ryu%
\BCBL {}\ \BBA {} Ludwig%
}{%
Wen%
\ \protect \BOthers {.}}{%
{\protect \APACyear {2016}}%
}]{%
Wen:2016inm}
\APACinsertmetastar {%
Wen:2016inm}%
\begin{APACrefauthors}%
Wen, X.%
, Ryu, S.%
\BCBL {}\ \BBA {} Ludwig, A\BPBI W\BPBI W.%
\end{APACrefauthors}%
\unskip\
\newblock
\APACrefYearMonthDay{2016}{}{}.
\newblock
{\BBOQ}\APACrefatitle {{Evolution operators in conformal field theories and conformal mappings: Entanglement Hamiltonian, the sine-square deformation, and others}} {{Evolution operators in conformal field theories and conformal mappings: Entanglement Hamiltonian, the sine-square deformation, and others}}.{\BBCQ}
\newblock
\APACjournalVolNumPages{Phys. Rev. B}{93}{23}{235119}.
\newblock
\begin{APACrefDOI} \doi{10.1103/PhysRevB.93.235119} \end{APACrefDOI}
\PrintBackRefs{\CurrentBib}

\bibitem [\protect \citeauthoryear {%
Wen%
\ \BBA {} Wu%
}{%
Wen%
\ \BBA {} Wu%
}{%
{\protect \APACyear {2018}}%
{\protect \APACexlab {{\protect \BCnt {1}}}}}]{%
Wen:2018agb}
\APACinsertmetastar {%
Wen:2018agb}%
\begin{APACrefauthors}%
Wen, X.%
\BCBT {}\ \BBA {} Wu, J\BHBI Q.%
\end{APACrefauthors}%
\unskip\
\newblock
\APACrefYearMonthDay{2018{\protect \BCnt {1}}}{4}{}.
\newblock
{\BBOQ}\APACrefatitle {{Floquet conformal field theory}} {{Floquet conformal field theory}}.{\BBCQ}
\newblock

\PrintBackRefs{\CurrentBib}

\bibitem [\protect \citeauthoryear {%
Wen%
\ \BBA {} Wu%
}{%
Wen%
\ \BBA {} Wu%
}{%
{\protect \APACyear {2018}}%
{\protect \APACexlab {{\protect \BCnt {2}}}}}]{%
Wen:2018vux}
\APACinsertmetastar {%
Wen:2018vux}%
\begin{APACrefauthors}%
Wen, X.%
\BCBT {}\ \BBA {} Wu, J\BHBI Q.%
\end{APACrefauthors}%
\unskip\
\newblock
\APACrefYearMonthDay{2018{\protect \BCnt {2}}}{}{}.
\newblock
{\BBOQ}\APACrefatitle {{Quantum dynamics in sine-square deformed conformal field theory: Quench from uniform to nonuniform conformal field theory}} {{Quantum dynamics in sine-square deformed conformal field theory: Quench from uniform to nonuniform conformal field theory}}.{\BBCQ}
\newblock
\APACjournalVolNumPages{Phys. Rev. B}{97}{18}{184309}.
\newblock
\begin{APACrefDOI} \doi{10.1103/PhysRevB.97.184309} \end{APACrefDOI}
\PrintBackRefs{\CurrentBib}

\bibitem [\protect \citeauthoryear {%
Witten%
}{%
Witten%
}{%
{\protect \APACyear {1998}}%
}]{%
Witten:1998qj}
\APACinsertmetastar {%
Witten:1998qj}%
\begin{APACrefauthors}%
Witten, E.%
\end{APACrefauthors}%
\unskip\
\newblock
\APACrefYearMonthDay{1998}{}{}.
\newblock
{\BBOQ}\APACrefatitle {{Anti-de Sitter space and holography}} {{Anti-de Sitter space and holography}}.{\BBCQ}
\newblock
\APACjournalVolNumPages{Adv. Theor. Math. Phys.}{2}{}{253--291}.
\newblock
\begin{APACrefDOI} \doi{10.4310/ATMP.1998.v2.n2.a2} \end{APACrefDOI}
\PrintBackRefs{\CurrentBib}

\bibitem [\protect \citeauthoryear {%
Witten%
}{%
Witten%
}{%
{\protect \APACyear {2023}}%
}]{%
Witten:2023qsv}
\APACinsertmetastar {%
Witten:2023qsv}%
\begin{APACrefauthors}%
Witten, E.%
\end{APACrefauthors}%
\unskip\
\newblock
\APACrefYearMonthDay{2023}{3}{}.
\newblock
{\BBOQ}\APACrefatitle {{Algebras, Regions, and Observers}} {{Algebras, Regions, and Observers}}.{\BBCQ}
\newblock

\PrintBackRefs{\CurrentBib}

\bibitem [\protect \citeauthoryear {%
Zhou%
\ \BBA {} Nahum%
}{%
Zhou%
\ \BBA {} Nahum%
}{%
{\protect \APACyear {2019}}%
}]{%
Zhou:2018myl}
\APACinsertmetastar {%
Zhou:2018myl}%
\begin{APACrefauthors}%
Zhou, T.%
\BCBT {}\ \BBA {} Nahum, A.%
\end{APACrefauthors}%
\unskip\
\newblock
\APACrefYearMonthDay{2019}{}{}.
\newblock
{\BBOQ}\APACrefatitle {{Emergent statistical mechanics of entanglement in random unitary circuits}} {{Emergent statistical mechanics of entanglement in random unitary circuits}}.{\BBCQ}
\newblock
\APACjournalVolNumPages{Phys. Rev. B}{99}{17}{174205}.
\newblock
\begin{APACrefDOI} \doi{10.1103/PhysRevB.99.174205} \end{APACrefDOI}
\PrintBackRefs{\CurrentBib}

\bibitem [\protect \citeauthoryear {%
Zhou%
\ \BBA {} Nahum%
}{%
Zhou%
\ \BBA {} Nahum%
}{%
{\protect \APACyear {2020}}%
}]{%
Zhou:2019pob}
\APACinsertmetastar {%
Zhou:2019pob}%
\begin{APACrefauthors}%
Zhou, T.%
\BCBT {}\ \BBA {} Nahum, A.%
\end{APACrefauthors}%
\unskip\
\newblock
\APACrefYearMonthDay{2020}{}{}.
\newblock
{\BBOQ}\APACrefatitle {{Entanglement Membrane in Chaotic Many-Body Systems}} {{Entanglement Membrane in Chaotic Many-Body Systems}}.{\BBCQ}
\newblock
\APACjournalVolNumPages{Phys. Rev. X}{10}{3}{031066}.
\newblock
\begin{APACrefDOI} \doi{10.1103/PhysRevX.10.031066} \end{APACrefDOI}
\PrintBackRefs{\CurrentBib}

\end{thebibliography}

\begin{appendices}
\chapter{Appendices to Chapter~\ref{Floquet}}
\section{Review of SSD CFT}\label{Review of SSD CFT}
In the main text, we introduced the cylinder $\mathbb{R}\times S^1$ with coordinate \eqref{Eucledian_w}.  
To study CFT$_2$, we map the cylinder to the $\emph{complex\ plane}$ $\mathbb{C}$ via the exponential map, 
\begin{equation}
\label{expMap}
    \left\{
    \begin{aligned}
            z&=e^{w}=e^{\tau+i x},\\
    \overline{z}&=e^{\overline{w}}=e^{\tau-i x}.
    \end{aligned}
    \right.
\end{equation}
$H_0$ is $\emph{dilation}$ on the $z$-plane. Analytically continuing to real time $\tau\to it$, we have
\begin{equation}
\label{Lorentzianz}
\left\{
    \begin{aligned}
    z&=\lambda e^{w}=e^{i(t+x)},\\
    \overline{z}&=\lambda e^{\overline{w}}=e^{i(t-x)}.
\end{aligned}
\right.
\end{equation}
where $\lambda=e^{\tau}$ in Euclidean signature. In Lorentzian signature, $z_{{\rm new}}$ and $\overline{z}_{{\rm new}}$ are on the $\emph{unit\ circle}$. 

\subsection{SSD CFT}\label{SSDreview}
\subsubsection{The action of $H_1$}
Under the exponential map \eqref{expMap}, the SSD Hamiltonian \eqref{1} becomes 
\begin{align}
    H_1&=2\int_0^{2\pi} dx\ \sin ^2\Big(\frac{x}{2}\Big)T_{00}(x)\\
    &=\oint \frac{dz}{2\pi i} \Big(-\frac{1}{2}+z-\frac{z^2}{2}\Big)T(z)-(z\to\bar{z})-\frac{c}{12}\label{H_1_z}\\
    &=L_0-\frac{L_{-1}+L_1}{2}+\bar{L}_0-\frac{\bar{L}_{-1}+\bar{L_1}}{2}-\frac{c}{12}.
\end{align}
While $H_0$ acts as dilation, the action of $H_1$ is a more general $SL(2)$ transformation. To see this, we perform a coordinate transformation $(z,\overline{z})\to(\chi,\overline{\chi})$: 
\es{H1start}{
    H_1&=\oint\frac{dz}{2\pi i}\ \Big(-\frac{1}{2}(z-1)^2\Big)T(x)+h.c\\
    &=\oint\frac{d\chi}{2\pi i}\ \frac{dz}{d\chi}\Big(-\frac{1}{2}(z-1)^2\Big)\Big[\Big(\frac{d\chi}{dz}\Big)^2 T(\chi)+\frac{c}{12}{\rm Sch}(\chi,z)\Big]+h.c\,, 
}
    where ${\rm Sch}(\chi,z)$ is the Schwarzian. By requiring that $H_1$ is dilation on the $\chi$-plane \cite{Fan:2019upv}, i.e.~that the  prefactor of $T(\chi)$ is equal to $\chi$, we obtain a differential equation with the solution:\footnote{The most general solutions take the form $\chi(z)=c \,e^{\frac{z+1}{z-1}}$ and $\overline{z}=\overline{c}\,e^{\frac{\overline{z}+1}{\overline{z}-1}}$; the constants $c$ and $\overline{c}$ have no effect on the coordinate transformations.} 
    \begin{align}
    \chi(z)=e^{\frac{z+1}{z-1}}\,, && \overline{\chi}(\overline{z})=e^{\frac{\overline{z}+1}{\overline{z}-1}}\,,\label{chi}
\end{align}
Plugging in this map into  \eqref{H1start}, we get
   \begin{align} 
      H_1 &=\oint\frac{d\chi}{2\pi i}\Big(\chi T_{00}(\chi)-\frac{c}{24}\frac{1}{\chi}\Big)+h.c\,,
\end{align}
and evaluating the contour integral (together with an identical contribution coming from the anti-holomorphic term) gives the Casimir energy $-c/12$.

To our knowledge, the map \eqref{chi} has not appeared in the literature, but we show around \eqref{MobiusDeformedCFT} that it can be obtained as a limit of maps known from prior work. Note that the maps contain essential singularities at $z=1$ and $\overline{z}=1$, respectively, which means that we need to exclude them in defining \eqref{chi}.  These are the fixed points of the $SL(2)$ transformations \eqref{znew} and \eqref{znewbar}, so we simply have $z_{{\rm new}}=z=1$ and $\overline{z}_{{\rm new}}=\overline{z}=1$ in these two excluded points. On the $\chi$ plane, we then have the Heisenberg evolution 
\begin{align}
    e^{\tau H_1 }O(\chi,\bar{\chi})e^{-\tau H_1 }=O(\lambda\chi,\lambda\bar{\chi}),
\end{align}
where $\lambda=e^{\tau}$. Back to the $z$-plane, the effect of $H_1$ is to shift the operator $O$ from $z$ to $z_{{\rm new}}$, where $z_{{\rm new}}$ is related to $z$ via \cite{Wen:2018vux}
\begin{align}
    \chi(z_{{\rm new}})&=\lambda\chi(z),
\end{align}
from which one can solve for $z_{{\rm new}}$: 
\begin{align}
    e^{\frac{z_{{\rm new}}+1}{z_{{\rm new}}-1}}&=e^{\tau}\cdot e^{\frac{z+1}{z-1}},\\
    \Rightarrow z_{{\rm new}}&=\frac{(1+\frac{\tau}{2})z-\frac{\tau}{2}}{\frac{\tau}{2}z+(1-\frac{\tau}{2})}\label{znew}.
\end{align}
For the anti-holomorphic part, we repeat the above calculation with $z\to\overline{z}$, arriving at
\begin{align}
    \overline{z}_{{\rm new}}&=\frac{(1+\frac{\tau}{2})\overline{z}-\frac{\tau}{2}}{\frac{\tau}{2}\overline{z}+(1-\frac{\tau}{2})}\label{znewbar}.
\end{align}
\eqref{znew} and \eqref{znewbar} are $SL(2)$ transformations on the complex plane. Transforming back to the cylinder via the inverse of \eqref{expMap}, and performing analytic continuation $\tau=it$ then yields \eqref{wnew} and \eqref{wnewbar} in the main text. 

The relation among the cylinder (left), complex plane (middle), and the $SL(2)$ transformed plane (right) are summarised as follows:
\begin{align}
    \begin{CD}
w,\overline{w}@>z=e^{w}>\overline{z}=e^{ \overline{w}}> z, \overline{z}@>\chi=e^{\frac{z+1}{z-1}}>\overline{\chi}=e^{\frac{\overline{z}+1}{\overline{z}-1}}>\chi, \overline{\chi}\\
@VH_1VV @VH_1VV @VH_1V{\rm Dilations}V\\
w_{{\rm new}}, \overline{w}_{{\rm new}} @>z_{{\rm new}}=e^{ w_{{\rm new}}}>\overline{z}_{{\rm new}}=e^{\overline{w}_{{\rm new}}}> z_{{\rm new}}, \overline{z}_{{\rm new}}@>\lambda\chi=e^{\frac{z_{{\rm new}}+1}{z_{{\rm new}}-1}}>\lambda\overline{\chi}=e^{\frac{\overline{z}_{{\rm new}}+1}{\overline{z}_{{\rm new}}-1}}> \lambda\chi, \lambda\overline{\chi}
\end{CD}
\label{CoordTab}
\end{align}

\subsubsection{Trigonometric forms of the coordinate transformations}
In Lorentzian signature, $|z_{{\rm new}}|=|\overline{z}_{{\rm new}}|=1$ as Mobius transformations preserve the unit circles \eqref{Lorentzianz} $z_{{\rm new}}$ and $\overline{z}_{{\rm new}}$ lives on. For technical conveniences in making Figure \ref{fig:Triangle} and \ref{fig:CutoffSurface3d}, we follow \cite{Goto:2021sqx} and write $z_{{\rm new}}$ and $\overline{z}_{{\rm new}}$ in trigonometric  forms: \eqref{znew} and \eqref{wnew} can be written as 
\begin{align}
    z_{{\rm new}}=\cos \ 2\varphi+i\ \sin \ 2\varphi, && w_{{\rm new}}=2i\varphi\,,
\end{align}
where 
\begin{align}
    \sin \varphi=\frac{\sin \big(\frac{x}{2}\big)}{\sqrt{1-t\ \sin x+t^2 \sin ^2\big(\frac{x}{2}\big)}}, && \cos \varphi=\frac{\cos \big(\frac{x}{2}\big)-t\ \sin \big(\frac{x}{2}\big)}{\sqrt{1-t\ \sin x+t^2 \sin ^2\big(\frac{x}{2}\big)}}\,.\label{znewphi}
\end{align}
We then have $\varphi\in[\frac{ x}{2},\pi]\in [0,\pi]$ \cite{Goto:2021sqx} as $t$ ranges from $[0,\infty)$. Similarly, \eqref{znewbar} and \eqref{wnewbar} can be written as  
\begin{align}
    \overline{z}_{{\rm new}}=\cos \ 2\overline{\varphi}+i\ \sin \ 2\overline{\varphi}, && \overline{w}_{{\rm new}}=2i\overline{\varphi}\,,
\end{align}
where 
\begin{align}
    \sin \overline{\varphi}=\frac{-\sin \big(\frac{x}{2}\big)}{\sqrt{1+t\ \sin x+t^2 \sin ^2\big(\frac{x}{2}\big)}}, && \cos \overline{\varphi}=\frac{\cos \big(\frac{x}{2}\big)+t\ \sin \big(\frac{x}{2}\big)}{\sqrt{1+t\ \sin x+t^2 \sin ^2\big(\frac{x}{2}\big)}}\,.\label{overlineznewphi}
\end{align}
We then have $\overline{\varphi}\in[-\frac{ x}{2},0]\in [-\pi,0]$ \cite{Goto:2021sqx} as $t$ ranges from $[0,\infty)$. Note that $\varphi$ and $\overline{\varphi}$ are in general $\emph{not}$ complex conjugates. 

The explicit expressions of $t_{{\rm new}}(t,x)$ and $x_{{\rm new}}(t,x)$ can be found from \eqref{wnew}, \eqref{wnewbar}, \eqref{tnewxnew}, \eqref{znewphi} and \eqref{overlineznewphi}
\begin{align}
    t_{{\rm new}}(t,x)&=\arccos\Big[\frac{\cos \big(\frac{x}{2}\big)-t\ \sin \big(\frac{x}{2}\big)}{\sqrt{1-t\ \sin x+t^2 \sin ^2\big(\frac{x}{2}\big)}}\Big]-\arccos\Big[\frac{\cos \big(\frac{x}{2}\big)+t\ \sin \big(\frac{x}{2}\big)}{\sqrt{1+t\ \sin x+t^2 \sin ^2\big(\frac{x}{2}\big)}}\Big]\label{tnewtrig},\\
    x_{{\rm new}}(t,x)&=\arccos\Big[\frac{\cos \big(\frac{x}{2}\big)-t\ \sin \big(\frac{x}{2}\big)}{\sqrt{1-t\ \sin x+t^2 \sin ^2\big(\frac{x}{2}\big)}}\Big]+\arccos\Big[\frac{\cos \big(\frac{x}{2}\big)+t\ \sin \big(\frac{x}{2}\big)}{\sqrt{1+t\ \sin x+t^2 \sin ^2\big(\frac{x}{2}\big)}}\Big]\,.\label{xnewtrig}
\end{align}

\subsubsection{SSD CFT as a CFT on curved spacetime}\label{SSDcurvedreview}

Under Weyl transformations, primary correlators transform as 
\es{WeylTf}{
\le\langle O(z,\bar z) \dots \ri\rangle_{ds'^2=\Om \bar{\Om} \,ds^2}=\Om(z)^{-h}\,\bar{\Om}(\bar z)^{-\bar h}\dots\le\langle O(z,\bar z) \dots \ri\rangle_{ds^2}.
}
Now let us choose $\Om=\big(\frac{\partial w_{{\rm new}}}{\partial (i x)}\big)^{-1}$ and similarly for $\bar \Om$. Then the second quantity in \eqref{3desc} becomes the RHS of the above equation. It only remains to compute metric $ds'^2$ in $\tau,x$ coordinates. To this end, we write:

\es{dsprime}{
ds'^2&=\Om \bar \Om\ dw_{{\rm new}}d\overline{w}_{{\rm new}}\\
&=\Big(\frac{\partial w_{{\rm new}}}{\partial (i x)}\Big)^{-1}\Big(\frac{\partial\overline{w}_{{\rm new}}}{\partial(-ix))}\Big)^{-1}\Big(\frac{\partial w_{{\rm new}}}{\partial \tau}d\tau+\frac{\partial w_{{\rm new}}}{\partial x}dx\Big)\Big(\frac{\partial \overline{w}_{{\rm new}}}{\partial \tau} d\tau+\frac{\partial \overline{w}_{{\rm new}}}{\partial x} dx\Big)\\
&=\Big({\partial_\tau w_{{\rm new}}\over \partial_x w_{{\rm new}}}\, d\tau+dx\Big)\Big({\partial_\tau \overline{w}_{{\rm new}}\over \partial_x \overline{w}_{{\rm new}}} \, d\tau+ dx\Big)\\
&=\Big(-iv_1(x)\, d\tau+dx\Big)\Big(iv_1(x)\, d\tau+ dx\Big)\\
&=v_1(x)^2d\tau^2+dx^2\,,
}
which is the curved space metric \eqref{SSDMetric3}.
Thus the Weyl transformation indeed directly gives the third quantity in \eqref{3desc}.

\subsubsection{$H_q$ with $q>1$}\label{SSDGeneralq}
As for $H_q$ with $q>1$, the conformal map $z=e^{qw}$ takes the cylinder to a $q$-sheet Riemann surface, where $H_q$ takes the same form as $H_1$
\begin{align}
     H_q&=2\int_0^{2\pi} dx\, \sin ^2\Big(q\cdot \frac{x}{2}\Big)T_{00}(x)\\
    &=q \oint \frac{dz}{2\pi i} \Big(-\frac{1}{2}+z-\frac{z^2}{2}\Big)T(z)-(z\to\bar{z})-q\frac{c}{12}.
\end{align}
The calculation then follows the same as above, with $2\pi\to 2\pi/q$. Namely, 
\begin{align}
    z_{{\rm new}}&=\frac{(1+q\frac{\tau}{2})z-q\frac{\tau}{2}}{q\frac{\tau}{2}z+(1-q\frac{\tau}{2})},\\
    ds^2&=q^2  \frac{e^{2q\tau}}{(1-\cos \ qx)^2}\Big[(1-\cos \ qx)^{2}d\tau^2+dx^2\Big].
\end{align}
Note that now $z$ and $\overline{z}$ are coordinates on the $q$-sheeted Riemann surface and there is a branch cut in the $z$-plane. The fact that the system is $q$ copies of identical theories transcribes to the crossing of sheets on the $z$-plane. 

\subsection{General Cases}
\cite{Wen:2018vux,Goto:2021sqx} also studied quantum quench with the one-parameter generalisation of the SSD Hamiltonian $H_1$, the Mobius Hamiltonian, 
The Mobius deformed Hamiltonian $H_\theta$ is given by 
\begin{align}
    H_\theta=L_0-{\rm tanh}\ 2\theta\ \frac{L_{-1}+L_1}{2}+\overline{L}_0-{\rm tanh}\ 2\theta\ \frac{\overline{L}_{-1}+\overline{L}_1}{2}-\frac{c}{12}\label{MobiusDeformedCFT}.
\end{align}
The $\theta\to 0$ and $\theta\to\infty$ limit of $H_\theta$ are $H_0$ and $H_1$, respectively. This family of Hamiltonians act as dilatations on the $\xi$ plane as $H_\theta^{(\xi)}={1\ov \cosh(2\theta)}\le(L^{(\xi)}_0+\bar{L}_0^{(\xi)}\ri)-{c\ov 12}$ with
\es{xiplane}{
\xi=-{\cosh \theta\, z-\sinh \theta\ov \sinh \theta\, z-\cosh \theta}\,.
}
By raising this map to the power $\cosh(2\theta)$, we obtain a properly normalised dilatation operator:\footnote{Through the exponential map we can see that raising the map to some power corresponds to rescaling time on the cylinder.} 
\es{chiThetaplane}{
\chi_\theta=\le(-\xi\ri)^{\cosh(2\theta)}=\le({\cosh \theta\, z-\sinh \theta\ov \sinh \theta\, z-\cosh \theta}\ri)^{\cosh(2\theta)}\,,
}
which has the the limit
\es{chiThetaLimit}{
\chi=\lim_{\theta\to \infty}\chi_\theta=e^{\frac{z+1}{z-1}} \,.
}
This agrees with \eqref{chi}.

The methodology of describing the non-trivial time evolution of inhomogeneous non-equilibrium CFT as coordinate transformations reviewed in Section \ref{SSDsetup} is valid beyond SSD (where only $SL(2)$ deformations are considered) and extends to deformations involving the $\emph{full}$ Virasoro algebras. In these cases, it is shown \cite{moosavi2021inhomogeneous} that the projective unitary representation of the universal cover of the orientation-preserving diffeomorphisms of the circle $\widetilde{{\rm Diff}}_+(S^1)$ allows one to flatten out the deformation $v(x)$ and map the expectations to ordinary equilibrium CFT under coordinate transformations. Like in the SSD case, these coordinate transformations capture the full information of the time evolution of the deformed CFT. In \cite{deBoer:2023lrd}, it is further demonstrated that just as in SSD CFT, these generally deformed CFTs can be understood as CFTs on curved spacetime as well. 

\section{Changing the cutoff in AdS/CFT}

\subsection{AdS/CFT with a nonuniform cutoff}\label{app:WigglyCutoff}

We will show on the example of a scalar field how the extrapolate dictionary \cite{Banks:1998dd} of AdS/CFT changes under changing the cutoff prescription. In the absence of a source of a scalar field, we have
\es{Falloff}{
\phi(\rho,y^\mu)=(2\Delta-d)\rho^{-\Delta}\, O(y^\mu)+\text{(subleading)}\,,
}
where $\rho$ is the bulk radial direction and $O(y^\mu)$ is the boundary CFT primary field. It is clear that if we choose two different bulk radial coordinates related by $r=\Om(y^\mu) r_\text{new}$ then the same bulk physics will give rise to boundary correlators that differ by some factors
\es{BulkFactors}{
\le\langle O(y^\mu)\dots\ri\rangle_\text{CFT, flat}&=\lim_{r_\text{new}\to \infty} {r_\text{new}^{\Delta}\ov2\Delta-d} \dots\le\langle \phi(r_\text{new},y^\mu) \dots \ri\rangle_\text{AdS}\\[4pt]
&=\Om(y^\mu)^{-\Delta}\dots \le\langle O(y^\mu)\dots\ri\rangle_\text{wiggly bdy}\,,
}
where the last object is the boundary correlation function that is computed from the nonuniform cutoff prescription on the gravity side
\es{cutoffagain}{
\epsilon={1 \ov R}={1\ov \Om(y^\mu) R_\text{new}} \,,
}
where in our application the choice $\Om=\sqrt{\frac{\partial w_{{\rm new}}}{\partial w}\frac{\partial\overline{w}_{{\rm new}}}{\partial\overline{w}}}$ is made.
$\le\langle O(y^\mu)\dots\ri\rangle_\text{wiggly bdy}$ is hence equal to the second object in \eqref{3desc}, as stated in the main text. 

It is also equal to the curved space CFT correlation function, as we show on the CFT side from applying a Weyl transformation and on the gravity side by adapting Fefferman-Graham coordinates to the nonuniform cutoff surface.

\subsection{The Fefferman-Graham expansion with a curved boundary}\label{app:FGBanados}

Let us change the spatial coordinate as
\es{xX}{
x=2\arccot X\,, 
}
then the boundary metric \eqref{SSDMetric3} takes the conformally flat form:
\es{confflat}{
ds_\text{bdy}^2={4\ov(1+X^2)^2}\le(-dt^2+dX^2\ri)\,.
}
The Fefferman-Graham expansion with this boundary metric takes the form:
\es{BanadosFG}{
ds^2=&{dr^2\ov r^2}-{(4r^2+X^2-(1+X^2)^2\,L(t+X))(4r^2+X^2-(1+X^2)^2\,R(t-X))\ov 4r^2(1+X^2)^2}\,dt^2\\
&+{(4r^2-1+(1+X^2)^2\,L(t+X))(4r^2-1+(1+X^2)^2\,R(t-X))\ov 4r^2(1+X^2)^2}\,dX^2\\
&+{(8r^2-1+X^2)(L(t+X)-R(t-X))\ov 4r^2}\,dt\,dX\,,
}
where $L(t+X)$ and $R(t-X)$ are arbitrary left and right moving functions corresponding to the holomorphic and anti-holomorphic stress tensor. The geometry \eqref{BanadosFG} first considered in~\cite{Skenderis:1999nb} is the generalisation of the well-known Banados geometries \cite{Banados:1998gg} appropriate for a flat boundary, and reduce to them in the limit of small $X$. We obtained these expressions by solving Einstein's equations in a $1/r$ series expansion, a method reviewed in \cite{Skenderis:1999nb,deHaro:2000vlm}. We can easily convert the metric to $x$ coordinates using the relation \eqref{xX}.

If we set $t=0$ and convert $X$ back to $x$, the resulting metric on the $t=0$ slice should match with the metric on the $t_\text{new}=0$ slice of the empty AdS$_3$ or BTZ metrics. Converting them to Fefferman-Graham form allows us to read off the functions $L$ and $R$. For the AdS$_3$ case, we get that $L=R=0$. In this case, a redefinition of $r$ reproduces the metric presented in \cite{MacCormack:2018rwq}.  For BTZ we find:
\es{LR}{
L(y)=R(y)={1+r_+^2\ov (1+y^2)^2}\,.
}

\section{Asymptotic symmetries of AdS$_3$}\label{AsymptoticSymmetries}
In this appendix we present the Killing vectors of global AdS$_3$ in \cite{Hartman}: 
\begin{align}
    \zeta_{-1}&=\frac{1}{2}e^{-i(t+x)}\Big(\frac{r}{\sqrt{r^2+1}}\partial_t+\frac{\sqrt{r^2+1}}{r}\partial_x+i\sqrt{r^2+1}\partial_r\Big),\\
    \zeta_{0}&=\frac{1}{2}(\partial_t+\partial_x),\\
    \zeta_{1}&=\frac{1}{2}e^{i(t+x)}\Big(\frac{r}{\sqrt{r^2+1}}\partial_t+\frac{\sqrt{r^2+1}}{r}\partial_x-i\sqrt{r^2+1}\partial_r\Big),\\
    \overline{\zeta_{-1}}&=\frac{1}{2}e^{-i(t-x)}\Big(\frac{r}{\sqrt{r^2+1}}\partial_t-\frac{\sqrt{r^2+1}}{r}\partial_x+i\sqrt{r^2+1}\partial_r\Big),\\
    \overline{\zeta_{0}}&=\frac{1}{2}(\partial_t-\partial_x),\\
    \overline{\zeta_{1}}&=\frac{1}{2}e^{i(t-x)}\Big(\frac{r}{\sqrt{r^2+1}}\partial_t-\frac{\sqrt{r^2+1}}{r}\partial_x-i\sqrt{r^2+1}\partial_r\Big),
\end{align}
where we dropped the "new" scripts for notation simplicity. These vectors are indeed obey the Killing equations
\begin{align}
    \nabla_{(\mu}(\zeta_i)_{\nu)}=0, && \nabla_{(\mu}(\overline{\zeta_i})_{\nu)}=0, && (i=0,\pm 1)\,.\label{KillingEqn}
\end{align}
The Killing vectors obey the $SL(2,\mathbb{R})_L\times SL(2,\mathbb{R})_R$ algebra:
\begin{align}
    i\{\zeta_{1},\zeta_{-1}\}=2\zeta_{0}, && i \{\zeta_{1},\zeta_{0}\}=\zeta_{1}, && i\{\zeta_{-1},\zeta_{0}\}=-\zeta_{-1}\label{4.9},\\    
    i \{\overline{\zeta_{1}},\overline{\zeta_{-1}}\}=2\overline{\zeta_{0}}, && i \{\overline{\zeta_{1}},\overline{\zeta_{0}}\}=\overline{\zeta_{1}}, && 
    i \{\overline{\zeta_{-1}},\overline{\zeta_{0}}\}=-\overline{\zeta_{-1}}\label{4.10},
\end{align}
where $\{.\,,.\}$ here denotes the Lie bracket and hence we have extra $i$'s on the LHS compared to when writing commutators.

\section{Geodesic equations}
In this appendix, we study null and spacelike geodesic equations, on pure AdS$_3$ and BTZ black holes, respectively. Our method is largely based on \cite{Hubeny:2012ry}. We will be working in geometries in the $(t_{{\rm new}},x_{{\rm new}},r_{{\rm new}})$ coordinate throughout this appendix, so we will drop the "new" subscript for notation simplicity. 

We consider Lorentzian (2+1)-dimensional metric: 
\begin{align}
    ds^2=-f(r)dt^2+\frac{dr^2}{f(r)}+r^2dx^2,
\end{align}
where $\lim_{r\to\infty}f(r)=r^2 $ so that the spacetime is asymptotically AdS$_3$. As the metric is independent of $t$ and $x$, there are two conserved quantities: 
\begin{align}
    E=f(r) \frac{dt}{d\lambda}, && L=r^2 \frac{dx}{d\lambda} \label{conservedQuantities},
\end{align}
which correspond to the energy and angular momentum, respectively. The norm of the tangent vector is: 
\begin{align}
    \kappa&=g_{\mu\nu}\frac{dx^{\mu}}{d\lambda}\frac{dx^{\nu}}{d\lambda}=-\frac{1}{f(r)}E^2+\frac{1}{f(r)}\Big(\frac{dr}{d\lambda}\Big)^2+\frac{1}{r^2}L^2\label{normTan},
\end{align}
where $\kappa=+1, 0,-1$ for spacelike, null, and timelike geodesics, respectively. We will focus on null and spacelike geodesics. Equation \eqref{normTan} can be put into the general form 
\begin{align}
    \Big(\frac{dr}{d\lambda}\Big)^2+V(r)=0, && V(r)=-E^2-\kappa f(r)+L^2\frac{f(r)}{r^2}\label{effPotential},
\end{align}
where $V(r)$ can be thought of as an effective potential. Geodesic motion requires $\big(\frac{dr}{d\lambda}\big)^2\ge 0$. Consequently, the condition for a geodesic to reach the boundary is $\lim_{r\to\infty}V(r)\leq 0$; for a geodesic emanating from a point on the boundary, the deepest distance, i.e.~the smallest $r$ it can reach $r_*$ is given by the largest positive zero of the effective potential $V(r)$ \cite{Hubeny:2012ry}.  
 
\subsection{Null geodesics}\label{Null Geodesics}
For null geodesic, we have $\kappa=0$ in \eqref{effPotential}, and the two conserved quantities can be combined to a single parameter $\ell=L/E$ by rescaling the affine parameter $\lambda$ \cite{Hubeny:2012ry}. \eqref{effPotential} then becomes  
\begin{align}
    \Big(\frac{dr}{d\lambda}\Big)^2+\ell^2\frac{f(r)}{r^2}-1=0\,.\label{nullGeod}
\end{align}
From \eqref{nullGeod}, it is clear that in order for the null geodesic to reach the boundary, one must have $\ell^2\leq 1$; $\ell^2=1$ corresponds to the case that the null geodesic stays on the boundary \cite{Hubeny:2012ry}. 
\subsubsection{Pure AdS$_3$}
\label{subsubsec:PureAdS}
For pure AdS$_3$, we have $f(r)=r^2+1$ in \eqref{nullGeod}, which integrates to 
\begin{align}
    r(\lambda)=\sqrt{(1-\ell^2)\ \lambda^2+\frac{\ell^2}{1-\ell^2}}  && \lambda\in \big[0,\infty\big)\label{AdSnullGeod}.
\end{align}
The null geodesic reaches the boundary, i.e.~$r\to\infty$ when $\lambda\to\infty$. 

Substituting \eqref{AdSnullGeod} into the two conserved quantities \eqref{conservedQuantities} with $f(r)=r^2+1$, one finds 
\begin{align}
    E&=(r^2+1) \frac{dt}{d\lambda}=\frac{(1-\ell^2)^2\ \lambda^2+1}{1-\ell^2} \frac{dt}{d\lambda}=1,\\
    L&=r^2 \frac{dx}{d\lambda}=\frac{(1-\ell^2)^2\ \lambda^2+\ell^2}{1-\ell^2} \frac{dx}{d\lambda}=\ell,
\end{align}
which integrate to 
\begin{align}
    t(\lambda)&={\rm arctan}[(1-\ell^2)\lambda]+C_t,\\
    x(\lambda)&={\rm arctan}\Big[\frac{1}{\ell}(1-\ell^2)\lambda\Big]+C_x.
\end{align}
When the null geodesic approaches the boundary 
\begin{align}
    \lim_{\lambda\to\infty}  t(\lambda) = C_t +\frac{\pi}{2} =y,&& \lim_{\lambda\to\infty}  x(\lambda) = C_x +\frac{\pi}{2}=y.
\end{align}
The other half of the triangle can be studied by reflection, i.e.~by performing a change of variable $x'=2\pi-x\in[0,\pi]$. The general expressions of null geodesics equations are therefore 
\begin{equation}
\left\{
    \begin{aligned}
    r(\lambda)&=\sqrt{(1-\ell^2)\ \lambda^2+\frac{\ell^2}{1-\ell^2}},\\
    t(\lambda)&={\rm arctan}[(1-\ell^2)\lambda]+y-\frac{\pi}{2},\\
    x(\lambda)&={\rm arctan}\Big[\frac{1}{\ell}(1-\ell^2)\lambda\Big]+y-\frac{\pi}{2}.
\end{aligned}
\right.
\end{equation}
From the expression of $x(\lambda)$, we can solve for $\ell$ 
\begin{align}
    -\cot (x-y)&=\frac{1}{\ell}(1-\ell^2)\lambda\label{B16},\\
    \ell&=\frac{r}{\sqrt{1+r^2+\cot ^2(x-y)}}\leq 1\label{ell}.
\end{align}
Plugging \eqref{B16}, \eqref{ell} and $r(\lambda)$ into $t(\lambda)$
\begin{align}
    t(r,x,y)&={\rm arctan}\Big[\ell\cdot\frac{1}{\ell}(1-\ell^2)\lambda\Big]+y-\frac{\pi}{2}\\
    &={\rm arctan}\Big[\frac{-r\cot (x-y)}{\sqrt{r^2+\csc ^2(x-y)}}\Big]+y-\frac{\pi}{2},
\end{align}
which is repeated  in the main text in \eqref{AdSNullGeod}. It is easy to see that $t$ increases monotonically with respect to $y$, 
\begin{align}
    \frac{dt}{dy}=1-\underbrace{\frac{r}{\sqrt{r^2+\csc ^2(x-y)}}}_{\leq 1}\ge 0\label{AdSMono}.
\end{align}
So $t$ maximises at $y=\pi$
\begin{align}
    t(r,x,y=\pi)=-{\rm arctan}\Big[\frac{r\cot x}{\sqrt{r^2+\csc ^2x}}\Big]+\frac{\pi}{2}.
\end{align}
Further simplifications depend on the range of $x$: 
\begin{itemize}
    \item For $\cot x>0$, i.e.~$0<x<\pi/2$, we have 
    \begin{align}
    t(r,x)={\rm arctan}\Big[\frac{\sqrt{r^2+\csc ^2x}}{r\cot x}\Big]=\arccos\Big[\frac{r}{\sqrt{r^2+1}}\cos x\Big].
\end{align}
    \item For $\cot x<0$, i.e.~$\pi/2<x<\pi$, we have
    \begin{align}
    t(r,x)&=-{\rm arctan}\Big[\frac{r\cot x}{\sqrt{r^2+\csc ^2x}}\Big]-\frac{\pi}{2}+\pi={\rm arctan}\Big[\frac{\sqrt{r^2+\csc ^2x}}{r\cot x}\Big]+\pi\nonumber\\
    &=\arccos\Big[\frac{r}{\sqrt{r^2+1}}\cos x\Big].
\end{align}
\end{itemize}
both of which are \eqref{AdSKilling2} in the main text. 

\subsubsection{BTZ}\label{BTZNullGeod}
\label{subsubsec:BTZ}
For BTZ, we have $f(r)=r^2-r_+^2$ in \eqref{nullGeod}, which integrates to 
\begin{align}
    r(\lambda)=\sqrt{(1-\ell^2)\ \lambda^2-\frac{\ell^2r_+^2}{1-\ell^2}},  && \lambda\in \Big[\frac{\ell r_+}{1-\ell^2},\infty\big)\label{BTZnullGeod}.
\end{align}
The null geodesic reaches the boundary, i.e.~$r\to\infty$ when $\lambda\to\infty$. 

Substituting \eqref{BTZnullGeod} into the two conserved quantities \eqref{conservedQuantities} with $f(r)=r^2-r_+^2$, we find 
\begin{align}
    E&=(r^2-r_+^2) \frac{dt}{d\lambda}=\frac{(1-\ell^2)^2\ \lambda^2-r_+^2}{1-\ell^2} \frac{dt}{d\lambda}=1,\\
    L&=r^2 \frac{dx}{d\lambda}=\frac{(1-\ell^2)^2\ \lambda^2-\ell^2r_+^2}{1-\ell^2} \frac{dx}{d\lambda}=\ell.
\end{align}
which outside the horizon, $r\ge r_+$ for integrate to:
    \begin{align}
        t(\lambda)&=\frac{1}{2 r_+}\log \frac{(1-\ell^2)\ \lambda-r_+}{(1-\ell^2)\ \lambda+r_+}+C_t\\
        x(\lambda)&=\frac{1}{2 r_+}\log \frac{(1-\ell^2)\ \lambda-\ell r_+}{(1-\ell^2)\ \lambda+\ell r_+}+C_x
    \end{align}
When the null geodesic approaches the boundary  
\begin{align}
    \lim_{\lambda\to\infty}  t(\lambda) =y= C_t, && \lim_{\lambda\to\infty}  x(\lambda) = y= C_x.
\end{align}
The general expressions of null geodesics equations outside the horizon are therefore 
\begin{equation}
\left\{
    \begin{aligned}
    r(\lambda)&=\sqrt{(1-\ell^2)\ \lambda^2-\frac{\ell^2r_+^2}{1-\ell^2}},\\
    t(\lambda)&=\frac{1}{2 r_+}\log \frac{(1-\ell^2)\ \lambda-r_+}{(1-\ell^2)\ \lambda+r_+}+y,\\
    x(\lambda)&=\frac{1}{2 r_+}\log \frac{(1-\ell^2)\ \lambda-\ell r_+}{(1-\ell^2)\ \lambda+\ell r_+}+y.
\end{aligned}
\right.
\end{equation}
From the expression of $x(\lambda)$, we can solve for $\ell$ 
\begin{align}
    \ell=\frac{(-1+e^{2r_+(x-y)})r}{\sqrt{(-1+e^{2r_+(x-y)})^2r^2+4r_+^2e^{2r_+(x-y)}}}\leq 1\label{B37}.
\end{align}
Plugging \eqref{B37} and $r(\lambda)$ into $t(\lambda)$ gives
\begin{align}
    t(r,x,y)=\frac{1}{2r_+}\log \frac{r\sqrt{1+r_+^2\sinh ^2(x-y)}-r_+\sqrt{1+r^2\sinh ^2(x-y)}}{r\sqrt{1+r_+^2\sinh ^2(x-y)}+r_+\sqrt{1+r^2\sinh ^2(x-y)}}+y,
\end{align}
which is \eqref{BTZNullGeod1} in the main text. It is easy to see that $t$ increases monotonically with respect to $y$ outside the BTZ horizon
\begin{align}
    \frac{dt}{dy}\propto 4r(r^2-r^2_+)\ \sinh (x-y)\ge0\label{BTZMono}.
\end{align}
So $t$ is maximised at $y=\pi$ 
\begin{align}
    t(r,x,y=\pi)=\frac{1}{2r_+}\log \frac{r\sqrt{1+r_+^2\sinh ^2(x-\pi)}-r_+\sqrt{1+r^2\sinh ^2(x-\pi)}}{r\sqrt{1+r_+^2\sinh ^2(x-\pi)}+r_+\sqrt{1+r^2\sinh ^2(x-\pi)}}+\pi,
\end{align}
which is \eqref{BTZNullGeod2} in the main text. 

\subsection{Spacelike geodesics}\label{SpacelikeGeodesics}
For spacelike geodesic, we have $\kappa=1$ in \eqref{normTan}
\begin{align}
    \Big(\frac{dr}{d\lambda}\Big)^2+V(r)=0, && V(r)=-f(r)-E^2+L^2\frac{f(r)}{r^2}=0\label{spacelikeGeod}.
\end{align}
To compute the holographic entanglement entropy, one needs to compute the length of the spacelike geodesic obeying the appropriate homology constraints. In our case, the spacelike geodesic equation is for non-equal time. The length of the spacelike geodesic is given in terms of the temporal and angular separation of the two points on the boundary on which our geodesic anchors. 

The length of the spacelike geodesic is given by 
\begin{align}
    \mathcal{L}=2\int_{r_*}^{\infty}d\lambda,
\end{align}
with the factor of two coming from symmetry. But this is of course infinite; to get a finite result, we impose the large-radius cutoffs $R_1$, $R_2$ at the two geodesic endpoints as in \cite{Hubeny:2012ry}. Note that from \eqref{effPotential}, one can change integration variable $d\lambda=\frac{dr}{\sqrt{-V(r)}}$, which is appropriate since $V(r)<0$ in $r\in(r_*,\infty)$. Thus, we have
\begin{align}
    \mathcal{L}_{{\rm reg}}=\Big(\int_{r_*}^{R_1}+\int_{r_*}^{R_2}\Big)\frac{dr}{\sqrt{-V(r)}}\label{lReg},
\end{align}
and the universal diverging piece is simply $\log 4R_1R_2$, since $V(r)\propto -r^2$ for spacelike geodesics as $r$ gets large in \eqref{spacelikeGeod}. 

The temporal and angular separation are then given by \cite{Hubeny:2012ry}
\begin{align}
    \Delta t&=2\int_{t_*}^{\infty}dt=2\int_{\lambda_*}^{\infty}\frac{dt}{d\lambda}d\lambda=2\int_{\lambda_*}^{\infty}\frac{E}{f(r)} d\lambda=2\int_{r_*}^{\infty}\frac{E}{f(r)\sqrt{-V(r)}}dr\label{Deltat},\\
    \Delta x&=2\int_{x_*}^{\infty}dx=2\int_{\lambda_*}^{\infty}\frac{dx}{d\lambda}d\lambda=2\int_{\lambda_*}^{\infty}\frac{L}{r^2} d\lambda=2\int_{r_*}^{\infty}\frac{L}{r^2\sqrt{-V(r)}}dr\label{Deltax},
\end{align}
where the conserved quantities \eqref{conservedQuantities} have been substituted in. 

\subsubsection{Pure AdS$_3$}\label{AdSSpaceLikeGeod}
For pure AdS$_3$, we have $f(r)=r^2+1$ in \eqref{spacelikeGeod}. The relevant quantities needed in our calculation have been worked out in \cite{Hubeny:2012ry}. The minimal radius the spacelike geodesic can reach is
\begin{align}
    r_*=\sqrt{\frac{1}{2}\Big[-(E^2-L^2+1)+\sqrt{(E^2-L^2+1)^2+4L^2}\Big]}\label{r*AdS}.
\end{align}
The regulated length of the spacelike geodesic is
\begin{align}
    \mathcal{L}_{\rm reg}=\Big(\int_{r_*}^{R_1}+\int_{r_*}^{R_2}\Big)\frac{dr}{\sqrt{-V(r)}}=-\log \sqrt{(E^2-L^2+1)^2+4L^2}+\log (4R_1R_2)\label{D45},
\end{align}
where we added the universal diverging term $\log (4R_1R_2)$ that \cite{Hubeny:2012ry} subtracted off, as it will play a role in our analysis later on. The temporal and angular distance between the two geodesic endpoints on the boundary are
\begin{align}
    \Delta t&=2\int_{r_*}^{\infty}\frac{E}{f(r)\sqrt{-V(r)}}dr=\frac{\pi}{2}+\arcsin \Big[\frac{E^2-L^2-1}{\sqrt{(E^2-L^2-1)^2+4E^2}}\Big],\\
    \Delta x&=2\int_{r_*}^{\infty}\frac{L}{r^2\sqrt{-V(r)}}dr=\frac{\pi}{2}+\arcsin \Big[\frac{E^2-L^2+1}{\sqrt{(E^2-L^2+1)^2+4L^2}}\Big],
\end{align}
which are \eqref{delta_tAdS} and \eqref{delta_xAdS} in the main text. Note that $(E^2-L^2-1)^2+4E^2=(E^2-L^2+1)^2+4L^2$. 

From \eqref{delta_tAdS} and \eqref{delta_xAdS}, we can solve for $E$ and $L$:
\begin{align}
    E=\frac{\sin \Delta t}{\cos \Delta t-\cos \Delta x}, && 
    L=\frac{\sin \Delta x}{\cos \Delta t-\cos \Delta x}\label{ELAdS}.
\end{align}
Plugging \eqref{ELAdS} into \eqref{D45}, we have, 
\begin{align}
    \mathcal{L}_{{\rm reg}}&=\log \Big[\frac{1}{2}\big|\cos \Delta t-\cos \Delta x\big|\Big]+\log (4R_{{1}}R_{{2}}),\\
    &=\log \Big[4\Big|\sinh 
    \Big(\frac{w_{1}-w_{2}}{2}\Big)\ \sinh \Big(\frac{\overline{w}_{1}-\overline{w}_{2}}{2}\Big)\Big|\ \Big]+\log (R_{{1}}R_{{2}}),
\end{align}
where we have used \eqref{7.1} in getting from the second to the third line. Restoring the "new" subscript then gives \eqref{7.7} in the main text. 

\subsubsection{BTZ}\label{BTZSpaceLikeGeod}
For BTZ, we have $f(r)=r^2-r_+^2$ in \eqref{spacelikeGeod}. Let us consider the cases in which the spacelike geodesic never enters the black hole. In such a case, $V(r)\ge 0$ for some $r>r_+$ \cite{Hubeny:2012ry}. To see what this indicates more concretely, we first note that $V(r)$ peaks at $r=\sqrt{L r_+}$: 
\begin{align}
    \frac{d V(r)}{dr}=-2r+\frac{2L^2 r_+^2}{r^3}, && \frac{d V(r)}{dr}=0\Rightarrow r=\sqrt{L r_+}, 
\end{align}
where one needs $\sqrt{L r_+}\ge r_+$, i.e.~$L\ge r_+$. The condition for the maximal value of $V(r)$, that is $V(\sqrt{L r_+})$, to be non-negative is 
\begin{align}
    V(\sqrt{L r_+})=-E^2+(L-r_+)^2\ge0 \qquad \Rightarrow \qquad L\ge E+r_+,
\end{align}
which is more stringent than $L\ge r_+$. Henceforth, the criterion for the spacelike geodesic not entering the BTZ horizon is  $L\ge E+r_+$. The limiting case $L=r_+$ when $E=0$ corresponds to the known case that the spacelike geodesic gets trapped in a circular orbit at the horizon, which is studied in \cite{Hubeny:2012ry}. At the minimal radius the spacelike geodesic can reach, $dr/d\lambda=0$, so the minimal radius is given by the larger positive zero of $V(r)$ 
\begin{align}
    0&=r^2-r_+^2+E^2-L^2\frac{r^2-r_+^2}{r^2},\\
    \Rightarrow r_*&=\sqrt{\frac{1}{2}\Big[L^2-E^2+r_+^2+\sqrt{(L^2-E^2+r_+^2)^2-4r_+^2L^2}\Big]}\label{r_*BTZ}.
\end{align}
One can verify that for $L= E+r_+$, $r_*=\sqrt{r_+(r_++E)}> r_+$; increasing $L$ results in increasing $r_*$. Hence $r_*> r_+$ for $L\ge E+r_+$. By setting $r_+^2=-1$ in \eqref{r_*BTZ}, we recover \eqref{r*AdS} in pure AdS$_3$. 

The regulated geodesic length is then given by integrating \eqref{lReg}
\begin{align}
    \mathcal{L}_{\rm reg}=\Big(\int_{r_*}^{R_1}+\int_{r_*}^{R_2}\Big)\frac{dr}{\sqrt{-V(r)}}=-\log \sqrt{(L^2-E^2+r_+^2)^2-4r_+^2L^2}+\log 4R_1R_2\label{D53},
\end{align}
Note that by setting the BTZ black hole radius $r_+^2=-1$ in \eqref{D53}, we recover \eqref{D45} in pure AdS$_3$. The temporal and angular distance between the two geodesic endpoints on the boundary are given by integrating  \eqref{Deltat} and \eqref{Deltax}, respectively
\begin{align}
    \Delta t&=2\int_{r_*}^{\infty}\frac{E}{f(r)\sqrt{-V(r)}}dr\\
    &=\frac{1}{r_+}\log \Big[\frac{\sqrt{(L^2-E^2+r_+^2)^2-4r_+^2E^2}}{(L-E-r_+)(L+E+r_+)}\big]\\
    &=\frac{1}{r_+}{\rm arccosh}\Big[\frac{L^2-E^2-r_+^2}{\sqrt{(L^2-E^2-r_+^2)^2-4r_+^2E^2}}\Big],\\
    \Delta x&=2\int_{r_*}^{\infty}\frac{L}{r^2\sqrt{-V(r)}}dr\\
    &=\frac{1}{r_+}\log \Big[\frac{\sqrt{(E^2-L^2-r_+^2)^2-4r_+^2L^2}}{(L+E-r_+)(L-E-r_+)}\big]\nonumber\\
    &=\frac{1}{r_+}{\rm arccosh}\Big[\frac{L^2-E^2+r_+^2}{\sqrt{(L^2-E^2+r_+^2)^2-4r_+^2L^2}}\Big]\label{D61},
\end{align}
which are \eqref{delta_tBTZ} and \eqref{delta_xBTZ} in the main text. Note that $(L^2-E^2+r_+^2)^2-4r_+^2L^2=(L^2-E^2-r_+^2)^2-4r_+^2E^2$. One can verify that when $E=0$, \eqref{D61} becomes 
\begin{align}
    \Delta x=\frac{1}{r_+}{\rm arccosh}\Big(\frac{L^2+r_+^2}{L^2-r_+^2}\Big)=\frac{2}{r_+}{\rm arctanh}\Big(\frac{r_+}{L}\Big),
\end{align}
which matches with a known result in section 2.3 of \cite{Hubeny:2012ry}.

From \eqref{delta_tBTZ} and \eqref{delta_xBTZ}, one can solve for $E$ and $L$
\begin{align}
    E=r_+\frac{\sinh (r_+\Delta t)}{\cosh  (r_+\Delta t)-\cosh  (r_+\Delta x)}, && 
    L=r_+\frac{\sinh (r_+\Delta x)}{\cosh  (r_+\Delta t)-\cosh  (r_+\Delta x)}\label{ELBTZ}.
\end{align}
Plugging \eqref{ELBTZ} into \eqref{D53}, we have,
\begin{align}
    \mathcal{L}_{{\rm reg}}
    &=\log \Big[\Big(\frac{2}{r_+}\Big)^2\frac{1}{2}\big|\cosh  (r_+\Delta t)-\cosh  (r_+\Delta x)\big|\Big]+\log (R_{{1}}R_{{2}})\label{7.22}\\
    &=\log \Big[\Big(\frac{2}{r_+}\Big)^2\Big|\sin 
    \Big(r_+\frac{w_{1}-w_{2}}{2}\Big)\ \sin \Big(r_+\frac{\overline{w}_{1}-\overline{w}_{2}}{2}\Big)\Big|\ \Big]+\log (R_{{1}}R_{{2}}),
\end{align}
where we have used \eqref{7.1} in getting from the second to the third line. This then gives \eqref{7.18} in the main text with the "new" subscript restored. 

\section{Time evolution of SSD horizons}\label{ComparisonRyu}
In this appendix, we show how the coordinates in Figure \ref{fig:HorCompare} are established. In \cite{Goto:2021sqx}, a redefinition of the radial coordinate 
\begin{align}
    r'=\sqrt{\frac{\partial w_{{\rm new}}}{\partial w}\frac{\partial\overline{w}_{{\rm new}}}{\partial\overline{w}}}r_{{\rm new}}=\frac{1}{\sqrt{4t^2\sin ^4\frac{ x}{2}+\Big(1- t^2 \sin ^2\frac{ x}{2}\Big)^2}}r_{{\rm new}}\label{r'},
\end{align}
was introduced, and the BTZ black hole horizon was plotted on constant $t$ slices in the $(t, x, r')$ coordinate. In order to compare with their result, we adopt the same $(t, x, r')$ coordinate when plotting Figure \ref{fig:HorCompare}. Notice that unlike \eqref{BanadosFG}, the metric in the $(t,x,r')$ coordinate is not in the Fefferman-Graham gauge \cite{Skenderis:1999nb,deHaro:2000vlm}, as $r'$ is $t$ and $x$ dependent, leading to non-zero $dr'dt$ and $dr'dx$ terms in the metric. Therefore quantitative comparisons with the boundary stress tensor are in general hard to achieve. 

\section{Comments on bulk reconstruction}\label{bulk reconstructions}
Horizons make information inaccessible.
The presence of the horizons \eqref{AdSKilling2} and \eqref{BTZNullGeod2} in Figure \ref{fig:AdSHor} and \ref{fig:BTZHor} makes us wonder if an information puzzle can arise in our setup: e.g.~a bulk excitation prepared on the $t=0$ would fall behind the horizon and become inaccessible leading to tension with unitary time evolution. Equivalently, one asks how to reconstruct local bulk operators (dressed to make them diffeomorphism invariant) inside or outside the new horizons, respectively. 

For bulk fields $\emph{inside}$ the causal wedge, i.e.~in the bulk region enclosed by the null hypersurfaces in Figure \ref{fig:AdSHor} or \ref{fig:BTZHor} and the quench surface at $t=0$, we can recover the information using HKLL-like reconstructions \cite{Banks:1998dd,Hamilton:2006az,Witten:2023qsv}. In our case, the $t_\text{new}>0$ part of this subregion is the triangle. (The $t_\text{new}<0$ part is a time strip of width $\pi/2$ for pure AdS and infinite for BTZ.)   The reconstruction formula is given by   
\begin{align}
    \phi(X)=\int_{{\rm triangle + strip}}dx\, K(X|x) O(x),
\end{align}
where $X=(t_{{\rm new}},r_{{\rm new}},x_{{\rm new}})$ and $x=(t_{{\rm new}},x_{{\rm new}})$ stand for the coordinates on the bulk and boundary, respectively. $K(X|x)$ is a smearing function adapted to the subregion. 

For bulk operators $\emph{outside}$ the causal wedge, i.e.~behind the horizon \eqref{AdSKilling2} and \eqref{BTZNullGeod2}, the aforementioned HKLL-like reconstruction does not apply anymore. However, information falling through the horizon \eqref{AdSKilling2} or \eqref{BTZNullGeod2} (but not through the BTZ black hole horizon in the case with \eqref{BTZNullGeod2}) remains reconstructable, by considering $\emph{modular}$ reconstruction \cite{Jafferis:2015del,Dong:2016eik}. In our case, the modular Hamiltonian is simply $H_0$ up to a coefficient $\frac{\beta}{2\pi}$,\footnote{In the vacuum since the density matrix is a projector, the modular Hamiltonian is not well defined, so we need to regulate the vacuum with a small temperature.} so the modular evolution is just $H_0$ time evolution with the modular parameter given by $t_{{\rm new}}=t$.\footnote{Note that when the Hamiltonian is $H_0$, we have $t_{{\rm new}}=t$; $t_{{\rm new}}$ and $t$ are different only when the driving Hamiltonian is $H_1$. } We would like to then consider the $\emph{modular\ flow}$ of local operators $O$ on the boundary, $O(x,t)=U^{-1}(t)O(x,0)U(t)$, where $U(t)=e^{-i \beta H_0 t/ 2\pi}$. Modular reconstruction then states that $\emph{modular\ flows\ equal\ bulk\ modular\ flows}$ \cite{Jafferis:2015del}. This conclusion allows us to recover the information of all bulk operators within the $\emph{entanglement\ wedges}$, which in our case is the entire spacetime for the ground state case and all the spacetime regions outside the BTZ black hole horizon for the thermal state case. In each case, the entanglement wedge is larger than the casual wedge and, importantly, includes the spacetime region that is behind the horizon \eqref{AdSKilling2} or \eqref{BTZNullGeod2} (but outside the BTZ horizon in the thermal state case). 

\section{Quenches governed by other $SL(2)$ Virasoro subalgebras}\label{SSDq}

It is straightforward to generalise the SSD story, by replacing $L_{\pm1}$ in the expression of $H_1$ in \eqref{1} with $L_{\pm q}$ with $q>1$.
This corresponds to the envelope function \cite{Fan:2019upv}
\begin{align}
    v_q(x)=2\sin ^2\Big(q\cdot \frac{x}{2}\Big), 
\end{align}
and the resulting deformed Hamiltonian $H_q$ is given by
\begin{align}
    H_q&=2\int_0^{2\pi} dx\ \sin ^2\Big(q\cdot \frac{x}{2}\Big)\,T_{00}(x)\\
    &=L_0-\frac{L_{-q}+L_q}{2}+\overline{L}_0-\frac{\overline{L}_{-q}+\overline{L_q}}{2}-q\cdot\frac{c}{12}.
\end{align}
In order for $\{L_0, L_q, L_{-q}\}$ to form the same algebra as $\{L_0, L_1, L_{-1}\}$, we have to $\emph{shift}$ the Virasoro generators $\{L_0, L_q, L_{-q}\}$,\footnote{There are subtleties regarding the difference between $L_q$ ($q>1$) and the global conformal charges $\{L_0, L_1, L_{-1}\}$, as the former generate transformations that are not globally well defined and are associated to conformal Killing vectors with poles. But these poles appear at $0$ or $\infty$ on the complex $z$-plane or equivalently $\pm \infty$ on the cylinder, where we will not perform calculations. Therefore, our calculation is unaffected by this subtlety. }
\begin{align}
    \mathcal{L}_0=\frac{1}{q}\big(L_0-\frac{c}{12}q(q^2-1)\big), && \mathcal{L}_{\pm q}=\frac{1}{q}L_{\pm q}.
\end{align}
Note that such shifts also define a new vacuum $|0\rangle_q$ for each $H_q$. $H_q$ is now given by 
\begin{align}
    H_q=q\Big(\mathcal{L}_0-\frac{\mathcal{L}_{-q}+\mathcal{L}_q}{2}+\overline{\mathcal{L}}_0-\frac{\overline{\mathcal{L}}_{-q}+\overline{\mathcal{L}}_q}{2}-\frac{c}{12}\Big)\label{Hq},
\end{align}
which follows the same algebraic commutation rules as that of $H_1$. The relevant coordinate transformations then take the same form as \eqref{wnew} and \eqref{wnewbar}, with the circumference $2\pi$ replaced by $2\pi/q$. The system can then be regarded as $q$ copies of identical theories driven by $H_1$, each with a shrunk length scale $2\pi/q$ living in a subregion on the cylinder (see Appendix \ref{SSDGeneralq} for more details).  
Hence the constant $t$ lines approach a $q$-fold triangle, i.e.~$q$ triangles, each living in a subregion $\big[(p-1)\frac{2\pi}{q},p\frac{2\pi}{q}\big]\ (p=1,2,...,q)$. The boundary conformal Killing vector is 
\begin{align}
    k_q^{\mu}=\Big(\cosh  (q\tau_{{\rm new}})\ \cos (qx_{{\rm new}})-1, \sinh (q\tau_{{\rm new}})\ \sin (qx_{{\rm new}})\Big)\label{qTriangles}.
\end{align}
The boundary conformal Killing horizon is then the aforementioned $q$-fold triangle. 

\begin{figure}[htbp]
\centering
\includegraphics[width=.55\textwidth]{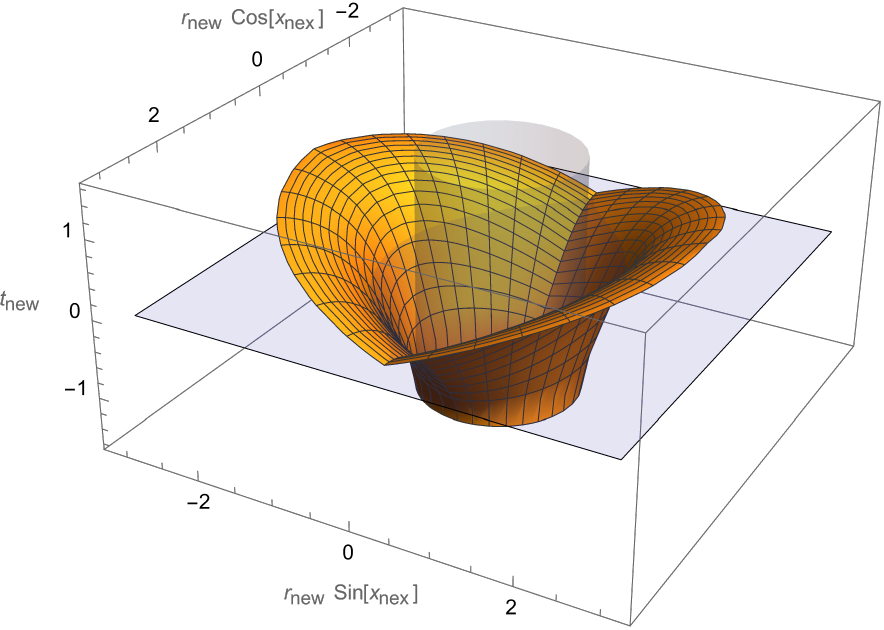}
\qquad
\caption{A plot of the new horizon in BTZ geometry with $H_2$ as driving Hamiltonian. The light blue surface is the quench surface $t_{{\rm new}}=t=0$. The BTZ black hole horizon is set to $r_+=1$ and plotted in light gray. The new horizon is larger than the BTZ black hole horizon. \label{fig:BTZHorq}}
\end{figure}

Generalisation of the new horizon with $H_q\ (q>1)$ as Hamiltonian is straightforward. From equation \eqref{qTriangles}, the boundary horizons when $H_q$ drives are $q$-triangles each in a subregion; according to the same maximisation procedure of null geodesics in \eqref{AdSNullGeod} and \eqref{BTZNullGeod1}, the new horizon then consists of $q$-copies of \eqref{AdSKilling2} or \eqref{BTZNullGeod2} when the CFT is in the ground and thermal state, respectively, each anchoring on one of the $q$ triangles on the boundary. An example for $q=2$ in the thermal state is plotted in Figure \ref{fig:BTZHorq}. Note that the bulk horizons for $H_q\ (q>1)$ in the ground state are no longer Killing horizons. (The naive generalisations of \eqref{BulkKillingVec} to $q>1$ do not obey the Killing equation \eqref{KillingEqn}. This is as anticipated as the bulk does not have an $SL(2)$ isometry for each $q$.)

\section{Two-point functions of twist operators}\label{Calc2pt}
In this appendix, we first evaluate the two-point function $\langle \mathcal{T}_{n}(w_{1,{{\rm new}}},\overline{w}_{1,{{\rm new}}})\mathcal{T}_{-n}(w_{2,{{\rm new}}},\overline{w}_{2,{{\rm new}}})\rangle$ for a CFT$_2$ on a compact space of length $2\pi$, and then evaluate the thermal two-point function $\langle \mathcal{T}_{n}(w_{1,{{\rm new}}},\overline{w}_{1,{{\rm new}}})\mathcal{T}_{-n}(w_{2,{{\rm new}}},\overline{w}_{2,{{\rm new}}})\rangle_{\beta}$ for a CFT$_2$ at temperature $T=\beta^{-1}$ on a non-compact space. Throughout this appendix, we will be working exclusively in $z_{{\rm new}}$ and $w_{{\rm new}}$ coordinate, so we drop the "new" subscript for notation simplicity. Our method is largely based on \cite{Calabrese:2004eu}. Note, however, that we are working in the coordinates 
\begin{align}
    w=\tau+i x, && \overline{w}=\tau-i x\label{holocoord},
\end{align}
on the Euclidean cylinder as is mentioned in the main text. This results in a different behaviour of the holomorphic and anti-holomorphic parts of the two-point function, in contrast to \cite{Calabrese:2004eu}. We can use the conformal transformation \eqref{expMap} that maps the Euclidean cylinder to the complex plane, where the two-point function of twist operator is simply
\begin{align}
    \langle \mathcal{T}_{n}(z_1,\overline{z}_1)\mathcal{T}_{-n}(z_2,\overline{z}_2)\rangle=\frac{1}{(z_1-z_2)^{2h}}\cdot \frac{1}{(\overline{z}_1-\overline{z}_2)^{2h}}.
\end{align}
For the holomorphic part, under a conformal transformation $z\to w$, 
\begin{align}
    \langle \mathcal{T}_{n}(z_1)\mathcal{T}_{-n}(z_2)\rangle=\frac{1}{(z_1-z_2)^{2h}}=\Big(\frac{\partial w_1}{\partial z_1}\Big)^h\Big(\frac{\partial w_2}{\partial z_2}\Big)^h\langle \mathcal{T}_{n}(w_1)\mathcal{T}_{-n}(w_2)\rangle
\end{align}
Substituting \eqref{expMap}, $\langle \mathcal{T}_{n}(w_1)\mathcal{T}_{-n}(w_2)\rangle$ on the Euclidean cylinder is then 
\begin{align}
    \langle \mathcal{T}_{n}(w_1)\mathcal{T}_{-n}(w_2)\rangle=\Big[2\sinh \Big(\frac{1}{2}(w_1-w_2)\Big)\Big]^{-2h}=\Big[2\sinh \Big(\frac{1}{2}((\tau_1-\tau_2)+i(x_1-x_2))\Big)\Big]^{-2h}\label{holo}.
\end{align}
The anti-holomorphic part is simply given by $w_i\to\overline{w}_i$, 
\begin{align}
    \langle \mathcal{T}_{n}(\overline{w}_1)\mathcal{T}_{-n}(\overline{w}_2)\rangle=\Big[2\sinh \Big(\frac{1}{2}(\overline{w}_1-\overline{w}_2)\Big)\Big]^{-2h}=\Big[2\sinh \Big(\frac{1}{2}((\tau_1-\tau_2)-i(x_1-x_2))\Big)\Big]^{-2h}\label{antiholo}.
\end{align}
Note that \eqref{holo} and \eqref{antiholo} are not equal due to \eqref{holocoord} in our setup. The two-point function $\langle \mathcal{T}_{n}(w_{1,{{\rm new}}},\overline{w}_{1,{{\rm new}}})\mathcal{T}_{-n}(w_{2,{{\rm new}}},\overline{w}_{2,{{\rm new}}})\rangle$ is then given by the product of \eqref{holo} and \eqref{antiholo} and restoring the "new" subscript, which is \eqref{2ptFuncNew} in the main text. 

To calculate the thermal two-point function, $\langle \mathcal{T}_{n}(w_{1,{{\rm new}}},\overline{w}_{1,{{\rm new}}})\mathcal{T}_{-n}(w_{2,{{\rm new}}},\overline{w}_{2,{{\rm new}}})\rangle_{\beta}$ on non-compact space, we simply perform a Wick rotation of the exponential map, and rescale the circumference of the cylinder from $2\pi$ to $\beta$ \cite{Calabrese:2004eu}, 
\begin{align}
    z=e^{\frac{2\pi}{\beta}iw}, && \overline{z}=e^{\frac{2\pi}{\beta}i\overline{w}},
\end{align}
this amounts to $w\to iw$ and $2\pi\to\beta$ in \eqref{holo} and \eqref{antiholo}, 
\begin{align}
    \langle \mathcal{T}_{n}(w_1)\mathcal{T}_{-n}(w_2)\rangle_{\beta}&=\Big[\frac{\beta}{\pi}\sin \Big(\frac{\pi}{\beta}(w_1-w_2)\Big)\Big]^{-2h}=\Big[\frac{\beta}{\pi}\sinh \Big(\frac{\pi}{\beta}((\tau_1-\tau_2)+i(x_1-x_2))\Big)\Big]^{-2h},\\
    \langle \mathcal{T}_{n}(\overline{w}_1)\mathcal{T}_{-n}(\overline{w}_2)\rangle_{\beta}&=\Big[\frac{_{\beta}}{\pi}\sin \Big(\frac{\pi}{\beta}(\overline{w}_1-\overline{w}_2)\Big)\Big]^{-2h}=\Big[\frac{_{\beta}}{\pi}\sinh \Big(\frac{\pi}{\beta}((\tau_1-\tau_2)-i(x_1-x_2))\Big)\Big]^{-2h}.
\end{align}
The thermal two-point function $\langle \mathcal{T}_{n}(w_{1,{{\rm new}}},\overline{w}_{1,{{\rm new}}})\mathcal{T}_{-n}(w_{2,{{\rm new}}},\overline{w}_{2,{{\rm new}}})\rangle_{\beta}$ is then given by the product of the two, and restoring the "new" subscript.

\chapter{Appendices to Chapter~\ref{Chap3}}
\section{Detailed analysis of solutions in the generalised membrane theory}\label{ApproxLagD}
\subsection{Displaced half-space}\label{ApproxLagDispHalf}
In this appendix, we solve the equations of motion arising from the effective action $\mathcal{L}_{{\rm eff}}$ \eqref{Leff} exactly. Our solution is still an approximation to the full solution presented in Appendix~\ref{Displaced half spaces}, since the effective action $\mathcal{L}_{{\rm eff}}$ \eqref{Leff} is derived under the scaling limit \eqref{AreaFunct3}. 

\begin{figure}[htbp]
\centering
\includegraphics[width=.4\textwidth]{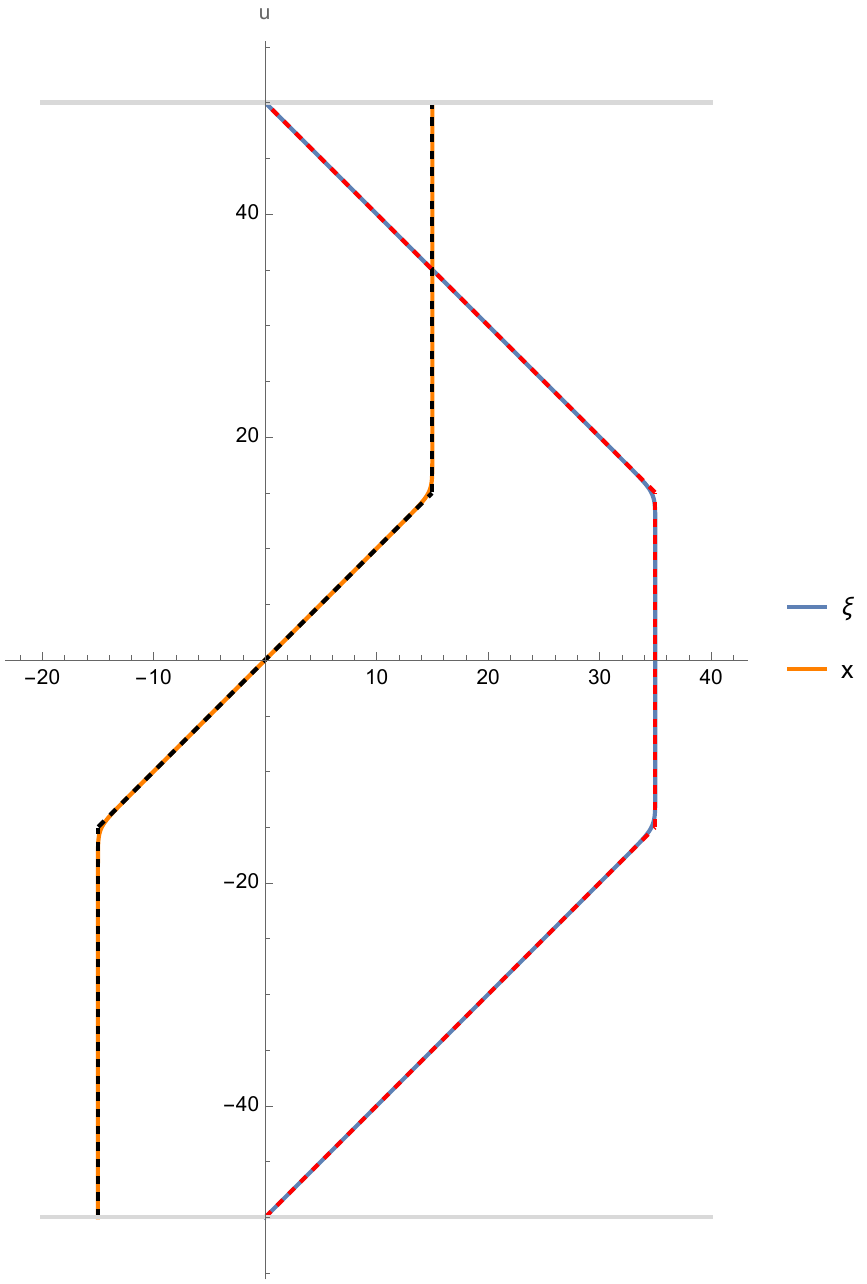}
\qquad
\caption{Plot of $\xi(u)$ (blue) \eqref{z(t)2} and $x(u)$ (orange) \eqref{x(t)2} when $T=50$ and $X=15$. The scaling approximation of $\xi(u)$ and $x(u)$ \eqref{FinalXi} are plotted as dashed in red and black, respectively. \label{fig:xiXApproxLag}}
\end{figure}

As \eqref{Leff} does not contain $u$ or $x$, it has two conserved quantities, the energy $E$ and momentum $p$, 
\begin{align}
    E&=\dot{\xi}\frac{\partial \mathcal{L}}{\partial \dot{\xi}}+\dot{x}\frac{\partial \mathcal{L}}{\partial \dot{x}}-\mathcal{L}=\frac{1}{2} e^{-2 \xi} \left(1+\dot{x}^2-\dot{\xi}^2\right)\label{simplifiedEBTZ},\\
    p&=\frac{\partial \mathcal{L}}{\partial \dot{x}}=e^{-2\xi}\dot{x}\label{simplifiedPBTZ}.
\end{align}
The generalised velocity $\dot{\xi}$ vanishes at the furthest point $\xi_0$ the geodesic can reach in the interior, $\dot{\xi}|_{\xi=\xi_0}=0$. Therefore, we can evaluate $E$ \eqref{simplifiedEBTZ} at $\xi_0$, which then allows us to solve for $\dot{\xi}$
\begin{align}
    e^{-2 \xi} \left(1+p^2e^{4\xi}-\dot{\xi}^2\right)&=e^{-2 \xi_0} \left(1+p^2e^{4\xi_0}\right),\\
    \Rightarrow \dot{\xi}^2&=(e^{2\xi_0}-e^{2\xi})(e^{-2\xi_0}-p^2e^{2\xi})\label{zdot},
\end{align}
where we have plugged in \eqref{simplifiedPBTZ}. We see that by making use of conserved quantity $E$ \eqref{simplifiedEBTZ}, we reduced the problem from a second order ODE \eqref{xiEOM} to a first order ODE \eqref{zdot}. Solving \eqref{zdot}, we find
\begin{align}
    \xi(u)=\log \left(\frac{e^{\text{$\xi $}_0} \text{sech}u}{\sqrt{1-e^{4 \text{$\xi $}_0} p^2
   \tanh ^2u}}\right)\label{z(t)}.
\end{align}
Then, plug \eqref{z(t)} into \eqref{simplifiedPBTZ} and integrate, we can further obtain $x(u)$
\begin{align}
    x(u)={\rm arctanh}(p e^{2\xi_0}\tanh u)\label{x(t)}.
\end{align}
From \eqref{x(t)}, we can find the total displacement $2X$ in terms of $p$ and $z_0$ by taking $|u|$ large, resulting in 
\begin{align}
    X={\rm arctanh}(Pe^{2\xi_0}) && \Rightarrow && p=e^{-2\xi_0}\tanh X\label{PwX}.
\end{align}
\eqref{PwX} indicates that $p\to e^{-2\xi_0}$ for large $X$. Plugging \eqref{PwX} into \eqref{z(t)}, we have 
\begin{align}
    \xi(u)=\log \left(\frac{e^{\text{$\xi $}_0} \text{sech}u}{\sqrt{1-\tanh
   ^2X \tanh ^2u }}\right)\label{z(t)3/2}.
\end{align}
This allows us to solve for $\xi_0$ in terms of $T$ and $X$ from \eqref{z(t)3/2}
\begin{align}
    \xi_0=\log \left(\sqrt{\frac{\cosh 2 T+\cosh 2 X}{\cosh 2 X +1}}\right)\label{xi0X},
\end{align}
which matches \eqref{z0X} obtained from the BTZ-AdS$_3$ mapping method. Plug \eqref{xi0X} and \eqref{PwX} into \eqref{z(t)} and \eqref{x(t)} we can express the geodesic as a parametric curve parameterised by the infalling time $u$
\begin{align}
    \xi(u)&=\frac{1}{2} \log \left(\frac{\cosh 2 T+\cosh 2 X}{\cosh 2 u+\cosh 2
   X}\right)\label{z(t)2},\\
    x(u)&={\rm arctanh}\ (\tanh X \tanh u)\label{x(t)2}.
\end{align}
A plot of \eqref{z(t)2} and \eqref{x(t)2} can be found in Fig.~\ref{fig:xiXApproxLag}. In the large $T$, large $X$ ($T>X$) limit, one can check that $\xi(u)$ \eqref{z(t)2} and $x(u)$ \eqref{x(t)2} match \eqref{FinalXi}.

 \subsection{Saturated EWCS}\label{Plateau+Null geodesic}
With the change of variable from $u$ to $x$ in \eqref{xiEOMx}, $E$ \eqref{simplifiedEBTZ} becomes 
\begin{align}
    E=\frac{1}{2} e^{-2 \xi } \left(1-e^{4 (\xi -\xi_p)} \left(\xi
   '^2-1\right)\right)\label{xixDiffEqn},
\end{align}
where we have plugged in $u'$ \eqref{xiEOMx}. Again, the conserved $E$ allows us to reduce the problem from a second-order ODE \eqref{xiEOMx} to a first-order ODE \eqref{xixDiffEqn}. Solving \eqref{xixDiffEqn} we find
\begin{align}
    \xi(x)=\frac{1}{2}\log\left(\frac{e^{4\xi_p}\left(1-E^2 e^{4\xi_p}\right) e^{-2 x} -e^{2
   x}+2 E e^{4\xi_p}}{2}\right)\label{xixv1}.
\end{align}
Notice that $\xi(-x)$ is also a solution to \eqref{xixDiffEqn}. This allows us to flip the sign of $u'$ \eqref{xiEOMx}.\footnote{It turned out that the effective theory allows both $\pm$ signs. To match with the exact solution, however, we choose $-$ sign here. } Plugging \eqref{xixv1} into \eqref{xiEOMx} and integrating, we have
\begin{align}
    u(x)=u_p+\ell_p-{\rm arctanh}(E e^{2\xi_p}-e^{2(x-\xi_p)})\label{txv1},
\end{align}
where for later convenience we write the integration constant as $u_p+\ell_p$. It can be easily observed from \eqref{txv1} that as $x\to-\infty$, $u(x)$ asymptotes a constant value. Denoting this value as $u_p$, we can solve for $E$ 
\begin{align}
    E=e^{-2\xi_p}\tanh\ell_p\label{ENullGeod}.
\end{align}
Notice the similarity between \eqref{ENullGeod} and \eqref{PwX}. Plugging \eqref{ENullGeod} into \eqref{xixv1}, one finds
\begin{align}
    \xi(x)=\frac{1}{2}\log\frac{1}{2}\Big(-e^{2x}+e^{-2(x-2\xi_p)}{\rm sech}^2\ell_p+2e^{2\xi_p}\tanh\ell_p\Big)\label{xixexact},
\end{align}
which, in the large $\ell_p$ limit, simplifies to 
\begin{equation}
\label{xixscaling}
    \xi(x)=
    \left\{
    \begin{aligned}
        &-x-\ell_p+2\xi_p+\frac{1}{2}\log2,&& x\in(-\infty,-\ell_p+\xi_p+\frac{1}{2}\log2],\\
        &\xi_p,&& x\in[-\ell_p+\xi_p+\frac{1}{2}\log2,\xi_p+\frac{1}{2}\log2).
    \end{aligned}
    \right.
\end{equation}
When $x\in[-\ell_p+\xi_p+\frac{1}{2}\log2,\xi_p+\frac{1}{2}\log2)$, the geodesic is on a $\emph{plateau}$ $\xi=\xi_p$. $\ell_p$ is therefore identified as the (horizontal) $\emph{length}$ of this plateau. When $x\in(-\infty,-\ell_p+\xi_p+\frac{1}{2}\log2]$, the geodesic shoots into the deep interior exponentially in $x$. Plugging \eqref{ENullGeod} into \eqref{txv1}, one finds 
\begin{align}
    u(x)=u_p+\ell_p+{\rm arctanh}(\tanh\ell_p-e^{2(x-\xi_p)})\label{txexact},
\end{align}
which, in the large $\ell_p$ limit, simplifies to 
\begin{equation}
\label{txscaling}
    u(x)=
    \left\{
    \begin{aligned}
        &u_p,&& x\in(-\infty,-\ell_p+\xi_p+\frac{1}{2}\log2],\\
        &x+u_p+\ell_p-\xi_p-\frac{1}{2}\log2,&& x\in[-\ell_p+\xi_p+\frac{1}{2}\log2,\xi_p+\frac{1}{2}\log2).
    \end{aligned}
    \right.
\end{equation}
Notice that when $(-\infty,-\ell_p+\xi_p+\frac{1}{2}\log2]$, the geodesic is on the $u=u_p$ slice. 

In general, one can shift $x\to x-x_c$ in \eqref{xixexact} and \eqref{txexact}. In the scaling limit, the shifted \eqref{xixexact} and \eqref{txexact} match the approximate solution \eqref{XiEWCS} in the main text with $x_0=x_c-\ell_p+\xi_p$. Therefore, one cannot fix all four parameters $(u_p,\ell_p,\xi_p,x_c)$ even solving the approximate equations of motion \eqref{xiEOMx} exactly. 

\begin{figure}[htbp]
\centering
\includegraphics[width=.6\textwidth]{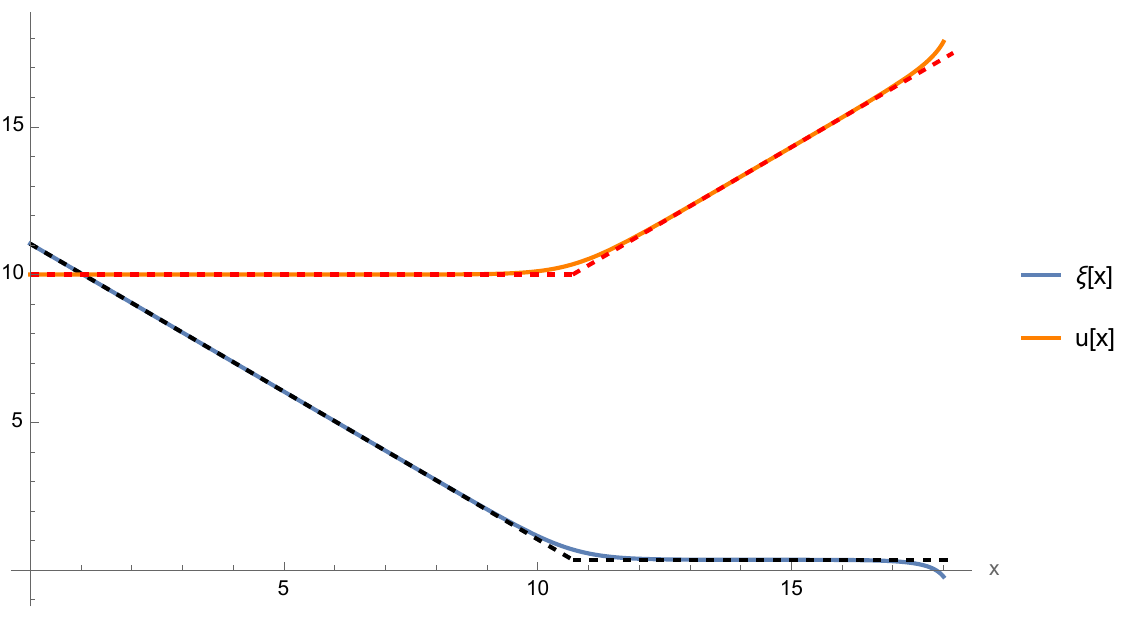}
\qquad
\caption{Plot of $\xi(x)$ \eqref{xixscaling} and $u(x)$ \eqref{txscaling} in \eqref{XiEWCS} when $u_p=10$, $\ell_p=7.5$, $x_c=17.5$ and $\xi_p=\log\sqrt{2}$ which is negligible in the scaling limit. The exact \eqref{xixexact} and scaling approximation \eqref{XiEWCS} of $\xi(x)$ are plotted in blue and dashed black, respectively;  the exact \eqref{txexact} and scaling approximation \eqref{XiEWCS} of $u(x)$ are plotted in orange and dashed red, respectively. The scaling approximations \eqref{xixscaling} and \eqref{txscaling} match \eqref{XiEWCS}. \label{fig:ApproxLagNullGeod}}
\end{figure}

In the exact solution in Appendix~\ref{SatEWCS}, we have $\xi_p=\frac{1}{2}\log2$ \eqref{zsqrt2} which is negligible, and $\ell_p=\frac{\ell}{2}$ \eqref{SatEWCSLen}, $u_p=T-\frac{1}{2}(\ell+\frac{D}{2})$ \eqref{BProjCoord}. Adapting to the coordinate system in Fig.~\ref{fig:ABProj} via a shift $x_c=\ell-\frac{D}{4}$, one finds that $\eqref{txscaling}$ becomes 
\begin{align}
    u(x)=x+T-\ell-\frac{D}{2}.
\end{align}
which matches the exact solution \eqref{x(t)SatEWCS} in the scaling limit. Notice that the effective theory in this section is technically much simpler than the exact solution in subsection~\ref{SatEWCS}. A plot of \eqref{xixexact}-\eqref{txscaling} with shift $x_c=\ell-\frac{D}{4}$ can be found in Fig.~\ref{fig:ApproxLagNullGeod}.

\section{Exact solutions of entanglement entropy in BTZ black hole} \label{app:AdStoBTZ}
In the main text as well as Appendix~\ref{ApproxLagD}, we have solved the equation of motions for various geodesics in planar BTZ black hole $\emph{approximately}$ in the scaling limit. These geodesic are in fact solvable $\emph{exactly}$, by virtue of the map between planar BTZ black hole and Poincare AdS$_3$~\cite{Hartman:2013qma}. We will see that all approximate solutions found earlier match with exact results detailed below.  

The planar BTZ black brane metric is given by
\begin{align}
    ds^2=\frac{1}{z^2}\Big(-f(z)dt^2+\frac{dz^2}{f(z)}+dx^2\Big)\label{BTZmetric},
\end{align}
where $f(z)=1-z^2$ for planar BTZ, and the BTZ horizon is at $z=1$. Unlike in the main text or Appendix~\ref{ApproxLagD}, for technical reasons we will work in Schwarzschild time $t$ in this appendix. The subtlety one needs to bear in mind is that in the interior, $t$ becomes complex~\cite{Fidkowski:2003nf}. We will use $\Tilde{t}=t+\frac{i\pi}{2}$ to denote the real part of the Schwarzschild time in the interior. As we will see in subsection~\ref{SchVSInfallingSec}, the differences between Schwarzschild time $t$ and infalling time $u$ turn out to be immaterial in the region $z\gg1$ that we are mostly interested in. 

\subsection{Difference between Schwarzschild and infalling time}\label{SchVSInfallingSec}
The infalling time $u$ is given by
\begin{align}
    du=dt-\frac{dz}{f(z)}\label{SchInf},
\end{align}
where $f(z)=1-z^2$. Outside the horizon $z<1$, one can simply integrate \eqref{SchInf} to obtain \eqref{u(z)2dExpansion}. In the interior $z>1$, notice that \eqref{SchInf} contains a simple pole at the horizon $z=1$. The resulting residue cancels the imaginary part of the interior Schwarzschild time $t$, and we have
\begin{align}
    u=\Tilde{t}-{\rm arccoth}z\approx\Tilde{t}-\frac{1}{z}\to\Tilde{t}\label{SchApproxInf}
\end{align}
for large $z$. Therefore, in the deep interior $z\gg1$, adopting Schwarzschild time $u$ or infalling time $t$ yields asymptotically the same result. 

\subsection{Map between BTZ and Poincare AdS$_3$}\label{BTZAdSMap}
\begin{figure}[htbp]
\centering
\includegraphics[width=.45\textwidth]{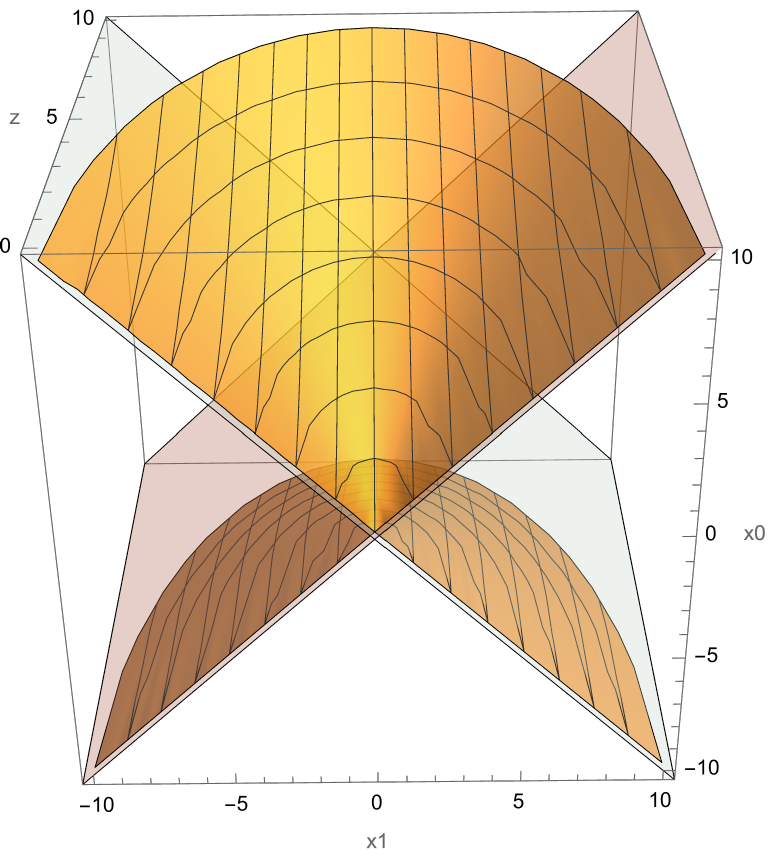}
\qquad
\caption{Map from BTZ black hole to Poincare AdS$_3$. The two boundaries on each side of the BTZ black hole becomes the two Rindler wedges in Poincare AdS$_3$. The BTZ black hole horizon is mapped to $|x_0|=|x_1|$, which are plotted in light blue. The exterior and interior regions are mapped to $x_0^2-x_1^2\leq 0$ and $0\leq x_0^2-x_1^2\leq Z^2$, respectively. In the interior, the hyperboloid $Z^2=x_0^2-x_1^2$ corresponds to $z\to\infty$ in BTZ. 
\label{fig:BTZAdS3Map}}
\end{figure}

BTZ black hole is a quotient of AdS$_3$~\cite{Banados:1992wn,Banados:1992gq}. We can therefore unwrap the BTZ angle and obtain Poincare AdS$_3$ as in~\cite{Hartman:2013qma}. In AdS$_3$ there is only one boundary, and the two boundaries on each side of the BTZ black hole become the two Rindler wedges in Poincare AdS$_3$ (See Fig.~\ref{fig:BTZAdS3Map})~\cite{Hartman:2013qma}. Notice that these two Rindler wedges do not cover the full Poincare AdS$_3$ boundary; the complementary regions are known as the two Milne patches. We would like to obtain from \eqref{BTZmetric} the Poincare AdS$_3$ metric 
\begin{align}
    ds^2=\frac{1}{Z^2}(-dx_0^2+dx_1^2+dZ^2)\label{AdSMetric2}.
\end{align}
It is helpful to discuss the exterior and interior regions of the BTZ black brane metric \eqref{BTZmetric}. To this end, we follow~\cite{Hartman:2013qma} and massage the BTZ radial coordinate $z$ via a coordinate transformation 
\begin{align}
    z={\rm sech}\rho\label{zRho},
\end{align}
where $\rho>0$ so that $z\in[0,1]$. The $\emph{exterior}$ part of the BTZ metric \eqref{BTZmetric} thus becomes
\begin{align}
    ds^2=- \sinh ^2\rho dt^2+d\rho ^2+\cosh ^2\rho dx^2\label{ExteriorMetric}.
\end{align}
The BTZ horizon is now at $\rho=0$. The coordinate time $t$ in \eqref{ExteriorMetric} diverges at the horizon $\rho=0$. The BTZ interior coordinate can be obtained via an analytic continuation on $\rho$ and adding an imaginary part in $t$~\cite{Hartman:2013qma}
\begin{align}
    \rho=i\alpha\ \Rightarrow\ z={\rm sec}\alpha\in[1,+\infty), && \Tilde{t}=t+\frac{i\pi}{2}. \label{zAlpha}
\end{align}
The $\emph{interior}$ part of the BTZ metric \eqref{BTZmetric} is then given by 
\begin{align}
    ds^2=\sin ^2\alpha d\Tilde{t}^2-d\alpha ^2+\cos ^2\alpha dx^2  \label{InteriorMetric}.
\end{align}
The interior time $\Tilde{t}$ in \eqref{InteriorMetric} diverges at the horizon $\alpha=0$. Notice that in the interior, $\alpha$ is timelike, and $\Tilde{t}$ is spacelike. 

Using the embedding coordinate in~\cite{Hartman:2013qma}, the coordinate transformation between BTZ exterior and Poincare AdS$_3$ are found to be
\begin{align}
    Z&=e^x{\rm sech}\rho\label{coordtransfExtZ},\\
    x_0&=e^x{\rm sinh}t\ {\rm tanh}\rho\label{coordtransfExtx0},\\
    x_1&=e^x{\rm cosh}t\ {\rm tanh}\rho\label{coordtransfExtx1}.
\end{align}
For the convenient of the readers, we also list the inverse of \eqref{coordtransfExtZ}-\eqref{coordtransfExtx1}
\begin{align}
    \rho(\lambda)&={\rm arcsinh}\frac{\sqrt{x_1(\lambda)^2-x_0(\lambda)^2}}{Z(\lambda)}\label{coordtransfExtZInv},\\
    t(\lambda)&={\rm arcsinh}\frac{x_0(\lambda)}{\sqrt{x_1(\lambda)^2-x_0(\lambda)^2}}\label{coordtransfExtx0Inv},\\
    x(\lambda)&=\frac{1}{2}\log\big(Z(\lambda)^2-x_0(\lambda)^2+x_1(\lambda)^2\big)\label{coordtransfExtx1Inv}.
\end{align}
The coordinate transformation between BTZ interior and Poincare AdS$_3$ are
\begin{align}
    Z&=e^x{\rm sec}\alpha\label{coordtransfIntZ},\\
    x_0&=e^x{\rm cosh}\Tilde{t}\ {\rm tan}\alpha\label{coordtransfIntx0},\\
    x_1&=e^x{\rm sinh}\Tilde{t}\ {\rm tan}\alpha\label{coordtransfIntx1},
\end{align}
the inverses of \eqref{coordtransfIntZ}-\eqref{coordtransfIntx1}
\begin{align}
    \alpha(\lambda)&={\rm arcsin}\frac{\sqrt{x_0(\lambda)^2-x_1(\lambda)^2}}{Z(\lambda)}\label{coordtransfIntZInv},\\
    \Tilde{t}(\lambda)&={\rm arcsinh}\frac{x_1(\lambda)}{\sqrt{x_0(\lambda)^2-x_1(\lambda)^2}}\label{coordtransfIntx0Inv},\\
    x(\lambda)&=\frac{1}{2}{\rm log}\big(Z(\lambda)^2-x_0(\lambda)^2+x_1(\lambda)^2\big)\label{coordtransfIntx1Inv}.
\end{align}
Notice that the expression of $x(\lambda)$ is the same in the exterior and interior. 

Under the above coordinate transformations, the BTZ black hole horizon $\rho=\alpha=0$ is mapped to $|x_0|=|x_1|$; the exterior region is mapped to $x_0^2-x_1^2\leq 0$; the BTZ black hole interior is mapped to the region $0\leq x_0^2-x_1^2\leq Z^2$ enclosed by the two planes $|x_0|=|x_1|$ and the hyperboloid $Z^2=x_0^2-x_1^2$. Notice that $z\to\infty$ in BTZ \eqref{BTZmetric} is mapped to the hyperboloid $Z^2=x_0^2-x_1^2$ in Poincare AdS$_3$ \eqref{AdSMetric2}. See Fig.~\ref{fig:BTZAdS3Map} for a demonstration. 

The above BTZ$\to$AdS     $_3$ mapping allows us to map the problem of interest in BTZ to that in AdS$_3$, where geodesic equations avoid the complexity due to the presence of horizons. In the next subsection, we will find the most general geodesic equations in Poincare AdS$_3$. 

\subsection{Spacelike Geodesics in Poincare AdS$_3$}\label{AdSGeod}
The Poincare AdS$_3$ metric \eqref{AdSMetric2} indicates that there are two conserved quantities, the Energy and Momentum of the geodesic 
\begin{align}
    E=\frac{1}{Z^2}\frac{dx_0}{d\lambda}, && P=\frac{1}{Z^2}\frac{dx_1}{d\lambda}\label{conservedQuantities2},
\end{align}
where $\lambda$ is the affine parameter along the geodesic. For spacelike geodesics, the norm of the tangent vector is 1, so we have
\begin{align}
    -\frac{1}{Z^2}\Big(\frac{dx_0}{d\lambda}\Big)^2+\frac{1}{Z^2}\Big(\frac{dZ}{d\lambda}\Big)^2+\frac{1}{Z^2}\Big(\frac{dx_1}{d\lambda}\Big)^2=1\label{normTan2}.
\end{align}
Plug \eqref{conservedQuantities2} into \eqref{normTan2} then yields the differential equation 
\begin{align}
    \Big(\frac{dZ}{d\lambda}\Big)^2=Z^2\big(1+(E^2-P^2)Z^2\big)\label{ZLambda0}.
\end{align}
We are interested in spacelike geodesics that reach some deepest points $Z_0$, where $\frac{dZ}{d\lambda}=0$. If $E^2-P^2>0$, then $\frac{dZ}{d\lambda}\neq 0$ for $\forall\ Z$; if instead $E^2-P^2<0$, then 
\begin{align}
    Z_0=\frac{1}{\sqrt{P^2-E^2}}, && Z\in [0,Z_0].
\end{align}
And we can solve the equation \eqref{ZLambda0}
\begin{align}
    \lambda(Z)=\pm{\rm arctanh}\sqrt{1-(P^2-E^2)Z^2}    \label{ZLambda2}.
\end{align}
At the AdS$_3$ boundary $Z=1$, $\lambda=\pm\infty$. We choose conventions such that $\lambda=+\infty$ and $\lambda=-\infty$ are at the right and left Rindler wedge boundary points, respectively. At the furthest point $Z_0$, $\lambda=0$. Inverting \eqref{ZLambda2} then gives 
\begin{align}
    Z(\lambda)=\frac{1}{\sqrt{P^2-E^2}}{\rm sech}\lambda\label{zLambda}.
\end{align}
Plug \eqref{zLambda} into \eqref{conservedQuantities2} and integrate, we have
\begin{align}
    x_0(\lambda)&=\frac{E}{P^2-E^2}{\rm tanh}\lambda+C_0\label{x0Lambda},\\
    x_1(\lambda)&=\frac{P}{P^2-E^2}{\rm tanh}\lambda+C_1\label{x1Lambda},
\end{align}
where $C_0$ and $C_1$ are integration constants that can be determined by the geodesic endpoints on the AdS$_3$ boundary. 

The separations in $x_0$ and $x_1$ determine $E$ and $P$ via 
\begin{align}
    \Delta x_0&=2\int_0^\infty\frac{dx_0}{d\lambda}d\lambda=2E\int_0^\infty Z^2 d\lambda=2E\int_0^{Z_0}\frac{ZdZ}{\sqrt{1-(P^2-E^2)Z^2}}=\frac{2E}{P^2-E^2}\label{Deltax0},\\
    \Delta x_1&=2\int_0^\infty\frac{dx_1}{d\lambda}d\lambda=2P\int_0^\infty Z^2 d\lambda=2P\int_0^{Z_0}\frac{ZdZ}{\sqrt{1-(P^2-E^2)Z^2}}=\frac{2P}{P^2-E^2}\label{Deltax1}.
\end{align}
Where in both equations, in the first step the 2 comes from $\lambda\to -\lambda$ symmetry, in the second step we plugged in \eqref{conservedQuantities2}, and in the third step we change the integration variable from $\lambda$ to $Z$ using \eqref{ZLambda0}. From \eqref{Deltax0} and \eqref{Deltax1}, one can then solve for $E$ and $P$ in terms of $\Delta x_0$ and $\Delta x_1$
\begin{align}
    E=\frac{2\Delta x_0}{\Delta x_1^2-\Delta x_0^2}, && P=\frac{2\Delta x_1}{\Delta x_1^2-\Delta x_0^2}\label{EPDelta}.
\end{align}
The two points $P_1$ and $P_2$ where the geodesic is anchored on the boundary therefore determine the value of $E$ and $P$. 

\subsection{Displaced half spaces}\label{Displaced half spaces}
In this subsection, we find the exact solution to the displaced half space geodesic considered in subsection~\ref{ApproxApproxLag} and Appendix~\ref{ApproxLagDispHalf}. 
\subsubsection{In Poincare AdS$_3$}
\begin{figure}[htbp]
\centering
\includegraphics[width=.5\textwidth]{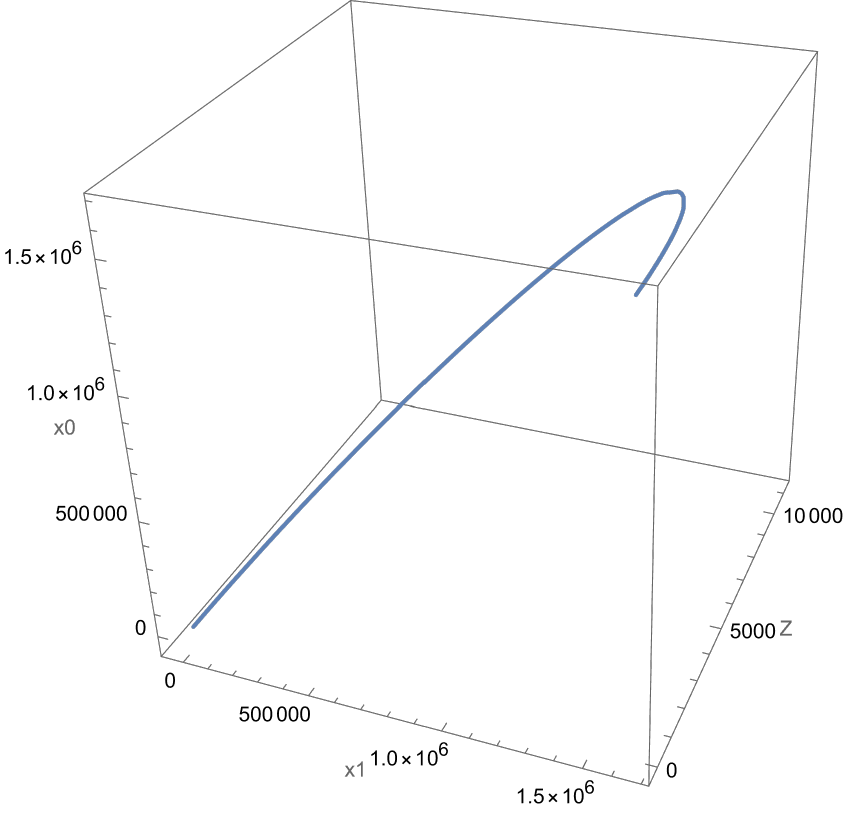}
\qquad
\caption{Spacelike geodesic in Poincare AdS$_3$, with coordinates $(Z,x_0,x_1)$. Here we plot for $T=10$ and $X=5$. \label{fig:PoincareAdS3Geodesic}}
\end{figure}

We have a spacelike geodesic anchoring on the two boundary points in AdS$_3$
\begin{align}
    P_1=(e^X{\rm sinh}T,\ e^X{\rm cosh}T), && P_2=(e^{-X}{\rm sinh}T,\ -e^{-X}{\rm cosh}T)\label{P1P2},
\end{align}
therefore, 
\begin{align}
    \Delta x_0=2\sinh X \sinh T, && \Delta x_1=2\cosh X \cosh T \label{deltax0x1}.
\end{align}
The explicit values of $E$ and $P$ can be determined by plugging \eqref{deltax0x1} into \eqref{EPDelta}. Further substituting these $E$ and $P$ into \eqref{zLambda}, \eqref{x0Lambda} and \eqref{x1Lambda}, fixes $Z(\lambda)$ and leaves $x_0(\lambda)$ and $x_1(\lambda)$ with $C_0$ and $C_1$ unknown. $C_0$ and $C_1$ can be evaluated by noting that the geodesic is anchored on the point $P_1$ on the AdS$_3$ boundary, 
\begin{align}
    x_0(+\infty)={\rm sinh}T, && x_1(+\infty)={\rm cosh}T.
\end{align}
Put everything together, we have the following parametric curves
\begin{align}
    Z(\lambda)&=\sqrt{\frac{\cosh 2T+\cosh 2X}{2}}\ {\rm sech}\lambda\label{Z(lambda)},\\
    x_0(\lambda)&=\sinh T(\sinh X\tanh\lambda+\cosh X)\label{x0(lambda)},\\
    x_1(\lambda)&=\cosh T(\cosh X\tanh\lambda+\sinh X)\label{x1(lambda)},
\end{align}
where $\lambda$ is the affine parameter along the geodesic. At $\lambda=0$, the geodesic reaches its deepest point $Z_0$; at $P_1$ and $P_2$, $\lambda=\pm\infty$, respectively. When the geodesic is at the horizon, $x_0=\pm x_1$, one finds $\lambda_{{\rm hor}}=\pm T-X$. Therefore, the interior range of the geodesic correspond to $\lambda\in(-T-X,T-X)$, and the exterior range $\lambda\in(-\infty,-T-X)\cup (T-X,+\infty)$. A plot of the parametric curve \eqref{Z(lambda)}, \eqref{x0(lambda)}, \eqref{x1(lambda)} in Poincare AdS$_3$ can be found in Fig.~\ref{fig:PoincareAdS3Geodesic}. 

When $X=0$, the above parametric curve \eqref{Z(lambda)}, \eqref{x0(lambda)} and \eqref{x1(lambda)}  becomes 
\begin{align}
    Z(\lambda)=\cosh T\ {\rm sech}\lambda, && x_0(\lambda)=\sinh T, && x_1(\lambda)=\cosh T\ \tanh \lambda \label{semicircleHM0},
\end{align}
which is the semi-circle 
\begin{align}
    Z^2+x_1^2=\cosh^2T, && x_0=\sinh T\label{semicircleHM}
\end{align}
studied in~\cite{Hartman:2013qma}. The furthest point this semi-circle can reach in Poincare AdS$_3$ is $Z_0=\cosh T$. 

\subsubsection{In planar BTZ}
By applying the inverse coordinate transformation \eqref{coordtransfIntZInv}-\eqref{coordtransfIntx1Inv}, one can map the spacelike geodesic \eqref{Z(lambda)}, \eqref{x0(lambda)} and \eqref{x1(lambda)} into the BTZ interior \footnote{\eqref{zBTZLambda} is obtained from 
\begin{align}
    \alpha(\lambda)&=\arcsin\left(\sqrt{\frac{\cosh 2 T-\cosh (2 (\lambda +X))}{\cosh 2 T+\cosh 2 X}}\right)\label{alphaLambda}
\end{align} using \eqref{zAlpha}.}
\begin{align}
    z(\lambda)&=\sqrt{\frac{\cosh 2 T+\cosh 2 X}{\cosh (2 (\lambda +X))+\cosh 2 X}}\label{zBTZLambda},\\
    \Tilde{t}(\lambda)&={\rm arcsinh} \left(\frac{\sqrt{2} \cosh T \sinh (\lambda +X)}{\sqrt{\cosh 2T-\cosh (2 (\lambda +X))}}\right)\label{TildeTLambda},\\
   x(\lambda)&=\frac{1}{2} \log (\text{sech}\lambda \cosh (\lambda +2 X))\label{xLambda}.
\end{align}
Notice that $\lambda$ is still the affine parameter in Poincare AdS$_3$, although the geodesic is mapped to BTZ. $z(\lambda)$ \eqref{zBTZLambda} peaks at $\lambda=-X$,
\begin{align}
    z_0=z(-X)=\sqrt{\frac{\cosh 2 T+\cosh 2 X}{1+\cosh 2 X}}\approx e^{T-X}\label{z0X},
\end{align}
where $\Tilde{t}=0$. Like in the radial $X=0$ case, $z_0$ grows exponentially with $T$ for large $T$. When $X=0$, \eqref{z0X} becomes $z_0={\rm cosh}T$, which is the same as $Z_0$ in AdS$_3$. From \eqref{xLambda} one have $x(+\infty)=X$ and $x(-\infty)=-X$. Notice that $x(\lambda)$ does not depend on $T$. 

The complicated parametric curves \eqref{zBTZLambda}, \eqref{TildeTLambda} and \eqref{xLambda} in the interior regions $-T-X<\lambda<T-X$  simplify greatly in the large $T$, large $X$ ($T>X$) limit. one can show that $\Tilde{t}$ is $\emph{linear}$ in $\lambda$, 
\begin{align}
    \Tilde{t}(\lambda)=\lambda+X\label{TildeTApprox},
\end{align}
and $z(\lambda)$ behaves as 
\begin{equation}
\label{zBTZApprox}
    z(\lambda)=
    \left\{
    \begin{aligned}
        &e^{T-X-\lambda},&&\lambda\in (0,T-X),\\
        &e^{T-X},&&\lambda\in (-2X,0),\\
        &e^{T+X+\lambda},&&\lambda\in (-T-X,-2X),
    \end{aligned}
    \right.
\end{equation}
which is at a $\emph{constant}$ value $z_0=e^{T-X}$ for $\lambda$ between $(-2X,0)$. This value is the $\emph{maximum}$ value of $z(\lambda)$, i.e. the deepest point $z_0$ \eqref{z0X} the geodesic can reach in the BTZ interior. Moreover, $x(\lambda)$ is $\emph{piecewise\ linear}$ in $\lambda$
\begin{equation}
\label{xApprox}
    x(\lambda)=
    \left\{
    \begin{aligned}
        &X,&&\lambda\in (0,T-X),\\
        &\lambda+X,&&\lambda\in (-2X,0),\\
        &-X,&&\lambda\in (-T-X,-2X),
    \end{aligned}
    \right.
\end{equation}
In other words, almost all the displacements in the $x$ direction occurred in the region $\lambda\in[-2X,0]$ linearly with $\lambda$, whereas in the complementary region of $\lambda$ (which includes all the exterior) there are nearly no displacements. 

From \eqref{zBTZApprox} and \eqref{xApprox}, it can be observed that when $\lambda\in(-2X,0)$, $z(\lambda)$ stays at $z_0$ while $x(\lambda)$ varies linearly; in the complementary interior region $\lambda\in(-T-X,-2X)\cup (0,T-X)$, $z(\lambda)$ varies non-trivially while $x(\lambda)$ stays at constant values. Combining with the fact that $\Tilde{t}(\lambda)=\lambda+X$ is linear in $\lambda$ for all interior values $\lambda\in(-T-X,T-X)$, one can eliminate $\lambda$ and write down the expression $z(\Tilde{t})$ and $x(\Tilde{t})$ in the scaling limit. Noticing that for large $z$ there is asymptotically no difference between Schwarzschild and infalling time \eqref{SchApproxInf}, and making the change of variable $z=e^{\xi}$, one arrives at \eqref{FinalXi} in the main text. 

By applying the inverse coordinate transformation \eqref{coordtransfExtZInv}-\eqref{coordtransfExtx1Inv}, one can map the spacelike geodesic \eqref{Z(lambda)}, \eqref{x0(lambda)} and \eqref{x1(lambda)} into the BTZ exterior 
\footnote{\eqref{zBTZLambdaExt} is obtained from 
\begin{align}
    \rho&={\rm arccosh}\left(\sqrt{\frac{\cosh (2 (\lambda +X))+\cosh 2 X}{\cosh 2 T+\cosh 2 X}}\right)
\end{align} using \eqref{zRho}.}
\begin{align}
    z(\lambda)&=\sqrt{\frac{\cosh 2 T+\cosh 2 X}{\cosh (2 (\lambda +X))+\cosh 2 X}}\label{zBTZLambdaExt},\\
    t(\lambda)&={\rm arcsinh} \left(\frac{\sqrt{2} \sinh T \cosh (\lambda +X)}{\sqrt{\cosh (2 (\lambda +X))-\cosh 2 T}}\right)\label{tLambdaExt},\\
    x(\lambda)&=\frac{1}{2} \log (\text{sech}\lambda \cosh (\lambda +2 X))\label{xLambdaExt}.
\end{align}
Notice that the expression of $z(\lambda)$ \eqref{zBTZLambdaExt} and $x(\lambda)$ \eqref{xLambdaExt} are the same as those in the BTZ black hole interior \eqref{zBTZLambda} and \eqref{xLambda}, respectively. It is just that the range of $\lambda$ are different. 

Under the scaling limit, the parametric curves \eqref{zBTZLambdaExt}, \eqref{tLambdaExt} and \eqref{xLambdaExt} in the exterior regions $\lambda\in(-\infty,-T-X)\cup (T-X,+\infty)$ also simplify greatly.  
\begin{equation}
\label{TApproxExt}
    t(\lambda)=
    \left\{
    \begin{aligned}
        &T,&&\lambda\in (T-X,+\infty),\\
        &-T,&&\lambda\in (-\infty,-T-X),
    \end{aligned}
    \right.
\end{equation}
where we have used the fact that for geodesics in the thermal double, i.e. when $\lambda\in(-\infty,-T-X)$, ${\rm Im}\ t=-i\pi$~\cite{Klebanov:1999tb}; we can therefore let $t\to t+i\pi$ in the thermal double, like introducing $\Tilde{t}=t+\frac{i\pi}{2}$ \eqref{zAlpha} in the interior. \eqref{TApproxExt} indicates that in the exterior regions, the HRT surface is on a $\emph{constant}$-$t$ $\emph{slice}$ $t=T$ ($t=-T$ in the thermal double) apart from the near-horizon divergence. For $z(\lambda)$ and $x(\lambda)$, we have 
\begin{equation}
\label{zBTZApproxExt}
    z(\lambda)=
    \left\{
    \begin{aligned}
        &e^{T-X-\lambda},&&\lambda\in (T-X,+\infty),\\
        &e^{T+X+\lambda},&&\lambda\in (-\infty,-T-X),
    \end{aligned}
    \right.
\end{equation}
\begin{equation}
\label{xApproxExt}
    x(\lambda)=
    \left\{
    \begin{aligned}
        &X,&&\lambda\in (T-X,+\infty),\\
        &-X,&&\lambda\in (-\infty,-T-X),
    \end{aligned}
    \right.
\end{equation}
which are the same as $z(\lambda)$ \eqref{zBTZApprox} and $x(\lambda)$ \eqref{xApprox} when $\lambda\in(0,T-X)$ and $\lambda\in(-T-X,-2X)$ in the interior, respectively.

When $T<X$. The expressions of $z(\lambda)$, $\Tilde{t}(\lambda)$, $t(\lambda)$ and $x(\lambda)$  are the same as \eqref{zBTZLambda}, \eqref{TildeTLambda}, \eqref{tLambdaExt}, and \eqref{xLambda}, respectively, but the scaling of $z(\lambda)$ is different. 
\begin{equation}
\label{zBTZApproxTleqX}
    z(\lambda)=
    \left\{
    \begin{aligned}
        &e^{-\lambda},&&\lambda\in (0,\infty),\\
        &1,&&\lambda\in (-2X,0),\\
        &e^{T+X+\lambda},&&\lambda\in (-\infty,-2X),
    \end{aligned}
    \right.
\end{equation}
\eqref{zBTZApproxTleqX} then indicates that $z$ reaches just to the horizon instead of going deeper into the interior. The geodesic moves transversely in $x$ when at the plateau $z=1$. 

\begin{figure}[htbp]
\centering
\includegraphics[width=.4\textwidth]{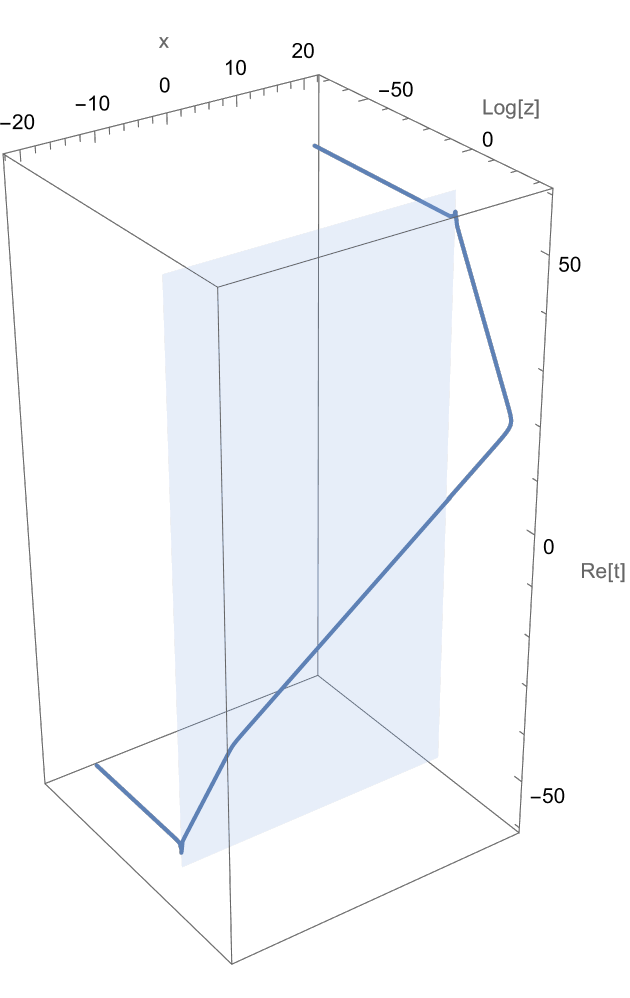}
\includegraphics[width=.42\textwidth]{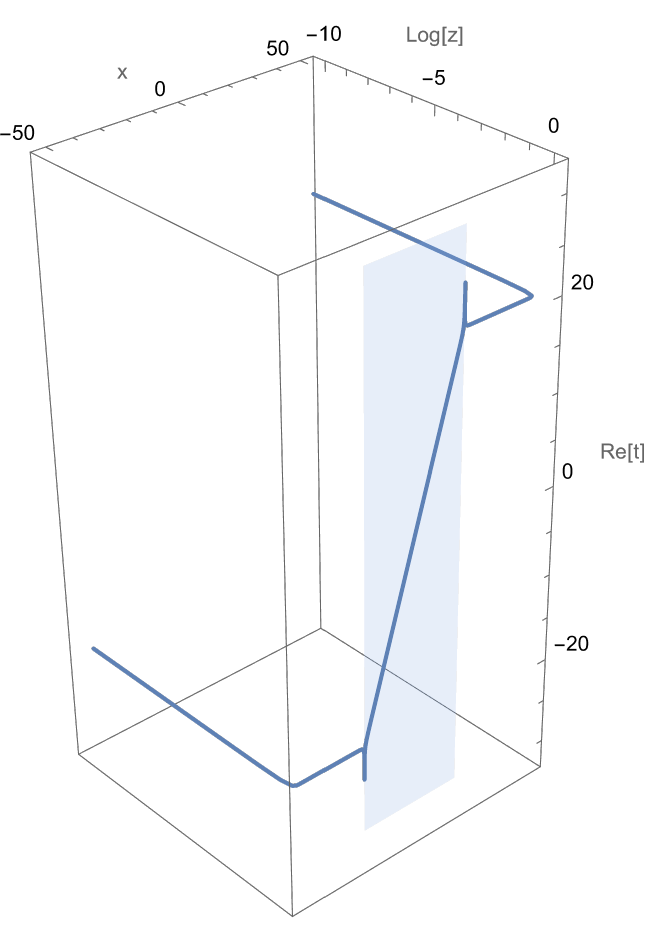}
\qquad
\caption{Spacelike geodesic in BTZ in the coordinates $(\log z,\Tilde{t},x)$. Here we plot for $T=50$ and $X=20$ in the left figure and $T=20$ and $X=50$ in the right figure. The light blue surface is the black hole horizon. \label{fig:BTZGeodesic}}
\end{figure}

\begin{figure}[htbp]
\centering
\includegraphics[width=.45\textwidth]{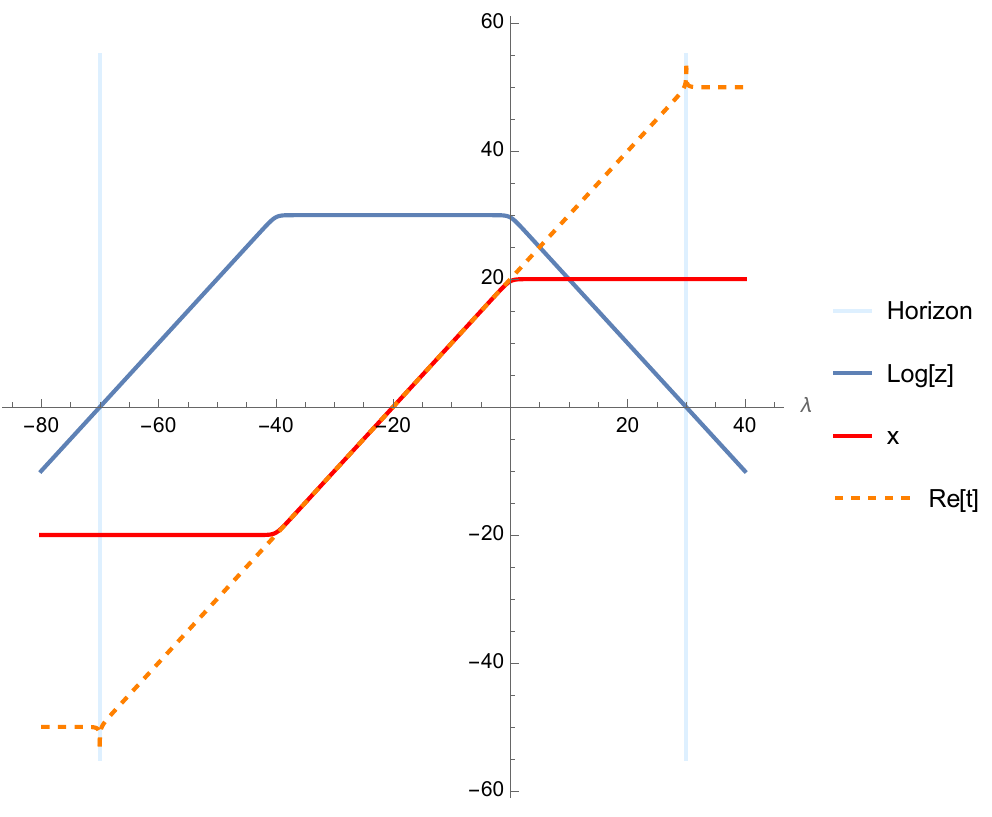}
\includegraphics[width=.45\textwidth]{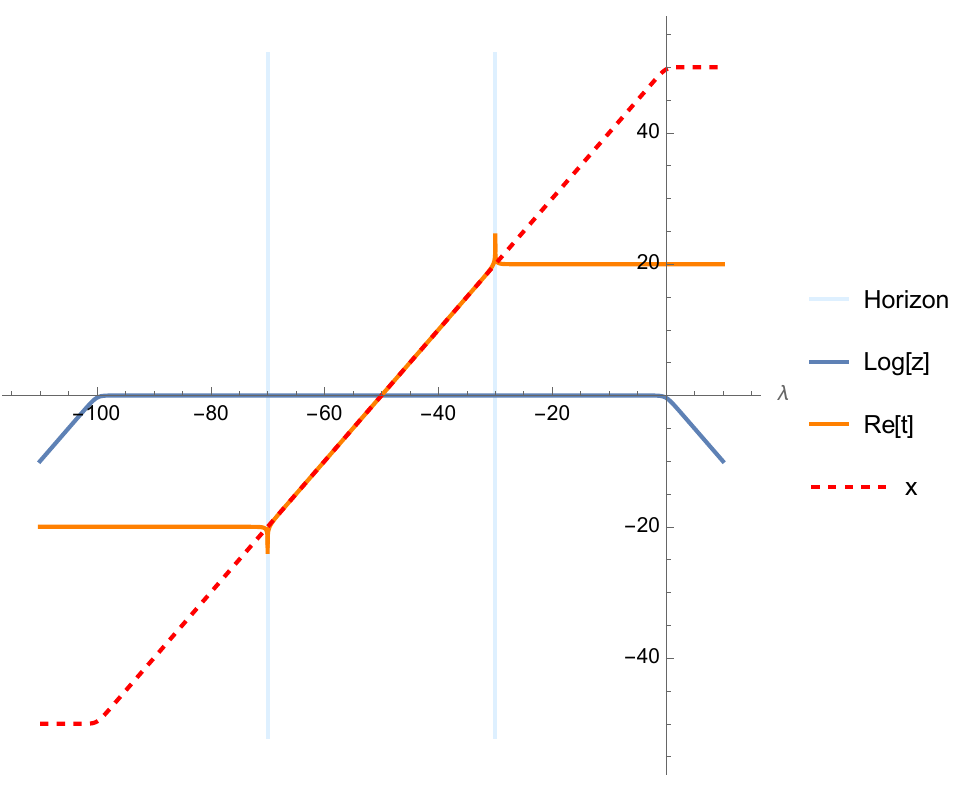}
\qquad
\caption{$\emph{Left}$: Plot of $\log z(\lambda)$ (blue), $\Tilde{t}(\lambda)$ or $t(\lambda)$ (dashed orange) and $x(\lambda)$ (red) when $T=50$ and $X=20$. $\emph{Right}$: Plot of $\log z(\lambda)$ (blue), $\Tilde{t}(\lambda)$ or $t(\lambda)$ (orange) and $x(\lambda)$ (dashed red) when $T=20$ and $X=50$. The light blue lines are the black hole horizons. \label{fig:ztx}}
\end{figure}

Plots of the parametric curves \eqref{zBTZLambda}-\eqref{xLambda}, \eqref{zBTZLambdaExt}-\eqref{xLambdaExt} and \eqref{zBTZApproxTleqX} can be found in Fig.~\ref{fig:BTZGeodesic} and~\ref{fig:ztx}. 

\subsection{Static geodesic}\label{Static geodesic}
To study the static geodesic only, it suffices to work on the $t=0$ slice in BTZ; we will, however, work on the $t=T$ slice for the need of subsection~\ref{SatEWCS}. 

\subsubsection{In Poincare AdS$_3$}
To find the expression of the spacelike geodesic with endpoints $P_2$ and $P_3$, with 
\begin{align}
    P_2=\Big(-\frac{D}{2},T\Big), && P_3=\Big(\frac{D}{2},T\Big)\label{P2P3},
\end{align}
 as a parametric curve in BTZ, we again map it to Poincare AdS$_3$ using the transformations in Appendix~\ref{BTZAdSMap}. The general expression of the spacelike geodesic in AdS$_3$ is the parametric curve \eqref{zLambda}-\eqref{x1Lambda}, but the boundary conditions are now given instead by $P_2$ and $P_3$ \eqref{P2P3} mapped to Poincare AdS$_3$ 
\begin{align}
    P_2=(e^{-\frac{D}{2}}\sinh T,\ e^{-\frac{D}{2}}\cosh T), && P_3=(e^{\frac{D}{2}}\sinh T,\ e^{\frac{D}{2}}\cosh T)\label{P2P3AdS},
\end{align}
which yields 
\begin{align}
    \Delta x_0=2\sinh \frac{D}{2} \sinh T, && \Delta x_1=2\sinh \frac{D}{2} \cosh T \label{deltax0x1Static}.
\end{align}
Plugging \eqref{deltax0x1Static} into \eqref{EPDelta}, and using the boundary condition \eqref{P2P3AdS}, one finds 
\begin{align}
    Z(\lambda)&=\sinh \frac{D}{2}\ {\rm sech}\lambda\label{Z(lambda)Static},\\
    x_0(\lambda)&=\sinh T\Big(\sinh \frac{D}{2}\tanh\lambda+\cosh \frac{D}{2}\Big)\label{x0(lambda)Static},\\
    x_1(\lambda)&=\cosh T\Big(\sinh \frac{D}{2}\tanh\lambda+\cosh \frac{D}{2}\Big)\label{x1(lambda)Static}.
\end{align}
Notice that $x_0(\lambda)< x_1(\lambda)$ due to $\sinh T<\cosh T$, indicating that the geodesic is always in the exterior region. 

\subsubsection{In planar BTZ}
Mapping the geodesic \eqref{Z(lambda)Static}-\eqref{x1(lambda)Static} to BTZ using \eqref{coordtransfExtZInv}-\eqref{coordtransfExtx1Inv}, one finds 
\begin{align}
    z(\lambda)&=\frac{\sinh \frac{D}{2}}{\sqrt{\cosh \lambda \cosh (D+\lambda )}}\label{zLambdaStatic},\\
    x(\lambda)&=\frac{1}{2} \log (\text{sech}\lambda  \cosh (D+\lambda ))\label{xLambdaStatic},
\end{align}
and $t=T$. In \eqref{zLambdaStatic} and \eqref{xLambdaStatic} $\lambda\in(-\infty,+\infty)$. $z(\lambda)$ \eqref{zLambdaStatic} peaks at $\lambda=-\frac{D}{2}$: $z(-\frac{D}{2})=\tanh \frac{D}{2}$. As $D$ gets large, $z(-\frac{D}{2})\to 1$. Notice that $x(\lambda)$ takes the same form as \eqref{xLambda} and \eqref{xLambdaExt} with $2 X\to D$. A plot of the full expression \eqref{zLambdaStatic}-\eqref{xLambdaStatic} can be found in Fig.~\ref{fig:StaticGeod}. 

\begin{figure}[htbp]
\centering
\includegraphics[width=.48\textwidth]{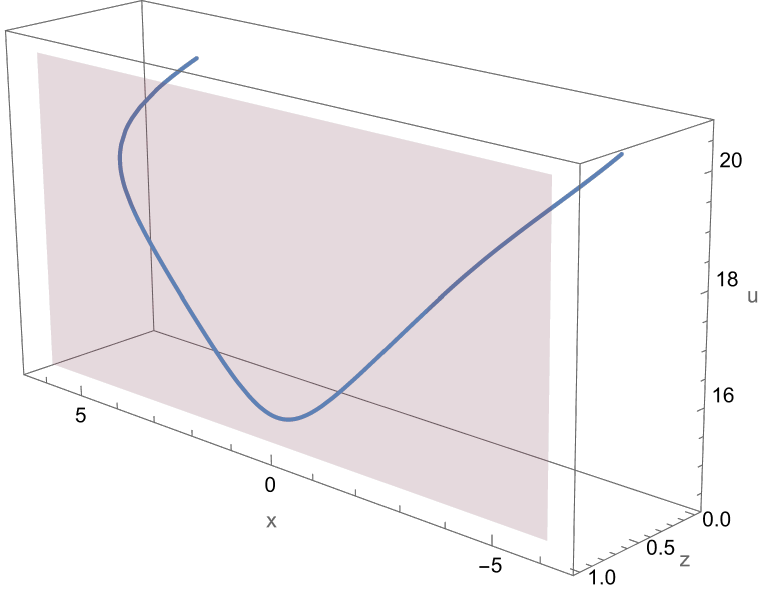}
\qquad
\caption{A 3d plot of the static geodesic given by \eqref{zLambdaStatic}-\eqref{xLambdaStatic} and shown in Fig.~\ref{fig:StaticGeod} in the $(u,z,x)$ coordinate, , for $T=20$ and $D=10$. The light blue surface denotes the black hole horizon $z=1$. The plateau part of the geodesic lies close to the $z=1$ plane.  \label{fig:RTInfalling}}
\end{figure}

As is explained in the main text, there is a finite difference between Schwarzschild and infalling time in the plateau regions due to \eqref{u(z)2dExpansion}. For example, at the deepest point the RT surfaces reach, we have $u(-\frac{D}{2})=T-\frac{D}{2}$. The plot of the static RT surface \eqref{zLambdaStatic}-\eqref{xLambdaStatic} in infalling coordinate can be found in Fig.~\ref{fig:RTInfalling}.

In the scaling limit, 
\begin{equation}
\label{zBTZStaticApprox}
    z(\lambda)=
    \left\{
    \begin{aligned}
        &e^{-\lambda},&&\lambda\in (0,+\infty),\\
        &1-\frac{1}{2}(e^{2\lambda}+e^{-2(\lambda+D)})\approx 1,&&\lambda\in (-D,0),\\
        &e^{\lambda+D},&&\lambda\in (-\infty,-D).
    \end{aligned}
    \right.
\end{equation}

One can then obtain \eqref{vBcone2dApprox} by plugging \eqref{zBTZStaticApprox} into \eqref{u(z)2dExpansion}, and using \eqref{xApprox} with $2 X\to D$ to eliminate $\lambda$. 

According to the BTZ-AdS$_3$ map \eqref{coordtransfExtZ}-\eqref{coordtransfExtx1}, a shift of $\ell+\frac{D}{2}$ in $x$ corresponds to a boost of $e^{\ell+\frac{D}{2}}$ to \eqref{Z(lambda)Static}-\eqref{x1(lambda)Static}, which are \eqref{AcoordZ}-\eqref{Acoordx1} in subsection~\ref{SatEWCS} of the main text. 

\subsection{Entanglement entropy of displaced half space}
In this appendix, we calculate the entanglement entropy computed by the displaced half space geodesic solution found in Appendix~\ref{Displaced half spaces}, using both holographic and CFT approaches. We will see that results from these two methods match each other. 
\subsubsection{Gravity calculation}\label{EEGravCalc}
The entanglement entropy is computed by the regulated length $A_{12}$ of the spacelike geodesic connecting $P_1$ and $P_2$. 
\begin{align}
    A_{12}=\int_{\lambda_2}^{\lambda_1} d\lambda = \lambda_1- \lambda_2,
\end{align}
where $\lambda_i$ ($i=1,2$) are the value of $\lambda$ corresponding to the near boundary cutoffs, $Z(\lambda_1)$ and $Z(\lambda_2)$. $\lambda_1\to +\infty$ and $\lambda_2\to -\infty$. In BTZ, the cutoff is at some large radius $\rho=\rho_{{\rm max}}$. Following~\cite{Hartman:2013qma} we introduce the notion $\epsilon=e^{-\rho_{{\rm max}}}$.  When mapped to Poincare AdS$_3$ using \eqref{coordtransfExtZ}, the near-boundary cutoffs $\rho=\rho_{{\rm max}}$ on the two sides are mapped to 
\begin{align}
    Z(\lambda_1)=e^{x(\lambda_1)}\ {\rm sech}{\rho_{{\rm max}}}= 2\epsilon e^{X}, && Z(\lambda_2)=e^{x(\lambda_2)}\ {\rm sech}{\rho_{{\rm max}}}=2\epsilon e^{-X},
\end{align}
where we have used $x(+\infty)=X$ and $x(-\infty)=-X$ from \eqref{xLambdaExt}. Therefore, from \eqref{Z(lambda)}, one finds 
\begin{align}
    \lambda_1&=-X+\log \Big(\frac{1}{\sqrt{2}\epsilon}\sqrt{{\rm cosh}2T+{\rm cosh}2X}\Big),\\
    -\lambda_2&= X+\log \Big(\frac{1}{\sqrt{2}\epsilon}\sqrt{{\rm cosh}2T+{\rm cosh}2X}\Big),
\end{align}
and the regulated geodesic length $A_{12}$ is given by 
\begin{align}
    A_{12}=\lambda_1-\lambda_2={\rm log}\Big(\frac{1}{2\epsilon^2}({\rm cosh}2T+{\rm cosh}2X)\Big)\label{A12}.
\end{align}
The entanglement entropy is thus 
\begin{align}
    S(T)=\frac{1}{4G_N}A_{12}=\frac{c}{6}\ {\rm log}\Big(\frac{1}{2\epsilon^2}({\rm cosh}2T+{\rm cosh}2X)\Big)\label{HEE}.
\end{align}
where we have used the Brown-Henneaux central charge~\cite{Brown:1986nw}. Notice that when $X=0$, $S(T)=\frac{c}{3}\ {\rm log}\big(\frac{{\rm cosh}T}{\epsilon}\big)$, which matches the result in~\cite{Hartman:2013qma}. In the scaling limit, 
\begin{equation}
\label{TApproxExt2}
    S(T)-S_{{\rm div}}=
    \left\{
    \begin{aligned}
        &\frac{c}{3}T,&& T>X,\\
        &\frac{c}{3}X,&& T<X,
    \end{aligned}
    \right.
\end{equation}
where $S_{{\rm div}}=\frac{c}{3}\log \frac{1}{\epsilon}$. \eqref{TApproxExt2} matches the operator entanglement entropy \eqref{OpEntFinal} ($T>X$) and \eqref{OpEntFinal3} ($T<X$) in the main text. For small $T$, 
\begin{align}
    S(T)\approx \frac{c}{3} \log \cosh X+\frac{c}{3 (\cosh 2X+1)}T^2+S_{{\rm div}},
\end{align}
indicating that the entanglement entropy grows quadratically with $T$, with a coefficient that depends on $X$. At $T=0$, one finds
\begin{align}
    S(T)\approx \frac{c}{3}X+S_{{\rm div}}
\end{align}
for large $X$. 

\subsubsection{CFT calculation}
\begin{figure}[htbp]
\centering
\includegraphics[width=.35\textwidth]{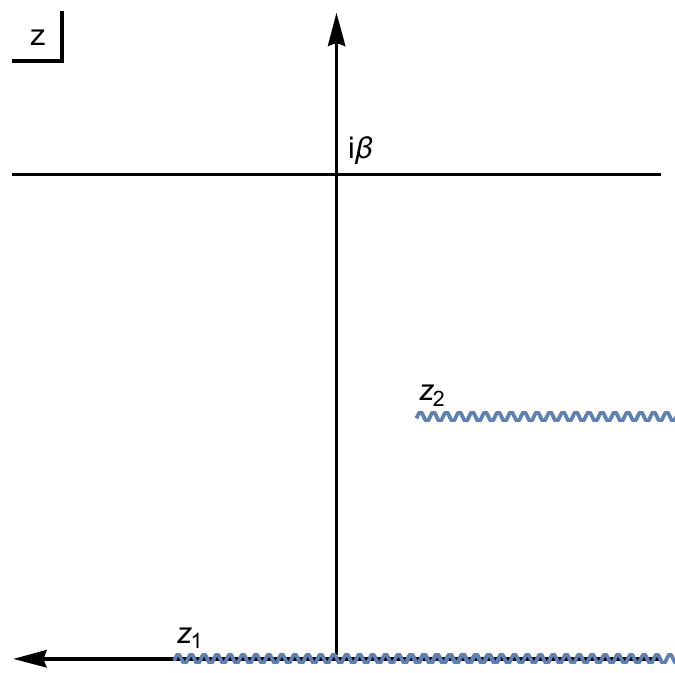}
\qquad
\caption{Insertion of twist operators at $z_1$ and $z_2$ on the Euclidean cylinder. \label{fig:TwistOp}}
\end{figure}

The CFT entanglement entropy is computed by the two-point function of twist operators~\cite{Calabrese:2004eu}
\begin{align}
    \langle \Phi_+(z_1,\overline{z}_1)\Phi_-(z_2,\overline{z}_2)\rangle\label{twopt}
\end{align}
with dimension $\Delta_n=\frac{c}{12}\Big(n-\frac{1}{n}\Big)$. The twist operators are inserted at 
\begin{align}
    z_1=\overline{z}_1=X, && z_2=2T-X+\frac{i\beta}{2}, && \overline{z}_2=-2T-X-\frac{i\beta}{2},
\end{align}
see Fig.~\ref{fig:TwistOp} for illustrations. 
Mapping to the complex plane, $w=e^\frac{2\pi z}{\beta}$, $\overline{w}=e^\frac{2\pi \overline{z}}{\beta}$, one finds the holomorphic and anti-holomorphic part of \eqref{twopt} are given by 
\begin{align}
    \langle \Phi_+(z_1)\Phi_-(z_2)\rangle&=\Big(\frac{\partial z_1}{\partial w_1}\Big)^{-\frac{\Delta_n}{2}}\Big(\frac{\partial z_2}{\partial w_2}\Big)^{-\frac{\Delta_n}{2}}\frac{1}{|w_1-w_2|}\nonumber\\
    &=\Big(\frac{\beta}{4\pi}\cosh\Big(\frac{2\pi}{\beta}(T-X)\Big)\Big)^{-\Delta_n}\label{holomorphic},\\
    \langle \overline{\Phi}_+(\overline{z}_1)\overline{\Phi}_-(\overline{z}_2)\rangle&=\Big(\frac{\partial \overline{z}_1}{\partial \overline{w}_1}\Big)^{-\frac{\Delta_n}{2}}\Big(\frac{\partial \overline{z}_2}{\partial \overline{w}_2}\Big)^{-\frac{\Delta_n}{2}}\frac{1}{|\overline{w}_1-\overline{w}_2|}\nonumber\\
    &=\Big(\frac{\beta}{4\pi}\cosh\Big(\frac{2\pi}{\beta}(T+X)\Big)\Big)^{-\Delta_n}\label{antiholomorphic}.
\end{align}
Taking the product of \eqref{holomorphic} and \eqref{antiholomorphic}, setting $\beta=2\pi$ and introducing the cutoff, one finds 
\begin{align}
    \langle \Phi_+(z_1,\overline{z}_1)\Phi_-(z_2,\overline{z}_2)\rangle=\Big(\frac{1}{2\epsilon^2}\big(\cosh 2T+\cosh 2X\big)\Big)^{-\Delta_n}.
\end{align}
The entanglement entropy is then obtained by $S(t)=-\partial_n I_n|_{n=1}$, which is precisely \eqref{HEE}. 

\section{Exact solutions of EWCS in BTZ black hole}\label{ExactEWCS}
In this Appendix, we find exact solutions to the dynamics of entanglement wedge cross sections in $d=2$, adopting the setup in section~\ref{sec:2dEWCS} in the main text. 
\subsection{Non-equilibrium EWCS}
\begin{figure}[htbp]
\centering
\includegraphics[width=.4\textwidth]{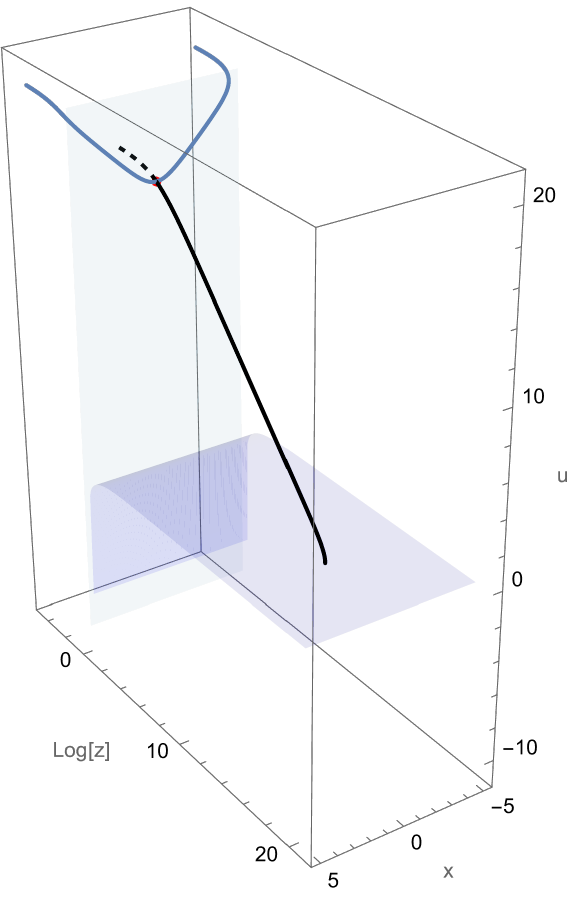}
\qquad
\caption{Plot of non-equilibrium EWCS in $d=2$ with BTZ coordinates $(u,\log(z),x)$, for $T=20$ and $D=10$. The static RT surface (blue) stays close to the horizon $z=1$ (vertical light green surface) before shooting exponentially towards the boundary $z=0$ (didn't plot). The saturated EWCS (black) starts from the static RT surface at the midpoint (red dot) of the latter, and then enters the interior where its length grows linearly according to \eqref{lEWCSNeqApprox}. The non-equilibrium EWCS ends perpendicularly on the EoW brane (light blue surface), whose location is given by $\tilde{t}=0$ or equivalently $u=-{\rm arccoth}z\to 0$ for large $z$. Notice that if one continues the EWCS to the boundary (black dashed line), it anchors on the $T-\frac{D}{2}$ line, instead of the $T$ line that the static RT surface ends. This $-\frac{D}{2}$ difference ensures that the EWCS growth is continuous in time in the scaling limit. \label{fig:NeqEWCS}}
\end{figure}

The time depending EWCS starts from the static RT surface $P_2P_3$ and ends on the end of the world brane. By symmetry, it starts from the mid-point $(t,x)=(T,\tanh\frac{D}{2})$ of $P_2P_3$. As all spacelike geodesics in AdS$_3$ reach the conformal boundary~\cite{Hubeny:2012ry}, we can imagine the EWCS as the part of a spacelike geodesic between $\big(T,\tanh\frac{D}{2}\big)$ and the EoW brane. Let us suppose that this fictitious geodesic starts from the $t=T'$ line on the boundary. As $P_1$ does not play any roles in this part of calculation, we can choose a coordinate such that this fictitious geodesic is at $x=0$. The exterior part of this fictitious geodesic is then given by \eqref{zBTZLambdaExt} and \eqref{xLambdaExt} with $X=0$
\begin{align}
    z(\lambda)&=\cosh T' \text{sech}\lambda,\\
    \sinh t(\lambda)&=\frac{\sinh T'}{\sqrt{1-\cosh ^2T' \text{sech}^2\lambda  }}.
\end{align}
As the fictitious geodesic passes the mid-point/furthest point of the static geodesic $\big(T,\tanh\frac{D}{2}\big)$, we have 2 equations
\begin{align}
    \tanh\frac{D}{2}&=\cosh T' \text{sech}\lambda\label{EqEWCS1},\\
    \sinh T&=\frac{\sinh T'}{\sqrt{1-\cosh ^2T' \text{sech}^2\lambda  }}\label{EqEWCS2},
\end{align}
which allows us to solve for $T'$ and $\lambda$ in terms of $T$ and $D$. Plugging \eqref{EqEWCS1} into \eqref{EqEWCS2}, we have 
\begin{align}
    T'&=\text{arcsinh}\Big(\sinh T \text{sech}\frac{D}{2}\Big)\label{tb'exact}\\
    &\approx T-\frac{D}{2}+\log 2\label{tb'},
\end{align}
where in getting \eqref{tb'}, we have taken the large $T$, large $D$ limit. This means that the fictitious geodesic emanates from the $t=T-\frac{D}{2}+\log 2$ line on the conformal boundary. Plugging \eqref{tb'exact} back into \eqref{EqEWCS1}, we can solve for $\lambda$. The EWCS intersects the EoW brane at the point of time reflection symmetry $\Tilde{t}=0$. According to \eqref{TildeTLambda} with $X=0$, this is at $\lambda=0$. Therefore, the length of EWCS is simply given by $\lambda$

\begin{align}
    \ell(\text{EWCS})&=\text{arccosh}\left(\coth \frac{D}{2} \sqrt{\frac{\cosh D+\cosh 2T}{\cosh D+1}}\right)\label{lEWCSNeq},\\
    &\approx T-\frac{D}{2}+\log 2\label{lEWCSNeqApprox},
\end{align}
which grows linearly with $T$ with slope $v_E=1$, see Fig.~\ref{fig:EWCSDyn}. Notice that the EWCS starts to be non-zero at $t={\rm arccosh}\sinh\frac{D}{2}\approx\frac{D}{2}$. Therefore, apart from the small ${\rm arccosh}\sqrt{2}$ factor that is negligible in the scaling limit, the EWCS growth is actually $\emph{continuous}$ in time. This is because the fictitious geodesic starts on the $t=T-\frac{D}{2}$ line \eqref{tb'} instead of $t=T$ line on the boundary. See Fig.~\ref{fig:NeqEWCS} for a plot of the non-equilibrium EWCS. 

As a consistency check, taking the $D\to 0$ limit of \eqref{lEWCSNeq}, one finds 
\begin{align}
    \ell(\text{EWCS})\approx\log\Big(\frac{4}{D}\cosh T\Big),
\end{align}
which reduces to the length of the spacelike geodesic computing the time-depending holographic entanglement entropy~\cite{Hartman:2013qma}. Here, the small $D$ can be interpreted as proportional to the UV cutoff $\epsilon$. 

\subsection{Saturated Entanglement Wedge Cross Section}\label{SatEWCS}
The saturated EWCS starts from the point $A$ on the static geodesic $P_2P_3$ and ends on the point $B$ on the radial geodesic ending on the EoW brane. Again, it is most straightforward to map this calculation to Poincare AdS$_3$. Notice that the static geodesic $P_2P_3$ is just the parametric curve \eqref{zLambdaStatic}-\eqref{xLambdaStatic} with a $\emph{shift}$ of $\ell+\frac{D}{2}$ in the $x$ coordinate \eqref{xLambdaStatic}; in AdS$_3$, this shift of $\ell+\frac{D}{2}$ in $x$ corresponds to a $\emph{boost}$ of $e^{\ell+\frac{D}{2}}$ to $\big(Z(\lambda),x_0(\lambda),x_1(\lambda)\big)$, see \eqref{coordtransfExtZ}-\eqref{coordtransfExtx1}. Therefore, in Poincare AdS$_3$, the coordinates of $A$ on the static geodesic $P_2P_3$ are given by (see Appendix~\ref{Static geodesic})
\begin{align}
    Z(\lambda)&=e^{\ell+\frac{D}{2}} \sinh \frac{D}{2}\  \text{sech}\lambda\label{AcoordZ},\\
    x_0(\lambda)&=e^{\ell+\frac{D}{2}} \sinh T\ \text{sech}\lambda \cosh \Big(\lambda +\frac{D}{2}\Big)\label{Acoordx0},\\
    x_1(\lambda)&=e^{\ell+\frac{D}{2}} \cosh T\ \text{sech}\lambda \cosh \Big(\lambda +\frac{D}{2}\Big)\label{Acoordx1}.
\end{align}
As $B$ is on a radial geodesic starts from the boundary point $P_1$ and ends on the EoW brane, its coordinates are simply given by \eqref{semicircleHM0}
\begin{align}
    Z(\mu)&=\cosh T\ \text{sech}\mu\label{BcoordZ},\\
    x_0(\mu)&=\sinh T\label{Bcoordx0},\\
    x_1(\mu)&=\cosh T\tanh\mu\label{Bcoordx1},
\end{align}
where $\mu$ is the affine parameter along the geodesic emanating from $P_1$ and ends on the EoW brane. 
\subsubsection{Length of saturated EWCS}
\begin{figure}[htbp]
\centering
\includegraphics[width=.6\textwidth]{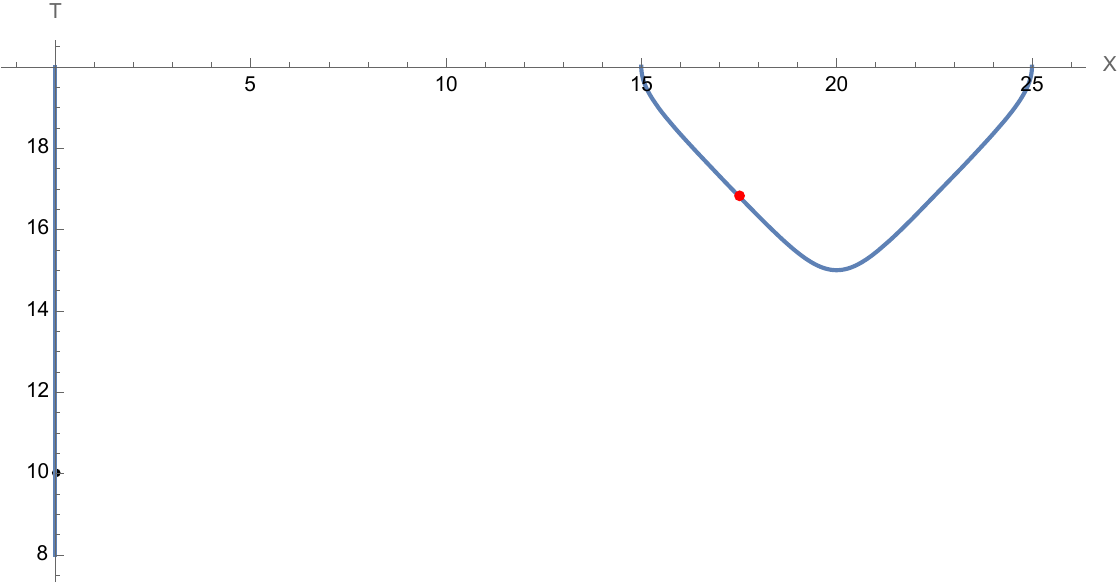}
\qquad
\caption{Locations of $A$ (red) and $B$ (black) upon projections to the boundary along constant infalling time, for $T=20$, $\ell=15$ and $D=10$. \label{fig:ABProj}}
\end{figure}

Now, we would like to calculate the geodesic distance between $A$ and $B$, and find its minimum with respect to the two variables $\lambda\in(-\infty,+\infty)$ and $\mu\in[0,+\infty)$.\footnote{This is because the radial geodesic $P_1B$ ends on the EoW brane, where $\mu=0$} In $2d$, one can find the geodesic distance between two arbitrary points using the geodesic distance formula in Poincare AdS$_3$ below, without knowing the explicit expressions of the geodesic connecting them
\begin{align}
    \cosh d(A,B)=T_1^AT_1^B+T_2^AT_2^B-X_1^AX_1^B-X_2^AX_2^B\label{geodDistAdS3},
\end{align}
where the embedding coordinates are given by 
\begin{align}
    T_1&=\frac{1+(Z^2-x_0^2+x_1^2)}{2 Z},\\
    T_2&=\frac{x_0}{Z},\\
    X_1&=\frac{x_1}{Z},\\
    X_2&=\frac{1-(Z^2-x_0^2+x_1^2)}{2 Z}.
\end{align}
Plugging in the coordinates of $A$ \eqref{AcoordZ}-\eqref{Acoordx1} and $B$ \eqref{BcoordZ}-\eqref{Bcoordx1} into the geodesic distance formula \eqref{geodDistAdS3}, one can find the geodesic distance $d(A,B)$. The explicit expression appears to be lengthy and uninspiring, so we will only present results in the scaling limit, 
\begin{align}
    T\to \Lambda \mathfrak{T}, && \ell\to \Lambda\mathfrak{l}, && d\to\Lambda \mathfrak{D}, && \lambda\to\Lambda s,  && \mu\to\Lambda m\label{scalingSatEWCS},
\end{align}
where $\Lambda$ is taken to be large. The EWCS length in the scaling limit \eqref{scalingSatEWCS} is  
\begin{align}
    d(A,B)\approx \frac{\ell}{2}+\log 2\label{SatEWCSLen},
\end{align}
which takes place at 
\begin{align}
    \lambda_A\approx -\frac{3}{4}D, && \mu_B\approx T-\frac{1}{2}\Big(\ell+\frac{D}{2}\Big)\label{lambdaAmuB},
\end{align}
here $\approx$ means up to exponentially small corrections. Henceforth, the ``equilibrium" EWCS is not truly in equilibrium, instead it asymptotes the extremal value \eqref{SatEWCSLen} exponentially slow. From \eqref{lambdaAmuB}, one finds that 
\begin{align}
    z(\lambda_A)\approx\frac{1}{\sqrt{1+e^{-\frac{D}{2}}}}, && u(\lambda_A)\approx T-\frac{D}{4}-\log 2, && x(\lambda_A)\approx \ell+\frac{D}{4}, && \label{AProjCoord}
\end{align}
indicating that $A$ is approximately at the $\emph{middle}$ point of the $v_B$ cone upon projection; and\footnote{Notice that as $z(\mu_B)\gg1$, the difference between Schwarzschild and infalling time at $B$ is negligible, i.e. $u(\mu_B)\approx\Tilde{t}(\mu_B)$.  One then have $\Tilde{t}(\mu_B)\approx\mu_B$ combining \eqref{AProjCoord} and \eqref{BProjCoord}, consistent with \eqref{TildeTApprox} in the radial, i.e. $X=0$ case. }
\begin{align}
    z(\mu_B)\approx e^{\frac{1}{2}\big(\ell+\frac{D}{2}\big)}\gg1, && u(\mu_B)\approx T-\frac{1}{2}\Big(\ell+\frac{D}{2}\Big), && x(\mu_B)=0 \label{BProjCoord}, 
\end{align}
implying that $B$ reaches deep into the interior. Notice that the location of $z(B)$ does not time evolve with $T$ to leading order. This is made possible by $B$ traveling towards the horizon since the geodesic goes deeper into the interior as time evolves, as can be seen from \eqref{lambdaAmuB}.\footnote{Notice that at where the geodesic meets the EoW brane, $\Tilde{t}=0$ and therefore $\mu=0$ according to \eqref{TildeTApprox} with $X=0$. Therefore, $\mu_B$ measures the distance between $B$ and the EoW brane along the geodesic. } A plot of the location of $A$ and $B$ upon projections can be found in Fig.~\ref{fig:ABProj}. 

\subsubsection{In Poincare AdS$_3$}
Having computed $l(EWCS)$ \eqref{SatEWCSLen} and the coordinates of $A$ and $B$ \eqref{lambdaAmuB}, we would like to now obtain the expression of the EWCS. Again, we imagine the EWCS as the part of a fictitious spacelike geodesic between $A$ and $B$. The general expression of a spacelike geodesic in AdS$_3$ is given as a parametric curve by \eqref{zLambda}-\eqref{x1Lambda}. Imposing the constraints that it passes point $A$ \eqref{AProjCoord} and $B$ \eqref{BProjCoord}, we have
\begin{align}
    Z(A)&=\frac{1}{\sqrt{\mathfrak{P}^2-\mathfrak{E}^2}}{\rm sech}\nu_A,\\
    x_0(A)&=\frac{\mathfrak{E}}{\mathfrak{P}^2-\mathfrak{E}^2}{\rm tanh}\nu_A+\mathfrak{C}_0,\\
    x_1(A)&=\frac{\mathfrak{P}}{\mathfrak{P}^2-\mathfrak{E}^2}{\rm tanh}\nu_A+\mathfrak{C}_1,\\
    Z(B)&=\frac{1}{\sqrt{\mathfrak{P}^2-\mathfrak{E}^2}}{\rm sech}\nu_B,\\
    x_0(B)&=\frac{\mathfrak{E}}{\mathfrak{P}^2-\mathfrak{E}^2}{\rm tanh}\nu_B+\mathfrak{C}_0,\\
    x_1(B)&=\frac{\mathfrak{P}}{\mathfrak{P}^2-\mathfrak{E}^2}{\rm tanh}\nu_B+\mathfrak{C}_1,
\end{align}
where $\nu$ is the affine parameter along the fictitious geodesic. Now, there are 6 variables $\mathfrak{E}$, $\mathfrak{P}$, $\mathfrak{C}_0$, $\mathfrak{C}_1$, $\nu_A$, and $\nu_B$, as well as 6 constraints \eqref{AProjCoord} and \eqref{BProjCoord}. To solve for these variables, we can proceed by first finding $\tanh \nu_A$ and $\tanh \nu_B$ from the $Z$ equations above.\footnote{It turned out that the physical range of parameters corresponds to $\nu_B<\nu_A<0$.} Substituting them into the $x_0$ and $x_1$ equations, and taking their differences $x_0(\nu_A)-x_0(\nu_B)$ and $x_1(\nu_A)-x_1(\nu_B)$, leaving us with two equations of $\mathfrak{E}$ and $\mathfrak{P}$. We can then solve for all 6 variables. The full expressions are too lengthy to be included in the paper, so we will only discuss their scaling limits \eqref{scalingSatEWCS} and show plots of this fictitious geodesic. 

An interesting feature of this fictitious geodesic is that it hits the hyperboloid $Z^2=x_0^2-x_1^2$ corresponding to the BTZ orbifold singularity $z=\infty$ (see Fig.~\ref{fig:BTZAdS3Map}), and then ends on the Milne patch! Of course, only the portion of the geodesic below the hyperboloid is physical when mapped to BTZ. In BTZ, the fictitious geodesic emanates from a point on the conformal boundary, passes $A$ and $B$, and eventually hits the BTZ singularity. 

In the scaling limit \eqref{scalingSatEWCS}, we found
\begin{align}
    \ell(\text{EWCS})=\nu_A-\nu_B\approx \frac{\ell}{2}+\log 2,
\end{align}
where $\nu$ is the affine parameter along the fictitious geodesic. This agrees with \eqref{SatEWCSLen}. 

\subsubsection{In planar BTZ}
\begin{figure}[htbp]
\centering
\includegraphics[width=.55\textwidth]{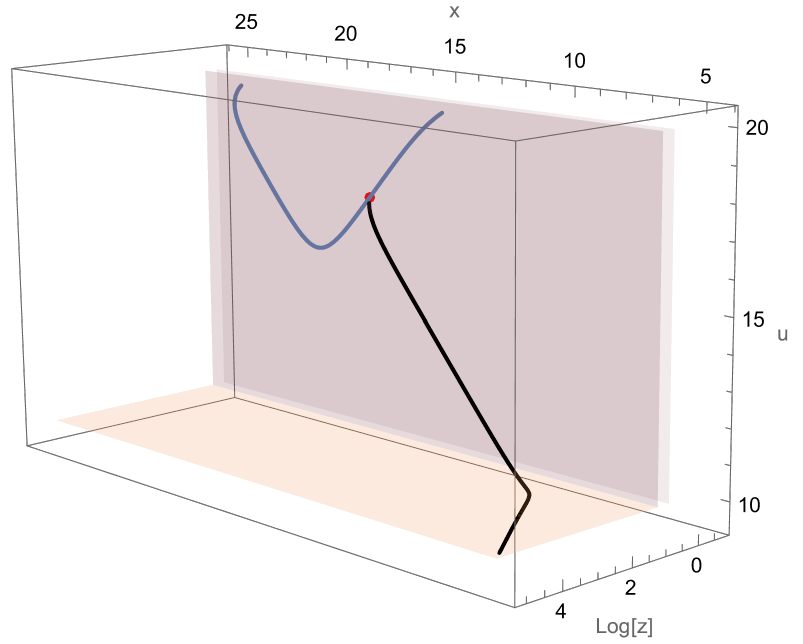}
\qquad
\caption{Plot of saturated EWCS in $d=2$ with BTZ coordinates $(u,\log(z),x)$, for $T=20$, $\ell=15$, and $d=10$. The static RT surface (blue) stays close to $z=1$ (vertical light green surface in the back) before shooting exponentially towards the boundary $z=0$ (didn't plot). The saturated EWCS (black) starts from the static RT surface at point $A$ (red dot), then enters the horizon and stays at $z=\sqrt{2}$ (vertical light blue surface in the front) before becoming almost null and shooting exponentially towards the BTZ singularity $z=\infty$ at constant interior Schwarzschild time $\Tilde{t}=T-\big(\ell+\frac{D}{2}\big)$ (horizontal light red surface). Notice that at large $z$ the difference between Schwarzschild and infalling time is negligible. \label{fig:SatEWCS}}
\end{figure}

When mapped to BTZ, we found that the saturated EWCS shows interesting behaviour as depicted in Fig.~\ref{fig:SatEWCS}: it consists of a plateau in the interior plus an almost null part that shoots exponentially towards the interior on a constant $\Tilde{t}$ slice. 

\begin{figure}[htbp]
\centering
\includegraphics[width=.35\textwidth]{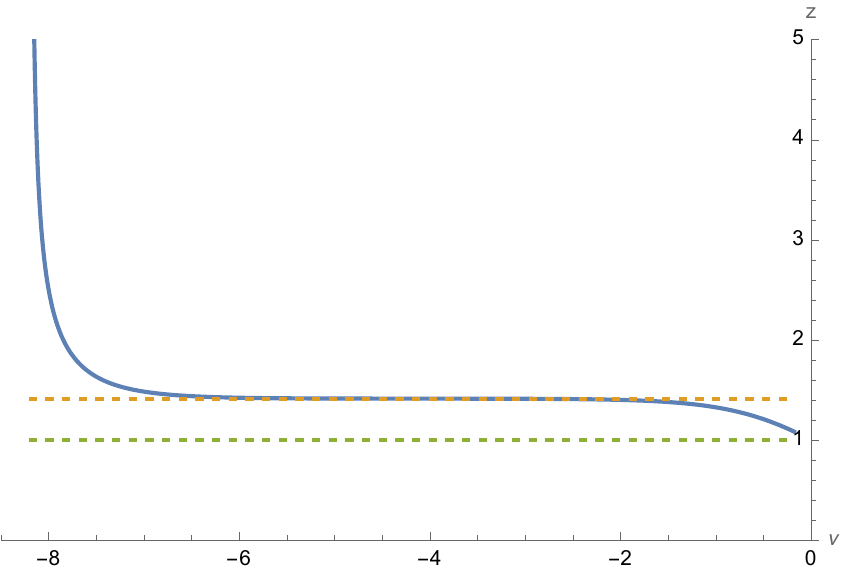}
\includegraphics[width=.28\textwidth]{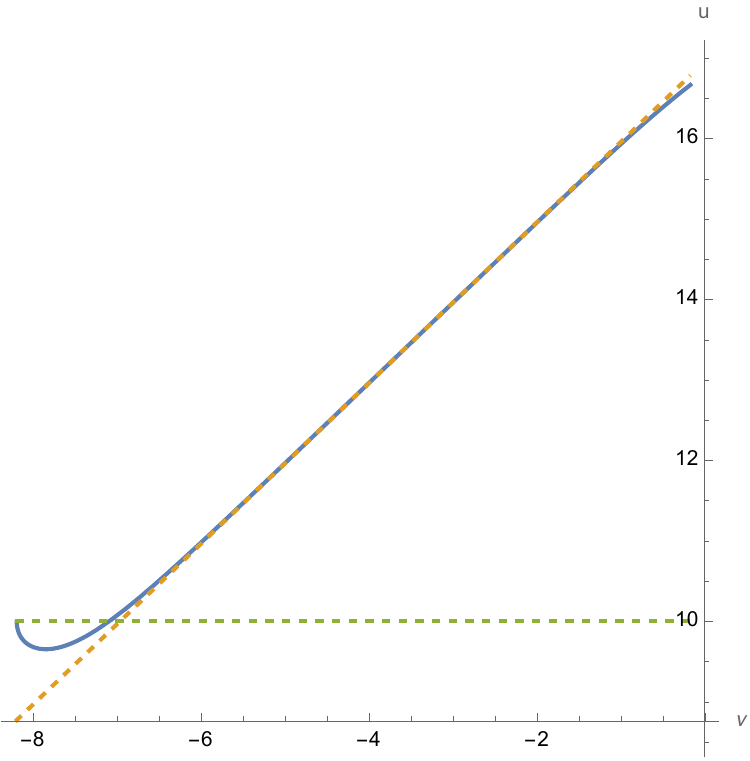}
\includegraphics[width=.28\textwidth]{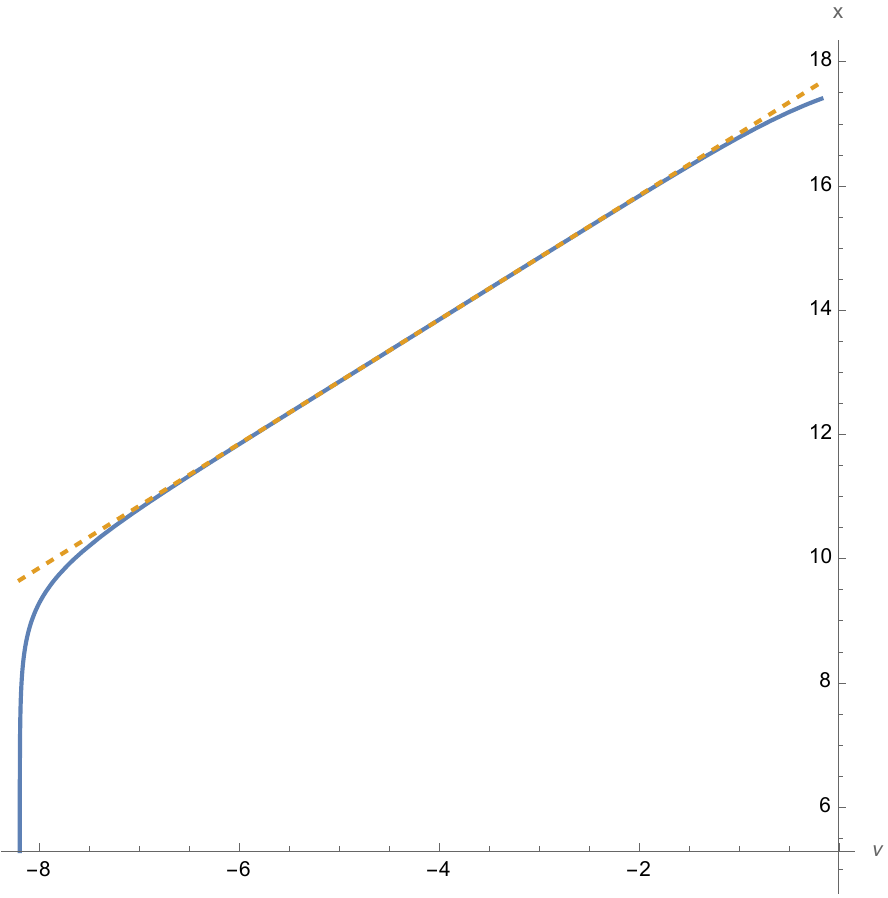}
\qquad
\caption{Plot of $z(\nu)$ (left), $u(\nu)$ (middle) and $x(\nu)$ (right) as well as their scaling limits (dashed), for $T=20$, $\ell=15$, and $D=10$. In the plot of $z(\nu)$, the two dashed lines represent the horizon $z=1$ and $z=\sqrt{2}$, respectively. In the plot of $u(\nu)$, the 45$^\circ$ and horizontal dashed lines stand for \eqref{unu} and the constant interior time slice $\Tilde{t}=T-\big(\ell+\frac{D}{2}\big)$ the null geodesic is on, respectively; the small lump near the horizontal dashed line is due to the difference between Schwarzschild and infalling time \eqref{SchInf} when $z$ is not yet large. In the plot of $x(\nu)$, the dashed line is \eqref{xnu}. \label{fig:zuxNu}}
\end{figure}

The plateau corresponds to $\nu\in(\nu_B,\nu_A]$, excluding a very small region near $\nu\to\nu_B^+$. In this regime we found 
\begin{align}
    z(\nu)&\approx\sqrt{2},\\
    \tilde{t}(\nu)&\approx \nu+T-\frac{D}{4}+\frac{1}{2}\log 2,\\
    x(\nu)&\approx \nu+\ell+\frac{D}{4}+\frac{1}{2}\log 2\label{xnu},
\end{align}
and therefore 
\begin{align}
    u(\nu)\approx\tilde{t}(\nu)-{\rm arccoth}z(\nu)=\tilde{t}(\nu)-{\rm arccoth}\sqrt{2}\label{unu}
\end{align}
according to \eqref{SchApproxInf}, see Fig.~\ref{fig:zuxNu}. Putting them together yields two equations \begin{align}
    z&\approx\sqrt{2}\label{zsqrt2},\\
    u(x)&\approx x+T-\ell-\frac{D}{2}-{\rm arccoth}\sqrt{2}\label{x(t)SatEWCS}
\end{align}
describing saturated EWCS. As the range of $\nu$ for the plateau region is almost the entire $[\nu_B,\nu_A]$, the EWCS in this region has length 
\begin{align}
    \ell({\rm Plateau})\approx \nu_A-\nu_B\approx \frac{\ell}{2}+\log 2\approx \ell(\text{EWCS}).
\end{align}
In other words, nearly $\emph{all}$ the length of EWCS \eqref{SatEWCSLen} comes from this plateau region. 

\begin{figure}[htbp]
\centering
\includegraphics[width=.5\textwidth]{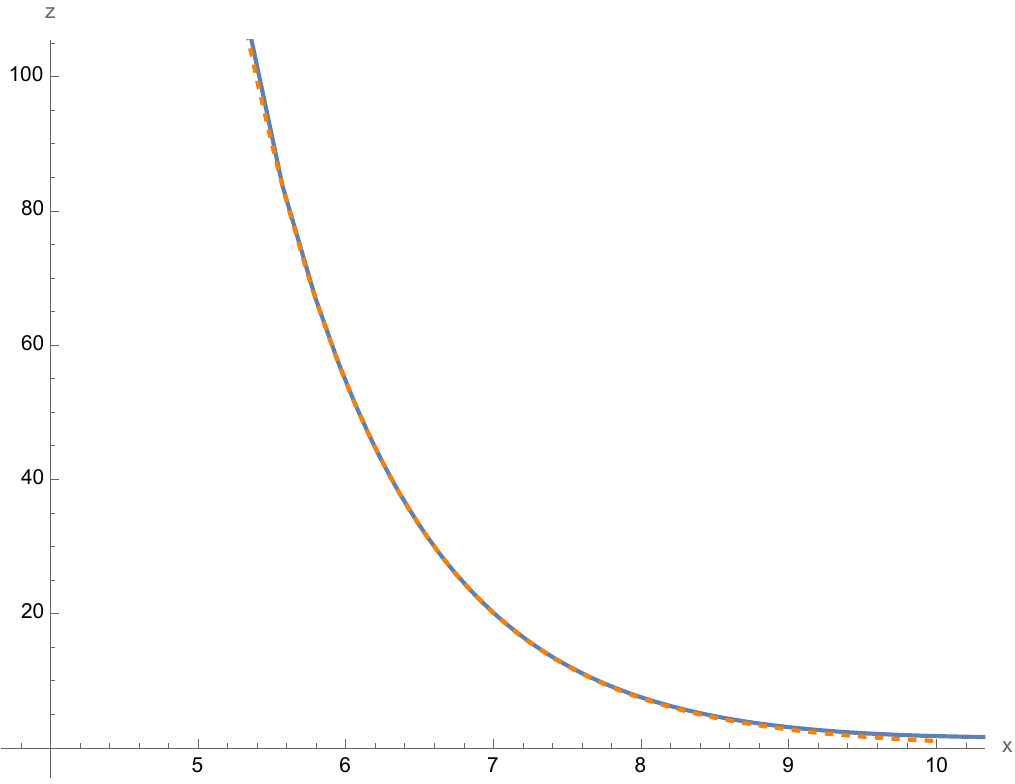}
\qquad
\caption{Exponential relation between $z$ and $|x|$ in the "null" part of EWCS, for $T=20$, $\ell=15$, and $D=10$. The dashed orange line represents $z\propto e^{-x}$. \label{fig:FlyingGeod}}
\end{figure}

The EWCS then transforms smoothly into a almost-$\emph{null}$ part that lies on the $\emph{constant}$ $\Tilde{t}\approx T-\frac{1}{2}\big(\ell+\frac{D}{2}\big)$ $\emph{slice}$ while shooting exponentially towards the BTZ singularity. In fact, the EWCS is a portion of a spacelike geodesic that $\emph{will}$ hit the BTZ singularity $z=\infty$! To understand the relation between $z$ and $x$ for the almost-null geodesic, we demand the metric \eqref{InteriorMetric} to be null on a constant-$\Tilde{t}$ slice, 
\begin{align}
    0=-d\alpha^2+\cos^2 \alpha dx^2 &&  \Rightarrow\ z=\cosh x \propto e^{|x|}\label{zxFlying},
\end{align}
where we have used \eqref{zAlpha}. The $\emph{exponential}$ relation between $z$ and $|x|$ (see Fig.~\ref{fig:FlyingGeod}) is consistent with the coordinates of $A$ and $B$, and the width of the plateau: we have $x(A)\approx \ell+\frac{D}{4}$ according to \eqref{AProjCoord}, and the plateau width is $\Delta x\approx \frac{\ell}{2}$. Therefore, at where the plateau ends and where the exponential growth of $z$ with respect to $x$ starts, we have $x\approx\frac{1}{2}\big(\ell+\frac{D}{2}\big)$. Thus, when the EWCS arrives at $B$, where $x=0$, one expects $z(B)\propto e^{\frac{1}{2}\big(\ell+\frac{D}{2}\big)}$ according to \eqref{zxFlying}. This is indeed the case as in \eqref{BProjCoord}.\footnote{Here we work in the scaling limit \eqref{scalingSatEWCS} and neglect small contributions such as factors of $\log 2$. } 

\begin{figure}[htbp]
\centering
\includegraphics[width=.6\textwidth]{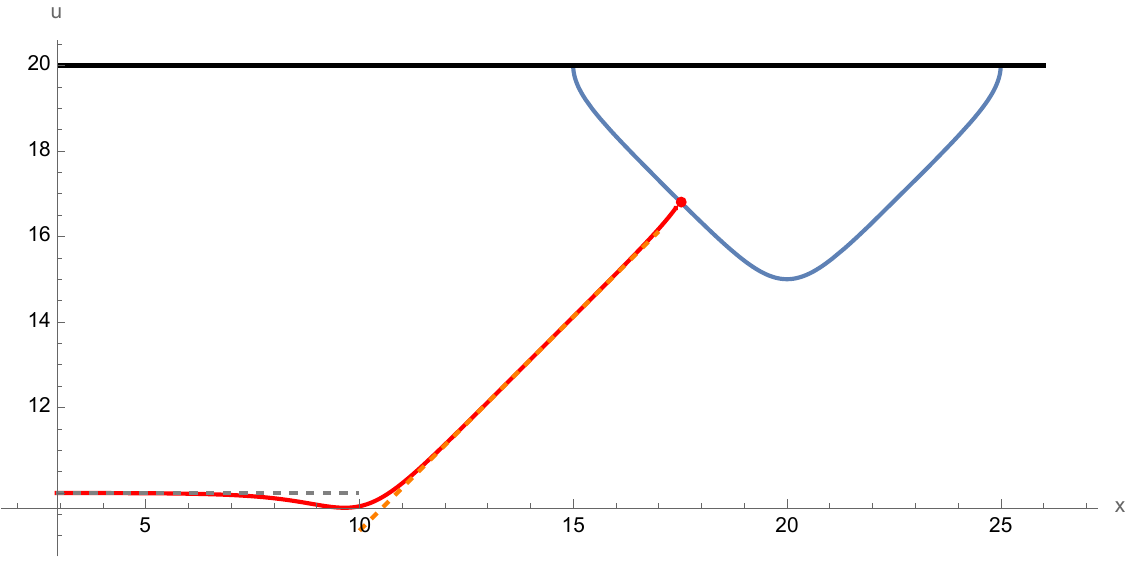}
\qquad
\caption{Projection of saturated EWCS (red) in $d=2$ along constant infalling time, for $T=20$, $\ell=15$, and $D=10$. The blue triangle is the projection of the static RT surface anchoring on $P_2$ and $P_3$ on the boundary. The dashed orange line is the plateau \eqref{x(t)SatEWCS}, which is responsible for almost all the length of the EWCS \eqref{SatEWCSLen}. The horizontal part of the projection is almost null on the 
 $\Tilde{t}=T-\frac{\ell}{2}\big(\ell+\frac{D}{2}\big)$ slice (dashed grey). The lump near the transition point is due to the difference between Schwarschild and infalling time \eqref{SchInf} when $z$ is not yet large.\label{fig:SatEWCSProj}}
\end{figure}

The projection of the saturated EWCS along constant infalling time can be found in Fig.~\ref{fig:SatEWCSProj}. Notice that $u(x)$ agrees with Fig.~\ref{fig:ApproxLagNullGeod} obtained from solving the approximate equations of motions in the scaling limit. 

\section{RT surfaces with spherical regions}
\label{sec:RTsphereBessel}
Recall that the static RT surface on the $t=T$ slice with a spherical entangling subregion has an area functional given by~\eqref{EEThmSphere}; in the large entangling subregion limit, this RT surface contains a plateau and two shooting pieces towards the boundary~\cite{Hubeny:2012ry,Mezei:2016zxg}. As such, we can introduce a near-horizon coordinate $s(r)$~\cite{Mezei:2016zxg}
\begin{align}
   z(r)=1-\epsilon s(r)^2,
\end{align}
where $\epsilon\ll 1$ denotes the closest distance the HRT surface is to the horizon. From \eqref{EEThmSphere}, one can obtain the Euler-Lagrange equation $s(r)$. Expanding for small $\epsilon$, and using $a(1)=0$, one finds the following second-order near-horizon wave equation~\cite{Mezei:2016zxg}
\begin{align}
   s''(r)+\frac{d-2}{r}s'(r)-M^2 s(r)=0, && M^2=-\frac{d-1}{2}a'(1)b(1)^2>0,\label{hyperbolicHelmholtz}
\end{align}
whose solution is given by modified Bessel function~\cite{Liu:2013una}
\begin{align}
   s(r)=C\frac{I_{\frac{d-3}{2}}(M r)}{r^\frac{d-3}{2}}\sim e^{M r}\label{BesselSln}.
\end{align}
At large $r$, the solutions \eqref{BesselSln} grows exponentially. From \eqref{BesselSln}, one can write 
\begin{align}
   z=1-e^{2M(r-r_0)}.\label{spherePlateau}
\end{align}
When $0<r\ll r_0$, $z\to 1^-$, the RT surface is a plateau skimming the horizon; when $r\to r_0^-$, $z$ quickly decays to 0, indicating that the RT surface shoots toward the conformal boundary. From \eqref{spherePlateau} we have
\begin{align}
   r-r_0\sim \frac{\log (1-z)}{b(1)\sqrt{-2(d-1) a'(1)}}\label{xzSphereApprox}.
\end{align}
Combining with \eqref{ExpandedInts} when $z\to 1^-$, we have 
\begin{align}
   r\sim \sqrt{\frac{a'(1)}{2(d-1)}}u
   = v_Bu \label{vBconeSphere},
\end{align}
where we again used~\eqref{vB} for $v_B$. \eqref{vBconeSphere} indicates that like in the strip case~\eqref{ExpandedInts2}, the projection of the RT surface in the large region limit along constant infalling time is a $v_B$ cone. 

We remark that the above calculations work for strip entangling subregions as well: the second-order near-horizon wave equation from~\eqref{EEThmStrip} is simply $(\frac{d^2}{dx^2}-M^2) s(x)^2=0$, with general solution $s(x)=c_1 e^{M x}+c_2 e^{-M x}$. $e^{\pm x}$ can be regarded as the $d>2$ analogue of $e^{\pm \lambda}$ in \eqref{zBTZStaticApprox}. 

\section{Scaling behaviour of EWCS in $d>2$}\label{App:ScalingECS}
In this appendix we compare the direct numerical evaluation of the reflected entropy and the corresponding membrane prediction. As described in the main text, these two results should agree in the scaling limit $D\rightarrow\infty$. However, we are constrained in our capacity to evaluate the EWCS within this limit. This problem arises from the numerical solution of the intersection equations between the EWCS and the HRT surfaces. Even for the relatively small values of $D$ we do have access to, we can see that the membrane theory is a surprisingly accurate description, able to capture all the main features of the reflected entropy.

To further test the validity of the membrane prediction, we can take a closer look at some of these key features. First, we can consider the time at which the reflected entropy begins to grow linearly and the size of the gap between the linear growth and the initial vanishing value.
\begin{figure}[h!]
\centering
\includegraphics[scale=0.35]{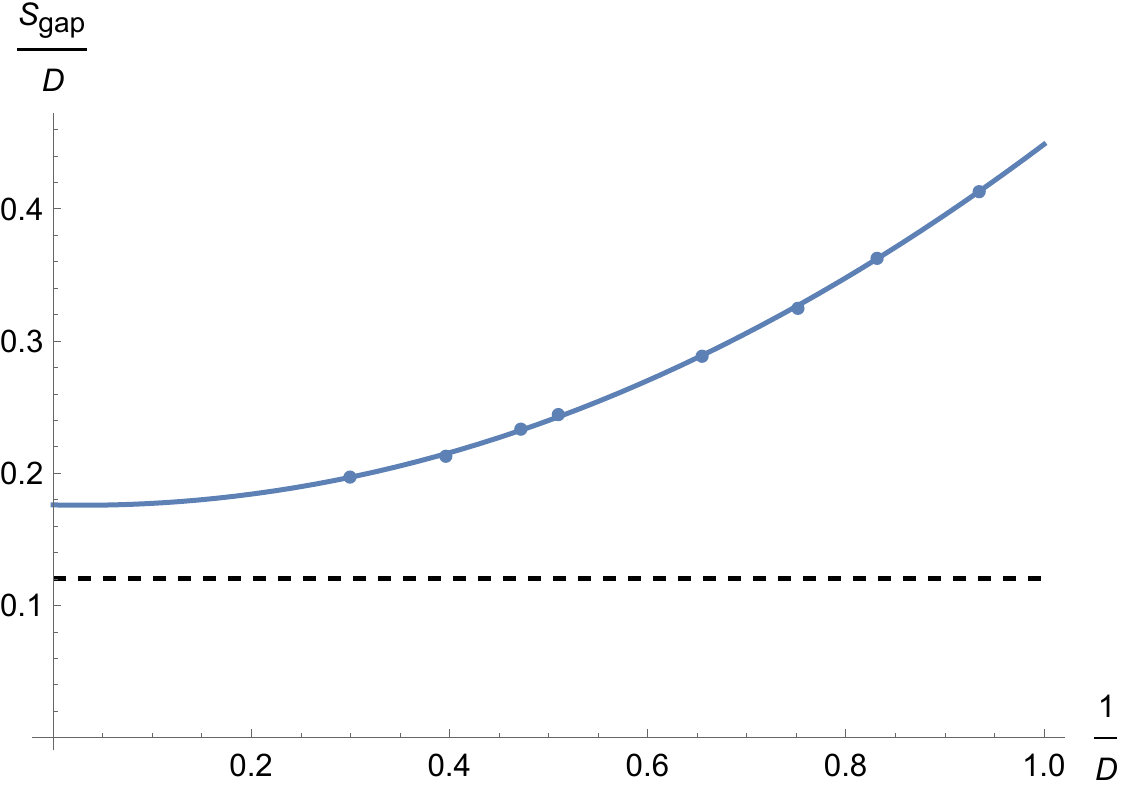}
\includegraphics[scale=0.35]{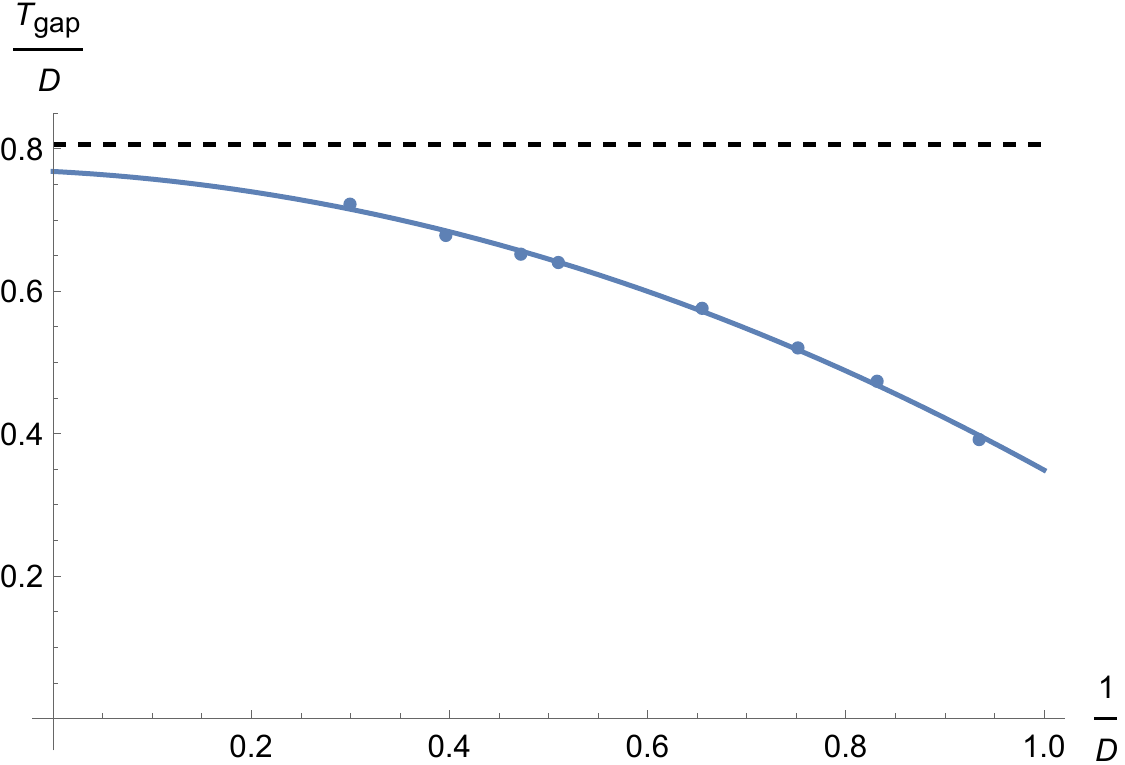}
\caption{Comparisson of the time of transition between vanishing EWCS and its regime of linear growth, as well as the size of the discontinuity, as functions of $1/D$. The membrane prediction is depicted in black.}
\end{figure}

While the discrepancy is somewhat worrying, we can provide an argument for this being simply a numerical issue. The transition between the vanishing and linearly growing regime is dictated not by the EWCS per se, but by the dynamics of entanglement entropy; namely, whether the entanglement wedge is connected or disconnected. This has been shown to be properly captured by the membrane prediction to a high degree of accuracy.

Another point we can compare is the transition between linear growth and the plateau. Once again we see a favorable trend in our evaluation, although precise numerical agreement remains elusive.
\begin{figure}[h!]
\centering
\includegraphics[scale=0.35]{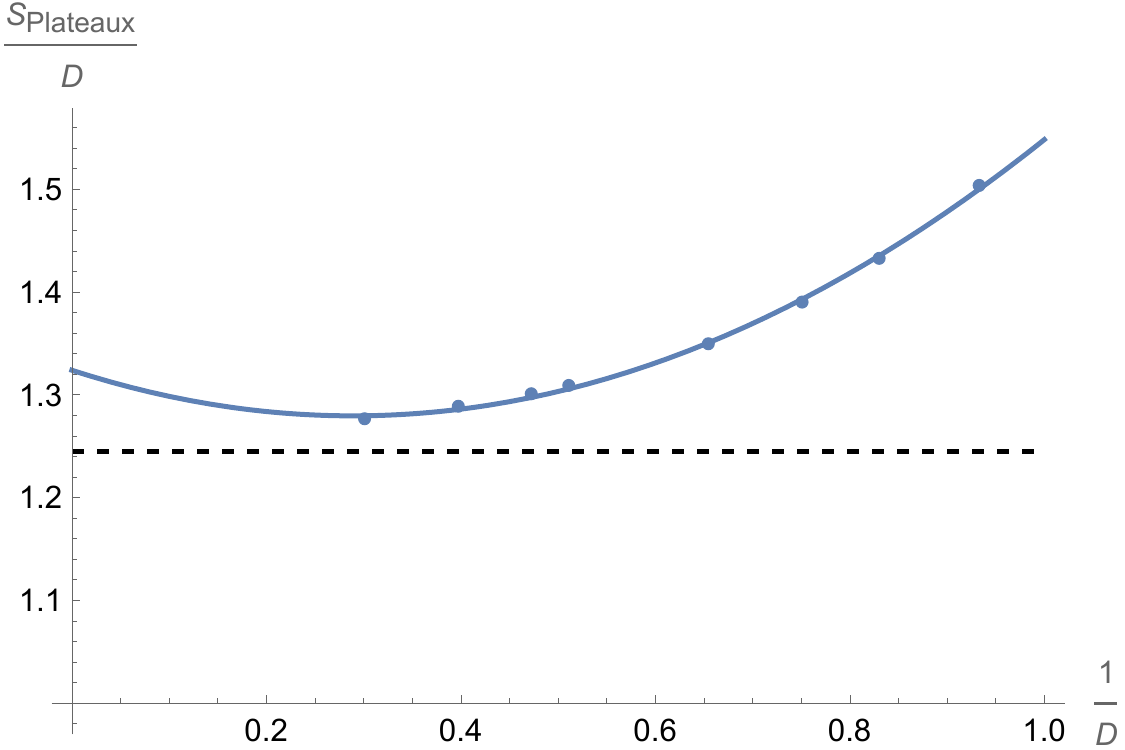}
\includegraphics[scale=0.35]{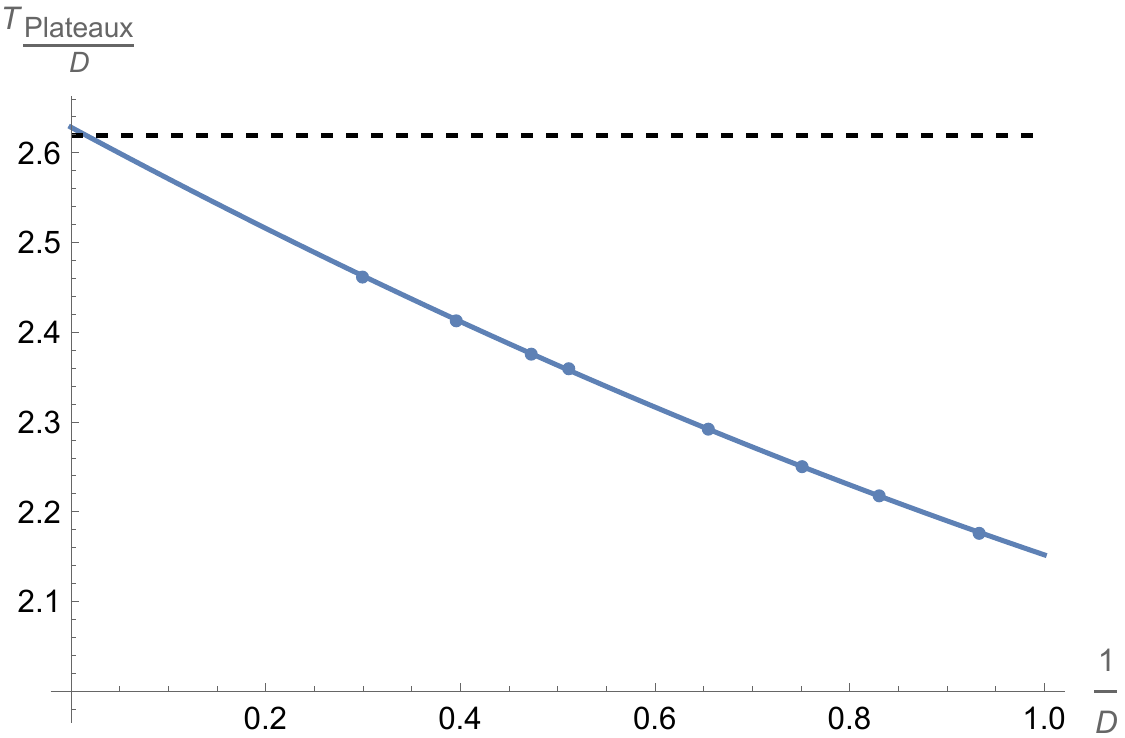}
\end{figure}

\section{Adding $\phi^4$-interactions}
An interesting extension of the story in section~\ref{smallPhiPertTh} is to study an RG flow that interpolates between a UV and an IR CFT. We expect that ${\cal E}(v)$ would be degenerate both in the UV and IR, but nontrivial in between. We could use the effective temperature of the state or operator whose entanglement dynamics we are studying as the RG scale. We can holographically model such an RG flow by
 adding a $\frac{\mu}{4!}\phi^4$ interaction to the bulk scalar fields in \eqref{action}:
\begin{align}
    \mathcal{L}=\frac{1}{2}(R+2)-\frac{1}{2}(g^{ab}\partial_a\phi \partial_b\phi+m^2\phi^2+\frac{\mu}{12}\phi^4)\label{actionphi4}.
\end{align}
The potential term in \eqref{actionphi4}, now containing both the mass term and the $\phi^4$ interaction, takes the form $V(\phi)=-\frac{3}{8}\phi^2+\frac{\mu}{24}\phi^4$. At zero temperature we would get a domain wall scalar profile that interpolates between $\phi=0$ at the boundary and the bottom of the potential $\phi_*={3\ov \sqrt{2\mu}}$, representing the IR fixed point \cite{Freedman:1999gp,Girardello:1998pd,Girardello:1999bd}.

The Einstein-scalar equations of motion in the presence of the $\phi^4$ interaction are 
\begin{align}
    \phi''+\Big(\frac{f'}{f}-\frac{1}{z}-\frac{\chi'}{2}\Big)\phi'+\frac{3}{4}\frac{\phi}{z^2f}-\frac{\mu}{6}\frac{\phi^3}{z^2f}&=0\,,\label{EoM1phi4}\\
    \chi'-2\frac{f'}{f}-\frac{3}{4}\frac{\phi^2}{zf}-\frac{4}{zf}+\frac{4}{z}+\frac{\mu}{12}\frac{\phi^4}{zf}&=0\,,\label{EoM2phi4}\\
    \chi'-z\phi'^2&=0\,.\label{EoM3phi4}
\end{align}
From \eqref{EoM1phi4}--\eqref{EoM3phi4}, we can obtain the shooting solution for $\phi^4$-interacting bulk scalars. A curious feature of these is that the scalar rolls through the minimum of the potential $\phi_*$ in the black brane interior and exhibits the same near-singularity behaviour in the presence of $\phi^4$  interactions as in their absence. We can understand this as follows: The terms in \eqref{EoM1phi4}--\eqref{EoM3phi4} that originate from $V(\phi)$ are 
\begin{align}
    \frac{3}{4}\frac{\phi}{z^2f}\,, && \frac{\mu}{6}\frac{\phi^3}{z^2f}\,, && \frac{3}{4}\frac{\phi^2}{zf}\,, && \frac{\mu}{12}\frac{\phi^4}{zf}\,,\label{VphiTemrs}
\end{align}
which upon substituting the asymptotic behaviour \eqref{Kasner}, all go to $0$ in the large $z$ limit. This is because $\phi$ and hence any polynomial of it only diverges $\emph{logarithmically}$, and the Kasner behaviour \eqref{Kasner} originates from the $\emph{kinetic}$ terms in the action \eqref{action} or \eqref{actionphi4}. To change \eqref{Kasner} near the singularity, one needs to introduce interactions that are (super-)exponential in $\phi$~\cite{Hartnoll:2022rdv}.  
This may make us doubt if the IR CFT leaves any imprint on entanglement dynamics that is determined by the interior. 

\begin{figure}[htbp]
\centering
\includegraphics[width=.49\textwidth]{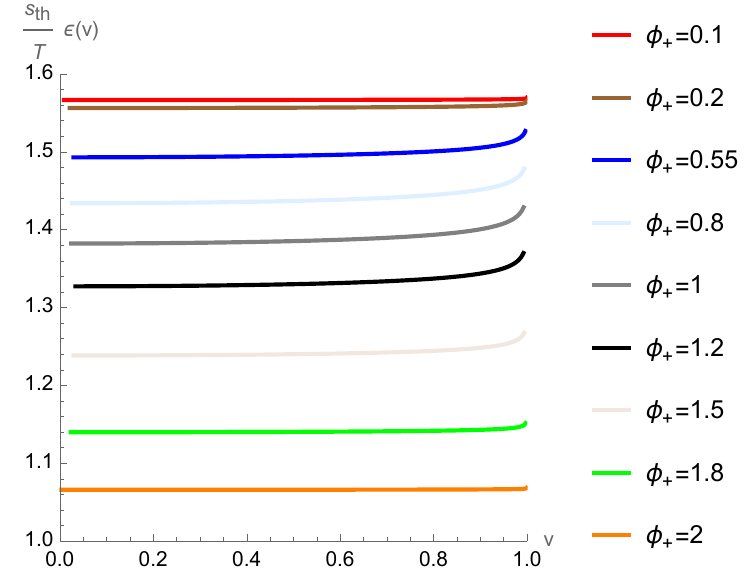}
\includegraphics[width=.49\textwidth]{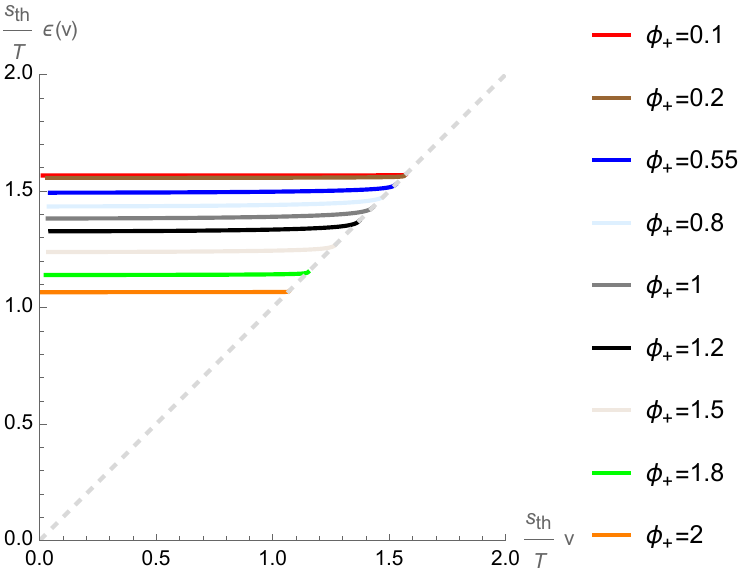}
\qquad
\caption{Plots of the membrane tension $\mathcal{E}(v)$ for a $\phi^4$-interacting bulk scalar with $\mu=1$ for the choices $\phi_+=0.1, 0.2, 0.55, 0.8, 1, 1.2, 1.5, 1.8$ and $2$. Note that the minimum of the potential that represents the IR fixed point is at $\phi_*=2.12$.\label{fig:EvPhi4}}
\end{figure}

Nevertheless, by tuning $\phi_+$, the horizon value of the scalar field, we can explore the RG flow, since  for  $\phi_+\to\phi_*$, i.e.~approaching the IR CFT,  can tune the Kasner exponent close to the BTZ value appropriate for the IR CFT. Our findings are fully consistent with RG intuition and we summarise them on two plots in Fig.~\ref{fig:EvPhi4}. Both show ${\cal E}(v)$ scaled differently. On the left we multiplied ${\cal E}(v)$ by the dimensionless  entropy density factor $s_\text{th}/ T$ that plays the role of the number of degrees of freedom. Recall that it is the combination $s_\text{th} {\cal E}(v)$ that shows up in the membrane action \eqref{MinMemb}. This combination indeed gives a monotonically decreasing  function  under RG flow, which is a satisfactory result on physical grounds. On the right we have also rescaled the horizontal axis by the same factor $s_\text{th}/ T$ to demonstrate that ${\cal E}(v)$ still is tangent to the $45^\circ$ line all along the RG flow. We note that we have found similar behaviour in the free scalar case as $\phi_+\to \infty$, but of course there was no IR CFT there.

\chapter{Appendices to Chapter~\ref{Chap4}}

\section{Generalised entanglement membrane for CFT$_2$}
\label{app:gen} 
So far, we have been studying multiparty entanglement signals in the membrane effective theory. For the special case of 2d CFT, however, that membrane theory becomes degenerate~\cite{Mezei:2018jco}, and the proper effective theory for entanglement dynamics is the generalised membrane theory introduced in Section~\ref{sec:2dEWCS}.

\subsection{$S_R$ in the case of adjacent intervals}
\begin{figure}[htbp]
\centering
\includegraphics[width=.4\textwidth]{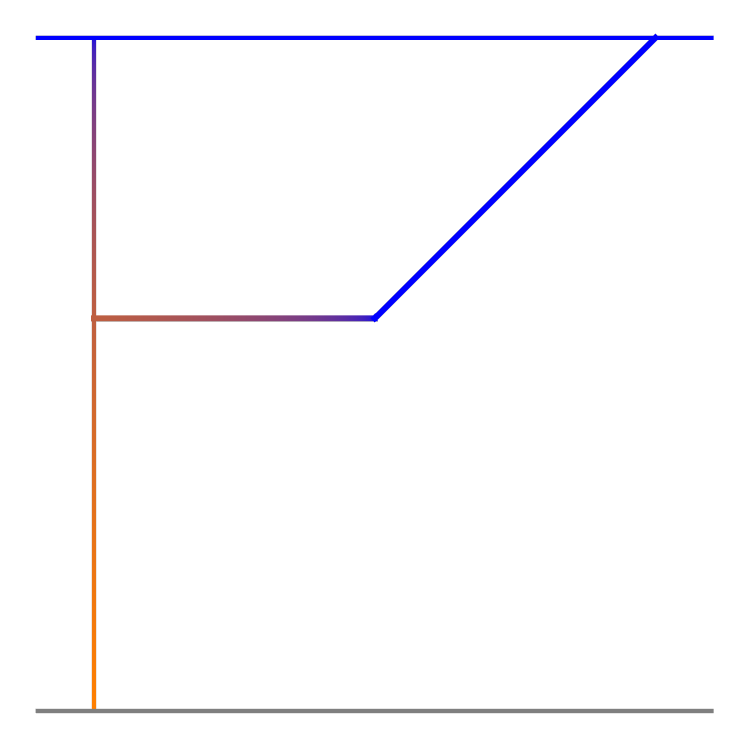}\hspace{0.5cm}
\qquad
\caption{The generalised membranes computing saturated $S_R$ (thick) in 2d CFT in the case of adjacent intervals. As in 2411.16542, we use colour to denote the additional degree of freedom $\xi$ on the membrane: the change from blue to orange denotes the increase in $\xi$.\label{fig:GenMembraneSRAdj}}
\end{figure}

The generalised membrane theory for reflected entropy~\cite{Jiang:2024tdj} was developed for the case depicted in Figure~\ref{fig:NeqEWCSProj} and~\ref{fig:EqEWCSProj} , where an interval of $O(\Lambda)$ size was inserted between $A$ and $B$. To make our multipartite entanglement signal calculations comparable with the ordinary membrane results in Sections~\ref{sec:r3} and~\ref{sec:I3}, we need to develop an effective description for adjacent $A$ and $B$. We assume that as in the disjoint interval case studied in~\cite{Jiang:2024tdj} and reviewed above, the generalised membrane contains a horizontal component that does not contribute to the reflected entropy in the scaling limit. In~\cite{Jiang:2024tdj}, the equations of motion for these generalised membranes parametrised by $(v(x),\xi(x))$\footnote{A change of variable from $v$ to $x$ has been made to describe the horizontal portion of the generalised membrane.} in the scaling limit were found to be
\es{xiEOMx2}{
 \xi''&=1-\xi'^2 - e^{4(\xi_p-\xi)}\,,\\
 v'&=e^{-2(\xi-\xi_p)}\,.
}
where $f'=\frac{df}{dx}$. The equations~\eqref{xiEOMx2} have the following approximate solutions in the scaling limit
\begin{equation}
    (v(x),\xi(x))=
    \left\{
    \begin{aligned}
        &(v_p,-x+x_0+\xi_p),&& 0<x<x_0,\\
        &(x+v_p-x_0,\xi_p),&& x_0<x,
    \end{aligned}
    \right.
\end{equation}
where $v_p$, $\xi_p$, and $x_0$ are constants, with $x_0$ labeling the location of the membrane's turning point. Like in ordinary membrane theory~\cite{Jonay:2018yei,Mezei:2018jco}, $v(x)$ describes the shape of the generalised membrane. To visualise the additional degree of freedom $\xi(x)$, we label it as colour on the generalised membrane as in~\cite{Jiang:2024tdj}. There are two boundary conditions for the EWCS: that $x=\ell_A$ at time $t$, and that it ends on the vertical membrane at $x=0$ perpendicularly. Notice that in imposing the boundary conditions for the EWCS in generalised membrane theory, we need to match not only the shape of the membrane $v$, but also the colour on the membrane $\xi$. Written explicitly, the two boundary conditions are 
\begin{align}
    &x_0+\xi_p=t-v_p\,&&\text{to match the vertical membrane HRT,}\label{matchingConds1}\\
    &(\ell_A+v_p-x_0, \xi_p)=(t, 0)\,&&\text{to match the connection point on the $t$ slice}\label{matchingConds2},
\end{align}
where the vertical membrane in~\eqref{matchingConds1} is described by~\cite{Jiang:2024tdj}
\begin{align}
    (x(v),\xi(v))=(0,t-v).
\end{align}
One can then solve for the 3 constants
\begin{align}
    \xi_p=0, && v_p=t-\frac{\ell_A}{2}, && x_0=\frac{\ell_A}{2}.
\end{align}
The length of the EWCS is thus 
\es{ECWS2}{
\ell(\text{EWCS})=x_A-x_0={\ell\ov 2}\,.
}
See Figure~\ref{fig:GenMembraneSRAdj} for the plot of this generalised membrane. Notice that the saturation value is half as big compared to that in ordinary membrane theory as shown in Figure~\ref{fig:SRABC} right. 

As such, the time evolution of $S_R(A:B)$ is given by\footnote{As in the $d>2$ case, when considering adjacent intervals $S_R$ is not UV finite~\cite{Kudler-Flam:2020url}, and we subtract the logarithmic UV divergences. }
\begin{equation}
\label{GenMembraneSRAdj}
\frac{S_R(A:B)}{2} = s_{\text{th}}
\begin{cases}
t \ &t < \frac{\ell_A}{2}\,, \\ 
\frac{\ell_A}{2}  \  &t > \frac{\ell_A}{2}\,.
\end{cases}
\end{equation}

As we will see shortly, the difference in the saturation of $S_R$  between the ordinary and generalised membrane theories,~\eqref{SRABSingReg} vs~\eqref{GenMembraneSRAdj}, leads to strikingly different behaviours of multipartite entanglement signals in 2d CFT and generic chaotic systems. This indicates that the infinite-dimensional conformal symmetry might have substantial effects on multiparty entanglement, a possibility that would be very interesting to investigate in the future. 

\subsection{Dynamics of $R_3$}
We use the generalised membrane theory for reflected entropy~\eqref{GenMembraneSRAdj} to study the dynamics of residual information in the setups considered in Section~\ref{sec:r3}. Quantities based solely on von Neumann entropy can be obtained from those in ordinary membrane theory in section~\ref{SingleFiniteReg} by simply setting $v_E=v_B=1$.\footnote{We note that while setting $v_E=v_B=1$ in ordinary membrane theory gives the correct generalised membrane theory results, the membranes computing entanglement entropy in generalised membrane theory have 2 degrees of freedom~\cite{Jiang:2024tdj}. However, the extra degree of freedom will not play any role in determining the value of entanglement entropy if we are not studying the saturation of $S_R$.} 

\subsubsection{A single finite interval}
We consider the setup of a single finite interval, $A = (-\infty, 0)$,  $B = (0, \ell)$, and  $C= (\ell, \infty)$, as is studied in section~\ref{SingleFiniteReg}. $I(A:B)$ and $I(A:C)$ can be simply obtained from~\eqref{IABSingleFiniteReg} and~\eqref{IACSingleFiniteReg} by setting $v_E=v_B=1$. Taking the difference between $S_R(A:B)$~\eqref{GenMembraneSRAdj} and $I(A:B)$, we find that
\begin{align}
    R_3(A:B)=0\ \ {\rm for}\ \ t\in[0,+\infty),
\end{align}
which is different from the ordinary membrane result~\eqref{R3ABSingInt} when setting $v_E=v_B=1$ due to the difference in saturation of $S_R$ in generalised membrane theory. 

$S_R(A:C)$ is computed by the same vertical membrane (with an additional degree of freedom) as is shown in Figure~\ref{fig:SRABCac}, which never saturates. As such, $S_R(A:C)$ is given by~\eqref{SRACSingleFiniteReg} with $v_E=v_B=1$. Taking the difference with $I(A:C)$, we have 
\begin{align}
    R_3(A:C)=0\ \ {\rm for}\ \ t\in[0,+\infty).
\end{align}
Notice that we have $R_3(A:C)=0$ when setting $v_E=v_B=1$ in~\eqref{R3ACSingInt}.  

As $R_3$ is only a signal of tripartite entanglement instead of a measure, $R_3(A:B)=R_3(A:C)=0$ in this setup does not necessarily indicate that there is only bipartite entanglement at any time. 

\subsubsection{Two finite regions}
We consider the setup of two finite regions, $A = (0,\ell_A)$, $B= (\ell_A, \ell_A + \ell_B)$, and $C = (-\infty,0) \cup (\ell_A + \ell_B,\infty)$, as  studied in Section~\ref{sec:r3twor}. Again, $I(A:B)$ and $I(A:C)$ can be simply obtained from~\eqref{IABTwoFiniteReg} and~\eqref{IACTwoFiniteReg} by setting $v_E=v_B=1$. Since we focus on the $\ell_B < \ell_A$ case, using~\eqref{GenMembraneSRAdj}, $S_R(A:B)$ is given by
\begin{equation}
\frac{S_R(A:B)}{2} = s_{\text{th}}
\begin{cases}
t \ &t < \frac{\ell_B}{2}\,, \\ 
\frac{\ell_B}{2}  \  &\frac{\ell_B}{2} < t <\frac{\ell_A+\ell_B}{2} \,,\\
0  \  &t > \frac{\ell_A+\ell_B}{2}\,.
\end{cases}
\end{equation}
Notice that when $AB$ saturates, $S_R(A:B)$ abruptly drops to 0. The corresponding cross-section is shown in Figure~\ref{fig:SatEWCSAdj}. Just as in the ordinary membrane theory case~\eqref{2srab}, this cross-section is no longer described by generalised membrane theory anymore, as it does not grow with $t$ or $x$ in the scaling limit. See Appendix~\ref{SatSRAdj} for details. The dynamics of $S_R$ in the case of two finite regions were also studied in~\cite{Kudler-Flam:2020url} from 2d CFT and holographic points of view. Our calculations from the generalised membrane effective theory agree with their results.

Taking the difference with $I(A:B)$ we find that 
\begin{equation}
R_3(A:B) = s_{\text{th}}
\begin{cases}
0 \ &t < \frac{\ell_A}{2}\,, \\ 
2t-\ell_A  \  &\frac{\ell_A}{2} < t <\frac{\ell_A+\ell_B}{2} \,,\\
0  \  &t > \frac{\ell_A+\ell_B}{2}\,.
\end{cases}
\end{equation}
Notice that ${\rm max}\{R_3(A:B)\}=\ell_B$, which is $\emph{half}$ the bound for $R_3(A:B)$~\cite{Balasubramanian:2024ysu}. As a comparison, in the same setup in ordinary membrane theory, $R_3(A:B)$~\eqref{R3ABTwoReg1} does saturate the $2\ell_B$ bound. 

$S_R(A:C)$ is computed by the same vertical membrane as is shown in Figure~\ref{fig:SRCABCac} (again, with an additional degree of freedom). Hence, we can set $v_E=v_B=1$ in~\eqref{2srac} to obtain $S_R(A:C)$ in generalised membrane theory. Taking the difference with $I(A:C)$, we arrive at 
\begin{equation}
\label{R3ACTwoReg2d}
R_3(A:C) = s_{\text{th}} 
\begin{cases}
0 &  t < \frac{\ell_A}{2} \,, \\
2  t - \ell_A  & \frac{\ell_A}{2} <  t < \frac{\ell_A}{2} +\frac{\ell_B}{4} \,,  \\
\ell_A + \ell_B - 2 t & \frac{\ell_A}{2} +\frac{\ell_B}{4}<  t < \frac{\ell_A+\ell_B}{2} \,,  \\
0 &   t > \frac{\ell_A+ \ell_B}{2} \,.
\end{cases}
\end{equation}
which is~\eqref{R3ACTwoReg} with $v_E=v_B=1$. We find that the initial plateau due to the saturation of $B$ in~\eqref{R3ACTwoReg} disappears in~\eqref{R3ACTwoReg2d}. The maximal value of $R_3(A:C)$ is ${\rm max}\{R_3(A:C)\}=\frac{\ell_B}{2}$, which reaches half the $R_3(A:C)$ bound. 

\subsection{Dynamics of $Q_4$}
We use the generalised membrane theory for reflected entropy~\eqref{GenMembraneSRAdj} to study the dynamics of $Q_4$ in the setup considered in Section~\ref{sec:I3}: $A = (0, \ell_A)$, $B = (\ell_A, \ell_A+\ell_B)$, $C= (\ell_A+\ell_B, \ell_A + \ell_B+\ell_C)$ and $D = (-\infty,0) \cup (\ell_A + \ell_B + \ell_C, \infty)$. Again, we assume $\ell_A=\ell_C$ for simplicity. 

The $\ell_A<\ell_B$ case is similar to that considered in section~\ref{sec:I3} except for the difference in saturation: the saturated EWCS ends on the vertical membrane on the left, but the saturation value is half as big as that in~\eqref{Q4DABCD1}. Therefore, we have  
\begin{equation}
\label{Q4DABCD2d1}
Q_4 = s_{\text{th}} 
\begin{cases}
0 \ &0<t < \frac{\ell_A+\ell_B}{2}\,, \\ 
2 \ell_A &\frac{\ell_A+\ell_B}{2}<t < \frac{\ell_A+\ell_B+\ell_C}{2}\,, \\ 
0\ &t>\frac{\ell_A+\ell_B+\ell_C}{2}\,.
\end{cases}
\end{equation}
Compared to the corresponding ordinary membrane theory result~\eqref{Q4DABCD1}, we see that ${\rm max}\{Q_4\}=2\ell_A$ reaches half the $2\ell_A$ bound.  

\begin{figure}[htbp]
\centering
\includegraphics[width=.31\textwidth]{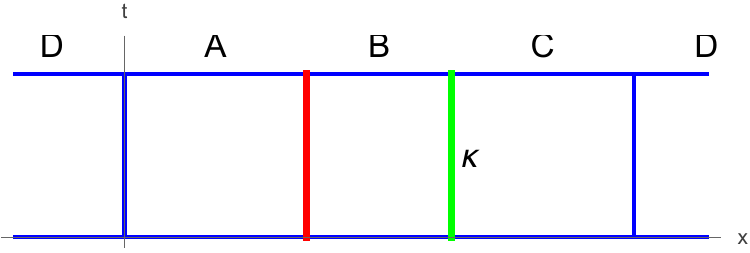}\hspace{3mm}
\includegraphics[width=.31\textwidth]{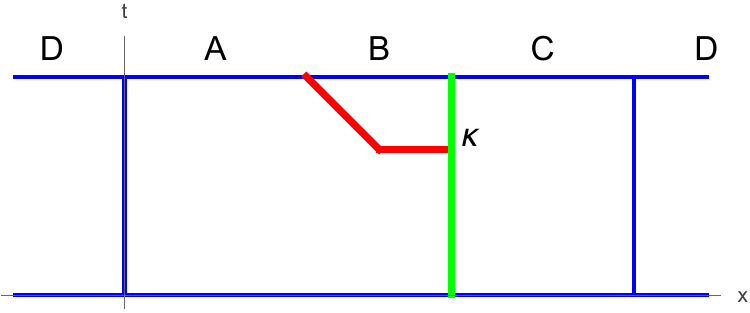}\hspace{3mm}
\includegraphics[width=.31\textwidth]{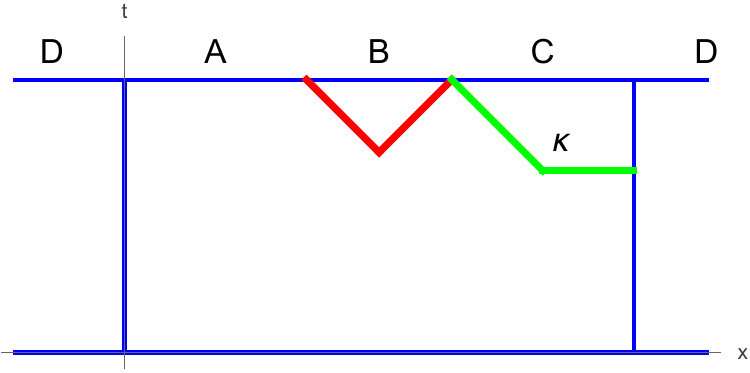}
\qquad
\caption{Minimal cross-sections computing $S_{AA^*A_*A^*_*}$ (red, thick) in $Q_4$ in the generalised membrane theory for CFT$_2$ when $\ell_A>\ell_B$. Notice that the saturation of $\kappa$ (green, thick) changes the shape as well as the value of the membrane computing $S_{AA^*A_*A^*_*}$. Here, we do not plot the additional degree of freedom $\xi$ as a colour on the membrane as in Figure~\ref{fig:GenMembraneSRAdj}, instead we follow the same colour conventions for $Q_4$ in the ordinary membrane cases in e.g. Figure~\ref{fig:kappa}. \label{fig:Q4DABCDLgA2d}}
\end{figure}

The $\ell_A>\ell_B$ case is more interesting: since the saturated EWCS now ends on $\kappa$, and the saturation value is half as big compared to the ordinary membrane result, the saturation of $\kappa$ does change the value of $S_{AA^*A_*A^*_*}$ by doubling its value, see Figure~\ref{fig:Q4DABCDLgA2d}. This is in contrast to the ordinary membrane theory case shown in Figure~\ref{fig:kappa}, where the saturation of $\kappa$ does not change the value of $S_{AA^*A_*A^*_*}$.

We thus have
\begin{equation}
\label{Q4DABCD2d2}
Q_4 = s_{\text{th}} 
\begin{cases}
0 \ &0<t < \frac{\ell_A+\ell_B}{2}\,, \\ 
2 \ell_B &\frac{\ell_A+\ell_B}{2}<t < \ell_A\,, \\ 
4 \ell_B &\ell_A<t < \frac{\ell_A+\ell_B+\ell_C}{2}\,, \\ 
0\ &t>\frac{\ell_A+\ell_B+\ell_C}{2}\,.
\end{cases}
\end{equation}
Notice that unlike the $\ell_A<\ell_B$ case, now ${\rm max}\{Q_4\}=4\ell_B$ does saturate the $4\ell_B$ bound. This indicates that the extensive part of the entanglement of $B$ is entirely multipartite. 

\section{Higher partite entanglement signals}
\label{sec:higher}
We can extend the above calculations to higher numbers of parties. The extension is conceptually straightforward, and does not seem to lead to qualitatively new features, but the calculations become more complicated due to the increasing number of regions. 

Consider for example the five-party entanglement signal $R_5$. Suppose we examine a system with subregions $A=(0,\ell)$, $B=(\ell,2\ell)$, $C=(2\ell,3\ell)$, $D=(3\ell,4\ell)$, $E=(-\infty,0)\cup (4\ell,+\infty)$, and trace over $E$. We have
\begin{equation}
R_5(A:B:C:D) = \frac{1}{2} I_4(AA^*:BB^*:CC^*:DD^*) - I_4(A:B:C:D),
\end{equation}
where
\begin{equation}
\label{R51}
    \begin{split}
   \frac{1}{2} I_4(AA^*:BB^*:CC^*:DD^*) =& I_3(AA^*:BB^*:CC^*) \\ =& S(AA^*) + S(BB^*)+ S(CC^*)+S(DD^*)\\ & - (S(AA^*BB^*) + S(AA^*CC^*) + S(BB^*CC^*)),     
    \end{split}
\end{equation}
and 
\begin{equation}
    \begin{split}
  I_4(A:B:C:D) =& S(A) + S(B)+ S(C)+ S(D) \\ &- (S(AB)+S(AC)+S(AD) + S(BC) + S(BD)+ S(CD) ) \\ &+ S(ABC) + S(ABD) + S(ACD) + S(BCD) - S(ABCD).          
    \end{split}
\end{equation}
%

\begin{figure}[htbp]
\centering
\includegraphics[width=.35\textwidth]{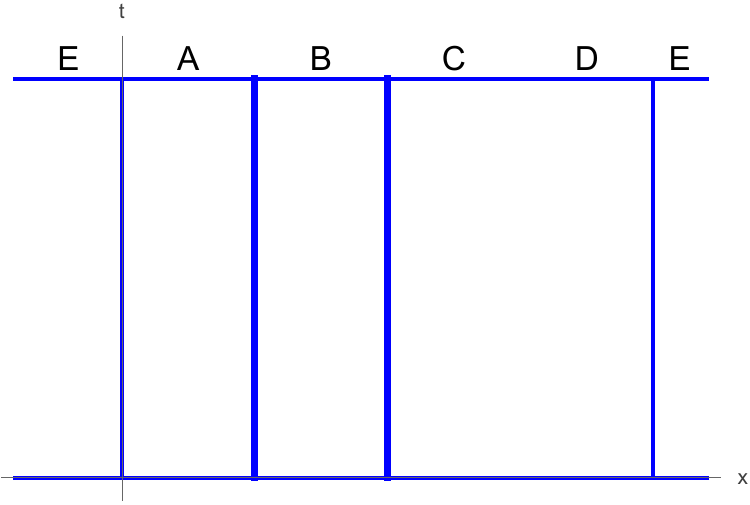}\hspace{3mm}
\includegraphics[width=.35\textwidth]{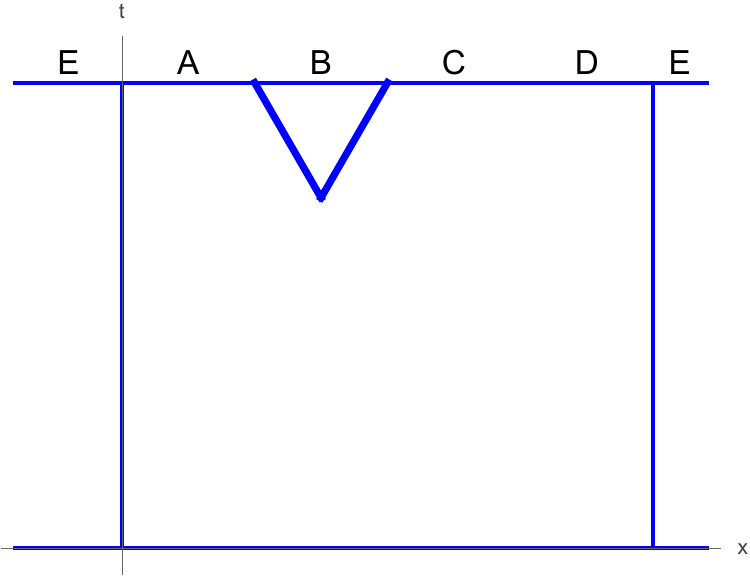}\vspace{5mm}
\includegraphics[width=.31\textwidth]{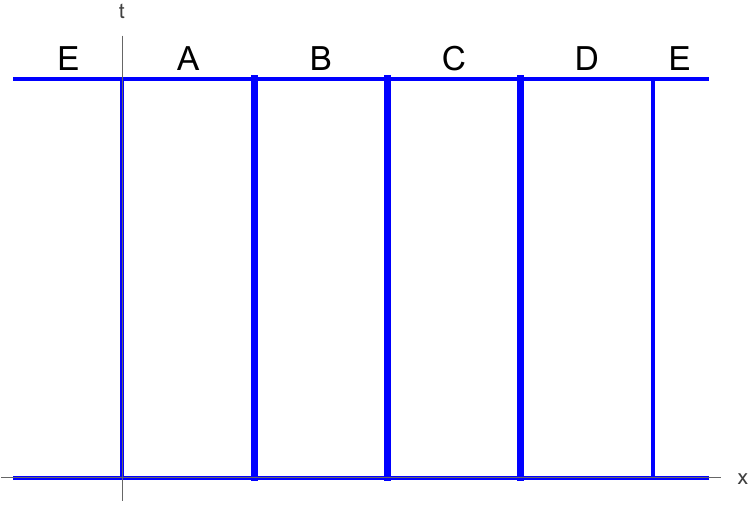}\hspace{3mm}
\includegraphics[width=.31\textwidth]{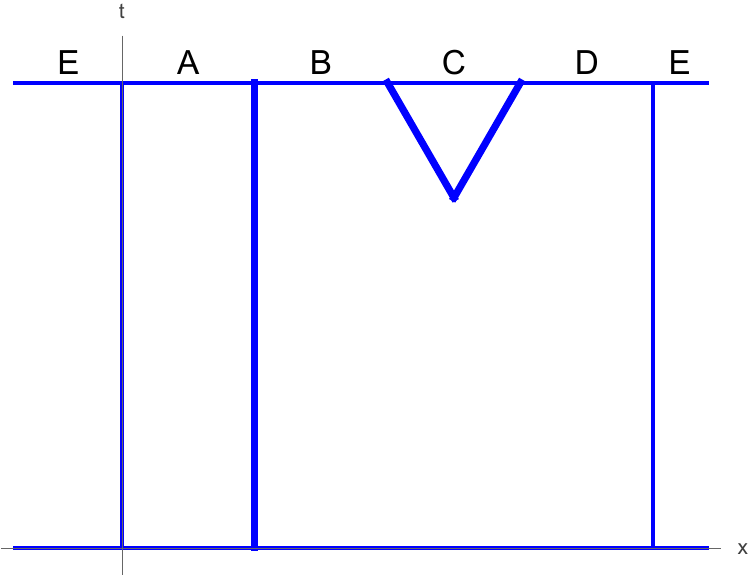}\hspace{3mm}
\includegraphics[width=.31\textwidth]{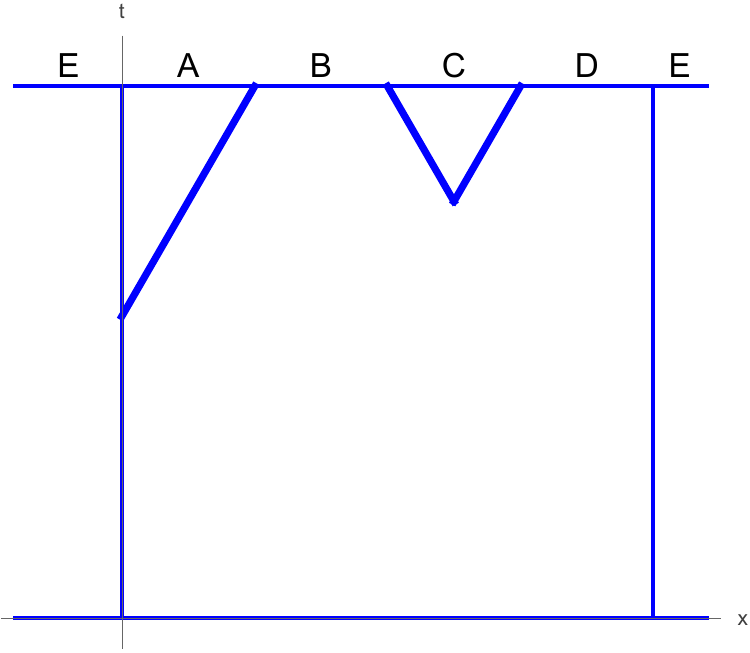}
\qquad
\caption{Membranes computing $S(BB^*)$ (top row) and $S(AA^*CC^*)$ (bottom row) in $R_5(A:B:C:D)$. We only plotted membranes before the saturation of the $ABCD$ EW, after which all the entanglement wedge cross-sections have vanishing area in the membrane description. \label{fig:R5}}
\end{figure}

We have $S(A) = S(B) = S(C) = S(D) = S_1$, $S(AB) = S(BC) = S(CD) = S_2$, $S(AC) = S(AD) = S(BD) = 2 S_1$, $S(ABC) = S(BCD) = S_3$, $S(ABD) = S(ACD) = S_2+S_1$, and $S(ABCD) = S_4$, where we use $S_n$ to denote the entanglement entropy of an interval of length $n \ell$. This gives
\begin{equation}
  I_4(A:B:C:D) = 2 S_3 - S_2 - S_4 = s_\mathrm{th}  \begin{cases} 0 & t < \ell , \\ 
  2t - 2\ell & \ell < t < \frac{3\ell}{2}, \\ 
  4 \ell-2t & \frac{3\ell}{2} < 2 \ell, \\
  0 & t > 2\ell. \end{cases}    
\end{equation}
The $ABCD$ entanglement wedge is always connected. Before the transition at $t = 2\ell$, it is bounded by two vertical surfaces (Figure~\ref{fig:R5}). The purification-based quantities $S(XX^*)$ in~\eqref{R51} are computed by the minimal cross-sections that bisect the entanglement wedge of $X$ and $ABCD/X$ (see Figure~\ref{fig:R5} for examples). As such, we have $S(AA^*) = S(DD^*) = S_2$, $S(BB^*) = S(CC^*) = 2 S_1$, $S(AA^*BB^*) = S_4$, $S(AA^*CC^*) = S_2 + 2 S_1$, and $S(BB^*CC^*) = 2S_2$. After the transition, the entanglement wedge only extends down to the horizon, and all the canonical purification quantities are given by vertical surfaces which extend down to the horizon (Figure~\ref{fig:SatEWCSAdj}), so $I_4(AA^*:BB^*:CC^*:DD^*) = 0$ for $t > 2\ell$. Thus
\begin{equation}
\frac{1}{2}   I_4(AA^*:BB^*:CC^*:DD^*) = s_\mathrm{th}  \begin{cases} 0 & t < \frac{\ell}{2} , \\ 
  2\ell - 4t & \frac{\ell}{2} < t < \ell, \\ 
  -2t & \ell < 2 \ell, \\ 
  0 & t > 2\ell. \end{cases}    
\end{equation}

This finally gives
\begin{equation}
\label{R5EABCDE}
R_5(A:B:C:D) = s_{\text{th}} 
\begin{cases}
0  &0<t < \frac{\ell}{2}\,, \\ 
2\ell-4 t &\frac{\ell}{2}<t < \frac{3\ell}{2}\,, \\ 
-4\ell &\frac{3\ell}{2}<t <2\ell\,, \\ 
0&t>2\ell\,.
\end{cases}
\end{equation}

\begin{figure}[htbp]
\centering
\includegraphics[width=.5\textwidth]{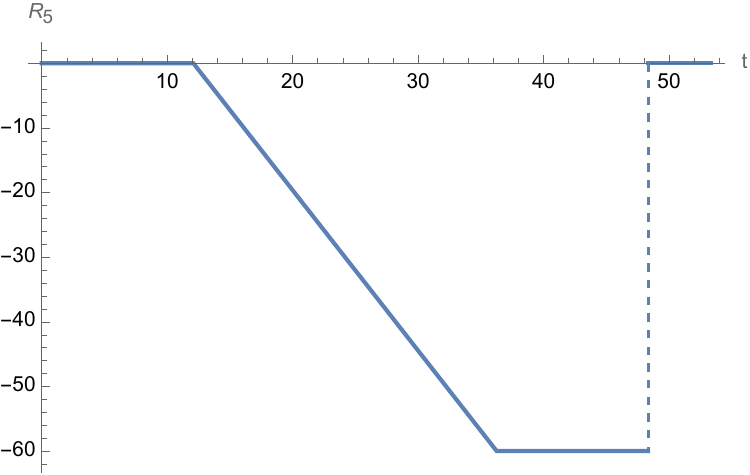}
\qquad
\caption{Time evolution of $R_5(A:B:C:D)$~\eqref{R5EABCDE}. Here, we plot for $\ell=15$. \label{fig:R5EABCDE}}
\end{figure}

This gives qualitatively similar behaviour to  $R_3(A:B)$ in Section \ref{sec:r3twor}. Notice that $R_5\leq 0$. In the vacuum AdS$_3$ case studied in~\cite{Balasubramanian:2024ysu}, $R_5$ is mostly negative, but with regions of the moduli where $R_5$ can be positive. It would be interesting to investigate the sign patterns of $R_5$ for more generic subsystem sizes here. 

We can generalise the $R_5$ calculation to $R_n$ for arbitrary odd $n$ ($n\geq 3$). Likewise, for arbitrary even $n$ ($n\geq 4$), we can generalise the $I_3=\frac{1}{2}I_4$ calculation in Section~\ref{sec:I3} to compute the $n$-partite entanglement signal $I_{n-1}=\frac{1}{2}I_n$. As $I_n$ are functions of von Neumann entropy, their time evolution will be continuous.

\section{Saturation of $S_R$ in finite adjacent intervals}\label{SatSRAdj}
In this Appendix, we use holography to verify that the saturated $S_R$ in the case of finite adjacent intervals as shown in Figure~\ref{fig:SatEWCSAdj} in the main text is indeed a vertical ``membrane'' that has vanishing extensive entropy in the scaling limit. 

We calculate saturated $S_R$ in holographic CFTs by examining the entanglement wedge cross-sections in the dual gravitational theories, where the planar AdS$_{d+1}$-Schwarzschild black brane metric is given by
\begin{align}
    ds^2=\frac{1}{z^2}\Big(-f(z)dt^2+\frac{dz^2}{f(z)}+dx^2+ d\vec{y}_{d-2}^2\Big), \ \ \ f(z)=1-z^d\label{AdSSchMetric2}.
\end{align}
The Schwarzschild time $t$ in~\eqref{AdSSchMetric2} is related to the infalling time $v$ in~\eqref{blackBraneInf} via
\begin{align}
\label{SchInf2}
    v(z)=t-\int_0^z\frac{dz'}{f(z')}=t-\frac{1}{d}B\left(z^d;\frac{1}{d},0\right),\ \ z\in [0,1),
\end{align}
where $B\left(z;a,b\right)$ is the incomplete Euler Beta function. The saturated EWCS lies on a constant Schwarzschild time slice. 

We are interested in the bulk RT surface homologous to a boundary subregion $AB$ of size $X=\ell_A+\ell_B$, see Figure~\ref{fig:SatEWCSAdj} (top). Like in section~\ref{sec:r3twor}, we consider the case $\ell_A>\ell_B$, with the boundary point $K$ between $A$ and $B$ located at $x=k>0$. We scale $\ell_A, \ell_B$ and $k$ to be large.

For simplicity, let us first consider the $d=2$ special case.\footnote{Notice that for saturated $S_R$, all the extremal surfaces involved are outside the black hole horizon. Hence, in this case the additional degree of freedom $\xi$ in generalised membrane theory is always of $O(1)$ value and can be scaled away. The results are thus the $v_E=v_B=1$ special case of planar AdS$_{d+1}$-Schwarzschild black holes.} In~\cite{Jiang:2024tdj}, it was shown that the static spacelike geodesic homologous to a finite boundary subregion is given by the parametric curve $\big(z(\lambda),x(\lambda)\big)$
\begin{align}
\begin{split}
\label{statGeod}
    z(\lambda)&=\frac{\sinh \frac{X}{2}}{\sqrt{\cosh \lambda \cosh (X+\lambda )}},\\
    x(\lambda)&=\frac{1}{2} \log (\text{sech}\lambda  \cosh (X+\lambda )),
\end{split}
\end{align}
where $\lambda\in(-\infty,+\infty)$. To find the minimal cross-section bisecting the $AB$ entanglement wedge, we minimise the geodesic distance between $K$ \footnote{To regularize the UV divergence, we consider the point $K$ at $z=\epsilon\ll 1$ in the asymptotic region.} and an arbitrary point $G$ on the parametric curve~\eqref{statGeod}. The geodesic distance formula in planar BTZ black brane is~\cite{Mezei:2016zxg,Shenker:2013pqa} 
\begin{align}
    \cosh d(K,G)=T_1^KT_1^G+T_2^KT_2^G-X_1^KX_1^G-X_2^KX_2^G\label{geodDistBTZ},
\end{align}
where the embedding coordinates are given by
\begin{align}
\begin{split}
    T_1&=\frac{\sqrt{1-z^2}}{z} \sinh t,\ \ \ \ T_2=\frac{1}{z}\cosh t,\\
    X_1&=\frac{\sqrt{1-z^2}}{z} \cosh t,\ \ \ X_2=\frac{1}{z}\sinh x.
\end{split}
\end{align}
The explicit expression for the geodesic distance between $K$ and $G$ is long and uninspiring. In the large region limit, we find that the minimal geodesic distance, after dropping the UV divergent $\log\frac{2}{\epsilon}$ term, is 
\begin{align}
    d_{{\rm reg}}=0 \ \ \ {\rm at} \ \ \ \lambda_*=-\frac{X}{2} + k.
\end{align}
At $\lambda_*$, one finds 
\begin{align}
    x(\lambda_*)=k.
\end{align}
In other words, the minimal geodesic between $K$ and $G$ is the $\emph{radial}$ one in the scaling limit. This result makes sense intuitively, as the geodesic~\eqref{statGeod} contains a plateau region where it skims the horizon almost transversely along the $x$ direction, and the EWCS needs to end on it perpendicularly. 

Now, let us consider saturated $S_R$ in the adjacent interval case in $d>2$ with strip entangling subregions. In the large region limit, the intersection point $z_0$ between the EWCS and the RT surface is close to the horizon, $z_0=1-\delta$ ($\delta\ll 1$). Following the intuition  above, the EWCS is also almost radial in the scaling limit when $d>2$.  Thus, it would be convenient to parametrise the RT surface by $x(z)$ (instead of $z(x)$). The area functional is given by
\begin{align}
    A=\int dz d^{d-2}\vec{y}\ \frac{1}{z^{d-1}}\sqrt{\frac{1}{f(z)}+x'(z)^2}.
\end{align}
Noting that $x(z)=k$ and $x'(z)=0$ for the saturated EWCS, the area functional becomes
\begin{align}
\label{AreaSatEWCSdGeq2}
    A&=\int_{\epsilon}^{z_0} dz  \frac{1}{z^{d-1}}\frac{1}{\sqrt{1-z^d}}=\frac{1}{d}B\left(z^d;\frac{2}{d}-1,\frac{1}{2}\right)\Big|_{\epsilon}^{z_0}\\
    &=\frac{\sqrt{\pi } \Gamma \left(\frac{2}{d}-1\right)}{d \Gamma \left(\frac{2}{d}-\frac{1}{2}\right)}-\frac{2}{\sqrt{d}}\sqrt{\delta}+\frac{1}{d-2}\frac{1}{\epsilon ^{d-2}},
\end{align}
where we have eliminated the $\int d^{d-2}\vec{y}={\rm Area}(\p A)$ infinite factor as in the main text after~\eqref{AreaFunct2}. Restoring this factor, we find that the last term diverges as $\frac{1}{d-2}\frac{{\rm Area}(\p A)}{\epsilon ^{d-2}}$, which is the same as the UV divergence of RT surfaces~\cite{Rangamani:2016dms}.\footnote{This result makes sense, as both the RT surface and the saturated EWCS end on the boundary perpendicularly, and are hence asymptotically the same.} Dropping this UV divergent $\frac{{\rm Area}(\p A)}{\epsilon ^{d-2}}$ term, we have  
\begin{align}
    A_{{\rm reg}}\sim \#\sqrt{\delta}+O(1)\to 0
\end{align}
in the scaling limit. Notice that when $d=2$, the area/length in~\eqref{AreaSatEWCSdGeq2} instead integrates to 
\begin{align}
    A=-\tanh ^{-1}\left(\sqrt{1-z^2}\right)\Big|_{\epsilon}^{z_0}=-\sqrt{2} \sqrt{\delta}+\log\frac{2}{\epsilon},
\end{align}
which agrees with our $d=2$ analysis above.

To sum up, we find that in $d\geq 2$ in the large region limit, the regularized area of the saturated EWCS is $0$. As these EWCS as shown in Figure~\ref{fig:SatEWCSAdj} (top) are radial and reach close to the horizon, their projections to the boundary along constant infalling time are vertical lines dangling from $K$, see Figure~\ref{fig:SatEWCSAdj} (bottom). This is due to the near-horizon logarithmic divergence of infalling time $v$~\eqref{SchInf2}~\cite{Jiang:2024tdj}. Nevertheless, these cross-sections are no longer described by membrane theory for holographic CFT~\cite{Mezei:2018jco,Jiang:2024tdj}. This is because to derive the membrane tension~\eqref{MinMemb} from the area of the HRT surface in the scaling limit, one needs to integrate out the radial bulk $z$ degree of freedom. However, from e.g.~\eqref{AreaSatEWCSdGeq2} above, it is clear that for the almost-radial cross-section shown in Figure~\ref{fig:SatEWCSAdj}, the radial bulk $z$ degree of freedom $\emph{cannot}$ be integrated out. The reason why this cross-section fails to be captured by membrane theory in other chaotic systems than holographic CFTs is technically less clear. Intuitively, one can understand the reason as cross-sections of this kind do not grow with $t$ or $x$ in the scaling limit. Hence, these vertical ``membranes'' should not be confused with the vertical membranes with tension $\mathcal{E}(0)=v_E$~\cite{Mezei:2018jco} obtained from projections of the HRT surfaces in the black hole interior~\cite{Hartman:2013qma,Liu:2013iza,Liu:2013qca}. The latter reaches the $t=0$ slice in membrane theory, computing entanglement entropy $S=s_{\rm th} v_E t$ that grows linearly with time.
\end{appendices}

\end{document}